%% file: main.tex
\PassOptionsToPackage{usenames,dvipsnames,rgb}{xcolor}
\documentclass[appendix,runningheads]{llncs}

\newif\ifaplas
\aplasfalse

\title{Typed Non-determinism in \\ Functional and Concurrent Calculi
}
\titlerunning{Typed Non-determinism in Functional and Concurrent Calculi}

\author{%
    Bas van den Heuvel\inst{1}\orcidID{0000-0002-8264-7371}
    \and
    Joseph W. N. Paulus\inst{1}\orcidID{0000-0002-1711-9361}
    \and \\
    Daniele Nantes-Sobrinho\inst{2,3}\orcidID{0000-0002-1959-8730}
    \and
    Jorge A. P\'erez\inst{1}\orcidID{0000-0002-1452-6180}%
}
\authorrunning{B.\,van den Heuvel et al.}
\institute{%
    University of Groningen, Groningen, The Netherlands
    \and
    University of Bras\'ilia, Bras\'ilia, Brazil
    \and
    Imperial College London, London, UK%
}

\usepackage{appendix}
\usepackage[pass]{geometry}
\usepackage{amsfonts}
\usepackage{amsmath,xspace}
\usepackage{amssymb}
\usepackage{bm}
\usepackage{mathpartir}
\usepackage{mathtools}
\usepackage{thm-restate}
\usepackage{relsize}
\usepackage{bussproofs}
\usepackage{stmaryrd}
\usepackage{tikz}
\usetikzlibrary{matrix}
\usetikzlibrary{decorations.pathreplacing,calligraphy}
\usetikzlibrary{decorations.pathmorphing}
\usepackage[framemethod=tikz]{mdframed}
\mdfsetup{
    innertopmargin=-5pt,
    innerleftmargin=4pt,
    innerrightmargin=4pt
}
\usepackage{centernot}
\usepackage{xcolor}
\usepackage{hyperref}
\usepackage{cleveref}
\usepackage{crossreftools}

\usepackage{thmtools}
\usepackage{amsthm}

\input{macros}

\spnewtheorem*{notation}{Notation}{\bfseries}{\itshape}

\usepackage{etoolbox}
\AtBeginEnvironment{example}{\pushQED{\qed}}
\AtEndEnvironment{example}{\popQED}

\newcommand{\myrevopt}[1]{}
\newcommand{\myrev}[1]{#1}
\newcommand{\mysmall}{\small}

\begin{document}

{\def\addcontentsline#1#2#3{}\maketitle}

\vspace{-5mm}
\begin{abstract}
    We study functional and concurrent calculi with \nond, along with  {type systems} to control resources based on  {linearity}.
The interplay between \nond and linearity is delicate: careless handling of branches can discard resources meant to be used exactly once.
Here we go beyond prior work by considering \nond in its standard sense: once a branch is selected, the rest are discarded.
Our technical contributions are three-fold.
    First, we introduce a $\pi$-calculus with non-de\-ter\-ministic choice, governed by session types.
    Second, we introduce a resource $\lambda$-calculus, governed by intersection types, in which \nond concerns fetching of resources from bags.
    Finally, we connect our two typed non-deterministic calculi via a correct translation.
    %
\end{abstract}

\section{Introduction}
\label{s:intro}

\myrev{In this paper, we present new formulations of typed programming calculi with \emph{non-determinism}.}
A classical ingredient of models of computation, \nond brings flexibility and generality in specifications.
In process calculi such as CCS and the $\pi$-calculus,  one source of non-determinism is {choice}, which is typically \emph{\nconf}: that is, given $P + Q$, we have either $P + Q \longrightarrow P$ or $P + Q \longrightarrow Q$.
Thus, committing to a branch entails discarding the rest.

\myrev{We study non-determinism as a way of increasing the expressivity of typed calculi in which resource control is based on \emph{linearity}.}
The interplay between non-determinism and linearity is delicate: a careless discarding of branches can jeopardize resources meant to be used exactly once.
On the concurrent side, we consider the $\pi$-calculus,
\myrev{the paradigmatic model of concurrency~\cite{DBLP:books/daglib/0004377}.}
We focus  on $\pi$-calculi with \emph{session types}~\cite{DBLP:conf/concur/Honda93,DBLP:conf/esop/HondaVK98}, in which linear logic principles ensure communication correctness:
here the resources are names that perform session protocols; they can be \emph{unrestricted} (used multiple times) and \emph{linear} (used exactly once).
To properly control resources, \nconf non-determinism is confined to unrestricted names; linear names can only perform {deterministic choices}.

In this context, considering \emph{confluent} forms of \nond can be appealing.
Intuitively, such formulations allow all branches to proceed independently:  given $P_1 \longrightarrow Q_1$ and $P_2 \longrightarrow Q_2$, then $P_1 + P_2 \longrightarrow  Q_1 + P_2$
and
$P_1 + P_2 \longrightarrow  P_1 + Q_2$.
Because confluent \nond does not discard branches, it is  compatible with a resource-conscious view of computation.

\Conf \nond has been studied mostly in the functional setting; it is present, e.g., in
Pagani and Ronchi della Rocca's {resource} $\lambda$-cal\-cu\-lus~\cite{PaganiR10}
and in
Ehrhard and Regnier's differential $\lambda$-cal\-culus~\cite{DBLP:journals/tcs/EhrhardR03}.
In~\cite{PaganiR10}, non-determinism resides in the application of a term $M$ to a \emph{bag} of available resources $C$; a $\beta$-reduction applies $M$ to a resource \emph{non-deterministically fetched} from $C$.
\Conf \nondt choice is also present in the session-typed $\pi$-calculus by Caires and P\'{e}rez~\cite{CairesP17}, where it expresses a choice between different implementations of the same session protocols, which are all \emph{non-deterministically available}---they may be available but may also \emph{fail}.
In their work, a Curry-Howard correspondence between linear logic and session types (`\emph{propositions-as-sessions}'~\cite{CairesP10,DBLP:conf/icfp/Wadler12}) ensures confluence,  protocol fidelity, and deadlock-freedom.
 Paulus et al.~\cite{DBLP:conf/fscd/PaulusN021} relate functional and concurrent calculi with confluent \nond: they give a translation
 of a resource $\lambda$-calculus into the session $\pi$-calculus from~\cite{CairesP17}, in the style of Milner's `\emph{functions-as-pro\-ce\-sses}'~\cite{Milner90}.

Although results involving confluent \nond are most significant, usual (\nconf) \nond remains of undiscussed convenience in formal modeling; consider, e.g., specifications of distributed protocols~\cite{DBLP:journals/entcs/BergerH00,DBLP:conf/concur/NestmannFM03} in which commitment is essential.
Indeed,  \nconf \nondt choice is commonplace in verification frameworks such as mCRL2~\cite{DBLP:books/mit/GrooteM2014}.
It is also  relevant in functional calculi; a well-known framework is De'Liguoro and Piperno's (untyped) non-deterministic
 $\lambda$-calculus~\cite{DBLP:journals/iandc/deLiguoroP95} (see also~\cite{DezaniCiancaglini:TR-TU-96} and references therein).

\smallskip
\myrev{
To further illustrate the difference between confluent and non-confluent \nond, we consider an example adapted from~\cite{CairesP17}: a movie server that offers a choice between buying a movie or watching its trailer.
In \clpi, the typed $\pi$-calculus that we present in this paper, this server can be specified as follows:
    \[
    \sff{Server}_s = \gsel{s} \left\{
        \begin{array}{@{}l@{}}
            \sff{buy} : \gname{s}{\textit{title}} ; \gname{s}{\textit{paym}} ; \pname{s}{\texttt{movie}} ; \pclose{s}
            ~,~
            \\
            \sff{peek} : \gname{s}{\textit{title}} ; \pname{s}{\texttt{trailer}} ;  \pclose{s}
        \end{array}
    \right\}
    \begin{array}{@{}l@{}}
        -\sff{Server}_s^{\sff{buy}}
        \\
        -\sff{Server}_s^{\sff{peek}}
    \end{array}
    \]
where $\gname{s}{-}$ and $\pname{s}{-}$ denote input and output prefixes on a name/channel $s$, respectively, and `\texttt{movie}' and `\texttt{trailer}' denote references to primitive data.
Also, the free names of a process are denoted with subscripts.
Process $\sff{Server}_s$ offers a choice on name~$s$ ($\gsel{s}\{-\}$) between labels \sff{buy} and \sff{peek}.
If \sff{buy} is received, process $\sff{Server}_s^{\sff{buy}}$ is launched: it receives the movie's title and a payment method, sends the movie, and closes the session on~$s$ ($\pclose{s}$).
If \sff{peek} is received, it proceeds as $\sff{Server}_s^{\sff{peek}}$: the server receives the title, sends the trailer, and closes the session.}

\myrev{Using the non-deterministic choice operator of \clpi, denoted `$\nd$', we can specify a process for  a client Alice who is interested in the movie `Jaws' but is undecided about buying the film or just watching its trailer for free:
\[
    \sff{Alice}_s := \begin{array}[t]{@{}r@{}lll@{}}
        &
        \psel{s}{\sff{buy}} ;
        &
        \pname{s}{\texttt{Jaws}} ; \pname{s}{\texttt{mcard}} ; \gname{s}{\textit{movie}} ; \gclose{s} ; \0
        &
        -\sff{Alice}_s^{\sff{buy}}
        \\
        {} \nd {} &
        \psel{s}{\sff{peek}} ;
        &
        \pname{s}{\texttt{Jaws}} ; \gname{s}{\textit{trailer}} ; \gclose{s} ; \0
        &
        -\sff{Alice}_s^{\sff{peek}}
    \end{array}
\]
If $\sff{Alice}_s$ selects the label $\sff{buy}$ ($\psel{s}{\sff{buy}}$), process $\sff{Alice}_s^{\sff{buy}}$ is launched: it sends title and payment method, receives the movie, waits for the session to close ($\gclose{s}$), and then terminates ($\0$).
If $\sff{Alice}_s$ selects \sff{peek}, process $\sff{Alice}_s^{\sff{peek}}$ is launched: it sends a title, receives the trailer, waits for the session to close, and terminates. Then, process $\sff{Sys} := \res{s} ( \sff{Server}_s \| \sff{Alice}_s )$ denotes the composition of client and server, connected  along $s$ (using $ \res{s}$).
Our semantics for  \clpi, denoted $\redtwo$, enforces non-confluent non-determinism, as $\sff{Sys}$ can reduce to separate processes, as expected:
\begin{mathpar}
    \sff{Sys} \redtwo \res{s} ( \sff{Server}_s^{\sff{buy}} \| \sff{Alice}_s^{\sff{buy}} )
    \and
    \text{and}
    \and
    \sff{Sys} \redtwo \res{s} ( \sff{Server}_s^{\sff{peek}} \| \sff{Alice}_s^{\sff{peek}} )
\end{mathpar}
}
\myrev{In contrast, the confluent non-deterministic choice  from \cite{CairesP17}, denoted~`$\oplus$', behaves differently: in their confluent semantics, $\sff{Sys}$ reduces to a  \emph{single} process including \emph{both alternatives}, i.e., $\sff{Sys} \red  \res{s} ( \sff{Server}_s^{\sff{buy}} \| \sff{Alice}_s^{\sff{buy}} ) \oplus \res{s} ( \sff{Server}_s^{\sff{peek}} \| \sff{Alice}_s^{\sff{peek}} )$.
}

\paragraph*{Contributions.}
We study new concurrent and functional  calculi with usual (\nconf) forms of \nond.
Framed in the typed (resource-conscious) setting,
we strive for definitions that do not exert a too drastic discarding of branches (as in the \nconf case) but also that do not exhibit a lack of commitment (as in the \conf case).
Concretely, we present:

\smallskip \noindent
(\secref{s:pi})
\clpi, a variant of the session-typed $\pi$-calculus in~\cite{CairesP17}, now with \nconf \nondt choice.
Its semantics adapts to the typed setting the usual semantics of non-deterministic choice in the untyped $\pi$-calculus~\cite{DBLP:books/daglib/0004377}.
Well-typed processes enjoy type preservation and deadlock-freedom (\Cref{t:srPi,t:dfPi}).

\smallskip \noindent
(\secref{s:lambda})
\lamcoldetshlin , a resource $\lambda$-calculus with   \nond, enhanced with constructs for expressing resource usage and failure.
Its  non-idempotent intersection type system provides a quantitative measure of the need/usage of resources.
Well-typed terms enjoy subject reduction and subject expansion (\Cref{t:lamSRShort,t:lamSEShort}).

\smallskip \noindent
(\secref{s:trans})
A typed translation of~\lamcoldetshlin into \clpi, which
provides further validation for our \nondt calculi, and casts them in the context of `{functions-as-processes}'.
We prove that our translation is \emph{correct}, i.e., it preserves types and satisfies tight operational correspondences (\Cref{def:encod_judge,t:correncLazy}).

\smallskip \noindent
Moreover, \secref{s:disc} closes by discussing related works.
\ifaplas
Omitted technical material
can be found in the full version of the paper~\cite{DBLP:journals/corr/abs-2205-00680}.
\else
Due to space limitations,
appendices contain (i)~omitted material (in particular, proofs of technical results); (ii)~an alternative \emph{eager} semantics for \clpi, which we  compare against the lazy semantics;
 and (iii)~extensions of \clpi and \lamcoldetshlin with  \emph{unrestricted} resources.
\fi

\section{A Typed \texorpdfstring{$\pi$}{Pi}-calculus with Non-deterministic Choice}
\label{s:pi}

We introduce $\clpi$, a session-typed $\pi$-calculus with   \nondt choice.
Following~\cite{CairesP17}, session types express protocols to be executed along channels.
These protocols can be \emph{non-deterministic}: sessions may succeed but also fail.
The novelty in $\clpi$ is the \nondt choice operator `$\!P \nd Q$', whose \emph{lazily committing semantics} is
compatible with linearity.
We prove that well-typed processes satisfy two key properties: \emph{type preservation} and  \emph{deadlock-freedom}.

\begin{figure}[t]
    \begin{mdframed} \mysmall
        \begin{align*}
            P,Q &::=
            \0 & \text{inaction}
            & \quad \sepr
            \pfwd{x}{y} & \text{forwarder}
            \\
            &~~~\sepr
            \res{x}(P \| Q) & \text{connect}
            & \quad \sepr
            P \nd Q & \text{non-determinism}
            \\ &~~~\sepr
            \pname{x}{y};(P \| Q) & \text{output}
            & \quad \sepr
            \gname{x}{y};P & \text{input}
            \\ &~~~\sepr
            \psel{x}{\ell};P & \text{select}
            & \quad \sepr
            \gsel{x}\{i:P\}_{i \in I} & \text{branch}
            \\ &~~~\sepr
            \pclose{x} & \text{close}
            & \quad \sepr
            \gclose{x};P & \text{wait}
            \\ &~~~\sepr
            \gsome{x}{w_1,\ldots,w_n};P & \text{expect}
            & \quad \sepr
            \psome{x};P & \text{available}
            \\ &~~~\sepr
            P \| Q & \text{parallel}
            & \quad \sepr
            \pnone{x} & \text{unavailable}
        \end{align*}

        \smallskip
        \dashes

        \vspace{-5ex}
        \begin{align*}
            P &\equiv P' ~ [P \equiv_\alpha P']
            &
            \pfwd{x}{y} &\equiv \pfwd{y}{x}
            &
            P \| \0 &\equiv P
            \\
            (P \| Q) \| R &\equiv P \| (Q \| R)
            &
            P \| Q &\equiv Q \| P
            &
            \res{x}(P \| Q) &\equiv \res{x}(Q \| P)
            \\
            P \nd P &\equiv P
            &
            P \nd Q &\equiv Q \nd P
            &
            (P \nd Q) \nd R &\equiv P \nd (Q \nd R)
        \end{align*}
        \vspace{-6ex}
        \begin{align*}
            \res{x}((P \| Q) \| R) &\equiv \res{x}(P \| R) \| Q
            &
            [ x \notin \fn{Q} ]
            \\
            \res{x}(\res{y}(P \| Q) \| R) &\equiv \res{y}(\res{x}(P \| R) \| Q)
            &
            [ x \notin \fn{Q}, y \notin \fn{R} ]
        \end{align*}
    \end{mdframed}
    \caption{ \clpi: syntax (top) and structural congruence (bottom).}\label{f:pilang}
\end{figure}

\subsection{Syntax and Semantics}

We use $P, Q, \ldots$ to denote processes, and $x,y,z,\ldots$ to denote \emph{names} representing  channels. 
\Cref{f:pilang} (top) gives the syntax of processes.
$P\{y/z\}$ denotes  the capture-avoiding substitution of $y$ for $z$ in  $P$.
Process~$\0$ denotes inaction, and
$\pfwd{x}{y}$ is a forwarder: a bidirectional link between $x$ and $y$.
Parallel composition appears in two forms:
while the process $P \| Q$ denotes communication-free concurrency,
process $\res{x}(P \| Q)$ uses restriction $\res{x}$ to express that $P$ and $Q$ implement complementary behaviors   on  $x$ and do not share any other names.

Process $P\nd Q$ denotes the non-deterministic choice between $P$ and $Q$: intuitively, if one choice can perform a synchronization, the other option may be discarded if it cannot.
Since $\nd$ is associative, we often omit parentheses. 
Also, we write $\bignd_{i \in I} P_i$ for the non-deterministic choice between each $P_i$ for $i \in I$.

Our output construct integrates parallel composition and restriction: process $\pname{x}{y};(P \| Q)$ sends a fresh name $y$ along $x$ and then continues as $P\| Q$.
The type system will ensure that behaviors on $y$ and $x$ are implemented by $P$ and $Q$, respectively, \myrev{which do not share any names---this separation defines communication-free concurrency and is key to ensuring deadlock-freedom}.
The  input process  $\gname{x}{y};P$ receives a name $z$ along $x$ and continues as $P\{z/y\}$, \myrev{which does not require the separation present in the output case. 
}
Process $\gsel{x}\{i:P_i\}_{i \in I}$  denotes a branch with labeled choices indexed by the finite set $I$: it awaits a choice on $x$ with continuation $ P_j$ for each $j \in I$.
The process $\psel{x}{\ell};P$ selects on $x$ the choice labeled $\ell$ before continuing as~$P$.
Processes $\pclose{x}$ and $\gclose{x}; P$ are dual actions for closing the session on $x$.
We omit replicated servers $\guname{x}{y};P$ and corresponding client requests $\puname{x}{y};P$, but they can be easily added 
\ifaplas
(cf.\ \cite{DBLP:journals/corr/abs-2205-00680}).
\else
(cf.~\secref{as:fullPi}).
\fi

The remaining constructs define non-deterministic sessions which
may provide a protocol or fail, following~\cite{CairesP17}.
    Process $\psome{x}; P$ confirms the availability of a session on $x$ and continues as $P$.
        Process $\pnone{x}$ signals the failure to provide the session on $x$.
    Process $\gsome{x}{w_1,\ldots,w_n}
    ; P$ specifies a dependency on a non-deterministic session on $x$ (names $w_1, \ldots , w_n$ implement sessions in~$P$). This
    process can either (i) synchronize with a `$\psome{x}$' and continue as $P$, or (ii) synchronize
    with a `$\pnone{x}$', discard $P$, and propagate the failure to $w_1, \ldots , w_n$.
To reduce eye strain, in writing $\gsome{x}{}$ we freely combine names and sets of names.
This way, e.g.,
we write $\gsome{x}{y,\fn{P},\fn{Q}}$ rather than $\gsome{x}{\{y\} \cup \fn{P} \cup \fn{Q}}$.

Name $y$ is bound in $\res{y}(P \| Q)$, $\pname{x}{y};(P \| Q)$, and $\gname{x}{y};P$.
\ifaplas
The set $\fn{P}$ includes the names in $P$ that are not bound.
\else
We write $\fn{P}$ and $\bn{P}$ for the free and bound names of $P$, respectively. 
\fi
We adopt Barendregt's convention.


\paragraph*{Structural Congruence.}
Reduction defines the steps that a process performs on its own.
It relies on \emph{structural congruence} ($\equiv$), the least congruence relation on processes induced by the rules in \Cref{f:pilang} (bottom).
Like the syntax of processes, the definition of $\equiv$ is aligned with the type system (defined next), such that $\equiv$ preserves typing (subject congruence, cf.\ \Cref{t:srPi}).
Differently from~\cite{CairesP17}, we do not allow distributing non-deterministic choice over parallel and restriction.
As shown in \ifaplas \cite{DBLP:journals/corr/abs-2205-00680}\else \secref{s:piEager}\fi, the position of a non-deterministic choice in a process determines how it may commit, so changing its position affects commitment.

\paragraph*{Reduction: Intuitions and Prerequisites.}
Barring non-deterministic choice, our reduction rules    arise as directed interpretations of proof transformations in the  underlying linear logic.
We follow Caires and Pfenning~\cite{CairesP10} and Wadler~\cite{DBLP:conf/icfp/Wadler12} in interpreting cut-elimination in linear logic as synchronization in $\clpi$.

Before delving into our reduction rules (\Cref{f:redtwo}), it may be helpful  to consider the usual reduction axiom for the (untyped) $\pi$-calculus (e.g.,~\cite{DBLP:journals/iandc/MilnerPW92a,DBLP:books/daglib/0004377}):
\begin{equation}
  (\pname{x}{z};P_1 + M_1) \|
(\gname{x}{y};P_2 + M_2)
\longrightarrow
P_1 \| P_2\{ z / y \}
\label{eq:usualpi}
\end{equation}
This axiom \myrev{captures the interaction of two (binary) choices: it} integrates the commitment of choice in  synchronization; after the reduction step,  \myrev{the two branches not involved in the synchronization, $M_1$ and $M_2$, are discarded}.
Our semantics of $\clpi$ is defined similarly: when a prefix within a branch of a choice synchronizes with its dual, that \myrev{branch reduces and the entire process commits to it}.

The key question at this point is: when and to which branches should we commit?
In~\eqref{eq:usualpi}, a communication commits to a single branch.
For \clpi,  we   define a \emph{lazy semantics} that minimizes commitment as much as possible.

The intuitive idea is that multiple branches of a choice may contain the same prefix, and so all these branches represent possibilities for synchronization (``possible branches'').
Other branches with different prefixes denote different possibilities (``impossible branches'').
When one synchronization is chosen, the possible branches are maintained while the impossible ones are discarded.

\begin{example}
\label{ex:possible}
    To distinguish possible and impossible branches, consider:
    \myrev{
    \begin{align*}
        P := \res{s} \mkern-1mu\big(\mkern-2mu \gsel{s}\{\sff{buy}:\myDots,\sff{peek}:\myDots\} \| ( \psel{s}{\sff{buy}}; \myDots \nd \psel{s}{\sff{buy}}; \myDots \nd \psel{s}{\sff{peek}} ; \myDots ) \mkern-3mu\big)
    \end{align*}
    The branch construct (case) provides the context for the non-deterministic choice.
    When the case synchronizes on the `\sff{buy}' label, the two branches prefixed by `$\psel{s}{\sff{buy}}$' are possible, whereas the branch prefixed by `$\psel{s}{\sff{peek}$}' becomes impossible, and can be discarded. The converse occurs when the  `\sff{peek}' label is selected.
    }
\end{example}

To formalize these intuitions, our reduction semantics (\Cref{f:redtwo})  relies on some auxiliary definitions. First, we define contexts.

\begin{definition}\label{d:ndctx}
    We define \emph{ND-contexts} ($\,\pctx{N},\pctx{M}$) as follows:
    \[
        \pctx{N},\pctx{M} ::= \hole \sepr \pctx{N} \| P \sepr \res{x}(\pctx{N} \| P) \sepr \pctx{N} \nd P
    \]
    The process obtained by replacing $\hole$ in $\pctx{N}$ with $P$ is denoted $\pctx{N}[P]$.
    We refer to ND-contexts that do not use the clause `\,$\pctx{N} \nd P$' as \emph{D-contexts}, denoted $\pctx{C},\pctx{D}$.
\end{definition}

Using D-contexts, we can express that, e.g.,  $\bignd_{i \in I} \pctx{C_i}[\pclose{x}]$ and $\bignd_{j \in J} \pctx{D_j}[\gclose{x};Q_j]$ should match.
To account for reductions with impossible branches, we define a precongruence on processes, denoted $\piprecong{S}$, where the parameter $S$ denotes the subject(s) of the prefix in the possible branches.
Our semantics is closed under $\piprecong{S}$.
Hence, e.g., anticipating a reduction on $x$, the possible branch $\pctx{C_1}[\gname{x}{y};P]$ can be extended with an  impossible branch to form $\pctx{C_1}[\gname{x}{y};P] \nd \pctx{C_2}[\gclose{z};Q]$.

Before defining $\piprecong{S}$ (\Cref{d:rpreone}), we first define prefixes (and their subjects).
\myrev{Below, we write $\widetilde{x}$ to denote a finite tuple of names $x_1, \ldots, x_k$.}

\begin{definition}\label{d:prefix}
    Prefixes are defined as follows:
    \[
        \alpha, \beta ::=
        \pname{x}{y} \sepr \gname{x}{y} \sepr \psel{x}{\ell} \sepr \gsel{x} 
        \sepr \pclose{x} \sepr \gclose{x} \sepr
        \psome{x} \sepr \pnone{x} \sepr \gsome{x}{\widetilde{w}} \sepr \pfwd{x}{y}
    \]
    The subjects of $\alpha$, denoted $\sub{\alpha}$, are $\{x,y\}$ in case of $\pfwd{x}{y}$, or $\{x\}$.
    By abuse of notation, we write $\alpha;P$ even when $\alpha$ takes no continuation (as in $\pclose{x}$, $\pnone{x}$, and $\pfwd{x}{y}$) and for $\pname{x}{y}$ which takes a parallel composition as continuation.
\end{definition}

\begin{definition}\label{d:rpreone}
    Let $\relalpha$ denote the least relation on prefixes (\defref{d:prefix}) defined by: \\
    (i)~$\pname{x}{y} \relalpha \pname{x}{z}$, (ii)~$\gname{x}{y} \relalpha \gname{x}{z}$, and (iii)~$\alpha \relalpha \alpha$ otherwise.

    \medskip
    Given a non-empty set $S\subseteq\{x,y\}$, the precongruence $P \piprecong{S} Q$ holds when both following conditions hold:
    \begin{enumerate}
        \item\label{i:pipreSingleton}
            $S = \{x\}$
            implies
            \\
            $P = \Big( \bignd_{i \in I} \pctx{C_i}[\alpha_i;P_i] \Big) \nd \Big( \bignd_{j \in J} \pctx{C_j}[\beta_j;Q_j] \Big)$
            and
            $Q = \bignd_{i \in I} \pctx{C_i}[\alpha_i;P_i]$,
            where
            \\
            (i)~$\forall i,i' \in I.\, \alpha_i \relalpha \alpha_{i'}$ and $\sub{\alpha_i} = \{x\} $,
            and \\
            (ii)~$\forall i \in I.\, \forall j \in J.\, \alpha_i \not\relalpha \beta_j \land x \in \fn{\beta_j;Q_j}$;

        \item\label{i:pipreForwarder}
            $S = \{x,y\}$
            implies
            \\
            $P = \Big( \bignd_{i \in I} \pctx{C_i}[\pfwd{x}{y}] \Big) \nd  \Big( \bignd_{j \in J} \pctx{C_j}[\pfwd{x}{z_j}] \Big) \nd \Big( \bignd_{k \in K} \pctx{C_k}[\alpha_k;P_k] \Big)$
            \\
            and
            $Q = \bignd_{i \in I} \pctx{C_i}[\pfwd{x}{y}]$,
            where
            \\
            (i)~$\forall j \in J.\, z_j \not = y$,
            and
            (ii)~$\forall k \in K.\ x \in \fn{\alpha_k;P_k} \land \forall z.\ \alpha_k \not\relalpha \pfwd{x}{z}$.
    \end{enumerate}
\end{definition}

\noindent
\myrev{
Intuitively, $\relalpha$ allows us to equate output/input prefixes with the same subject (but different object).
The rest of \Cref{d:rpreone} accounts for two kinds of reduction, using $S$ to discard ``impossible'' branches.
In case $S$ is $\{x\}$ (\Cref{i:pipreSingleton}), it concerns a synchronization on $x$; in case $S$ is $\{x,y\}$, it concerns forwarding on $x$ and $y$ (\Cref{i:pipreForwarder}).
In both cases, $P$ and $Q$ contain matching prefixes on $x$, while $P$ may contain additional branches with different or blocked prefixes on $x$; $x$ must appear in the hole of the contexts in the additional branches in $P$ (enforced with $x \in \fn{\ldots}$), to ensure that no matching prefixes are discarded.
}

\begin{example}
\label{ex:piprecongr}
    Recall process $P$ from \Cref{ex:possible}.
    \myrev{%
    To derive a synchronization with the `\sff{buy}' alternative of the case, we can use $\piprecong{S}$ to discard the `\sff{peek}' alternative, as follows:
    $
        \psel{s}{\sff{buy}} ; \myDots \nd \psel{s}{\sff{buy}} ; \myDots \nd \psel{s}{\sff{peek}} ; \myDots ~\piprecong{s}~ \psel{s}{\sff{buy}} ; \myDots \nd \psel{s}{\sff{buy}} ; \myDots
    $  
    }%
\end{example}

\begin{figure}[!t]
    \begin{mdframed} \mysmall
        \begin{align*}
            \mathsmaller{\rredtwo{\scc{Id}}}~
            &
            \res{x} \Big(\!\! \bignd_{i \in I}\pctx[\big]{C_i}[\pfwd{x}{y}] \| Q\Big)
            \redtwo_{x,y}
            \bignd_{i \in I} \pctx{C_i}[Q\{ y / x \}]
            \\
            \mathsmaller{\rredtwo{\tensor \parr}}~
            &
            \res{x} \Big(\!\! \bignd_{i \in I} \pctx{C_i}[\pname{x}{y_i};(P_i \| Q_i)] \| \bignd_{j \in J} \pctx{D_j}[x(z);R_j] \Big)
            \\ &
            \redtwo_x
            \bignd_{i \in I} \pctx[\Big]{C_i}[\! \res{x} \big(  Q_i \|  \res{w}( P_i\{w/y_i\} \| \bignd_{j \in J} \pctx{D_j}[R_j\{w/z\}] ) \big) \!]
            \\
            \mathsmaller{\rredtwo{\oplus \with}}~
            &
            \res{x} \Big(\!\! \bignd_{i \in I} \pctx{C_i}[\psel{x}{k'};P_i]  \| \bignd_{j \in J} \pctx{D_j}[\gsel{x}\{k:Q_j^k\}_{k \in K}] \Big)
            \\ &
            \redtwo_{x}
            \res{x} \Big( \bignd_{i \in I} \pctx{C_i}[P_i]  \| \bignd_{j \in J} \pctx{D_j}[ {Q_j^{k'}}  ]  \Big)
            \qquad [k' \in K]
            \\
            \mathsmaller{\rredtwo{\1 \bot}}~
            &
            \res{x} \Big(\!\! \bignd_{i \in I} \pctx{C_i}[ \pclose{x} ] \|  \bignd_{j \in J} \pctx{D_j}[ \gclose{x};Q_j ] \Big)
            \redtwo_{x}
            \bignd_{i \in I} \pctx{C_i}[ \0 ]  \| \bignd_{j \in J} \pctx{D_j}[Q_j]
            \\
            \mathsmaller{\rredtwo{\some}}~
            &
            \res{x} \Big(\!\! \bignd_{i \in I} \pctx{C_i}[ \psome{x};P_i ]  \| \bignd_{j \in J} \pctx{D_j}[\gsome{x}{w_1, \ldots, w_n};Q_j] \Big)
            \\ &
            \redtwo_{x}
            \res{x} \Big( \bignd_{i \in I} \pctx{C_i}[P_i]  \| \bignd_{j \in J} \pctx{D_j}[ {Q_j}  ]  \Big)
            \\
            \mathsmaller{\rredtwo{\none}}~
            &
            \res{x} \Big(\!\! \bignd_{i \in I} \pctx{C_i}[ \pnone{x} ] \| \bignd_{j \in J} \pctx{D_j}[\gsome{x}{w_1, \ldots, w_n};Q_j] \Big)
            \\ &
            \redtwo_{x}
            \bignd_{i \in I} \pctx{C_i}[\0]  \| \bignd_{j \in J} \pctx{D_j}[ \pnone{w_1} \| \ldots \| \pnone{w_n}  ]
        \end{align*}
        \begin{mathpar}
            \mathsmaller{\rredtwo{\piprecong{S}}}~~
            \inferrule{
                x \in S
                \\
                P \piprecong{S} P'
                \\
                Q \piprecong{S} Q'
                \\
                \res{x}(P' \| Q') \redtwo_S R
            }{
                \res{x}(P \| Q) \redtwo_S R
            }
                        \and
            \mathsmaller{\rredtwo{\nu\nd}}~~
            \inferrule{
                \res{x}(P \| \pctx[\big]{N}[\pctx{C}[Q_1] \nd \pctx{C}[Q_2]])\redtwo_S R
            }{
                \res{x}(P \| \pctx[\big]{N}[\pctx{C}[Q_1 \nd Q_2]])\redtwo_S R
            }
                        \and
            \mathsmaller{\rredtwo{\equiv}}~~
            \inferrule{
                P \equiv P'
                \\
                P' \redtwo_S Q'
                \\
                Q' \equiv Q
            }{
                P \redtwo_S Q
            }
            \and
            \mathsmaller{\rredtwo{\nu}}~~
            \inferrule{
                P \redtwo_S P'
            }{
                \res{x}(P \| Q) \redtwo_S \res{x}(P' \| Q)
            }
            \and
            \mathsmaller{\rredtwo{\|}}~~
            \inferrule{
                P \redtwo_S P'
            }{
                P \| Q \redtwo_S P' \| Q
            }
            \and
            \mathsmaller{\rredtwo{\nd}}~~
            \inferrule{
                P \redtwo_S P'
            }{
                P \nd Q \redtwo_S P' \nd Q
            }
        \end{mathpar}
    \end{mdframed}
    \caption{
       Reduction semantics for \clpi.
    }
    \label{f:redtwo}
\end{figure}

\paragraph*{Reduction Rules.}
\Cref{f:redtwo} gives the rules for the (lazy) reduction semantics, denoted $\redtwo_S$, where the set
$S$  contains the names involved in the interaction. We  omit the curly braces in this annotation; this way, e.g., we write `$\redtwo_{x,y}$' instead of `$\redtwo_{\{x,y\}}$'.
\myrev{
Also, we write $\redtwo_S^k$ to denote a sequence of $k \geq 0$ reductions.
}

The first six rules in \Cref{f:redtwo} formalize forwarding and communication: they are defined on choices containing different D-contexts (cf. \Cref{d:ndctx}), each with the same prefix but possibly different continuations; these rules preserve the non-deterministic choices.
Rule~$\rredtwo{\scc{Id}}$ fixes $S$ to the forwarder's two names, and the other rules fix $S$ to the one involved name.
\myrev{In particular, Rule~$\rredtwo{{\tensor}{\parr}}$ formalizes name communication: it involves multiple senders and multiple receivers (grouped in choices indexed by $I$ and $J$, respectively). Because they proceed in lock-step, reduction leads to substitutions involving the same (fresh) name $w$; also, the scopes of the choice and the contexts enclosing the senders is extended.}

Rule~$\rredtwo{\piprecong{S}}$ is useful to derive a synchronization that discards groups of  choices.
\myrev{
Rule~$\rredtwo{\nu\nd}$ allows inferring reductions when non-deterministic choices are not top-level: e.g., $\res{x} \big( \pclose{x} \| \res{y} ( ( \gclose{x} ; Q_1 \nd \gclose{x} ; Q_2 ) \| R ) \big) \redtwo_x \res{y} ( Q_1 \| R ) \nd \res{y} ( Q_2 \| R )$.
}
The last four rules formalize that reduction is closed under structural congruence, restriction, parallel composition, and non-deterministic choice.

\myrev{As mentioned earlier, a key motivation for our work is to have \nondt choices that effectively enforce commitment, without a too drastic discarding of alternatives.
Next we illustrate this intended form of \emph{gradual commitment}.}
\begin{example}[A Modified Movie Server]\label{ex:eve}
    \myrev{
    Consider the following variant of the movie server from the introduction, where the handling of the payment   is now modeled as a branch:
    \[
        \sff{NewServer}_s := \gname{s}{\textit{title}} ; \gsel{s} \left\{
            \begin{array}{@{}l@{}}
                \sff{buy} : \gsel{s}\left\{
                    \begin{array}{@{}l@{}}
                        \sff{card} : \gname{s}{\textit{info}} ; \pname{s}{\texttt{movie}} ; \pclose{s} ,
                        \\
                        \sff{cash} : \pname{s}{\texttt{movie}} ; \pclose{s}
                    \end{array}
                \right\}
                ,
                \\
                \sff{peek} : \pname{s}{\texttt{trailer}} ; \pclose{s}
            \end{array}
        \right\}
    \]
    Consider a client, Eve, who cannot decide between buying `Oppenheimer' or watching its trailer.
    In the former case, she has two options for payment method:
    \[
        \sff{Eve}_s := \pname{s}{\texttt{Oppenheimer}} ; \left(
            \begin{array}{@{}rl@{}}
                &
                \psel{s}{\sff{buy}} ; \psel{s}{\sff{card}} ; \pname{s}{\texttt{visa}} ; \gname{s}{\textit{movie}} ; \gclose{s} ; \0
                \\ {\nd} &
                \psel{s}{\sff{buy}} ; \psel{s}{\sff{cash}} ; \gname{s}{\textit{movie}} ; \gclose{s} ; \0
                \\ {\nd} &
                \psel{s}{\sff{peek}} ; \gname{s}{\textit{link}} ; \gclose{s} ; \0
            \end{array}
        \right)
    \]
    }
    \myrev{
    Let $\sff{Sys}^* := \res{s} ( \sff{NewServer}_s \| \sff{Eve}_s )$.
    After sending the movie's title, Eve's choice (buying or watching the trailer) enables gradual commitment. We have:
    \begin{mathpar}
        \sff{Sys}^* \redtwo_s^2 \res{s} \big( \gsel{s}\{ \sff{card} : \ldots , \sff{cash} : \ldots \} \| ( \psel{s}{\sff{card}} ; \ldots \nd \psel{s}{\sff{cash}} ; \ldots ) \big) =: \sff{Sys}^*_1
        \\
        \text{and}
        \and
        \sff{Sys}^* \redtwo_s^2 \res{s} ( \pname{s}{\texttt{trailer}} ; \ldots \| \gname{s}{\textit{trailer}} ; \ldots ) =: \sff{Sys}^*_2
    \end{mathpar}
    Process $\sff{Sys}^*_1$ represents the situation for Eve after selecting $\sff{buy}$, in which case the third alternative ($\psel{s}{\sff{peek}}; \ldots$) can be discarded as an impossible branch.
    Process $\sff{Sys}^*_2$ represents the dual situation. 
    From~$\sff{Sys}^*_1$, the selection of payment method completes the commitment to one alternative; we have:
    $\sff{Sys}^*_1 \redtwo_s \res{s} ( \gname{s}{\textit{info}} ; \myDots \| \pname{s}{\texttt{visa}} ; \myDots )$
    and
    $\sff{Sys}^*_1 \redtwo_s \res{s} ( \pname{s}{\texttt{movie}} ; \myDots \| \gname{s}{\textit{movie}} ; \myDots )$.
    }
\end{example}


\ifaplas
\else
In~\Cref{s:piEager}
 we discuss an alternative \emph{eager} semantics that commits to a single branch upon communication, as in~\eqref{eq:usualpi}.
\fi

\subsection{Resource Control for \texorpdfstring{\clpi}{spi+} via Session Types}
\label{ss:piTypeSys}

We define a session type system for \clpi\kern-.7ex, following `propositions-as-sessions'~\cite{CairesP10,DBLP:conf/icfp/Wadler12}.
As already mentioned, in a session type system,
resources are names that perform protocols:
the \emph{type assignment} $x:A$ says that $x$ should conform to the protocol specified by the session type $A$.
We give the syntax of types:
\begin{align*}
    A,B &::= \1 \sepr \bot \sepr A \tensor B \sepr A \parr B \sepr {\oplus}\{i:A\}_{i \in I} 
    \sepr {\with}\{i:A\}_{i \in I} \sepr {\with}A \sepr {\oplus}A
\end{align*}
The units $\1$ and $\bot$ type closed sessions.
$A \tensor B$ types a name that first outputs a name of type~$A$ and then proceeds as   $B$.
Similarly, $A \parr B$ types a name that inputs a name of type $A$ and then proceeds as~$B$.
Types ${\oplus}\{i:A_i\}_{i \in I}$ and  ${\with}\{i:A_i\}_{i \in I}$ are given to names that can select and offer a labeled choice, respectively.
 Then, ${\with}A$ is the type of a name that \emph{may produce} a behavior of type $A$, or fail; dually, ${\oplus}A$ types a name that \emph{may consume} a behavior of type $A$.


For any type $A$ we denote its \emph{dual} as $\ol{A}$.
Intuitively,  dual types serve to avoid communication errors: the type at one end of a channel is the dual of the type at the opposite end.
Duality is an involution, defined as follows:
\begin{align*}
    \ol{\1} &= \bot
    & \ol{A \tensor B} &= \ol{A}\parr\ol{B}
    & \ol{{\oplus} \{i:A_i\}_{i\in I}} &= {\with}\{i:\ol{A_i}\}_{i\in I}
    & \ol{{\with}A} &= {\oplus}\ol{A}
    \\
    \ol{\bot} &= \1
    & \ol{A \parr B} &= \ol{A}\tensor \ol{B}
    & \ol{{\with}\{i:A_i\}_{i\in I}} &= {\oplus}\{i:\ol{A_i} \}_{i\in I}
        & \ol{{\oplus}A} &= {\with}\ol{A}
\end{align*}

Judgments are of the form $P\vdash \Gamma$, where $P$ is a process and $\Gamma$ is a context, a collection of type assignments.
In writing $\Gamma, x:A$, we assume $x \notin \dom{\Gamma}$.
We write $\dom{\Gamma}$ to denote the set of names appearing in $\Gamma$.
We write $\with \Gamma$ to denote that $\forall x:A \in \Gamma.~ \exists A'.~ A = \with  A'$.

\begin{figure}[!t]
    \begin{mdframed} \mysmall
        \begin{mathpar}
            {\ttype[\scriptsize]{cut}}~
            \inferrule{
                P \vdash \Gamma, x{:}A
                \\
                Q \vdash \Delta, x{:}\ol{A}
            }{
                \res{x}(P \| Q) \vdash \Gamma, \Delta
            }
            \and
            \ttype[\scriptsize]{mix}~
            \inferrule{
                P \vdash \Gamma
                \\
                Q \vdash \Delta
            }{
                P \| Q \vdash \Gamma, \Delta
            }
            \and
            \ttype[\scriptsize]{$\nd$}~
            \inferrule{
                P \vdash \Gamma
                \\
                Q \vdash \Gamma
            }{
                P \nd Q \vdash \Gamma
            }
            \and
            \ttype[\scriptsize]{empty}~
            \inferrule{ }{
                \0 \vdash \emptyset
            }
            \and
            \ttype[\scriptsize]{id}~
            \inferrule{ }{
                \pfwd{x}{y} \vdash x{:}A, y{:}\ol{A}
            }
            \and
            \ttype[\scriptsize]{$\1$}~
            \inferrule{ }{
                \pclose{x} \vdash x{:}\1
            }
            \and
            \ttype[\scriptsize]{$\bot$}~
            \inferrule{
                P \vdash \Gamma
            }{
                \gclose{x};P \vdash \Gamma, x{:}\bot
            }
            \and
            \ttype[\scriptsize]{$\tensor$}~
            \inferrule{
                P \vdash \Gamma, y{:}A
                \\
                Q \vdash \Delta, x{:}B
            }{
                \pname{x}{y};(P \| Q) \vdash \Gamma, \Delta, x{:}A \tensor B
            }
            \and
            \ttype[\scriptsize]{$\parr$}~
            \inferrule{
                P \vdash \Gamma, y{:}A, x{:}B
            }{
                \gname{x}{y}; P \vdash \Gamma, x{:}A \parr B
            }
            \and
            \ttype[\scriptsize]{$\oplus$}~
            \inferrule{
                P \vdash \Gamma, x{:}A_j
                \\
                j \in I
            }{
                \psel{x}{j};P \vdash \Gamma, x{:}{\oplus}\{i:A_i\}_{i \in I}
            }
            \and
            \ttype[\scriptsize]{$\with$}~
            \inferrule{
                \forall i \in I.~ P_i \vdash \Gamma, x{:}A_i
            }{
                \gsel{x}\{i:P_i\}_{i \in I} \vdash \Gamma, x{:}{\with}\{i:A_i\}_{i \in I}
            }
            \and
            \ttype[\scriptsize]{${\with}\some$}~
            \inferrule{
                P \vdash \Gamma, x{:}A
            }{
                \psome{x};P \vdash \Gamma, x{:}{\with}A
            }
            \and
            \ttype[\scriptsize]{${\with}\none$}~
            \inferrule{ }{
                \pnone{x} \vdash x{:}{\with}A
            }
            \and
            \ttype[\scriptsize]{${\oplus}\some$}~
            \inferrule{
                P \vdash {\with}\Gamma,  x{:}A
            }{
                \gsome{x}{\dom{\Gamma}};P \vdash {\with}\Gamma,   x{:}{\oplus}A
            }
        \end{mathpar}
    \end{mdframed}
    \caption{Typing rules for \clpi.}
    \label{fig:trulespi}
\end{figure}

\Cref{fig:trulespi} gives the typing rules: they correspond to the rules in Curry-Howard interpretations of classical linear logic as session types (cf.\ Wadler~\cite{DBLP:conf/icfp/Wadler12}), with the rules for ${\with}A$ and ${\oplus}A$ extracted from~\cite{CairesP17}, and the additional Rule~\ttype{$\nd$} for non-confluent non-deterministic choice, which modifies the confluent rule in~\cite{CairesP17}.

Most rules follow~\cite{DBLP:conf/icfp/Wadler12}, so we focus on those related to non-determinism.
Rule~\ttype{${\with}\some$} types a process with a name whose behavior can be provided, while Rule~\ttype{${\with}\none$} types a name whose behavior cannot.
Rule~\ttype{${\oplus}\some$} types a process with a name $x$ whose   behavior may not be available.
If the behavior is not available, all  the sessions in the process must be canceled; hence, the rule requires all names to be typed under the ${\with}A$ monad.

Rule~\ttype{$\nd$} types our new non-deterministic choice operator;  the branches must be typable under the same typing context.
Hence, all branches denote the same sessions, which may be implemented differently.
In context of a synchronization, branches that are kept are able to synchronize, whereas the discarded branches are not; nonetheless, the remaining branches still represent different implementations of the same sessions.
Compared to the rule for non-determinism in~\cite{CairesP17}, we do not require processes to be typable under the ${\with}A$ monad.
   \myrev{
   \begin{example}
   Consider again process $\sff{Eve}_s$ from \Cref{ex:eve}.
    The three branches of the non-deterministic choice give \emph{different implementations of the same session}: 
    assuming primitive, self-dual data types $\mathtt{C}$, $\mathtt{M}$, and $\mathtt{L}$,
    all three branches on $s$ are typable by $\oplus \big\{ \sff{buy} : \oplus \{ \sff{card} : \mathtt{C} \tensor \mathtt{M} \parr \bot , \sff{cash} : \mathtt{M} \parr \bot \} , \sff{peek} : \mathtt{L} \parr \bot \big\}$.
       \end{example}
    }

    \myrev{
\begin{example}[Unavailable Movies]\label{x:unavailableMovies}
    Consider now a modified movie server, 
   which offers movies that may not be yet available.
    We specify this server using non-deterministic choice and non-deterministically available sessions:
    \[
        \sff{BuyServ}_s := \gname{s}{\textit{title}} ; ( \pnone{s} \nd \psome{s} ; \gname{s}{\textit{paym}} ; \pname{s}{\texttt{movie}} ; \pclose{s} ) \vdash s:\mathtt{T} \parr \big( \with ( \mathtt{P} \parr \mathtt{M} \tensor \1 ) \big),
    \]
    where $\mathtt{T},\mathtt{P},\mathtt{M}$ denote primitive, self-dual data-types.
    While the branch `$\pnone{s}$' signals that the movie is not available, the branch `$\psome{s} ; ...$' performs the expected protocol.
    We now define a client Ada who buys a movie for Tim, using session $s$; Ada only forwards it to him (using session $u$) if it is actually available:
    \begin{align*}
        \sff{Ada}_{s,u} &:= \pname{s}{\texttt{Barbie}} ; \gsome{s}{u} ; \pname{s}{\texttt{visa}} ; \gname{s}{\textit{movie}} ; \gclose{s} ; \psome{u} ; \pname{u}{\textit{movie}} ; \pclose{u}
        \\
        &\hphantom{:= } \vdash s:\mathtt{T} \tensor \big( \oplus ( \mathtt{P} \tensor \mathtt{M} \parr \bot ) \big) , u:\with  ( \mathtt{M} \tensor \1 )
        \\
        \sff{Tim}_u &:= \gsome{u}{} ; \gname{u}{\textit{movie}} ; \gclose{u} ; \0 \vdash u:\oplus ( \mathtt{M} \parr \1 )
    \end{align*}
    Let $\sff{BuySys} := \res{s} \big( \sff{BuyServ}_s \| \res{u} ( \sff{Ada}_{s,u} \| \sff{Tim}_u ) \big)$.
    Depending on whether the server has the movie ``Barbie'' available, we have the following reductions:
    \[
        \sff{BuySys} \redtwo_s^2 \res{u} ( \pnone{u} \| \sff{Tim}_u \mkern-2mu )
        ~~
        \text{or}
        ~~
        \sff{BuySys} \redtwo_s^5 \res{u} ( \psome{u} {;} \mkern-2mu \myDots \| \sff{Tim}_u \mkern-2mu )
        \tag*{\qedhere}
    \]
\end{example}
    }

Our type system ensures \emph{session fidelity} and \emph{communication safety}, but not confluence:
the former says that processes correctly follow their ascribed session protocols, and the latter that no communication errors/mismatches occur.
Both properties follow from the fact that typing is consistent across structural congruence and reduction.
\ifaplas
See~\cite{DBLP:journals/corr/abs-2205-00680} for details.
\else
See~\Cref{ss:TPLazy} for details.
\fi

\begin{theorem}[Type Preservation]
\label{t:srPi}
    If $P \vdash \Gamma$, then both $P \equiv Q$ and $P \redtwo_S Q$ (for any $Q$ and $S$) imply $Q \vdash \Gamma$.
\end{theorem}


Another important, if often elusive, property in session types is \emph{dead\-lock-free\-dom}, which ensures that processes can reduce as long as they are not inactive.
Our type system satisfies deadlock-freedom for processes with fully connected names, i.e., typable under the empty context.
\ifaplas
See~\cite{DBLP:journals/corr/abs-2205-00680} for details.
\else
See \Cref{ss:DFLazy} for details.
\fi

\begin{theorem}[Deadlock-freedom]\label{t:dfPi}
    If $P \vdash \emptyset$ and $P \not\equiv \0$, then there are $Q$ and $S$ such that $P \redtwo_S Q$.
\end{theorem}


\section{A Non-deterministic Resource \texorpdfstring{$\lambda$}{Lambda}-calculus}\label{s:lambda}

We present \lamcoldetshlin, a resource $\lambda$-calculus
with   \nond  and lazy evaluation.
In \lamcoldetshlin,   \nond is  \col and \emph{implicit}, as it arises from the fetching of terms from {bags} of \emph{linear} resources.
This is different from  $\clpi$, where the choice operator `$\nd$'
specifies \nond \emph{explicitly}.
A mismatch between the number of variable occurrences and the size of the bag induces \emph{failure}.

In $\lamcoldetshlin$, the \emph{sharing} construct  $M\sharing{x_1,\ldots, x_n}{x}$,
expresses that
$x$  may be used in $M$ under ``aliases'' $x_1,\ldots, x_n$.
Hence, it atomizes $n$ occurrences of   $x$   in~$M$, via an explicit pointer to $n$ variables.
This way, e.g.,
the $\lambda$-term $\lambda x.(x\ x)$ is expressed in $\lamcoldetshlin$ as $\lambda x. (x_1 \bag{x_2}\sharing{x_1,x_2}{x})$, where $\bag{x_2}$ is a bag containing~$x_2$.

\begin{figure}[t]
    \begin{mdframed} \mysmall
        \begin{align*}
            M,N,L ::=~
            & x & \text{variable}
            & \quad
            {}\sepr M \esubst{C}{x} & \text{intermediate subst.}
            \\
            \sepr~
            & (M\ C) & \text{application}
            & \quad
            {}\sepr M \linexsub{C/\widetilde{x}} & \text{explicit subst.}
            \\
            \sepr~
            & \lambda x.M & \text{abstraction}
            & \quad
            {}\sepr \fail^{\tilde{x}} & \text{failure}
            \\
            \sepr~
            & M\sharing{\widetilde{x}}{x} & \text{sharing}
            \\
            C,D ::=~
            & \oneb \sepr \bag{M} \cdot\, C
            \span\span
            & \text{ bag}
            \\
            \lctx{C} ::=~
            & \hole \sepr \lctx{C}\sharing{\widetilde{x}}{x} \sepr (\lctx{C}\ C) \sepr \lctx{C} \linexsub{C/\widetilde{x}}
            \span\span
            & \text{context}
        \end{align*}
    \end{mdframed}
    \caption{Syntax of $\lamcoldetshlin$: terms, bags, and contexts.}
    \label{f:lambdalin}
\end{figure}

\subsection{Syntax and Reduction Semantics}

\paragraph*{Syntax.}
We use $x,y,z,\ldots $ for variables, and write $\widetilde{x}$ to denote a finite sequence of pairwise distinct $x_i$'s, with length $|\widetilde{x}|$.
\Cref{f:lambdalin} gives the syntax of terms ($M,N,L$) and bags~($C, D$).
The empty bag is denoted $\oneb$.
We use $C_i$ to denote the $i$-th term in $C$, and $\size{C}$ denotes the number of elements in $C$.
To ease readability, we often write, e.g., $\bag{N_1, N_2}$
as a shorthand notation for
$\bag{N_1} \cdot \bag{N_2}$.

In  $M\sharing{\widetilde{x}}{x}$, we say that $\widetilde{x}$ are the \emph{shared variables}  and that $x$  is the \emph{sharing variable}.
We require for each $x_i \in \widetilde{x}$: (i)~$x_i$ occurs exactly once in $M$; (ii)~$x_i$ is not a sharing variable.
The sequence $\widetilde{x}$ can be empty:
$M\sharing{}{x}$ means that $x$ does not share any variables in $M$.
Sharing binds the shared variables in the term.

An abstraction $\lambda x. M$ binds
occurrences  of  $x$ in $M$.
Application $(M\ C)$ is as usual.
The term  $M \linexsub{C /  \widetilde{x}}$  is the \emph{explicit substitution}   of a bag $C$ for $\widetilde{x}$ in  $M$.
We require $\size{C} = |\widetilde{x}|$ and for each $x_i \in \widetilde{x}$: (i)~$x_i$ occurs in $M$; (ii)~$x_i$ is not a sharing variable; (iii)~$x_i$ cannot occur in another explicit substitution in $M$.
The term $M\esubst{ C }{ x }$ denotes an intermediate explicit substitution that does not (necessarily) satisfy the conditions for explicit substitutions.

The term $\fail^{\tilde{x}}$ denotes failure; the variables in $\widetilde{x}$ are ``dangling'' resources, which cannot be accounted for after failure.
We write $\lfv{M}$ to denote the free variables of $M$, defined as expected.
Term $M$ is \emph{closed} if $\lfv{M} = \emptyset$.

\begin{figure}[t]
    \begin{mdframed} \mysmall
        \begin{mathpar}
            \inferrule[$\redlab{RS{:}Beta}$]{ }{
                (\lambda x . M) C  \red M\esubst{ C }{ x }
            }
            \and
            \inferrule[$\redlab{RS{:}Ex \dash Sub}$]{
                \size{C} = |\widetilde{x}|
                \\
                M \not= \fail^{\tilde{y}}
            }{
                (M\sharing{\widetilde{x}}{x})\esubst{ C }{ x } \red  M\linexsub{C  /  \widetilde{x}}
            }
            \and
                        \inferrule[$\redlab{RS:TCont}$]{
                M \red    N
            }{
                \lctx{C}[M] \red   \lctx{C}[N]
            }
            \and
            \inferrule[$\redlab{RS{:}Fetch^{\ell}}$]{
                \headf{M} =  {x}_j
                \\
                0 < i \leq \size{C}
            }{
                M \linexsub{C /  \widetilde{x}, x_j} \red  M \headlin{ C_i / x_j }  \linexsub{(C \setminus C_i ) /  \widetilde{x}  }
            }
            \and
            \inferrule[$\redlab{RS{:}Fail^{\ell}}$]{
                \size{C} \neq |\widetilde{x}|
                \\
                \widetilde{y} = (\lfv{M} \setminus \{  \widetilde{x}\} ) \cup \lfv{C}
            }{
                (M\sharing{\widetilde{x}}{x}) \esubst{C }{ x }  \red  \fail^{\tilde{y}}
            }
            \and
            \inferrule[$\redlab{RS{:}Cons_1}$]{
                \widetilde{y} = \lfv{C}
            }{
                \fail^{\tilde{x}}\ C  \red  \fail^{\tilde{x} \cup \widetilde{y}}
            }
            \and
            \inferrule[$\redlab{RS{:}Cons_2}$]{
                \size{C} =   |  {\widetilde{x}} |
                \\
                \widetilde{z} = \lfv{C}
            }{
                (\fail^{ {\tilde{x}} \cup \tilde{y}} \sharing{\widetilde{x}}{x})\esubst{ C }{ x }  \red  \fail^{\tilde{y} \cup \widetilde{z}}
            }
            \and
            \inferrule[$\redlab{RS{:}Cons_3}$]{
                \widetilde{z} = \lfv{C}
            }{
                \fail^{\tilde{y}\cup \tilde{x}} \linexsub{C /  \widetilde{x}} \red  \fail^{\tilde{y} \cup \widetilde{z}}
            }
        \end{mathpar}
        where $\headf{M}$ is defined as follows:
        \begin{align*}
            \headf{ x } &= x
            &
            \headf{ \lambda x.M } &= \lambda x.M
            &
            \headf{ (M\ C) } &= \headf{M}
            \\
            \headf{ \fail^{\widetilde{x}} } &= \fail^{\widetilde{x}}
            &
            \headf{ M \esubst{ C }{ x } } &= M \esubst{ C }{ x }
            &
            \headf{ M \linexsub{C / \widetilde{x}} } &= \headf{M}
            \\
            \headf{ M\sharing{\widetilde{x}}{x} } = \begin{cases}
                x & \text{$\headf{ M } = y$ and $y \in  {\widetilde{x}}$}
                \\
                \headf{ M } & \text{otherwise}
            \end{cases}
            \span\span\span\span
        \end{align*}
    \end{mdframed}
    \caption{Reduction rules for $\lamcoldetshlin$.}
    \label{fig:reduc_intermlin}\label{f:lambda_redlin}
\end{figure}

\paragraph*{Semantics.}
\Cref{fig:reduc_intermlin} gives the reduction semantics, denoted $\red$, and  the \emph{head variable} of term $M$, denoted $\headf{M}$.
 Rule~$\redlab{RS:Beta}$ induces an intermediate substitution.
Rule~$\redlab{RS:Ex{\dash}Sub}$  reduces an intermediate substitution to an explicit substitution, provided the size of the bag equals the number of shared variables.
In case of a mismatch, the term evolves into failure via Rule~$\redlab{RS:Fail^\ell}$.

An explicit substitution $M \linexsub{C/\widetilde{x}}$, where the head variable of $M$ is $x_j \in \widetilde{x}$, reduces via Rule~$\redlab{R:Fetch^\ell}$.
The rule extracts a $C_i$ from $C$ (for some $0 < i \leq \size{C}$) and substitutes it for $x_j$ in $M$; this is how fetching induces a non-deterministic choice between $\size{C}$ possible reductions.
Rules~$\redlab{RS:Cons_j}$ for $j \in \{1,2,3\}$ consume terms when they meet failure.
Finally, Rule~$\redlab{RS:TCont}$ closes reduction under contexts.
The following example illustrates reduction.

\begin{example}\label{ex:syntax}\label{ex:lambdaRed}
    Consider the term $M_0 = ( \lambda x. x_1 \bag{x_2 \bag{x_3\ \oneb} } \sharing{ \widetilde{x} }{x} )\ \bag{\fail^{\emptyset} , y ,I\,}$,
    where $I = \lambda x. (x_1 \sharing{x_1}{x})$ and $\widetilde{x} = x_1 , x_2 ,x_3$.
    First, $M_0$ evolves into an intermediate substitution~\eqref{eq:lin_cons_sub1}.
    The bag can provide for all shared variables, so it then evolves into an explicit substitution~\eqref{eq:lin_cons_sub2}:
    \begin{align}
        M_0
        &\red  (x_1 \!\bag{\!x_2 \!\bag{\!x_3\ \oneb\!} \!}  \sharing{\widetilde{x}}{x}) \esubst{ \bag{\fail^{\emptyset} , y , I}   }{x}
        \label{eq:lin_cons_sub1}
        \\
        &\red
        (x_1 \!\bag{\!x_2 \!\bag{\!x_3\ \oneb\!}  \!}) \linexsub{\! \bag{\!\fail^{\emptyset} , y , I\!} \!/ \widetilde{x} } = M
        \label{eq:lin_cons_sub2}
    \end{align}
    Since $ \headf{M} = x_1$, one of the three elements of the bag will be substituted.
    $M$ represents a non-deterministic choice between the following three reductions:
    \begin{align*}
        \mathbin{\rotatebox[origin=r]{30}{$\red$}} &~ (\fail^{\emptyset} \bag{x_2 \bag{x_3\ \oneb}  }) \linexsub{  \bag{y, I} /  x_2,x_3  }
        = N_1
        \\[-5pt]
        M~
        \red &~ (y \bag{x_2 \bag{x_3\ \oneb}  }) \linexsub{ \bag{\fail^{\emptyset} , I} /  x_2,x_3  }
        = N_2
        \\[-3pt]
        \mathbin{\rotatebox[origin=r]{-30}{$\red$}} &~ (I \bag{x_2 \bag{x_3\ \oneb}}) \linexsub{ \bag{\fail^{\emptyset} , y}  /  x_2,x_3  }
        = N_3
        \tag*{\qedhere}
    \end{align*}
\end{example}

\subsection{Resource Control for \texorpdfstring{\lamcoldetshlin}{Lambda} via Intersection Types}\label{sec:lamTypes}

Our type system for   $\lamcoldetshlin$ is based on non-idempotent intersection types.
As in prior works~\cite{PaganiR10,DBLP:conf/birthday/BoudolL00}, intersection types account for available resources in bags, which are unordered and have all the same type.
Because we admit the term $\fail^{\tilde x}$ as typable, we say that our system enforces \emph{well-formedness} rather than \emph{well-typedness}.
As we will see, well-typed terms form the sub-class of well-formed terms that does not include $\fail^{\tilde x}$ (see the text after \Cref{t:lamSRShort}).

Strict types ($\sigma, \tau, \delta$) and multiset types ($\pi, \zeta$) are defined as follows:
\begin{align*}
    \sigma, \tau, \delta ::=~ &
    \unit \sepr \arrt{ \pi }{\sigma}
   \qquad  \pi, \zeta ::=
    \bigwedge_{i \in I} \sigma_i \sepr \omega
\end{align*}
\noindent
Given a non-empty $I$, multiset types $\bigwedge_{i\in I}\sigma_i$ are given to bags of size $|I|$.
This operator is associative, commutative,  and non-idempotent (i.e., $\sigma\wedge\sigma\neq \sigma$), with identity $\omega$.
Notation $\sigma^k$ stands for $\sigma \wedge \cdots \wedge \sigma$ ($k$ times, if $k>0$) or $\omega$ (if $k=0$).

Judgments have the form $\Gamma\wfdash M:\tau$, with contexts defined as follows:
\begin{align*}
    \Gamma,\Delta &::= \dash \sepr \Gamma , x:\pi  \sepr \Gamma, x:\sigma
\end{align*}
where $\dash$ denotes the empty context.
We write $\dom{\Gamma}$ for the set of variables in~$\Gamma$.
For $\Gamma, x:\pi$, we assume $x \not \in \dom{\Gamma}$.
To avoid ambiguities, we write $x:\sigma^1$  to denote that the assignment involves a multiset type, rather than a strict~type.
Given $\Gamma$, its \emph{core context} $\core{\Gamma}$ concerns variables with types different from $\omega$; it is defined as
$\core{\Gamma} = \{ x:\pi \in \Gamma \,|\, \pi \not = \omega\}$.

\begin{definition}[Well-formedness in $\lamcoldetshlin$]
    A  term $M$ is \emph{well-formed} if there exists a context $\Gamma$ and a  type  $\tau$ such that the rules in \Cref{fig:wfsh_ruleslin} entail $\Gamma \wfdash  M : \tau $.
\end{definition}

\begin{figure}[t]
    \begin{mdframed} \mysmall
        \begin{mathpar}
            \mprset{sep=1.8em}
            \inferrule[$\redlab{FS{:}var^{\ell}}$]{ }{
                 {x}: \sigma \wfdash  {x} : \sigma
            }
            \and
            \inferrule[$\redlab{FS{:}\oneb^{\ell}}$]{ }{
                \dash \wfdash \oneb : \omega
            }
            \and
            \inferrule[$\redlab{FS{:}bag^{\ell}}$]{
                 \Gamma \wfdash N : \sigma
                \\
                 \Delta \wfdash C : \sigma^k
            }{
                 \Gamma , \Delta \wfdash \bag{N}\cdot C:\sigma^{k+1}
            }
            \and
            \inferrule[$\redlab{FS{:}fail}$]{
                \dom{\core{\Gamma}} = \widetilde{x}
            }{
                \core{\Gamma} \wfdash  \fail^{\widetilde{x}} : \tau
            }
            \and
            \inferrule[$\redlab{FS{:}weak}$]{
                 \Gamma  \wfdash M : \tau
            }{
                \Gamma ,  {x}: \omega \wfdash M\sharing{}{x}: \tau
            }
            \and
            \inferrule[$\redlab{FS{:}shar}$]{
                \Gamma ,  {x}_1: \sigma, \ldots,  {x}_k: \sigma \wfdash M : \tau
                \\
                k \not = 0
            }{
                \Gamma ,  {x}: \sigma^{k} \wfdash M \sharing{{x}_1 , \ldots ,  {x}_k}{x}  : \tau
            }
            \and
            \inferrule[$\redlab{FS{:}abs\dash sh}$]{
                \Gamma ,  {x}: \sigma^k \wfdash M\sharing{\widetilde{x}}{x} : \tau
            }{
                 \Gamma \wfdash \lambda x . (M\sharing{\widetilde{x}}{x})  : \sigma^k \rightarrow \tau
            }
            \and
            \inferrule[$\redlab{FS{:}app}$]{
                \Gamma \wfdash M : \sigma^{j} \rightarrow \tau
                \\
                \Delta \wfdash C : \sigma^{k}
                \\
            }{
                 \Gamma , \Delta \wfdash M\ C : \tau
            }
            \and
            \inferrule[$\redlab{FS{:}Esub}$]{
                \Gamma ,  {x}: \sigma^{j} \wfdash M\sharing{\widetilde{x}}{x} : \tau
                \\\\
                \Delta \wfdash C : \sigma^{k}
            }{
                \Gamma , \Delta \wfdash (M\sharing{\widetilde{x}}{x})\esubst{ C }{ x }  : \tau
            }
            \and
            \inferrule[$\redlab{FS{:}Esub^{\ell}}$]{
                \Gamma  ,  x_1:\sigma, \cdots , x_k:\sigma \wfdash M : \tau
                \\\\
                \Delta \wfdash C : \sigma^k
            }{
                \Gamma , \Delta \wfdash M \linexsub{C /  x_1, \cdots , x_k} : \tau
            }
        \end{mathpar}
    \end{mdframed}
    \caption{Well-Formedness Rules for $\lamcoldetshlin$.}
    \label{fig:wfsh_ruleslin}
\end{figure}

\noindent
In \Cref{fig:wfsh_ruleslin},
Rule~$\redlab{FS:var^{\ell}}$ types variables.
Rule~$\redlab{FS:\oneb^{\ell}}$ types the empty bag with $\omega$.
Rule~$\redlab{FS:bag^{\ell}}$ types the concatenation of bags.
Rule~$\redlab{FS:\fail}$ types the term $\fail^{\widetilde{x}}$ with a strict type $\tau$, provided that the domain of the core context coincides with $\widetilde{x}$ (i.e., no  variable in $\widetilde{x}$ is typed with $\omega$).
Rule~$\redlab{FS:weak}$ types $M\sharing{}{x}$ by weakening the context with $x:\omega$.
Rule~$\redlab{FS:shar}$ types $M\sharing{\widetilde{x}}{x}$ with $\tau$, provided that there are assignments to the shared variables in~$\widetilde{x}$.

Rule~$\redlab{FS:abs{\dash}sh}$ types an abstraction $\lambda x.( M\sharing{\widetilde{x}}{x})$ with  $\sigma^k\to \tau$, provided that   $M\sharing{\widetilde{x}}{x}:\tau$ can be entailed from an assignment  $x:\sigma^k$.
Rule~\redlab{FS:app}  types $(M\ C)$, provided that $M$ has type $\sigma^j \to \tau$ and  $C$ has type $\sigma^k$. Note that, unlike usual intersection type systems, $j$ and $k$ may differ.
Rule~$\redlab{FS:Esub}$ types the intermediate substitution of a bag $C$ of type $\sigma^k$, provided that $x$ has type $\sigma^j$; again, $j$ and $k$ may differ.
Rule~$\redlab{FS:Esub^{\ell}}$  types  $M\linexsub{C/\widetilde{x}}$ as long as $C$ has type $\sigma^{|\widetilde{x}|}$,  and  each $x_i \in \widetilde{x}$ is of type $\sigma$.

Well-formed terms satisfy subject reduction (SR), whereas \emph{well-typed} terms, defined below, satisfy also subject expansion (SE).
\ifaplas
See~\cite{DBLP:journals/corr/abs-2205-00680} for details.
\else
See \secref{a:lamTypes} and \secref{a:subexpan} for details.
\fi

\begin{theorem}[SR in \lamcoldetshlin]\label{t:lamSRShort}
    If $ \Gamma \wfdash M:\tau$ and $M \red M'$, then $ \Gamma \wfdash M' :\tau$.
\end{theorem}


From our system for well-formedness we can extract a system for \emph{well-typed} terms, which do not include $\fail^{\tilde x}$.
Judgments for well-typedness are denoted $ \Gamma \wtdash M:\tau$, with rules copied from \Cref{fig:wfsh_ruleslin} (the rule name prefix \texttt{FS} is replaced with \texttt{TS}), with the following modifications:
(i)~Rule~$\redlab{TS{:}fail}$ is removed; (ii)~Rules~$\redlab{TS{:}app}$ and $\redlab{TS{:}Esub}$ are modified to disallow a mismatch between variables and resources, i.e., multiset types should match in size.
Well-typed terms are also well-formed, and thus satisfy SR.
Moreover, \myrev{as a consequence of adopting (non-idempotent) intersection types,} they also satisfy~SE:

\begin{theorem}[SE in $\lamcoldetshlin$]\label{t:lamSEShort}
    If $ \Gamma \wtdash M':\tau$ and $M \red M'$, then $ \Gamma \wtdash M :\tau$.
\end{theorem}


\section{A Typed Translation of \texorpdfstring{\lamcoldetshlin}{Lambda} into \texorpdfstring{\clpi}{Pi}
}
\label{s:trans}

While \clpi features \nondt choice,
\lamcoldetshlin
\myrev{is a prototypical programming language in which implicit non-determinism}
implements  fetching of resources.
Resources are controlled using different type systems  (session types in \clpi, intersection types  in \lamcoldetshlin).
To reconcile these differences and
\myrev{illustrate the potential of \clpi to precisely model non-determinism as found in realistic programs/protocols},
we give a translation of  \lamcoldetshlin into \clpi. 
This translation 
preserves types (\thmref{def:encod_judge}) and respects well-known  criteria for dynamic correctness~\cite{DBLP:journals/iandc/Gorla10,DBLP:phd/dnb/Peters12a,DBLP:journals/corr/abs-1908-08633} (\thmref{t:correncLazy}).

\paragraph*{The Translation.}
Given a \lamcoldetshlin-term \(M\), its translation into \clpi is denoted \(\piencod{M}_{u}\) and given in \Cref{fig:encodinglin}.
As usual, every variable \(x\) in $M$ becomes a name \(x\) in process \(\piencod{M}_{u}\), where
name \(u\) provides the behavior of \(M\).
A peculiarity is that, to handle failures in \lamcoldetshlin, \(u\) is a non-deterministically available session: the translated term can be available or not, as signaled by prefixes \(\psome{u}\) and \(\pnone{u}\), respectively.
As a result, reductions from \(\piencod{M}_{u}\) include synchronizations that codify $M$'s behavior but also   synchronizations that confirm a session's availability.

\begin{figure*}[t]
    \begin{mdframed} \mysmall
        \begin{align*}
            \piencodf{ {x}}_u
            &= \psome{x}; \pfwd{x}{u}
            \\
            \piencodf{\lambda x.M}_u
            &= \psome{u};\gname{u}{x}; \piencodf{M}_u
            \\
            \piencodf{(M\,C)}_u
            &=
            \res{v} (\piencodf{M}_v \| \gsome{v}{u , \llfv{C}};\pname{v}{x}; ( \piencodf{C}_x   \| \pfwd{v}{u}  ) )
            \\
            \piencodf{ M \esubst{ C }{ x} }_u
            &=  \res{x}( \piencodf{ M}_u \| \piencodf{ C }_x )
            \\
            \piencodf{\bag{N} \cdot~ C}_{{x}}
            &=
            \gsome{{x}}{\lfv{C}, \lfv{N} }; \gname{x}{y_i}; \gsome{{x}}{y_i, \lfv{C}, \lfv{N}}; \psome{{x}}; \pname{{x}}{z_i};
            \\ &\qquad
            ( \gsome{z_i}{\lfv{N}}; \piencodf{N}_{z_i}
              \| \piencodf{C}_{{x}} \| \pnone{y_i} )
            \\
            \piencodf{{\oneb}}_{{x}}
            &=
            \gsome{{x}}{\emptyset};\gname{x}{y_n};  ( \psome{ y_n}; \pclose{y_n}  \| \gsome{{x}}{\emptyset}; \pnone{{x}} )
            \\
            \piencodf{M \!\linexsub{ \!\bag{ \!N_1 , N_2\! } \!/ x_1,x_2 }}_u
            &=
            \res{z_1}( \gsome{z_1}{\lfv{N_{1}}};\piencodf{ N_{1} }_{ {z_1}} \| \res{z_2} (
                \gsome{z_2}{\lfv{N_{2}}};\piencodf{ N_{2} }_{ {z_2}}
            \\ & \qquad
                \| \bignd_{x_{i} \in \{ x_1 , x_2  \}} \bignd_{x_{j} \in \{ x_1, x_2 \setminus x_{i}  \}} \piencodf{ M }_u \{ z_1 / x_{i} \} \{ z_2 / x_{j} \} ) )
            \\
            \piencodf{M\sharing{}{x}}_u
            &=
            \psome{{x}}; \pname{{x}}{y_i}; ( \gsome{y_i}{ u , \lfv{M} }; \gclose{ y_{i} } ;\piencodf{M}_u \| \pnone{ {x} } )
            \\
            \piencodf{M\sharing{\widetilde{x}}{x}}_u
            &=
                \psome{{x}}; \pname{{x}}{y_i}; \big( \gsome{y_i}{ \emptyset }; \gclose{ y_{i} } ; \0
                \| \psome{{x}}; \gsome{{x}}{u, \lfv{M} \setminus  \widetilde{x} };
                \\ & \qquad
            \bignd_{x_i \in \widetilde{x}} \,\gname{{x}}{{x}_i};\piencodf{M\sharing{(\widetilde{x} {\setminus} x_i)}{x}}_u \big)
            \\
            \piencodf{\fail^{x_1, \ldots, x_k}}_u
            &= \pnone{ u}  \| \pnone{ x_1} \| \ldots \| \pnone{ x_k}
        \end{align*}
    \end{mdframed}
    \caption{Translation of $\lamcoldetshlin$ into $\clpi$.}\label{fig:encodinglin}
\end{figure*}
At its core, our translation  follows Milner's.
This way, e.g., the process \(\piencodf{(\lambda x.M)\ C}_u\)  enables synchronizations between \(\piencodf{\lambda x.M}_v\) and \(\piencodf{C}_{x}\) along
  name~$v$, resulting in the translation of an intermediate substitution.
The \emph{key novelty} is the role and treatment of   non-determinism. Accommodating \col \nond is non-trivial, as it entails translating explicit substitutions and sharing in $\lamcoldetshlin$ using the \nondt choice operator $\nd$ in $\clpi$.
Next we discuss these novel aspects, while highlighting differences with respect to a translation by Paulus et al.~\cite{DBLP:conf/fscd/PaulusN021}, which is given in the confluent setting (see \secref{s:disc}).

   In \Cref{fig:encodinglin}, non-deterministic choices occur in the translations of   $M\linexsub{C/\widetilde{x}}$ (explicit substitutions) and  $M\sharing{\widetilde{x}}{x}$ (non-empty sharing).
    Roughly speaking, the position  of $\nd$  in the translation of $M\linexsub{C/\widetilde{x}}$  represents the most desirable way of mimicking the fetching of terms from a bag.
This use of $\nd$ is a central idea in our translation: as we explain below,  it allows for appropriate commitment in  \nondt choices, but also for \emph{delayed} commitment when necessary.

For simplicity, we consider explicit substitutions \(M\mkern-3mu\linexsub{C/\widetilde{x}}\) where $C=\bag{\mkern-2mu N_1{,}N_2}$ and \(\widetilde{x}=x_1,x_2\).
The translation \(\piencodf{M \linexsub{C/\widetilde{x}}}_{u}\) uses the processes \(\piencodf{N_i}_{z_i}\), where each $z_i$ is fresh.
First, each bag item confirms its behavior.
Then, a variable~\(x_{i} \in \widetilde{x}\) is chosen non-deterministically;
we ensure that these choices consider all variables.
Note that writing \(\bignd_{x_{i} \in \{x_1,x_2 \}}\bignd_{x_{j}\in \{x_1,x_2\}\setminus x_{i} }\) is equivalent to non-de\-ter\-mi\-nis\-tic\-ally assigning \(x_{i},x_{j} \) to each permutation of \(x_1,x_2\).
The resulting choice involves \(\piencodf{M}_{u}\) with $x_{i}, x_j$ substituted by $z_1, z_2$.
Commitment here is triggered only via synchronizations along $z_1$ or $z_2$; synchronizing with
$\gsome{z_i}{\lfv{N_{i}}};\piencodf{N_{i} }_{ {z_i}}$ then represents fetching   $N_{i}$ from the bag.
\myrev{The size of the translated term $\piencodf{M \linexsub{C/\widetilde{x}}}_{u}$ is exponential with respect to the size of $C$.}

The process
\(\piencodf{M\sharing{\widetilde{x}}{x}}_{u}\)
proceeds as follows.
First, it confirms its behavior along \(x\).
Then it sends a name \(y_i\) on \(x\), on which a failed reduction may be handled.
Next,
the translation confirms again its behavior along \(x\) and non-deterministically receives a reference to an $x_i \in \widetilde{x}$.
Each branch consists of $\piencodf{M\sharing{(\widetilde{x} {\setminus} x_i)}{x}}_u$.
The possible choices are permuted, represented by~\(\bignd_{x_i\in \widetilde{x}}\).
{Synchronizations with $\piencodf{M\sharing{(\widetilde{x} {\setminus} x_i)}{x}}_u$ and bags delay commitment in this choice (we return to this point below).}
The process \(\piencodf{M\sharing{}{x}}_{u}\) is similar but simpler: here the  name \(x\) fails, as it cannot take further elements to substitute.

In case of a failure (i.e., a mismatch between the size of the bag~\(C\) and the number of variables in $M$), our translation ensures that the confirmations of \(C\) will not succeed.
This is how failure in \lamcoldetshlin is correctly translated to failure in~\clpi.

\paragraph*{Translation Correctness.}
The translation is typed: intersection types in \lamcoldetshlin are translated into session types in \clpi (\Cref{fig:enc_typeslin}).
This translation of types abstractly describes how non-deterministic fetches are codified as non-deterministic session protocols.
{
    It is worth noting that this translation of types is the same as in~\cite{DBLP:conf/fscd/PaulusN021}.
   This is not surprising: as we have seen, session types effectively abstract away from the behavior of processes, as all branches of a non-deterministic choice use the same typing context.
    Still, it is pleasant that the translation of types remains unchanged across different translations with
        our (non-confluent) non-determinism (in \Cref{fig:encodinglin})
        and with
    confluent non-determinism (in~\cite{DBLP:conf/fscd/PaulusN021}).
}

To state \emph{static} correctness,
we require the following definition:

\begin{definition}\label{def:enc_sestypfaillin}
    Let
    $
        \Gamma =  {{x}_1: \sigma_1}, \ndots,  {{x}_m : \sigma_m},  {{v}_1: \pi_1} , \ndots ,  {v}_n: \pi_n
    $
    be a context.
    The translation  $\piencodf{\cdot}_{\_}$  in~\Cref{fig:enc_typeslin} extends to contexts
    as follows:
    \begin{align*}
        \piencodf{\Gamma} = {} & {x}_1 : \with \overline{\piencodf{\sigma_1}} , \cdots ,   {x}_m : \with \overline{\piencodf{\sigma_m}} ,
         {v}_1:  \overline{\piencodf{\pi_1}_{(\sigma, i_1)}}, \cdots ,  {v}_n: \overline{\piencodf{\pi_n}_{(\sigma, i_n)}}
    \end{align*}
\end{definition}

\begin{figure}[t]
    \begin{mdframed} \mysmall
        \begin{align*}
            \piencodf{\unit} &= \with \onef
            &
            \qquad \qquad
            \piencodf{\sigma^{k}   \rightarrow \tau} &= \with( \dual{\piencodf{ \sigma^{k}  }_{(\sigma, i)}} \ampy \piencodf{\tau})
            \\
            \piencodf{ \sigma \wedge \pi }_{(\tau, i)} = \oplus(( \with \onef) \ampy ( \oplus  \with (( \oplus \piencodf{\sigma} ) \otimes (\piencodf{\pi}_{(\tau, i)}))))
            \span\span\span
            \\
            \piencodf{\omega}_{(\sigma, i)}
            &= \begin{cases}
                \oplus ((\with \1) \parr (\oplus \with \1))
                & \text{if $i = 0$}
                \\
                \oplus ((\with \1) \parr (\oplus\, \with ((\oplus \piencodf{\sigma}) \tensor (\piencodf{\omega}_{(\sigma, i-1)}))))
                & \text{if $i > 0$}
            \end{cases}
            \span\span
        \end{align*}
    \end{mdframed}
    \caption{Translation of intersection types into session types  (cf.\ \defref{def:enc_sestypfaillin}).}
    \label{fig:enc_typeslin}
\end{figure}

\noindent
Well-formed terms translate into well-typed processes:

\begin{theorem}
    \label{def:encod_judge}
    If
    $ \Gamma \wfdash {M} : \tau$,
    then
    $
        \piencodf{{M}}_u \vdash
        \piencodf{\Gamma},
        u : \piencodf{\tau}
    $.
\end{theorem}


To state \emph{dynamic} correctness, we rely on established  notions that (abstractly) characterize \emph{correct translations}.
A language \({\cal L}=(L,\to)\)  consists of a set of terms $L$ and a reduction relation $\to$ on $L$.
Each language ${\cal L}$ is assumed to contain a success constructor \(\sucs{}\).
A term \(T \in  L\) has {\em success}, denoted \(\succp{T}{\sucs{}}\), when there is a sequence of reductions (using \(\to\)) from \(T\) to a term satisfying success criteria.

Given ${\cal L}_1=(L_1,\to_1)$ and ${\cal L}_2=(L_2,\to_2)$, we seek translations \(\encod{\cdot }{}: {L}_1 \to {L}_2\) that are correct: they satisfy well-known correctness criteria~\cite{DBLP:journals/iandc/Gorla10,DBLP:phd/dnb/Peters12a,DBLP:journals/corr/abs-1908-08633}.
We state the set of correctness criteria that determine the correctness of a translation.

\begin{definition}[Correct Translation]\label{d:encCriteria}
    Let $\mathcal{L}_1= (\mathcal{M}, \shred_1)$ and  $\mathcal{L}_2=(\mathcal{P}, \shred_2)$ be two languages.
    Let $ \asymp_2 $ be an equivalence 
  over $\mcl{L}_2$.
    We use $M,M' $ (resp.\ $P,P' $) to range over terms in $\mcl{M}$  (resp.\ $\mcl{P}$).
  Given a translation $\encod{\cdot }{}: {\cal M}\to {\cal P}$, we define:
    \begin{description}
        \item \textbf{Completeness:}
            For every ${M}, {M}' $ such that ${M} \shred_1^\ast {M}'$, there exists $ P $ such that $ \encod{{M}}{} \shred_2^\ast P \asymp_2 \encod{{M}'}{}$.

        %
        \item \textbf{Weak Soundness:}
            For every $M$ and $P$ such that $\encod{M}{} \shred_2^\ast P$, there exist  $M'$, $P' $ such that $M \shred_1^\ast M'$ and $P \shred_2^\ast P' \asymp_2 \encod{M'}{}  $.

        %
        \item \textbf{Success Sensitivity:}
            For every ${M}$, we have $\succp{M}{\sucs{}}$ if and only if $\succp{\encod{M}{}}{\sucs{}}$.
    \end{description}
    %
\end{definition}

Let us write $\Lambda$ to denote the set of well-formed \lamcoldetshlin terms, and $\Pi$ for the set of all well-typed \clpi processes, both including $\sucs{}$.
We have 
our final  result:

\begin{theorem}[Translation correctness under $\redtwo$]
    \label{t:correncLazy}
    The translation $\piencod{\cdot}_{\_}: (\Lambda, \red) \to (\Pi, \redtwo)$ is 
    correct  (cf.\ \Cref{d:encCriteria}) using equivalence $\equiv$ (\Cref{f:pilang}).
\end{theorem}

\noindent
The proof of \Cref{t:correncLazy} involves instantiating/proving each of the parts of \mbox{\defref{d:encCriteria}}.
%
%
%
%
%
%
%
%
%
%
Among these, \emph{weak soundness} is the most  challenging to prove.
Prior work on translations of typed $\lambda$ into $\pi$ with confluent non-determinism~\cite{DBLP:conf/fscd/PaulusN021} rely critically on
confluence to match a behavior in $\pi$ with a corresponding behavior in $\lambda$.
Because in our setting  confluence is lost,  we must resort to a different proof.

As already discussed, our translation makes the implicit non-determinism in a $\lamcoldetshlin$-term $M$ explicit by adding non-deterministic choices in key points of $\piencodf{M}_u$.
  Our reduction $\redtwo$ preserves those branches that simultaneously have the same prefix available (up to $\relalpha$). In proving
  weak soundness,
  we exploit the fact that reduction entails delayed commitment.
    To see this, consider the following terms:
\begin{eqnarray}
    \res{x}(  (\alpha_1 ; P_1 \nd \alpha_2 ; P_2) \| Q)  \label{ex:sound1}
    \\
    \res{x}(  \alpha_1 ; P_1 \| Q) \nd \res{x}(\alpha_2 ; P_2 \| Q) \label{ex:sound2}
\end{eqnarray}
In \eqref{ex:sound1}, commitment to a choice
relies on whether $\alpha_1 \relalpha\alpha_2$ holds (cf. \Cref{d:rpreone}).
If $\alpha_1 \not \relalpha \alpha_2$, a choice is made; otherwise,   commitment is delayed, and depends on $P_1$ and $P_2$.
Hence, in \eqref{ex:sound1} the possibility of committing to either branch is kept open.
In contrast, in \eqref{ex:sound2} commitment to a choice is independent of   $\alpha_1 \relalpha\alpha_2$.

Our translation exploits the delayed commitment of non-determinism illustrated by \eqref{ex:sound1} to mimic commitment to non-deterministic choices in \lamcoldetshlin, which manifests in fetching resources from bags.
The fact that this delayed commitment preserves information about the different branches (e.g., $P_1$ and $P_2$ in \eqref{ex:sound1}) is essential to establish
weak soundness,
i.e., to match a behavior in \clpi with a corresponding  step in \lamcoldetshlin.
In contrast, forms of non-determinism in $\piencodf{N}_u$ that resemble \eqref{ex:sound2} are useful to characterize behaviors  different from fetching.

\section{Summary and Related Work}
\label{s:disc}

We studied the interplay between resource control and non-determinism in typed  calculi.
We introduced \clpi and \lamcoldetshlin, two calculi with \col \nond, both with type systems for resource control.
Inspired by the untyped $\pi$-calculus,  \nond in \clpi is lazy and explicit, with  session types defined following `propositions-as-sessions'~\cite{CairesP17}.
In \lamcoldetshlin,   \nond  arises in the fetching of resources, and is regulated by intersection types.
A correct translation of \lamcoldetshlin into \clpi precisely connects their different forms of  \nond. 


\paragraph{Related Work}
Integrating (\col) \nond within session types is non-trivial, as carelessly discarding  branches would break  typability.
Work by Caires and P\'erez~\cite{CairesP17}, already mentioned, develops a confluent semantics by requiring that non-determinism is only used inside the monad $\with A$; our non-confluent semantics drops this requirement.
This allows us to consider non-deterministic choices not possible in~\cite{CairesP17}, such as, e.g., selections of different labels.
We stress that linearity is not jeopardized: the branches of `$\nd$' do not represent \emph{different sessions}, but \emph{different implementations} of the same sessions.

Atkey et al.~\cite{DBLP:conf/birthday/AtkeyLM16} and Kokke et al.~\cite{journal/lmcs/KokkeMW20} extend `pro\-po\-si\-tions-as-sessions' with \nond. Their approaches are very different (conflation of the additives and bounded linear logic, respectively) and
  support non-determinism for {unrestricted names only}.
\myrevopt{This is a major difference:  
\Cref{ex:piprecongr,ex:piprecongr_red,ex:safe_or_risk_pro}
can only be supported in \cite{DBLP:conf/birthday/AtkeyLM16,journal/lmcs/KokkeMW20} by dropping linearity, which is important there.}
Also, \cite{DBLP:conf/birthday/AtkeyLM16,journal/lmcs/KokkeMW20} do not connect with typed $\lambda$-calculi, as we do. 
Rocha and Caires also consider non-determinism, relying on confluence in~\cite{conf/icfp/RochaC21} and on unrestricted names in~\cite{conf/esop/RochaC23}.
Casal et al.~\cite{DBLP:journals/tcs/CasalMV22,DBLP:conf/esop/VasconcelosCAM20} develop a type system for \emph{mixed sessions} (sessions with mixed choices), which can express non-determinism but does not ensure deadlock-freedom.
Ensuring deadlock-freedom by typing is a key feature of the `propositions-as-sessions' approach that we adopt for \clpi.

Our language \lamcoldetshlin is most related to  calculi by Boudol~\cite{DBLP:conf/concur/Boudol93}, Boudol and Laneve~\cite{DBLP:conf/birthday/BoudolL00}, and by Pagani and {Ronchi Della Rocca}~\cite{PaganiR10}.
Non-determinism  in the calculi in~\cite{DBLP:conf/concur/Boudol93,DBLP:conf/birthday/BoudolL00}   is committing and implicit; their linear resources can be consumed \emph{at most} once, rather than \emph{exactly} once.
The work~\cite{PaganiR10} considers non-committing \nond that is both implicit (as in \lamcoldetshlin) and explicit (via a sum operator on terms).
Both~\cite{DBLP:conf/concur/Boudol93,PaganiR10} develop (non-idempotent) intersection type systems to regulate resources.
In our type system,  all terms in a bag have the same type; the system in~\cite{PaganiR10} does not enforce this condition.
Unlike these type systems, our system for well-formedness can type terms with a lack or an excess of resources.

 Boudol and Laneve~\cite{DBLP:conf/birthday/BoudolL00} and Paulus et al.~\cite{DBLP:conf/fscd/PaulusN021} give translations of resource $\lambda$-calculi into $\pi$.
The translation in~\cite{DBLP:conf/birthday/BoudolL00}
is used to study the semantics induced upon $\lambda$-terms by a translation into \(\pi\); unlike ours, it does not consider types.
As already mentioned in~\secref{s:trans},  Paulus et al.~\cite{DBLP:conf/fscd/PaulusN021} relate calculi with \emph{confluent} \nond: a resource $\lambda$-calculus with sums on terms, and the session $\pi$-calculus from~\cite{CairesP17}.
Our translation of terms and that in~\cite{DBLP:conf/fscd/PaulusN021} are very different:
while here we use non-deterministic choice to mimic the sharing construct, the translation in~\cite{DBLP:conf/fscd/PaulusN021} uses it to translate bags.
Hence, our \Cref{t:correncLazy} cannot be derived from~\cite{DBLP:conf/fscd/PaulusN021}.

\myrevopt{Milner's  seminal work~\cite{Milner90} connects
untyped, deterministic $\lambda$ and $\pi$.
Sangiorgi studies the behavioral equivalences that Milner's translations induce on $\lambda$-terms~\cite{DBLP:phd/ethos/Sangiorgi93,DBLP:journals/iandc/Sangiorgi94,DBLP:journals/mscs/Sangiorgi99};   in particular, the work~\cite{DBLP:journals/iandc/Sangiorgi94} considers an untyped, non-de\-ter\-minis\-tic $\lambda$-calculus.
Sangiorgi and Walker~\cite{DBLP:books/daglib/0004377} offer a unified presentation of (typed) translations of $\lambda$ into $\pi$: 
they consider simply-typed $\lambda$-calculi and $\pi$-calculi with input-output types;  non-deterministic choices are not considered.}

The last decade of work on `prop\-o\-sitions-as-sessions' has delivered   insightful connections with typed $\lambda$-calculi---see, e.g.,~\cite{DBLP:conf/icfp/Wadler12,DBLP:conf/fossacs/ToninhoCP12,DBLP:conf/esop/ToninhoY18}.
Excepting~\cite{DBLP:conf/fscd/PaulusN021}, already discussed,
none of these works consider non-deterministic $\lambda$-calculi.

\paragraph{Acknowledgments}
We are grateful to
the anonymous reviewers for useful comments on previous versions of this paper.
We are also grateful to
Mariangiola Dezani
for her encouragement and suggestions.
This research has been supported by
the Dutch Research Council (NWO) under project No.\ 016.Vidi.189.046
(`Unifying Correctness for Communicating Software')
and the EPSRC Fellowship `VeTSpec: Verified Trustworthy Software Specification' (EP/R034567/1).

\addcontentsline{toc}{section}{References}
\bibliography{references}

\ifaplas
\end{document}
\else

\newpage
\newgeometry{margin=1.5cm}
\tableofcontents
\appendix


\section{Full \texorpdfstring{\clpi}{Pi}: Replicated Servers and Clients}
\label{as:fullPi}

Full \clpi includes unrestricted session behaviors (replicated servers and clients), not presented in \Cref{s:pi}.
Here we discuss how to add these omitted unrestricted sessions to the system described in \Cref{s:pi}.
The proofs of Type Preservation (\Cref{t:srPi}) and Deadlock-freedom (\Cref{t:dfPi}) in \Cref{s:piProofs} concern Full \clpi.
\begin{itemize}
    \item
        To the syntax of processes in \Cref{f:pilang} (top) we add two prefixes, $\puname{x}{y};P$ and $\guname{x}{y};P$, for client requests and server definitions, respectively.
        Both prefixes bind $y$ in $P$.

    \item
        Now $\fn{P}$ denotes the set of free names of $P$, including those used for unrestricted sessions.
        We write $\fln{P}$ to denote the set of free linear names of $P$, and $\fpn{P}$ for the set of free non-linear names.
        Note that $\fpn{P} = \fn{P} \setminus \fln{P}$.

    \item
        To the structural congruence in \Cref{f:pilang} (bottom) we add a rule that cleans up unused servers:
        \[
            \res{x}(\guname{x}{y};P \| Q) \equiv Q \quad \text{(if $x \notin \fn{Q}$)}.
        \]

    \item
        To the lazy semantics in \Cref{f:redtwo} we add the following rule that initiates a session between a client and a copy of a server:
        \begin{align*}
            \mathsmaller{\rredtwo{{?}{!}}}\quad
            & \res{x} \Big( \bignd_{i \in I} \pctx{C_i}[\puname{x}{y_i};P_i] \| \bignd_{j \in J} \pctx{D_j}[\guname{x}{z};Q_j] \Big)
            \\
            & \redtwo_x
            \bignd_{j \in J} \pctx[\Big]{D_j}[ \res{x} \Big( \res{w} \Big( \bignd_{i \in I} \pctx{C_i}[P_i\{w/z\}] \| Q_j\{w/z\} \Big) \| \guname{x}{z};Q_j \Big) ]
        \end{align*}


    \item
        Writing ${?}\Gamma$ to denote that $\forall x \in \Gamma.~ \exists A.~ \Gamma(x) = {?}A$, to the typing rules in \Cref{fig:trulespi} we add:
        \begin{mathpar}
            \ttype[\scriptsize]{$?$}~
            \inferrule{
                P \vdash \Gamma, y:A
            }{
                \puname{x}{y};P \vdash \Gamma, x:{?}A
            }
            \and
            \ttype[\scriptsize]{$!$}~
            \inferrule{
                P \vdash {?}\Gamma, y:A
            }{
                \guname{x}{y};P \vdash {?}\Gamma, x:{!}A
            }
            \and
            \ttype[\scriptsize]{weaken}~
            \inferrule{
                P \vdash \Gamma
            }{
                P \vdash \Gamma, x:{?}A
            }
            \and
            \ttype[\scriptsize]{contract}~
            \inferrule{
                P \vdash \Gamma, x:{?}A, x':{?}A
            }{
                P\{x/x'\} \vdash \Gamma, x:{?}A
            }
        \end{mathpar}
        Moreover, we replace Rule~$\ttype{${\oplus}\some$}$ with the following:
        \[
            \ttype{${\oplus}\some$}~
            \inferrule{
                P \vdash {\with}\Gamma, {?}\Delta, x:A
            }{
                \gsome{x}{\dom{\Gamma}};P \vdash {\with}\Gamma, {?}\Delta, x:{\oplus}A
            }
        \]
\end{itemize}

\section{An Alternative Eager Semantics for \texorpdfstring{\clpi}{spi+}}\label{s:piEager}
Let us consider a variant of \clpi in which syntax, typing, and structural congruence are as in \secref{s:pi}, but with an \emph{eagerly committing} semantics.
The idea is simple: we fully commit to a non-deterministic choice once a prefix synchronizes.

\begin{figure}[t]
    \begin{mdframed} \mysmall
        \begin{mathpar}
            \mathsmaller{\rredone{\scc{Id}}}~~
            \inferrule{}{
                \res{x}(\pctx[\big]{N}[\pfwd{x}{y}] \| Q) \redone \pctx*{N}[Q\{y/x\}]
            }
            \and
            \mathsmaller{\rredone{\1 \bot}}~~
            \inferrule{}{
                \res{x}(\pctx{N}[\pclose{x}] \| \pctx{N'}[\gclose{x};Q]) \redone \pctx*{N}[\0] \| \pctx*{N'}[Q]
            }
            \and
            \mathsmaller{\rredone{\tensor \parr}}~~
            \inferrule{}{
                \res{x}(\pctx{N}[\pname{x}{y};(P \| Q)] \| \pctx{N'}[\gname{x}{z};R]) \redone \pctx*[\big]{N}[\res{x}(Q \| \res{y}(P \| \pctx*{N'}[R\{y/z\}]))]
            }
            \and
            \mathsmaller{\rredone{\oplus \with}}~~
            \inferrule{}{
                \forall k' \in K.~ \res{x}(\pctx{N}[\psel{x}{k'};P] \| \pctx{N'}[\gsel{x}\{k:Q^k\}_{k \in K}]) \redone \res{x}(\pctx*{N}[P] \| \pctx*{N'}[Q^{k'}])
            }
            \and
            \mathsmaller{\rredone{{?}{!}}}~~
            \inferrule{}{
                \res{x}(\pctx{N}[ \puname{x}{y};P ] \| \pctx{N'}[ \guname{x}{z};Q ])
                \redone \pctx*[\big]{N'}[ \res{x}\big( \res{y}( \pctx*{N}[P] \| Q\{y/z\} ) \| \guname{x}{z};Q \big) ]
            }
            \and
            \mathsmaller{\rredone{\some}}~~
            \inferrule{}{
                \res{x}(\pctx{N}[\psome{x};P] \| \pctx{N'}[\gsome{x}{w_1, \ldots, w_n};Q]) \redone \res{x}(\pctx*{N}[P] \| \pctx*{N'}[Q])
            }
            \and
            \mathsmaller{\rredone{\none}}~~
            \inferrule{}{
                \res{x}(\pctx{N}[\pnone{x}] \| \pctx{N'}[\gsome{x}{w_1, \ldots, w_n};Q]) \redone \pctx*{N}[\0] \| \pctx*{N'}[\pnone{w_1} \| \ldots \| \pnone{w_n}]
            }
            \and
            \mathsmaller{\rredone{\equiv}}~~
            \inferrule{
                P \equiv P'
                \\
                P' \redone Q'
                \\
                Q' \equiv Q
            }{
                P \redone Q
            }
            \and
            \mathsmaller{\rredone{\nu}}~~
            \inferrule{
                P \redone P'
            }{
                \res{x}(P \| Q) \redone \res{x}(P' \| Q)
            }
            \and
            \mathsmaller{\rredone{\|}}~~
            \inferrule{
                P \redone P'
            }{
                P \| Q \redone P' \| Q
            }
            \and
            \mathsmaller{\rredone{\nd}}~~
            \inferrule{
                P \redone P'
            }{
                P \nd Q \redone P' \nd Q
            }
        \end{mathpar}
    \end{mdframed}
    \caption{
        Eager reduction semantics for \texorpdfstring{\clpi}{spi+}.
    }
    \label{f:eagerReductions}
\end{figure}

The eager reduction semantics, denoted $\redone$, is given in \Cref{f:eagerReductions}.
This semantics implements the full commitment of non-deterministic choices by committing ND-contexts to D-contexts as follows:

\begin{definition}\label{d:ncoll}
    The \emph{commitment} of an ND-context $\pctx{N}$, denoted $\D{\pctx{N}}$, is defined as follows:
    \begin{align*}
        \D{\hole} &:= \hole
        & \D{\pctx{N} \|  P} &:= \D{\pctx{N}} \| P
        & \D{\res{x}(\pctx{N} \| P)} &:= \res{x}(\D{\pctx{N}} \| P)
        & \D{\pctx{N} \nd P} &:= \D{\pctx{N}}
    \end{align*}
\end{definition}

\begin{proposition}\label{p:ncoll}
    For any ND-context $\pctx{N}$, the context $\D{\pctx{N}}$ is a D-context.
\end{proposition}

Just as \clpi  with lazy semantics,  \clpi with $\redone$ satisfies type preservation and deadlock-freedom.
See \Cref{ss:proofsEager} for details.

\begin{theorem}[Type Preservation: Eager Semantics]\label{t:typePresEager}
    If $P \vdash \Gamma$, then both $P \equiv Q$ and $P \redone Q$ (for any $S$) imply $Q \vdash \Gamma$.
\end{theorem}

\begin{proof}[Proof (Sketch)]
    If $P \equiv Q$, the thesis follows directly from \Cref{t:srPi}.
    If $P \redone Q$, we apply induction on the derivation of the reduction.
    In each case, we show that the commitment of ND-contexts (\Cref{d:ncoll}) preserves typing.
\end{proof}

\begin{restatable}[Deadlock-freedom: Eager Semantics]{theorem}{thmDlfreeOne}\label{t:dlfreeOne}
    If $P \vdash \emptyset$ and $P \not\equiv \0$, then there is $R$ such that $P \redone R$.
\end{restatable}

\begin{proof}[Proof (Sketch)]
    First, we write $P$ in such a way that we can access all its unblocked prefixes.
    Then we inductively show that there must be at least one pair of such prefixes that are connected by a restriction.
    Hence, these prefixes are duals and thus the process can reduce.
\end{proof}

\begin{figure*}[!t]
    \begin{mdframed} \mysmall
        \begin{align*}
            \mathbin{\rotatebox[origin=r]{30}{$\red$}} &~ (\fail^{\emptyset} \bag{x_2 \bag{x_3\ \oneb}  }) \linexsub{  \bag{y, I} /  x_2,x_3  }
            = N_1
            \\[-5pt]
            M
            =
            (x_1 \!\bag{\!x_2 \!\bag{\!x_3\ \oneb\!}  \!}) \linexsub{\! \bag{\!\fail^{\emptyset} , y , I\!} \!/x_1,x_2,x_3 }~
            \red &~ (y \bag{x_2 \bag{x_3\ \oneb}  }) \linexsub{ \bag{\fail^{\emptyset} , I} /  x_2,x_3  }
            = N_2
            \\[-3pt]
            \mathbin{\rotatebox[origin=r]{-30}{$\red$}} &~ (I \bag{x_2 \bag{x_3\ \oneb}}) \linexsub{ \bag{\fail^{\emptyset} , y}  /  x_2,x_3  }
            = N_3
        \end{align*}
        \vspace{-4ex}
        \begin{align*}
            \piencodf{M}_u
            =
            &~
            \res{z_1}(
            \gsome{z_1}{\emptyset}; \piencodf{ \fail^{\emptyset} }_{z_1}
            \| \res{z_2}(
            \gsome{z_2}{y}; \piencodf{ y }_{z_2}
            \| \res{z_3}(
            \gsome{z_3}{\emptyset}; \piencodf{ I }_{z_3}
            \\
            &~ \quad {}
            \| \bignd_{(x_i,x_j,x_k) \in \pi(\{z_1,z_2,z_3\})}
            \res{v}(
            \psome{x_i}; \pfwd{x_i}{v} \| \gsome{v}{v,x_j,x_k}; \pname{v}{z}; (
            \piencodf{ \bag{ x_j \bag{ x_k \ \oneb } } }_z
            \| \pfwd{v}{u}
            ) ) ) ) )
        \end{align*}

        \newlength{\sibldist}\setlength{\sibldist}{1.3cm}
            \mbox{}\hfill
            \begin{tikzpicture}
            [
                level 1/.style = {sibling distance = \sibldist},
                level 2/.style = {sibling distance = 0.5cm},
                level distance = 2.2cm,
                squigly/.style = {decorate,decoration={snake,amplitude=1pt,segment length=6pt,pre length=0pt,post length=2pt}}
            ]

                \node {$\piencodf{M}_u$}
                    child { node {$\piencodf{N_1}_u$}
                        edge from parent [->, squigly, cblGreen] node [above, xshift=-.5ex, pos=1.0, text=black] {\scriptsize $\ast$}}
                    child { node {$\piencodf{N_2}_u$}
                        edge from parent [->, squigly, cblGreen] node [left, yshift=1.0ex, xshift=.1ex, pos=1.0, text=black] {\scriptsize $\ast$}}
                    child { node {$\piencodf{N_3}_u$}
                        edge from parent [->, squigly, cblGreen] node [above, xshift=.5ex, pos=1.0, text=black] {\scriptsize $\ast$}};
            \end{tikzpicture}
            \hfill
            \textcolor{gray}{\vrule}
            \hfill
            \begin{minipage}[t]{11.6cm}
                \centering
                \begin{tikzpicture}
                [
                    level 1/.style = {sibling distance = 2.0cm},
                    level 2/.style = {sibling distance = 0.5cm},
                    level distance = 1.1cm
                ]

                    \node {$\piencodf{ M }_u$ }
                        child { node {$P_1(z_2,z_3)$}
                            child { node [text=black] {$\piencodf{N_1}_u$}
                                edge from parent [draw=none] node [draw=none] {$\premattwo$}}
                            edge from parent [->, cblRed] node [above, pos=1.0, text=black] {\scriptsize $\ast$}}
                        child {node {$P_1(z_3,z_2)$}
                            child { node [text=black] {$\piencodf{N_1}_u$}
                                edge from parent [draw=none] node [draw=none] {$\premattwo$}}
                            edge from parent [->, cblRed] node [above, pos=1.0, text=black] {\scriptsize $\ast$}}
                        child {node {$P_2(z_1,z_3)$}
                            child { node [text=black] {$\piencodf{N_2}_u$}
                                edge from parent [draw=none] node [draw=none] {$\premattwo$}}
                            edge from parent [->, cblRed] node [above, xshift=-.5ex, pos=1.0, text=black] {\scriptsize $\ast$}}
                        child {node {$P_2(z_3,z_1)$}
                            child { node [text=black] {$\piencodf{N_2}_u$}
                                edge from parent [draw=none] node [draw=none] {$\premattwo$}}
                            edge from parent [->, cblRed] node [above, xshift=.5ex, pos=1.0, text=black] {\scriptsize $\ast$}}
                        child {node {$P_3(z_1,z_2)$}
                            child { node [text=black] {$\piencodf{N_3}_u$}
                                edge from parent [draw=none] node [draw=none] {$\premattwo$}}
                            edge from parent [->, cblRed] node [above, pos=1.0, text=black] {\scriptsize $\ast$}}
                        child {node {$P_3(z_2,z_1)$}
                            child { node [text=black] {$\piencodf{N_3}_u$}
                                edge from parent [draw=none] node [draw=none] {$\premattwo$}}
                            edge from parent [->, cblRed] node [above, pos=1.0, text=black] {\scriptsize $\ast$}};
                \end{tikzpicture}

                The processes above are as follows:
                \begin{align*}
                    Q(a,b)
                    &=
                    \gsome{v}{u,a,b}; \pname{v}{z}; ( \piencodf{ \bag{a \ \bag{ b \ \oneb}} }_z \| \pfwd{v}{u} )
                    \\[-1mm]
                    P_1(a,b)
                    &=
                    \res{z_2}( \gsome{z_2}{y}; \piencodf{ y }_{z_2} \| \res{z_3}( \gsome{x_3}{\emptyset}; \piencodf{ I }_{z_3} \| \res{v}( \piencodf{ \fail^{\emptyset} }_{v} \| Q(a,b) ) ) )
                    \\[-1mm]
                    P_2(a,b)
                    &=
                    \res{z_1}( \gsome{z_1}{\emptyset}; \piencodf{ \fail^{\emptyset} }_{z_1} \| \res{z_3}( \gsome{x_3}{\emptyset}; \piencodf{ I }_{z_3} \| \res{v}( \piencodf{ y }_{v} \| Q(a,b) ) ) )
                    \\[-1mm]
                    P_3(a,b)
                    &=
                    \res{z_1}( \gsome{z_1}{\emptyset}; \piencodf{ \fail^{\emptyset} }_{z_1} \| \res{z_2}( \gsome{z_2}{y}; \piencodf{ y }_{z_2} \| \res{v}( \piencodf{ I }_{v} \| Q(a,b) ) ) )
                \end{align*}
            \end{minipage}
            \hfill\mbox{}
    \end{mdframed}
    \caption{\Cref{ex:looseTight}: Reductions of $M$ and of $\piencodf{M}_u$ under lazy and eager semantics. In $\piencodf{M}_u$, we write `$\pi(X)$' for the permutations of finite set $X$.}
    \label{fig:eager_m_red}\label{f:eager_v_lazy}
\end{figure*}

It is insightful to formally contrast our lazy and eager semantics.
We discuss two different ways.

\paragraph*{Lazy vs Eager, Part I: Translating \lamcoldetshlin.}
One way of comparing $\redtwo$ and $\redone$ is to establish the correctness of the translation
$\piencod{\cdot}_{-}$ (\Cref{fig:encodinglin}) but now in the eager case.
It turns out that the eager semantics leads to a \emph{strictly weaker} form of correctness, whereby completeness and weak soundness (cf.\ \Cref{d:encCriteria}) hold up to a precongruence $\premat$  instead of an equivalence (as in \Cref{t:correncLazy}).

The  precongruence $\premat$ is defined as follows:
\begin{mathpar}
    \inferrule{ }{
        P \premat P
    }
    \and
    \inferrule{
        P_i \premat P'_i
        \\
        {\scriptstyle i \in \{1,2\}}
    }{
        P_1 \nd P_2 \premat P'_i
    }
    \and
    \inferrule{
        P \premat P'
        \\
        Q \premat Q'
    }{
        P   \| Q \premat P'   \| Q'
    }
    \and
    \inferrule{
        P \premat P'
    }{
        \res{ x }   P \premat  \res{ x }  P'
    }
\end{mathpar}
Intuitively, $P \premat Q$ says that $P$ has at least as many branches as $Q$.
Translation correctness up to $\premat$ thus means that $\redone$ is ``{too eager}'', as it \emph{prematurely commits} to branches.

\appref{a:tight} proves that the translation under the eager semantics satisfies such   criteria.
Before discussing the corresponding completeness and soundness results, we present an example.

\begin{example}
\label{ex:looseTight}
    To contrast commitment in eager and lazy semantics (and their effect on the translation's correctness), recall from \Cref{ex:lambdaRed} the term $M$~\eqref{eq:lin_cons_sub2}:
    \[
        M = \big(x_1\ \bag{x_2\ \bag{x_3\ \oneb}}\big) \linexsub{ \bag{\fail^{\emptyset} , y , I} / x_1,x_2,x_3 }
    \]
    \Cref{f:eager_v_lazy} recalls the three branching reductions from $M$ to $N_1$, $N_2$ and $N_3$.
    It also depicts a side-by-side comparison of the reductions of   $\piencodf{M}_u$ under the lazy ($\redtwo$) and eager ($\redone$) semantics.
    In the figure, $\redtwo^\ast$ and $\redone^\ast$ denote the reflexive, transitive closures of $\redtwo$ and $\redone$, respectively.

    Under \redtwo there are three different reduction paths, each resulting directly in the translation of one of $N_1,N_2,N_3$: after the first choice, the following choices are preserved.
    In contrast, under \redone  there are six different reduction paths, each resulting in a process that relates to the translation of one of $N_1,N_2,N_3$ through $\premat$: after the first choice for an item from the bag is made, the semantics commits to choices for the other items.
\end{example}

The correctness properties induced by $\redone$ are \emph{loose} (rather than \emph{tight}, as in \secref{s:trans}; cf.\ \Cref{d:encCriteria}):

\begin{restatable}[Loose Completeness (Under $\redone$)]{theorem}{thmEncLWCompl}\label{thm:opcompletenessweak}
    If $ {N}\red {M}$ for a well-formed closed $\lamcoldetsh$-term $N$, then there exists $Q$ such that $\piencodf{{N}}_u \redone^* Q$ and $\piencodf{{M}}_u \premat Q $.
\end{restatable}

\begin{proof}[Proof (Sketch)]
    By induction on reductions.
    See \secref{a:looscompleteness} for details.
\end{proof}

\begin{restatable}[Loose Weak Soundness (Under $\redone$)]{theorem}{thmEncLWSound}\label{thm:opsoundweak}
    If  $\piencodf{{N}}_u \redone^* Q$ for a well-formed closed $\lamcoldetsh$-term $N$, then there exist ${N}'$ and $Q'$ such that
    (i)~${N} \redone^* {N}'$
    and
    (ii)~$Q \redone^* Q' $ with
    $ \piencodf{{N}'}_u \premat Q'$.
\end{restatable}

While \redtwo reduces multiple branches of a choice in lockstep, \redone  reduces only one branch and discards the rest.
Accordingly, weak soundness under \redtwo (\Cref{t:soundnesstwounres}) relates a sequence of lazy reductions to a sequence of   reductions in \lamcoldetshlin.
In contrast, \Cref{thm:opsoundweak} is weaker: it relates a sequence of eager reductions to a subset of branches, as some branches may have been eagerly discarded.
Hence, the proof of \Cref{thm:opsoundweak} has the added complexity of showing that every branch that is eagerly discarded must also be precongruent to a source reduction. This makes it difficult to apply induction directly, as we do not know which branches have been discarded in $\clpi$.

More in details, to prove \Cref{thm:opsoundweak}, we first show that $\premat$ is stable under reductions.
We need a way of denoting all possible reductions from a process.
We define $P \redone \{P_i\}_{i \in I}$, for a fixed (maximal) finite set $I = \{ i \mid P \redone P_i \}$.
Similarly, we define $P \redone^* \{P_i\}_{i \in I}$ inductively: if $P \redone^* \{P_i\}_{i \in I} $ and $P_i \redone \{P_j\}_{j \in J_i}$ for each $i \in I$, then $P \redone^* \{P_j\}_{j \in J} $ with $J = \bigcup_{ i \in I} J_i  $.
We then have the following:
\begin{restatable}{proposition}{propSoundextra}\label{prop:soundextra}
    If $P \premat Q$ and $P \redone^* \{P_i\}_{i \in I}$,
    then there exist $J$ and $ \{Q_j \}_{j \in J}$ such that $Q \redone^* \{Q_j \}_{j \in J} $, $ J \subseteq I $, and for each $j \in J$, $P_j \premat Q_j$.
\end{restatable}

\begin{proof}[Proof (Sketch)]
   {By induction on the derivation rules of the precongruence $\premat$.
    See \secref{a:loossoundness} for details.}
\end{proof}

Then, using \Cref{prop:soundextra} we prove the following:
\begin{restatable}{lemma}{thmOpsoundone}\label{thm:opsoundone}
    Let ${N}$ be a well-formed closed term.
    If  $\piencodf{{N}}_u \redone^* Q$, then there exist ${N}'$ and $\{Q_i\}_{i \in I}$ such that
    (i)~${N} \red^* {N}'$
    and
    (ii)~$Q \redone^* \{Q_i\}_{i \in I}$ where for each $j \in I$, $ \piencodf{{N}'}_u \premat Q_j$.
\end{restatable}

 \Cref{thm:opsoundone} ensures that the translation does not add behaviors not present in the source term. \Cref{thm:opsoundweak} then follows directly from \Cref{thm:opsoundone} by taking an arbitrary $Q_i$.

\paragraph*{Lazy vs Eager, Part II: Behavioral Equivalence.}
One may ask if the differences between \redtwo and \redone are confined to the ability to correctly translate \lamcoldetshlin, or, relatedly, whether \lamcoldetshlin's formulation is responsible for these differences.

We now compare \redtwo and \redone  \emph{independently from \lamcoldetshlin} by resorting to \emph{behavioral equivalences}.
We define a simple behavioral notion of equivalence on \clpi processes, parametric in \redtwo or \redone; then, we prove that there are classes of processes that are equal with respect to \redtwo, but incomparable with respect to \redone (\Cref{t:piBisim}).
A key ingredient is the following notion of observable on processes:

\begin{definition}\label{d:readyPrefix}
    A process $P$ has a \emph{ready-prefix} $\alpha$, denoted $P \readyPrefix{\alpha}$, if and only if there exist $\pctx{N},P'$ such that $P \equiv \pctx{N}[\alpha; P']$.
\end{definition}

We may now define:

\begin{definition}[Ready-Prefix Bisimilarity] \label{d:readyPrefixBisim}
    A relation $\mathbb{B}$ on \clpi processes is a \emph{(strong) ready-prefix bisimulation with respect to \redtwo} if and only if, for every $(P,Q) \in \mathbb{B}$,
    \begin{enumerate}
        \item
            For every $P'$ such that $P \redtwo P'$, there exists $Q'$ such that $Q \redtwo Q'$ and $(P',Q') \in \mathbb{B}$;

        \item
            For every $Q'$ such that $Q \redtwo Q'$, there exists $P'$ such that $P \redtwo P'$ and $(P',Q') \in \mathbb{B}$;

        \item
            For every $\alpha \relalpha \beta$, $P \readyPrefix{\alpha}$ if and only if $Q \readyPrefix{\beta}$.
    \end{enumerate}
    $P$ and $Q$ are \emph{ready-prefix bisimilar with respect to \redtwo}, denoted $P \readyPrefixBisim{L} Q$, if there exists a relation $\mathbb{B}$ that is a
    ready-prefix bisimulation with respect to \redtwo such that $(P,Q) \in \mathbb{B}$.

    A \emph{(strong) ready-prefix bisimulation with respect to \redone} is defined by replacing every occurrence of `$\redtwo$' by `$\redone$' in the definition above.
    We write $P \readyPrefixBisim{E} Q$ if $P$ and $Q$ are \emph{ready-prefix bisimilar with respect to \redone}.
\end{definition}

Ready-prefix bisimulation can highlight a subtle but significant difference between the behavior induced by our lazy and eager semantics.
To demonstrate this, we consider session-typed implementations of a vending machine.
\newcommand{\myMachine}{VM}
\newcommand{\myInterface}{IF}

\begin{example}[Two Vending Machines]\label{ex:prefixReadyBisim}
    Consider vending machines $\sff{\myMachine}_1$ and $\sff{\myMachine}_2$ consisting of three parts:
    (1)~an interface, which interacts with the user to send money and choose between coffee ($\sff{c}$) and tea ($\sff{t}$);
    (2)~a brewer, which produces either beverage;
    (3)~a system, which collects the money and forwards the user's choice to the brewer.
    A \clpi specification follows (below $\textup{€}$ and $\textup{€}2$ stand for names):
    \begin{align*}
        \sff{\myMachine}_1 &:= \res{x} \big( \sff{\myInterface}_1 \| \res{y} ( \sff{Brewer} \| \sff{System} ) \big)
        \\ \displaybreak[1]
        \sff{\myMachine}_2 &:= \res{x} \big( \sff{\myInterface}_2 \| \res{y} ( \sff{Brewer} \| \sff{System} ) \big)
            \\ \displaybreak[1]
        \sff{\myInterface}_1 &:= \pname{x}{\textup{€}2}; \big( \pclose{\textup{€}2} \| ( \psel{x}{\sff{c}}; \pclose{x} \nd \psel{x}{\sff{t}}; \pclose{x} ) \big)
        \\ \displaybreak[1]
        \sff{\myInterface}_2 &:= \pname{x}{\textup{€}2}; ( \pclose{\textup{€}2} \| \psel{x}{\sff{c}}; \pclose{x} ) \nd \pname{x}{\textup{€}2}; ( \pclose{\textup{€}2} \| \psel{x}{\sff{t}}; \pclose{x} )
                    \\ \displaybreak[1]
        \sff{System} &:= \gname{x}{\!\textup{€}}; \gsel{x} \!\left\{\!
            \begin{array}{@{}l@{}}
                \sff{c} : \psel{y}{\sff{c}}; \gclose{x}; \gclose{\textup{€}}; \pclose{y},
                \\
                \sff{t} : \psel{y}{\sff{t}}; \gclose{x}; \gclose{\textup{€}}; \pclose{y}
            \end{array}
        \!\right\}
        \\ \displaybreak[1]
        \sff{Brewer} &:= \gsel{y} \{ \sff{c} : \gclose{y}; \sff{Brew}_{\sff{c}}, \sff{t} : \gclose{y}; \sff{Brew}_{\sff{t}} \}
    \end{align*}
    where $\sff{Brew}_{\sff{c}} \vdash \emptyset$, $\sff{Brew}_{\sff{t}} \vdash \emptyset$, such that $\sff{\myMachine}_1 \vdash \emptyset$, $\sff{\myMachine}_2 \vdash \emptyset$.

    We give two implementations of the interface:
    $\sff{\myInterface}_1$ sends the money and then chooses coffee or tea;
    $\sff{\myInterface}_2$ chooses sending the money and then requesting coffee, or sending the money and then requesting tea.
    Then, $\sff{\myInterface}_1$ and $\sff{\myInterface}_2$ result in two different vending machines, $\sff{\myMachine}_1$ and $\sff{\myMachine}_2$.

    We have $\sff{\myMachine}_1 \nreadyPrefixBisim{E} \sff{\myMachine}_2$: the eager semantics distinguishes between the implementations; e.g., $\sff{\myInterface}_1$ has a single money slot, a button for coffee, and another button for tea, whereas $\sff{\myInterface}_2$ has two money slots, one for coffee, and another for tea.
    In contrast, under the lazy semantics, these machines are indistinguishable:
    $\sff{\myMachine}_1 \readyPrefixBisim{L} \sff{\myMachine}_2$.
\end{example}

\Cref{ex:prefixReadyBisim} highlights a difference in behavior between \redtwo and \redone when a moment of choice is subtly altered.
The following theorem captures this distinction (see \secref{a:piBisim}):

\begin{restatable}{theorem}{thmPiBisim}\label{t:piBisim}
    Take $R \equiv \pctx{N}[\alpha_1; (P \nd Q)] \vdash \emptyset$ and $S \equiv \pctx{N}[\alpha_2; P \nd \alpha_3; Q] \vdash \emptyset$, where $\alpha_1 \relalpha \alpha_2 \relalpha \alpha_3$ and $\alpha_1,\alpha_2,\alpha_3$ require a continuation.
    Suppose that $P \nreadyPrefixBisim{L} Q$ and $P \nreadyPrefixBisim{E} Q$.
    Then (i)~$R \readyPrefixBisim{L} S$ but (ii)~$R \nreadyPrefixBisim{E} S$.
\end{restatable}

Processes~\eqref{ex:sound1} and~\eqref{ex:sound2} from \secref{s:trans} (Page~\pageref{ex:sound1}) provide another example of a change in the moment of choice, different from the one discussed above.
In~\eqref{ex:sound1}, the choice depends on $\alpha_1$ and $\alpha_2$.
In~\eqref{ex:sound2}, the choice does not depend on $\alpha_1$ or $\alpha_2$, but  on choices made in the context in which the process resides.
Though the choice is subtly changed between~\eqref{ex:sound1} and~\eqref{ex:sound2}, the impact is significant: these processes are not ready-prefix bisimilar, with respect to neither semantics.
This is because, under both lazy and eager semantics, in~\eqref{ex:sound1} the two branches evolve in lockstep, whereas in~\eqref{ex:sound2} they evolve independently.


\section{Beyond Linear Resources}
Our results extend to  the language  $\lamcoldetsh$, an extension of $\lamcoldetshlin$ with a more general bag which includes unrestricted resources: resources that may be used zero or more times.
\Cref{f:lambda} gives the syntax of \lamcoldetsh-terms, bags and contexts.

A key difference with \lamcoldetshlin is that variables $x,y,z,\ldots $ have linear and unrestricted occurrences.
Notation $x[\mtt{l}]$ denotes a \emph{linear} occurrence of $x$; we often omit the annotation `$[\mtt{l}]$', and a sequence $\widetilde{x}$ always involves linear occurrences.
Notation~$x[i]$ denotes an \emph{unrestricted} occurrence of $x$, explicitly referencing the $i$-th element of an unrestricted (ordered) bag.
The structure of a bag is now split into a linear and an unrestricted component: as in $\lamcoldetshlin$, linear resources in bags cannot be duplicated, but unrestricted resources are always duplicated when consumed.
The empty unrestricted bag is denoted $\unvar{\oneb}$.
Notation $U_i$ denotes the singleton bag at the $i$-th position in $U$; if there is no $i$-th position in $U$, then $U_i$ defaults to $\unvar{\oneb}$.
We use `$\bagsep$' to combine a linear and an unrestricted bag, and unrestricted bags are joined via the non-commutative `$\concat$'.

To account for the explicit distinction between linear and unrestricted occurrences of variables, we now have two forms of explicit substitution:
\begin{itemize}
    \item $M \linexsub{C /  x_1 , \ldots , x_k}$, a linear substitution as in $\lamcoldetshlin$;
    \item $M \unexsub{U / \unvar{x}}$,  an unrestricted substitution   of an unrestricted  bag $U$ for an unrestricted variable $x^!$ in  $M$, with the assumption that $x^!$ does not appear in another unrestricted substitution in $M$.
\end{itemize}

\begin{figure}[t]
    \begin{mdframed} \mysmall
        \begin{align*}
            M,N,L ::=~
            & x[*] & \text{variable}
            & \quad
            {}\sepr M[\widetilde{x} \leftarrow x] & \text{sharing}
            \\
            \sepr~
            & \lambda x.M & \text{abstraction}
            & \quad
            {}\sepr M \esubst{B}{x} & \text{intermediate substitution}
            \\
            \sepr~
            & (M\ B) & \text{application}
            & \quad
            {}\sepr M \linexsub{C/x_1,\ldots,x_k} & \text{linear substitution}
            \\
            \sepr~
            & \fail^{\tilde{x}} & \text{failure}
            & \quad
            {}\sepr M \unexsub{U/\lunvar{x}} & \text{unrestricted substitution}
            \\
            [*] ::=~
            & [\mtt{l}]
            \sepr
            [i] ~~ i \in \mathbb{N} & \text{annotations}
            & \quad\hphantom{{}\sepr{}}
            A,B ::=~
            C \bagsep U & \text{bag}
            \\
            U,V ::=~
            & \unvar{\oneb} \sepr {\unvar{\bag{M}}} \sepr {U \concat V} & \text{unrestricted bag}
            & \quad\hphantom{{}\sepr{}}
            C,D ::= \oneb \sepr \bag{M} \cdot\, C & \text{linear bag}
            \\
            \lctx{C} ::=~
            & \hole \sepr (\lctx{C}\ B) \sepr \lctx{C} \linexsub{C/\widetilde{x}} \sepr \lctx{C} \unexsub{U/\lunvar{x}} \sepr \lctx{C}[\widetilde{x} \leftarrow x]
            \span\span
            & \text{context}
        \end{align*}
    \end{mdframed}
    \caption{Syntax of $\lamcoldetsh$.}
    \label{f:lambda}
\end{figure}


The reduction rules for \lamcoldetsh, extend the rules given in \Cref{f:lambda_redlin} with some modifications to accomodate the two-component format of bags and dedicated rules for the new constructs. Here, we describe the most interesting new rules: Rule~$\redlab{RS{:}Ex \dash Sub}$ for explicit substitution, and also Rules~$\redlab{R:Fetch^!}$ and $\redlab{R:Fail^!}$ for unrestricted substitution.

\begin{mathpar}
      \inferrule[$\redlab{RS{:}Ex \dash Sub}$]{
                \size{C} = |\widetilde{x}|
                \\
                M \not= \fail^{\tilde{y}}
            }{
                (M\sharing{\widetilde{x}}{x}) \esubst{ C \bagsep U }{ x } \red  M \linexsub{C  /  \widetilde{x}} \unexsub{U / \lunvar{x} }
            }
\and
  \inferrule[$\redlab{RS{:} Fetch^!}$]{
                \headf{M} = {x}[i]
                \\
                U_i = \unvar{\bag{N}}
            }{
                M \unexsub{U / \lunvar{x}} \red  M \headlin{ N /{x}[i] }\unexsub{U / \lunvar{x}}
            }
            \and
            \inferrule[$\redlab{RS{:}Fail^!}$]{
                \headf{M} = {x}[i]
                \\
                U_i = \unvar{\oneb}
            }{
                M \unexsub{U / \lunvar{x} } \red M \headlin{ \fail^{\emptyset} /{x}[i] } \unexsub{U / \lunvar{x} }
            }
\end{mathpar}
An explicit substitution $(M\sharing{\widetilde{x}}{x}) \esubst{ C \bagsep U }{ x }$ reduces to a term in which the linear and unrestricted parts of the bag are separated into their own explicit substitutions $M\linexsub{C  /  \widetilde{x}} \unexsub{U / \lunvar{x} }$, if successful. The fetching of linear/unrestricted resources from their corresponding bags is done by the appropriated fetch rules.
The reduction of an unrestricted substitution $M \unexsub{U/\lunvar{x}}$, where the head variable of $M$ is $x[i]$, depends on $U_i$:
\begin{itemize}
    \item
        If $U_i = \unvar{\bag{N}}$, then the term reduces via Rule~$\redlab{R:Fetch^!}$ by substituting the head occurrence of $x[i]$ in $M$ with $N$, denoted $M\headlin{N/x[i]}$; note that $U_i$ remains available after this reduction.

    \item
        If $U_i = \unvar{\oneb}$, the head variable is instead substituted with failure via Rule~$\redlab{R:Fail^!}$.
\end{itemize}
The definition of $\head{M}$ is as in \Cref{f:lambda_redlin} (bottom), extended with $\head{x[i]} = x[i]$ and $\head{M \unexsub{U/x}} = \head{M}$. The complete set of rules is in\appref{ss:lambdaUnresBags}.


\paragraph{Well-typedness and well-formedness.}

Types for $\lamcoldetsh$ extend the types for \lamcoldetshlin in \Cref{sec:lamTypes} with:
\begin{align*}
    \sigma, \tau, \delta ::=~ &
    \unit \sepr \arrt{ (\pi , \eta) }{\sigma}
\end{align*}
\begin{align*}
    \eta, \epsilon  &::=
    \sigma  \sepr \epsilon \concat \eta
    \qquad \text{list}
    &
    ( \pi , \eta)
    \qquad \text{tuple}
\end{align*}

The list type $\epsilon\concat \eta$ types the concatenation of unrestricted bags.
It can be recursively unfolded into a finite composition of strict types $\sigma_1\concat \ldots\concat \sigma_n$, for some $n\geq 1$, with length $n$ and $\sigma_i$ its $i$-th strict type ($1\leq i\leq n$).
We  write $\unvar{x}:\eta $ to denote for $x[1]:\eta_1 , \ldots , x[k]:\eta_k$ where $\eta$ has length $k$.
The tuple type $(\pi,\eta)$ types concatenation of a linear bag of type $\pi$ with an unrestricted bag of type $\eta$. Finally strict types are amended to allow for unrestricted functional types which go from tuple types to strict types \arrt{ (\pi , \eta) }{\sigma} rather then multiset types to strict types.

We separate contexts into two parts: linear ($\Gamma, \Delta,\ldots$) and unrestricted ($\Theta,\Upsilon,\ldots$):
\begin{align*}
    \Gamma,\Delta &::= \dash \sep \Gamma , x:\pi  \sepr \Gamma, x:\sigma
    &
    \Theta,\Upsilon &::= \dash \sepr \Theta, x^!:\eta
\end{align*}
Both linear and unrestricted occurrences of variables may occur at most once in a context.
Judgments have the form $\Theta;\Gamma\wfdash M:\tau$.
We write $\wfdash M:\tau$ to denote $\dash;\dash\wfdash M:\tau$.

Well-formedness rules for $\lamcoldetsh$ extend the rules given in \Cref{fig:wfsh_ruleslin} with specific rules to handle unrestricted resources, among those, we select  Rules \redlab{FS:var^!}, \redlab{FS{:}abs\dash sh}, \redlab{FS{:}Esub^!} and
\redlab{FS{:} bag^{!}}, for typing unrestricted occurrences of a variable, abstraction of a sharing term, explicit substitution and unrestricted bags, respectively.

\begin{mathpar}
   \inferrule[$\redlab{FS{:}var^!}$]{
                \Theta , x^!: \eta;  {x}: \eta_i , \Delta \wfdash  {x} : \sigma
            }{
                \Theta ,  x^!: \eta; \Delta \wfdash {x}[i] : \sigma
            }
            \and
             \inferrule[$\redlab{FS{:}abs\dash sh}$]{
                \Theta , x^!:\eta ; \Gamma ,  {x}: \sigma^k \wfdash M\sharing{\widetilde{x}}{x} : \tau
                \\
                {x} \notin \dom{\Gamma}
            }{
                \Theta ; \Gamma \wfdash \lambda x . (M[ {\widetilde{x}} \leftarrow  {x}])  : (\sigma^k, \eta )  \rightarrow \tau
            }
               \and
            \inferrule[$\redlab{FS{:}Esub^!}$]{
                \Theta , x^! {:} \eta; \Gamma  \wfdash M : \tau
                \\
                \Theta ; \dash \wfdash U : \epsilon
                \\
                \eta \relunbag \epsilon
            }{
                \Theta ; \Gamma \wfdash M \unexsub{U / \unvar{x}}  : \tau
            }
              \and
            \inferrule[$\redlab{FS{:} bag^{!}}$]{
                \Theta ; \dash \wfdash U : \epsilon
                \\
                \Theta ; \dash \wfdash V : \eta
            }{
                \Theta ; \dash  \wfdash U \concat V :\epsilon \concat \eta
            }
\end{mathpar}

In Rule \redlab{FS{:}Esub!},  $\eta \relunbag \epsilon$ denotes the fact that  $\epsilon=\epsilon_1\concat \ldots \concat \epsilon_k\concat \ldots \concat \epsilon_n$ and $\eta=\eta_1\concat \ldots \concat \eta_k$ are two list types, with $n\geq k$, such that the following hold:   for all $i$, $1\leq i \leq k$, $\epsilon_i = \eta_i $. The complete set of well-formedness rules for $\lamcoldetsh$ is in\appref{ss:lambdaUnresBags}.


\paragraph{Extended translation.}
To extend the translation
in \figref{fig:encodinglin} to \lamcoldet:
The access and use of unrestricted resources in \lamcoldet is codified in   \clpi by combining labeled choices and clients/servers (\appref{ss:transUnres}).

Note: for the sake of generality the proofs in the appendices concern \lamcoldet.


\input{appendix/unres-extentions.tex}

\input{appendix/pi-proofs.tex}

\input{appendix/lambda-correct.tex}

\input{appendix/lazy-encod.tex}
\input{appendix/eager-encod.tex}
\input{appendix/pi-comparison.tex}

\end{document}
\fi

%% file: macros.tex
\definecolor{cblRed}{rgb}{.60,.00,.01}
\definecolor{cblBlue}{rgb}{.16,.23,.42}
\definecolor{cblBlueLt}{rgb}{.00,.56,.61}
\definecolor{cblYellow}{rgb}{1.00,.69,.00}
\definecolor{cblPink}{rgb}{.85,.00,.42}
\definecolor{cblPurple}{rgb}{.10,.00,.54}
\definecolor{cblPurpleLt}{rgb}{.76,.39,.89}
\definecolor{cblOrange}{rgb}{.74,.29,.00}
\definecolor{cblGreen}{rgb}{.38,.61,.27}

\hypersetup{
    hypertexnames=false,
    colorlinks=true,
    final,
    citecolor=cblGreen,
    linkcolor=cblRed,
    urlcolor=cblRed,
    bookmarksnumbered=true,
    bookmarksopen=true,
    bookmarksopenlevel=1,
    pdfpagemode=UseOutlines
}

\newcommand{\clrone}[1]{\textcolor{cblRed}{#1}}
\newcommand{\clrtwo}[1]{\textcolor{cblGreen}{#1}}
\newcommand{\pctxcolor}{cblBlueLt}

\newcommand{\nconf}{non-con\-flu\-ent\xspace}

\newcommand{\conf}{con\-flu\-ent\xspace}
\newcommand{\Conf}{Con\-flu\-ent\xspace}
\newcommand{\col}{\nconf}

\newcommand{\nond}{non-de\-ter\-min\-ism\xspace}
\newcommand{\nondt}{non-de\-ter\-min\-is\-tic\xspace}

\newcommand{\lamcoldet}{\ensuremath{\lambda_{\mathtt{C}}}\xspace}
\newcommand{\lamcoldetsh}{\ensuremath{\lambda_{\mathtt{C}}}\xspace}
\newcommand{\lamcoldetshlin}{\ensuremath{\lambda_{\mathtt{C}}^{\ell}}\xspace}
\newcommand{\core}[1]{#1^{\downarrow}}

\newcommand{\linvar}[1]{ #1 ^{\ell}}
\newcommand{\unvar}[1]{ #1 ^{!}  }
\newcommand{\lunvar}[1]{#1}
\newcommand{\bagsep}{\star}

\newcommand{\sep}{\ | \ } 
\newcommand{\bag}[1]{\lbag #1 \rbag}
\newcommand{\oneb}{\mathtt{1}}
\newcommand{\perm}[1]{\ensuremath{\mathsf{PER}(#1)}}
\newcommand{\dom}[1]{\mathit{dom}(#1)}
\newcommand{\arrt}[2]{\ensuremath{#1 \rightarrow #2}}
\newcommand{\fail}{\mathtt{fail}}

\newcommand{\concat}{\ensuremath{\diamond}}
\newcommand{\relunbag}{\ensuremath{\sim}}
\newcommand{\sharing}[2]{[#1 \leftarrow #2]}

\newcommand{\contexcat}{\ensuremath{\wedge}}

\newcommand{\size}[1]{\mathsf{size}(#1)} 
\newcommand{\headf}[1]{\ensuremath{\mathsf{head}(#1)}}

\newcommand{\dash}{\text{-}}

\def\esubst#1#2{\langle\!\langle \raisebox{.5ex}{\small$#1$}\! / \mbox{\small$#2$}\rangle\!\rangle} 

\newcommand{\headlin}[1]{  \{ #1 \} } 
\newcommand{\ltalltriangle}{{\langle\!|}}
\newcommand{\rtalltriangle}{{|\!\rangle}}
\newcommand{\linexsub}[1]{\ltalltriangle #1 \rtalltriangle}
\newcommand{\unexsub}[1]{\llfloor #1 \rrceil }

\newcommand{\lfv}[1]{\mathsf{fv}(#1)}
\newcommand{\llfv}[1]{\mathsf{lfv}(#1)}

\newcommand{\unit}{\mathbf{unit}}
\newcommand{\head}[1]{\mathsf{head}(#1)} 

\newcommand{\red}{\mathbin{\longrightarrow}}
\newcommand{\redone}{\ensuremath{\mathbin{\clrone{\longrightarrow}}}\xspace}
\newcommand{\redtwo}{\ensuremath{\mathbin{\clrtwo{\leadsto}}}\xspace}

\newcommand{\readyPrefixBisimSub}[1]{\mkern-1mu\mathtt{#1}}
\newcommand{\readyPrefixBisim}[1]{\sim_{\readyPrefixBisimSub{#1}}}
\newcommand{\nreadyPrefixBisim}[1]{\not\sim_{\readyPrefixBisimSub{#1}}}

\newcommand{\succone}{\mathbin{\clrone{\Downarrow}}}
\newcommand{\succtwo}{\mathbin{\clrtwo{\Downarrow}}}

\newcommand{\redlab}[1]{\ensuremath{\mathtt{[#1]} }}

\newcommand{\equivlam}{\equiv_\lambda}

\newcommand{\secref}[1]{$\S$\,\labelcref{#1}\xspace}
\newcommand{\figref}[1]{Fig.\ \labelcref{#1}\xspace}
\newcommand{\defref}[1]{Def.\ \labelcref{#1}\xspace}
\newcommand{\appref}[1]{\appendixref*{#1}}
\newcommand{\thmref}[1]{\Cref{#1}\xspace}

\newcommand{\wfdash}{\vDash}
\newcommand{\wtdash}{\vdash}

\newcommand{\some}{\mathtt{some}}
\newcommand{\none}{\mathtt{none}}

\newcommand{\zero}{{\bf 0}}

\newcommand{\fn}[1]{\mathit{fn}(#1)}
\newcommand{\fln}[1]{\mathit{fln}(#1)}

\newcommand{\ampy}{\mathbin{\bindnasrepma}}
\newcommand{\with}{{\binampersand}}
\newcommand{\onef}{\mathbf{1}}
\newcommand{\dual}[1]{\overline{#1}}

\newcommand{\encod}[2]{\llbracket#1\rrbracket_{#2}} 

\newcommand{\piencodf}[1]{{\textcolor{cblPink}{\llbracket}#1\textcolor{cblPink}{\rrbracket}}}
\newcommand{\piencod}[1]{\piencodf{#1}}
\newcommand{\piencodfscale}[1]{\color{cblPink}\left\llbracket \normalcolor #1 \color{cblPink}\right\rrbracket \normalcolor}

\newcommand{\succp}[2]{\ensuremath{{#1} \Downarrow {#2}}}

\newcommand{\premat}{\succeq_{\mkern-2mu \scalebox{.8}{$\nd$}} }
\newcommand{\premattwo}{\preceq_{\nd} }

\newcommand{\relalpha}{ \bowtie }

\newcommand{\readyPrefix}[1]{\downarrow_{#1}}
\newcommand{\nreadyPrefix}[1]{\centernot{\downarrow}_{\mkern-7mu#1}}
\newcommand{\dualPrefix}{\mathbin{\dual{\relalpha}}}

\newcommand{\bignd}{\scalebox{1.5}{$\nd\mkern-3mu$}}

\newcommand{\ol}[1]{\overline{#1}}
\newcommand{\0}{\bm{0}}
\newcommand{\fwd}{\mathbin{\leftrightarrow}}
\newcommand{\sepr}{\mathbin{\scalebox{1.3}{$\mid$}}}

\newcommand{\rredone}[1]{[\clrone{\shortrightarrow}_{#1}]}
\newcommand{\rredtwo}[1]{[\clrtwo{\rightsquigarrow}_{#1}]}
\newcommand{\ttype}[2][]{\text{\normalfont#1[\scc{T#2}]}}

\newcommand{\sff}[1]{\relax\ifmmode\mathsf{#1}\else\textsf{#1}\fi}
\newcommand{\mathsc}[1]{\text{\normalfont\scshape#1}}
\newcommand{\scc}[1]{\relax\ifmmode\mathsc{#1}\else\textsc{#1}\fi}

\newcommand{\tensor}{\ensuremath{\mathbin{\otimes}}}
\newcommand{\parr}{\mathbin{\rott{\with}}}
\newcommand{\rott}[1]{\mathpalette\rot{#1}}
\newcommand{\rot}[2]{\rotatebox[origin=c]{180}{$#1{#2}$}}

\renewcommand{\gets}{\mathbin{\triangleright}}

\newcommand{\subj}{\mathit{subj}}

\newcommand{\D}[1]{\textcolor{\pctxcolor}{\llparenthesis}\mkern1mu #1 \mkern1mu\textcolor{\pctxcolor}{\rrparenthesis}}

\newcommand{\hole}{[\cdot]}

\let\obigcup\bigcup
\renewcommand{\bigcup}{{\mathchoice{\textstyle}{}{}{}\obigcup}}
\let\osum\sum
\renewcommand{\sum}{{\mathchoice{\textstyle}{}{}{}\osum}}
\let\oprod\prod
\renewcommand{\prod}{\mathchoice{\textstyle}{}{}{}{\oprod}}

\newcommand{\prlplus}{\hbox{\rlap{$\mkern-.4mu{}\mathbin{|\mkern-3mu|}{}$}$-$}}
\newcommand{\nd}{\mathbin{\prlplus}}

\newcommand{\piprecong}[1]{\mathrel{\succeq_{#1}}}
\newcommand{\sub}[1]{\ensuremath{\mathit{subj}\{#1\}}}

\newcommand{\infAss}[1]{\AxiomC{#1}}

\newcommand{\infUn}[2]{\RightLabel{#2} \UnaryInfC{#1}}

\newcommand{\mtt}[1]{\mathtt{#1}}
\let\oprl\|
\renewcommand{\|}{\mathbin{|}}
\newcommand{\pclose}[1]{\ol{#1}[]}
\newcommand{\gclose}[1]{#1()}
\newcommand{\psome}[1]{\ol{#1}.\some}
\newcommand{\pnone}[1]{\ol{#1}.\none}
\newcommand{\gsome}[2]{{#1.\some}_{#2}}
\newcommand{\pname}[2]{\ol{#1}[#2]}
\newcommand{\puname}[2]{{?}\ol{#1}[#2]}
\newcommand{\gname}[2]{#1(#2)}
\newcommand{\guname}[2]{{!}#1(#2)}
\newcommand{\psel}[2]{\ol{#1}.#2}
\newcommand{\gsel}[1]{#1.\mtt{case}}
\newcommand{\pfwd}[2]{[#1 \fwd #2]}
\newcommand{\res}[1]{(\bm{\nu} #1)}
\newcommand{\1}{\bm{1}}
\newcommand{\clpi}{\ensuremath{{{\mathsf{s}\pi\mkern-2mu}^{+}\mkern-1mu}}\xspace}
\newenvironment{tarr}[2][t]{\begin{array}[#1]{@{}#2@{}}}{\end{array}}
\newcommand{\ih}[1]{IH\textsubscript{#1}\xspace}
\newcommand{\inferruleDbl}[3][]{\inferrule*[#1,fraction={===}]{#2}{#3}}
\newcommand{\bn}[1]{\mathit{bn}(#1)}
\newcommand{\fpn}[1]{\mathit{fpn}(#1)}
\newcommand{\disj}{\mathbin{\#}}
\newcommand{\refitem}[3]{\Cref{#1:#2}.\labelcref{i:#2#3}}
\renewcommand{\flat}[1]{\mathit{flat}(#1)}
\newcommand{\dashes}{\vspace{-1.25em}\hbox to \textwidth{\leaders\hbox to 3pt{\hss . \hss}\hfil}\smallskip}
\newcommand{\mcl}[1]{\mathcal{#1}}

\newcommand{\lctx}[1]{\mcl{#1}}
\NewDocumentCommand \pctx {s O{} m o} {\IfBooleanT{#1}{\textcolor{\pctxcolor}{\llparenthesis}\mkern1mu}\textcolor{\pctxcolor}{\mathtt{#3}}\IfBooleanT{#1}{\mkern1mu\textcolor{\pctxcolor}{\rrparenthesis}} \IfValueT{#4}{\textcolor{\pctxcolor}{#2[}#4\textcolor{\pctxcolor}{#2]}}}
\newcommand{\sucs}[1]{{\checkmark\mkern-4mu}_{#1}}

\def\shred{\mathbin{\shortrightarrow}}
\def\ndots{\hbox to 1em{.\hss.\hss.}}

\newcommand{\myDots}{\ifmmode\mathinner{\ldotp\kern-0.2em\ldotp\kern-0.2em\ldotp}\else.\kern-0.13em.\kern-0.13em.\fi}


%% file: appendix/unres-extentions.tex
\section{Extensions in Detail: Eager Semantics and Unrestricted Resources}
\label{s:fullAccounts}

\subsection{\texorpdfstring{$\lamcoldetsh$}{Unrestricted Lambda}:  An Extension of \texorpdfstring{\lamcoldetshlin}{Lambda}  with Unrestricted Bags}
\label{ss:lambdaUnresBags}

\Cref{f:lambda_llfv} defines the free linear variables of a term, denoted $\llfv{M}$.
\Cref{f:lambda_red} gives the reduction rules for \lamcoldetsh, extending the rules given in \Cref{f:lambda_redlin} with Rules~$\redlab{R:Fetch^!}$ and $\redlab{R:Fail^!}$ for unrestricted substitution.
%

\begin{figure}[t]
    \begin{mdframed} \mysmall
        \begin{align*}
            \llfv{x}&=\{x\}
            & \llfv{(M\ B)} &= \llfv{M}\cup \llfv{B}\\
            \llfv{x[i]}&=\emptyset
           & \llfv{M\sharing{\widetilde{x}}{x}}&=(\llfv{M}\setminus\{\widetilde{x}\})\cup \{x\} \\
            \llfv{\oneb}&=\emptyset
            & \llfv{\lambda x. (M\sharing{\widetilde{x}}{x})}&=\llfv{M\sharing{\widetilde{x}}{x}}\setminus \{x\}\\
            \llfv{\bag{M}}&=\llfv{M}
            & \llfv{M\linexsub{C/x_1,\ldots, x_k}}&=(\llfv{M}\setminus \{x_1,\ldots, x_k\})\cup \llfv{C}\\
            \llfv{C\bagsep  U}&=\llfv{C}
            & \llfv{ M\esubst{ B }{ x }}&= (\llfv{M}\setminus \{x\})\cup\llfv{B}\\
            \llfv{\fail^{\widetilde{x}}}&=\{\widetilde{x}\}
            & \llfv{\bag{M}\cdot C}&=\llfv{M}\cup \llfv{C}
        \end{align*}
    In the case $\llfv{M}=\emptyset$, the term $M$ is called \emph{linearly closed}.
    \end{mdframed}
    \caption{Free Linear Variables}
    \label{f:lambda_llfv}
\end{figure}

\begin{figure}[t]
    \begin{mdframed} \mysmall
        \begin{mathpar}
            \inferrule[$\redlab{RS{:}Beta}$]{ }{
                (\lambda x . M)\ B  \red M \esubst{ B }{ x }
            }
            \and
            \inferrule[$\redlab{RS{:}Ex \dash Sub}$]{
                \size{C} = |\widetilde{x}|
                \\
                M \not= \fail^{\tilde{y}}
            }{
                (M\sharing{\widetilde{x}}{x}) \esubst{ C \bagsep U }{ x } \red  M \linexsub{C  /  \widetilde{x}} \unexsub{U / \lunvar{x} }
            }
            \and
            \inferrule[$\redlab{RS{:}Fetch^{\ell}}$]{
                \headf{M} =  {x}_j
                \\
                0 < i \leq \size{C}
            }{
                M \linexsub{C /  \widetilde{x}, x_j} \red  (M \headlin{ C_i / x_j })  \linexsub{(C \setminus C_i ) /  \widetilde{x}  }
            }
            \and
            \inferrule[$\redlab{RS{:}Fail^{\ell}}$]{
                \size{C} \neq |\widetilde{x}|
                \\
                \widetilde{y} = (\llfv{M} \setminus \{  \widetilde{x}\} ) \cup \llfv{C}
            }{
                (M\sharing{\widetilde{x}}{x}) \esubst{C \bagsep U}{ x }  \red  \fail^{\widetilde{y}}
            }
            \and
            \inferrule[$\redlab{RS{:} Fetch^!}$]{
                \headf{M} = {x}[i]
                \\
                U_i = \unvar{\bag{N}}
            }{
                M \unexsub{U / \lunvar{x}} \red  M \headlin{ N /{x}[i] }\unexsub{U / \lunvar{x}}
            }
            \and
            \inferrule[$\redlab{RS{:}Fail^!}$]{
                \headf{M} = {x}[i]
                \\
                U_i = \unvar{\oneb}
            }{
                M \unexsub{U / \lunvar{x} } \red M \headlin{ \fail^{\emptyset} /{x}[i] } \unexsub{U / \lunvar{x} }
            }
            \and
            \inferrule[$\redlab{RS{:}Cons_1}$]{
                \widetilde{y} = \llfv{C}
            }{
                \fail^{\widetilde{x}}\ (C \bagsep U) \red  \fail^{\widetilde{x} \ \widetilde{y}}
            }
            \and
            \inferrule[$\redlab{RS{:}Cons_2}$]{
                \size{C} =   |  {\widetilde{x}} |
                \\
                \widetilde{z} = \llfv{C}
            }{
                (\fail^{ {\widetilde{x}} \cup \widetilde{y}} \sharing{\widetilde{x}}{x}) \esubst{ C \bagsep U }{ x }  \red  \fail^{\widetilde{y} \cup \widetilde{z}}
            }
            \and
            \inferrule[$\redlab{RS{:}Cons_3}$]{
                \widetilde{z} = \llfv{C}
            }{
                \fail^{\widetilde{y}\cup \widetilde{x}} \linexsub{C /  \widetilde{x}} \red  \fail^{\widetilde{y} \cup \widetilde{z}}
            }
            \and
            \inferrule[$\redlab{RS{:}Cons_4}$]{ }{
                \fail^{\widetilde{y}} \unexsub{U / \lunvar{x}}  \red  \fail^{\widetilde{y}}
            }
            \and
            \inferrule[$\redlab{RS:TCont}$]{
                M \red    N
            }{
                \lctx{C}[M] \red   \lctx{C}[N]
            }
        \end{mathpar}
    \end{mdframed}
    \caption{Reduction rules for \texorpdfstring{\lamcoldetsh}{lambda}.}
    \label{fig:reduc_interm}\label{f:lambda_red}
\end{figure}

\paragraph{Well-typedness and well-formedness.}

Types for $\lamcoldetsh$ extend the types for \lamcoldetshlin in \Cref{sec:lamTypes} with:
\begin{align*}
    \sigma, \tau, \delta ::=~ &
    \unit \sepr \arrt{ (\pi , \eta) }{\sigma}
\end{align*}
\begin{align*}
    \eta, \epsilon  &::=
    \sigma  \sepr \epsilon \concat \eta
    \qquad \text{list}
    &
    ( \pi , \eta)
    \qquad \text{tuple}
\end{align*}

The list type $\epsilon\concat \eta$ types the concatenation of unrestricted bags.
It can be recursively unfolded into a finite composition of strict types $\sigma_1\concat \ldots\concat \sigma_n$, for some $n\geq 1$, with length $n$ and $\sigma_i$ its $i$-th strict type ($1\leq i\leq n$).
We  write $\unvar{x}:\eta $ to denote for $x[1]:\eta_1 , \ldots , x[k]:\eta_k$ where $\eta$ has length $k$.
The tuple type $(\pi,\eta)$ types concatenation of a linear bag of type $\pi$ with an unrestricted bag of type $\eta$. Finally strict types are amended to allow for unrestricted functional types which go from tuple types to strict types \arrt{ (\pi , \eta) }{\sigma} rather then multiset types to strict types.

\begin{definition}[$\eta \relunbag \epsilon$]\label{not:ltypes}
    Let $\epsilon$ and $\eta$ be two list types, with the length of $\epsilon$ greater or equal to that of $\eta$.
    We say that $\epsilon$ \emph{embraces} $\eta$, denoted $\eta \relunbag \epsilon$, whenever there exist $  \epsilon'$ and $ \epsilon''$   such that: i) $ \epsilon = \epsilon' \concat \epsilon''$; ii)  the size of $\epsilon' $ is that of $\eta$; iii)  for all $i$, $\epsilon'_i = \eta_i $.

\end{definition}

We separate contexts into two parts: linear ($\Gamma, \Delta,\ldots$) and unrestricted ($\Theta,\Upsilon,\ldots$):
\begin{align*}
    \Gamma,\Delta &::= \dash \sep \Gamma , x:\pi  \sepr \Gamma, x:\sigma
    &
    \Theta,\Upsilon &::= \dash \sepr \Theta, x^!:\eta
\end{align*}
Both linear and unrestricted occurrences of variables may occur at most once in a context.

Judgments have the form $\Theta;\Gamma\wfdash M:\tau$.
We write $\wfdash M:\tau$ to denote $\dash;\dash\wfdash M:\tau$.

\begin{definition}[Well-formedness in $\lamcoldetsh$]
    A $\lamcoldetsh$-term $M$ is \emph{well-formed} if there exists a context $\Theta$ and $\Gamma$ and a  type  $\tau$ such that the rules in \Cref{fig:wfsh_rulesunres} entail $\Theta ; \Gamma \wfdash  M : \tau $.
\end{definition}

\begin{figure}[ht]
    \begin{mdframed} \mysmall
        \begin{mathpar}
            \inferrule[$\redlab{FS{:}var^{\ell}}$]{ }{
                \Theta ;  {x}: \sigma \wfdash  {x} : \sigma
            }
            \and
            \inferrule[$\redlab{FS{:}var^!}$]{
                \Theta , x^!: \eta;  {x}: \eta_i , \Delta \wfdash  {x} : \sigma
            }{
                \Theta ,  x^!: \eta; \Delta \wfdash {x}[i] : \sigma
            }
            \and
            \inferrule[$\redlab{FS{:}\oneb^{\ell}}$]{ }{
                \Theta ; \dash \wfdash \oneb : \omega
            }
            \and
            \inferrule[$\redlab{FS{:}bag^{\ell}}$]{
                \Theta ; \Gamma \wfdash M : \sigma
                \\
                \Theta ; \Delta \wfdash C : \sigma^k
            }{
                \Theta ; \Gamma , \Delta \wfdash \bag{M}\cdot C:\sigma^{k+1}
            }
            \and
            \inferrule[$\redlab{FS{:}\oneb^!}$]{ }{
                \Theta ;  \dash  \wfdash \unvar{\oneb} : \sigma
            }
            \and
            \inferrule[$\redlab{FS{:} bag^{!}}$]{
                \Theta ; \dash \wfdash U : \epsilon
                \\
                \Theta ; \dash \wfdash V : \eta
            }{
                \Theta ; \dash  \wfdash U \concat V :\epsilon \concat \eta
            }
            \and
            \inferrule[$\redlab{FS{:}bag}$]{
                \Theta ; \Gamma\wfdash C : \sigma^k
                \\
                \Theta ;\dash \wfdash  U : \eta
            }{
                \Theta ; \Gamma \wfdash C \bagsep U : (\sigma^{k} , \eta  )
            }
            \and
            \inferrule[$\redlab{FS{:}fail}$]{
                \dom{\core{\Gamma}} = \widetilde{x}
            }{
                \Theta ; \core{\Gamma} \wfdash  \fail^{\widetilde{x}} : \tau
            }
            \and
            \inferrule[$\redlab{FS{:}weak}$]{
                \Theta ; \Gamma  \wfdash M : \tau
            }{
                \Theta ; \Gamma ,  {x}: \omega \wfdash M\sharing{}{x}: \tau
            }
            \and
            \inferrule[$\redlab{FS{:}shar}$]{
                \Theta ;  \Gamma ,  {x}_1: \sigma, \cdots,  {x}_k: \sigma \wfdash M : \tau
                \\
                {x}\notin \dom{\Gamma}
                \\
                k \not = 0
            }{
                \Theta ;  \Gamma ,  {x}: \sigma^{k} \wfdash M\sharing{{x}_1 , \ldots ,  {x}_k}{x}  : \tau
            }
            \and
            \inferrule[$\redlab{FS{:}abs\dash sh}$]{
                \Theta , x^!:\eta ; \Gamma ,  {x}: \sigma^k \wfdash M\sharing{\widetilde{x}}{x} : \tau
                \\
                {x} \notin \dom{\Gamma}
            }{
                \Theta ; \Gamma \wfdash \lambda x . (M[ {\widetilde{x}} \leftarrow  {x}])  : (\sigma^k, \eta )  \rightarrow \tau
            }
            \and
            \inferrule[$\redlab{FS{:}app}$]{
                \Theta ;\Gamma \wfdash M : (\sigma^{j} , \eta ) \rightarrow \tau
                \\
                \Theta ;\Delta \wfdash B : (\sigma^{k} , \epsilon )
                \\
                \eta \relunbag \epsilon
            }{
                \Theta ; \Gamma , \Delta \wfdash (M\ B) : \tau
            }
            \and
            \inferrule[$\redlab{FS{:}Esub}$]{
                \Theta , x^! : \eta ; \Gamma ,  {x}: \sigma^{j} \wfdash M[ {\widetilde{x}} \leftarrow  {x}] : \tau
                \\
                \Theta ; \Delta \wfdash B : (\sigma^{k} , \epsilon )
                \\
                \eta \relunbag \epsilon
            }{
                \Theta ; \Gamma , \Delta \wfdash (M[ {\widetilde{x}} \leftarrow  {x}])\esubst{ B }{ x }  : \tau
            }
            \and
            \inferrule[$\redlab{FS{:}Esub^{\ell}}$]{
                \Theta ; \Gamma  ,  x_1:\sigma, \cdots , x_k:\sigma \wfdash M : \tau
                \\
                \Theta ; \Delta \wfdash C : \sigma^k
            }{
                \Theta ; \Gamma , \Delta \wfdash M \linexsub{C /  x_1, \cdots , x_k} : \tau
            }
            \and
            \inferrule[$\redlab{FS{:}Esub^!}$]{
                \Theta , x^! {:} \eta; \Gamma  \wfdash M : \tau
                \\
                \Theta ; \dash \wfdash U : \epsilon
                \\
                \eta \relunbag \epsilon
            }{
                \Theta ; \Gamma \wfdash M \unexsub{U / \unvar{x}}  : \tau
            }
        \end{mathpar}
    \end{mdframed}
    \caption{Well-Formedness Rules for $\lamcoldetsh$.}
    \label{fig:wfsh_rulesunres}
\end{figure}

\paragraph{A congruence.}
Some terms, though syntactically different, display the same behavior.
For example, assuming $x \notin \lfv{M}$, this holds for $M \unexsub{U/x^!}$ and $M$: the former describes a substitution that ``does nothing'' and would result in $M$ itself.
This notion is formalized through a \emph{congruence} ($\equivlam$) closed under syntax, given in \Cref{fig:rsPrecongruencefailure}.

\begin{figure}[t]
\begin{mdframed}
\small
\begin{mathpar}
    \inferrule{
        \widetilde{x} \disj \lfv{B}
    }{
        (M\ B) \linexsub{C/\widetilde{x}}  \equivlam (M \linexsub{C/\widetilde{x}})\ B
    }
    \and
    \inferrule{
        x \not \in \lfv{M}
    }{
        M \unexsub{U/x^!} \equivlam M
    }
    \and
    \inferrule{
        x \not \in \lfv{B}
    }{
        (M\ B)  \unexsub{U/x^!}  \equivlam (M \unexsub{U/x^!})\ B
    }
    \and
    \inferrule{
        \widetilde{x} \disj \lfv{A}
    }{
        \big((M\ A)\sharing{\widetilde{x}}{x}\big) \esubst{B}{x} \equivlam \big((M\sharing{\widetilde{x}}{x}) \esubst{B}{x}\big)\ A
    }
    \and
    \inferrule{
        \widetilde{x} \disj \lfv{A}
        \\
        \widetilde{y} \disj \lfv{B}
    }{
        \Big(\big((M\sharing{\widetilde{y}}{y}) \esubst{A}{y}\big)\sharing{\widetilde{x}}{x}\Big) \esubst{B}{x} \equivlam \Big(\big(M\sharing{\widetilde{y}}{y}) \esubst{B}{y}\big)\sharing{\widetilde{x}}{x}\Big) \esubst{A}{x}
    }
    \and
    \inferrule{
        \widetilde{x} \disj \lfv{B}
        \\
        \widetilde{y} \disj \lfv{C}
    }{
        \big((M\sharing{\widetilde{y}}{y}) \esubst{B}{y}\big) \linexsub{C/\widetilde{x}}  \equivlam \big((M \linexsub{C/\widetilde{x}})\sharing{\widetilde{y}}{y}\big) \esubst{B}{y}
    }
    \and
    \inferrule{
        x \not \in \lfv{B}
        \\
        \widetilde{y} \disj \lfv{U}
    }{
        \big((M\sharing{\widetilde{y}}{y}) \esubst{B}{y}\big)  \unexsub{U/x^!}  \equivlam \big((M \unexsub{U/x^!})\sharing{\widetilde{y}}{y}\big) \esubst{B}{y}
    }
    \and
    \inferrule{
        \widetilde{x} \disj \lfv{C_2}
        \\
        \widetilde{y} \disj \lfv{C_1}
    }{
        (M \linexsub{C_2/\widetilde{y}}) \linexsub{C_1/\widetilde{x}} \equivlam (M \linexsub{C_1/ \widetilde{x} }) \linexsub{C_2/\widetilde{y}}
    }
    \and
    \inferrule{
        \widetilde{x} \disj \lfv{U}
        \\
        y \not \in \lfv{C}
    }{
        (M\linexsub{C/\widetilde{x}}) \unexsub{U/y^!} \equivlam (M  \unexsub{U/y^!})   \linexsub{C/\widetilde{x}}
    }
    \and
    \inferrule{
        x \not \in \lfv{U}
        \\
        y \not \in \lfv{V}
    }{
        (M\unexsub{V/\widetilde{x}}) \unexsub{U/y^!} \equivlam (M  \unexsub{U/y^!})   \unexsub{V/{x}^!}
    }
\end{mathpar}
\end{mdframed}
\caption{Congruence in \lamcoldetsh.}\label{fig:rsPrecongruencefailure}
\end{figure}

\subsection{Translating \texorpdfstring{\lamcoldetsh}{Unrestricted Lambda} into (Full) \texorpdfstring{\clpi}{Pi}}
\label{ss:transUnres}

\Cref{fig:encodinglin} gives the translation of \lamcoldetshlin into \clpi.
Here we extend this translation to consider the extended calculus \lamcoldetsh.
The key differences are in the translation of  unrestricted variable occurrences, intermediate substitution, abstraction, and the new structure of bags.
\Cref{fig:encoding} gives the translation that maps terms in \lamcoldetsh into processes in full \clpi, denoted $\piencod{\cdot}_u$.

The translation of an unrestricted variable $x[j]$ first connects to a server along channel $x$ via a request $ \puname{\unvar{x}}{{x_i}}$ followed by a selection on $ \psel{x_i}{j}$.

Process $\piencodf{\lambda x. (M\sharing{\widetilde{x}}{x})}_u$ first confirms its behavior  along $u$, followed by the reception of a channel $x$.
The channel $x$ provides a linear channel $\linvar{x}$ and an unrestricted channel $\unvar{x}$ for dedicated substitutions of the linear and unrestricted bag components.
This separation is also present in the translation of $ \piencodf{ M\esubst{B}{x}}_u $, for the same reason.

Process $\piencodf{M\, (C \bagsep U)_u}$ consists of synchronizations between the translation of $\piencod{M}_v$ and
$\piencodf{C\bagsep U}_x$:  the translation
of  $C \bagsep U$ evolves when $M$ is an abstraction, say
${\lambda x . (M'\sharing{\widetilde{x}}{x})}$.
The channel $ \linvar{x}$ provides the linear behavior of the bag $C$ while $\unvar{x}$ provides the behavior of $U$; this is done by guarding the translation of $U$ with a server connection, such that every time a channel synchronizes with it a fresh copy of $U$ is spawned.

Process $\piencodf{ M \unexsub{U / \unvar{x}}}_u $ consists of  the composition of the translation of $M$ and a server guarding the translation of $U$: in order for $\piencodf{M}_u$ to gain access to $\piencodf{U}_{x_i}$ it must first synchronize with the server channel $\unvar{x}$ to spawn a fresh copy of the translation of $U$.

\begin{figure}[t]
    \begin{mdframed} \mysmall
        \begin{align*}
            \piencodf{ {x}}_u & = \psome{x}; \pfwd{x}{u} \hspace{1cm}
            \piencodf{{x}[j]}_u  =   \puname{\unvar{x}}{{x_i}}; \psel{x_i}{j}; \pfwd{x_i}{u}
            \\[1mm]
            \piencodf{\lambda x.M}_u & = \psome{u};\gname{u}{x};  \psome{x};\gname{x}{\linvar{x}}; \gname{x}{\unvar{x}};  \gclose{x} ; \piencodf{M}_u
            \\[1mm]
            \piencodf{ M \esubst{ C \bagsep U }{ x} }_u & =  \res{x}( \psome{x}; \gname{x}{\linvar{x}}; \gname{x}{\unvar{x}};  \gclose{x} ;\piencodf{ M}_u \| \piencodf{ C \bagsep U}_x )
            \\[1mm]
            \piencodf{M (C \bagsep U)}_u & = \res{v} (\piencodf{M}_v \| \gsome{v}{u , \llfv{C}};\pname{v}{x}; ( \piencodf{C \bagsep U}_x   \| \pfwd{v}{u}  ) )
            \\[1mm]
            \piencodf{ C \bagsep U }_x & = \gsome{x}{\llfv{C}};  \pname{x}{\linvar{x}}; \big( \piencodf{ C }_{\linvar{x}} \|  \pname{x}{\unvar{x}}; ( \guname{\unvar{x}}{x_i}; \piencodf{ U }_{x_i}  \| \pclose{x} ) \big)
            \\[1mm]
            \piencodf{\bag{M_j} \cdot~ C}_{\linvar{x}}  &=
            \begin{array}{@{}l@{}}
                \gsome{\linvar{x}}{\llfv{C} }; \gname{x}{y_i}; \gsome{\linvar{x}}{y_i, \llfv{C}}; \psome{\linvar{x}}; \\
                \quad \pname{\linvar{x}}{z_i}; ( \gsome{z_i}{\llfv{M_j}};  \piencodf{M_j}_{z_i} \| (\piencodf{(C \setminus M_j)}_{\linvar{x}} \| \pnone{y_i} ))
            \end{array}
            \\[1mm]
            \piencodf{{\oneb}}_{\linvar{x}} & = \gsome{\linvar{x}}{\emptyset};\gname{x}{y_n};  ( \psome{ y_n}; \pclose{y_n}  \| \gsome{\linvar{x}}{\emptyset}; \pnone{\linvar{x}} )
            \\[1mm]
            \span
            \piencodf{\unvar{\oneb}}_{x}  = \pnone{x}
            \qquad\qquad
            \piencodf{\unvar{\bag{N}}}_{x}  =  \piencodf{N}_{x}
            \qquad\qquad
            \piencodf{ U }_{x}  = \gsel{x}\{i:\piencodf{ U_i }_{x} \}_{U_i \in U}
            \\[1mm]
            \piencodfscale{
            M \ltalltriangle \bag{M_1} \cdot \bag{M_2}
                    {} /  x_1, x_2 \rtalltriangle
            }_u    &= \begin{array}{@{}l@{}}
                \res{z_1}( \gsome{z_1}{\llfv{M_{1}}};\piencodf{ M_{1} }_{ {z_1}} \| \res{z_2} ( \gsome{z_2}{\llfv{M_{2}}};\piencodf{ M_{2} }_{ {z_2}} \\
                \qquad \quad {} \| {} {} \bignd_{x_{i_1} \in \{ x_1 , x_2  \}} \bignd_{x_{i_2} \in \{ x_1, x_2 \setminus x_{i_1}  \}} \piencodf{ M }_u \{ z_1 / x_{i_1} \} \{ z_2 / x_{i_2} \} ) \ldots )
            \end{array}
            \\[1mm]
            \piencodf{ M \unexsub{U / \lunvar{x}}  }_u   & =  \res{\unvar{x}} ( \piencodf{ M }_u \|   \guname{\unvar{x}}{x_i}; \piencodf{ U }_{x_i} )
            \\[1mm]
            \piencodf{M[  \leftarrow  {x}]}_u & = \psome{\linvar{x}}; \pname{\linvar{x}}{y_i}; ( \gsome{y_i}{ u , \llfv{M} }; \gclose{ y_{i} } ;\piencodf{M}_u \| \pnone{ \linvar{x} } )
            \\[1mm]
            \piencodf{M[ \widetilde{x} \leftarrow  {x}]}_u & = \begin{array}{@{}l@{}}
                \psome{\linvar{x}}; \pname{\linvar{x}}{y_i}; \big( \gsome{y_i}{ \emptyset }; \gclose{ y_{i} } ; \0 \\
                \quad {} \| \psome{\linvar{x}}; \gsome{\linvar{x}}{u, \llfv{M} \setminus  \widetilde{x} }; \bignd_{x_i \in \widetilde{x}} \gname{\linvar{x}}{{x}_i};\piencodf{M[ (\widetilde{x} \setminus x_i ) \leftarrow  {x}]}_u \big)
            \end{array}
            \\[1mm]
            \piencodf{\fail^{x_1, \ldots, x_k}}_u & = \pnone{ u}  \| \pnone{ x_1} \| \ldots \| \pnone{ x_k}
        \end{align*}
    \end{mdframed}
    \caption{Translation of \texorpdfstring{\lamcoldetsh}{lambda} into \texorpdfstring{\clpi}{spi+}.}\label{fig:encoding}
\end{figure}

To complete this section, \Cref{fig:enc_typesunres} gives the translation of intersection types for \lamcoldetsh to session types for full \clpi.

\begin{definition}\label{def:enc_sestypfailunres}
    The translation  $\piencodf{\cdot}_{\_}$  in \Cref{fig:enc_typesunres} extends to a linear context
    \[
        \Gamma =  {{x}_1: \sigma_1} , \ldots, {{x}_m : \sigma_m} , {{v}_1: \pi_1} , \ldots ,  {v}_n: \pi_n
    \]
    and an unrestricted context $\Theta =\unvar{x}[1] : \eta_1 , \ldots , \unvar{x}[k] : \eta_k$  as follows:
    \begin{align*}
        \piencodf{\Gamma} &= {x}_1 : \with \overline{\piencodf{\sigma_1}} , \ldots ,   {x}_m : \with \overline{\piencodf{\sigma_m}} ,
        {v}_1:  \overline{\piencodf{\pi_1}_{(\sigma, i_1)}}, \ldots ,  {v}_n: \overline{\piencodf{\pi_n}_{(\sigma, i_n)}}\\
        \piencodf{\Theta}&=\unvar{x}[1] : \dual{\piencodf{\eta_1}} , \ldots , \unvar{x}[k] : \dual{\piencodf{\eta_k}}
    \end{align*}
\end{definition}

\begin{figure}[t]
    \begin{mdframed} \mysmall
        \begin{align*}
            \piencodf{\unit} &= \with \onef
            &
            \piencodf{ \eta } &= {!} \with_{\eta_i \in \eta} \{ i : \piencodf{\eta_i} \}
            \\
            \piencodf{(\sigma^{k} , \eta )   \rightarrow \tau} &= \with( \dual{\piencodf{ (\sigma^{k} , \eta  )  }_{(\sigma, i)}} \ampy \piencodf{\tau})
            &
            \piencodf{ (\sigma^{k} , \eta  )  }_{(\sigma, i)} &= \oplus( (\piencodf{\sigma^{k} }_{(\sigma, i)}) \otimes (( \piencodf{\eta}) \otimes (\onef))  )
            \\[1mm]
            \piencodf{ \sigma \wedge \pi }_{(\sigma, i)} & = \oplus(( \with \onef) \ampy ( \oplus  \with (( \oplus \piencodf{\sigma} ) \otimes (\piencodf{\pi}_{(\sigma, i)}))))
            \span \span
            \\
            \piencodf{\omega}_{(\sigma, i)} & = \begin{cases}
                \oplus ((\with \1) \parr (\oplus \with \1))
                &  \text{if $i = 0$}
                \\
                \oplus ((\with \1) \parr (\oplus\, \with ((\oplus \piencodf{\sigma}) \tensor (\piencodf{\omega}_{(\sigma, i-1)}))))
                & \text{if $i > 0$}
            \end{cases}
            \span \span
        \end{align*}
    \end{mdframed}
    \caption{Translation of intersection types into session types  (cf.\ \defref{def:enc_sestypfailunres}).}
    \label{fig:enc_typesunres}
\end{figure}

%% file: appendix/pi-proofs.tex
\section{Proofs of Type Preservation and Deadlock-freedom for (Full) \texorpdfstring{\clpi}{Pi}}
\label{s:piProofs}

Here we prove \Cref{t:srPi,t:dfPi} (type preservation and deadlock-freedom for the lazy semantics, respectively), as well as the analogue results for the eager semantics.
In fact, deadlock-freedom for the lazy semantics follows from deadlock-freedom for the eager semantics, so we present the proofs for the eager semantics first.

\subsection{Eager Semantics}
\label{ss:proofsEager}

\subsubsection{Subject Congruence}

\begin{theorem}\label{t:subcong}
    If $P \vdash \Gamma$ and $P \equiv Q$, then $Q \vdash \Gamma$.
\end{theorem}

\begin{proof}
    By induction on the derivation of the structural congruence.
    We first detail the base cases:
    \begin{itemize}
        \item
            $P \equiv_\alpha P' \implies P \equiv P'$.
            Since alpha-renaming only affects bound names, it does not affect the names in $\Gamma$, so clearly $P' \vdash \Gamma$.

        \item
            $\pfwd{x}{y} \equiv \pfwd{y}{x}$.
            \begin{mathpar}
                \inferrule{ }{
                    \pfwd{x}{y} \vdash x{:}A, y{:}\ol{A}
                }
                \equiv
                \inferrule{ }{
                    \pfwd{y}{x} \vdash x{:}A, y{:}\ol{A}
                }
            \end{mathpar}

        \item
            $P \| Q \equiv Q \| P$.
            \begin{mathpar}
                \inferrule{
                    P \vdash \Gamma
                    \\
                    Q \vdash \Delta
                }{
                    P \| Q \vdash \Gamma, \Delta
                }
                \equiv
                \inferrule{
                    Q \vdash \Delta
                    \\
                    P \vdash \Gamma
                }{
                    Q \| P \vdash \Gamma, \Delta
                }
            \end{mathpar}

        \item
            $(P \| Q) \| R \equiv P \| (Q \| R)$.
            \begin{mathpar}
                \inferrule{
                    \inferrule*{
                        P \vdash \Gamma
                        \\
                        Q \vdash \Delta
                    }{
                        P \| Q \vdash \Gamma, \Delta
                    }
                    \\
                    R \vdash \Lambda
                }{
                    (P \| Q) \| R \vdash \Gamma, \Delta, \Lambda
                }
                \equiv
                \inferrule{
                    P \vdash \Gamma
                    \\
                    \inferrule*{
                        Q \vdash \Delta
                        \\
                        R \vdash \Lambda
                    }{
                        Q \| R \vdash \Delta, \Lambda
                    }
                }{
                    P \| (Q \| R) \vdash \Gamma, \Delta, \Lambda
                }
            \end{mathpar}

        \item
            $P \| \0 \equiv P$
            \begin{mathpar}
                \inferrule{
                    P \vdash \Gamma
                    \\
                    \inferrule*{ }{
                        \0 \vdash \emptyset
                    }
                }{
                    P \| \0 \vdash \Gamma
                }
                \equiv
                P \vdash \Gamma
            \end{mathpar}

        \item
            $\res{x}(P \| Q) \equiv \res{x}(Q \| P)$.
            \begin{mathpar}
                \inferrule{
                    P \vdash \Gamma, x{:}A
                    \\
                    Q \vdash \Delta, x{:}\ol{A}
                }{
                    \res{x}(P \| Q) \vdash \Gamma, \Delta
                }
                \equiv
                \inferrule{
                    Q \vdash \Delta, x{:}\ol{A}
                    \\
                    P \vdash \Gamma, x{:}A
                }{
                    \res{x}(Q \| P) \vdash \Gamma, \Delta
                }
            \end{mathpar}

        \item
            $x \notin \fn{Q} \implies \res{x}(\res{y}(P \| Q) \| R) \equiv \res{y}(\res{x}(P \| R) \| Q)$.
            \begin{mathpar}
                \inferrule{
                    \inferrule*{
                        P \vdash \Gamma, y{:}A, x{:}B
                        \\
                        Q \vdash \Delta, y{:}\ol{A}
                    }{
                        \res{y}(P \| Q) \vdash \Gamma, \Delta, x{:}B
                    }
                    \\
                    R \vdash \Lambda, x{:}\ol{B}
                }{
                    \res{x}(\res{y}(P \| Q) \| R) \vdash \Gamma, \Delta, \Lambda
                }
                \equiv
                \inferrule{
                    \inferrule*{
                        P \vdash \Gamma, y{:}A, x{:}B
                        \\
                        R \vdash \Lambda, x{:}\ol{B}
                    }{
                        \res{x}(P \| R) \vdash \Gamma, \Lambda, y{:}A
                    }
                    \\
                    Q \vdash \Delta, y{:}\ol{A}
                }{
                    \res{y}(\res{x}(P \| R) \| Q) \vdash \Gamma, \Delta, \Lambda
                }
            \end{mathpar}

        \item
            $x \notin \fn{Q} \implies \res{x}((P \| Q) \| R) \equiv \res{x}(P \| R) \| Q$.
            \begin{mathpar}
                \inferrule{
                    \inferrule*{
                        P \vdash \Gamma, x{:}A
                        \\
                        Q \vdash \Delta
                    }{
                        P \| Q \vdash \Gamma, \Delta, x{:}A
                    }
                    \\
                    R \vdash \Lambda, x{:}\ol{A}
                }{
                    \res{x}((P \| Q) \| R) \vdash \Gamma, \Delta, \Lambda
                }
                \equiv
                \inferrule{
                    \inferrule*{
                        P \vdash \Gamma, x{:}A
                        \\
                        R \vdash \Lambda, x{:}\ol{A}
                    }{
                        \res{x}(P \| R) \vdash \Gamma, \Lambda
                    }
                    \\
                    Q \vdash \Delta
                }{
                    \res{x}(P \| R) \| Q \vdash \Gamma, \Delta, \Lambda
                }
            \end{mathpar}

        \item
            $x \notin \fn{Q} \implies \res{x}(\guname{x}{y};P \| Q) \equiv Q$.
            \begin{mathpar}
                \inferrule{
                    \inferrule*{
                        P \vdash {?}\Gamma, y{:}A
                    }{
                        \guname{x}{y};P \vdash {?}\Gamma, x{:}{!}A
                    }
                    \\
                    \inferrule*{
                        Q \vdash \Delta
                    }{
                        Q \vdash \Delta, x{:}{?}\ol{A}
                    }
                }{
                    \res{x}(\guname{x}{y};P \| Q) \vdash {?}\Gamma, \Delta
                }
                \equiv
                \inferruleDbl[vcenter]{
                    Q \vdash \Delta
                }{
                    Q \vdash {?}\Gamma, \Delta
                }
            \end{mathpar}

        \item
            $P \nd Q \equiv Q \nd P$.
            \begin{mathpar}
                \inferrule{
                    P \vdash \Gamma
                    \\
                    Q \vdash \Gamma
                }{
                    P \nd Q \vdash \Gamma
                }
                \equiv
                \inferrule{
                    Q \vdash \Gamma
                    \\
                    P \vdash \Gamma
                }{
                    Q \nd P \vdash \Gamma
                }
            \end{mathpar}

        \item
            $(P \nd Q) \nd R \equiv P \nd (Q \nd R)$
            \begin{mathpar}
                \inferrule{
                    \inferrule*{
                        P \vdash \Gamma
                        \\
                        Q \vdash \Gamma
                    }{
                        P \nd Q \vdash \Gamma
                    }
                    \\
                    R \vdash \Gamma
                }{
                    (P \nd Q) \nd R \vdash \Gamma
                }
                \equiv
                \inferrule{
                    P \vdash \Gamma
                    \\
                    \inferrule*{
                        Q \vdash \Gamma
                        \\
                        R \vdash \Gamma
                    }{
                        Q \nd R \vdash \Gamma
                    }
                }{
                    P \nd (Q \nd R) \vdash \Gamma
                }
            \end{mathpar}

        \item
            $P \nd P \equiv P$.
            \begin{mathpar}
                \inferrule{
                    P \vdash \Gamma
                    \\
                    P \vdash \Gamma
                }{
                    P \nd P \vdash \Gamma
                }
                \equiv
                P \vdash \Gamma
            \end{mathpar}
    \end{itemize}
    The inductive cases follow from the IH straightforwardly.
    Note that the rules for parallel composition do not apply directly behind the output prefix and restriction.
\end{proof}

\subsubsection{Subject Reduction}

\begin{lemma}\label{l:ctxType}
    Suppose $P \vdash \Gamma, x{:}A$.
    \begin{enumerate}
        \item\label{i:ctxTypePclose}
            If $P = \pctx{N}[\pclose{x}]$, then $A = \1$.

        \item\label{i:ctxTypeGclose}
            If $P = \pctx{N}[\gclose{x};P']$, then $A = \bot$.

        \item\label{i:ctxTypePname}
            If $P = \pctx{N}[\pname{x}{y};(P' \| P'')]$, then $A = B \tensor C$.

        \item\label{i:ctxTypeGname}
            If $P = \pctx{N}[\gname{x}{y};P']$, then $A = B \parr C$.

        \item\label{i:ctxTypePsel}
            If $P = \pctx{N}[\psel{x}{j};P']$, then $A = {\oplus}\{i:B_i\}_{i \in I}$ where $j \in I$.

        \item\label{i:ctxTypeGsel}
            If $P = \pctx{N}[\gsel{x}\{i:P'_i\}_{i \in I}]$, then $A = {\with}\{i:B_i\}_{i \in I}$.

        \item\label{i:ctxTypePsome}
            If $P = \pctx{N}[\psome{x};P']$, then $A = {\with}B$.

        \item\label{i:ctxTypePnone}
            If $P = \pctx{N}[\pnone{x}]$, then $A = {\with}B$.

        \item\label{i:ctxTypeGsome}
            If $P = \pctx{N}[\gsome{x}{w_1,\ldots,w_n}]$, then $A = {\oplus}B$.

        \item\label{i:ctxTypePuname}
            If $P = \pctx{N}[\puname{x}{y};P']$, then $A = {?}B$.

        \item\label{i:ctxTypeGuname}
            If $P = \pctx{N}[\guname{x}{y};P']$, then $A = {!}B$.
    \end{enumerate}
\end{lemma}

\begin{proof}
    Each item follows by induction on the structure of the ND-context.
    The base case follows by inversion of typing, and the inductive cases follow from the IH straightforwardly.
\end{proof}

\begin{lemma}\label{l:ctxRedOne}
    For each of the following items, assume $\Gamma \disj \Delta$.
    \begin{enumerate}
        \item\label{i:ctxRedOneFwd}
            If $\pctx[\big]{N}[\pfwd{x}{y}] \vdash \Gamma, x{:}A$ and $Q \vdash \Delta, x{:}\ol{A}$, then $\pctx*{N}[Q \{y/x\}] \vdash \Gamma, \Delta$.

        \item\label{i:ctxRedOnePclose}
            If $\pctx{N}[\pclose{x}] \vdash \Gamma, x{:}\1$, then $\pctx*{N}[\0] \vdash \Gamma$.

        \item\label{i:ctxRedOneGclose}
            If $\pctx{N}[\gclose{x};Q] \vdash \Gamma, x{:}\bot$, then $\pctx*{N}[Q] \vdash \Gamma$.

        \item\label{i:ctxRedOneName}
            If $\bn{\pctx{N}} \disj \fn{\pctx{N}'}$ and $\pctx{N}[\pname{x}{y};(P \| Q)] \vdash \Gamma, x{:}A \tensor B$ and $\pctx{N'}[\gname{x}{z};R] \vdash \Delta, x{:}\ol{A} \parr \ol{B}$, then $\pctx*{N}[\res{x}(Q \| \res{y}(P \| \pctx*{N'}[R\{y/z\}]))] \vdash \Gamma, \Delta$.

        \item\label{i:ctxRedOnePsel}
            If $\pctx{N}[\psel{x}{j}; P] \vdash \Gamma, x{:}{\oplus}\{i: A_i\}_{i \in I}$ and $j \in I$, then $\pctx*{N}[P] \vdash \Gamma, x{:}A_j$.

        \item\label{i:ctxRedOneGsel}
            If $\pctx{N}[\gsel{x}\{i: P_i\}_{i \in I}] \vdash \Gamma, x{:}\with\{i: A_i\}_{i \in I}$, then $\pctx*{N}[P_i] \vdash \Gamma, x{:}A_i$ for every $i \in I$.

        \item\label{i:ctxRedOnePsome}
            If $\pctx{N}[\psome{x}; P] \vdash \Gamma, x{:}{\with}A$, then $\pctx*{N}[P] \vdash \Gamma, x{:}A$.

        \item\label{i:ctxRedOnePnone}
            If $\pctx{N}[\pnone{x}] \vdash \Gamma, x{:}{\with}A$, then $\pctx*{N}[\0] \vdash \Gamma$.

        \item\label{i:ctxRedOneGsome}
            If $\pctx{N}[\gsome{x}{w_1,\ldots,w_n}; P] \vdash \Gamma, x{:}{\oplus} A$, then $\pctx*{N}[P] \vdash \Gamma, x{:}A$ and $\pctx*{N}[\pnone{w_1} \| \ldots \| \pnone{w_n}] \vdash \Gamma$.

        \item\label{i:ctxRedOneUname}
            If $\bn{\pctx{N'}} \disj \fn{\pctx{N}}$ and $\pctx{N}[\puname{x}{y}; P] \vdash \Gamma, x{:}{?}A$ and $\pctx{N}'[\guname{x}{y}; Q] \vdash \Delta, x{:}{!}\ol{A}$, then $\pctx*{N'}\big[\res{x}(\res{y}(\pctx*{N}[P] \| Q\{y/z\}) \| \guname{x}{z};Q)\big]$.
    \end{enumerate}
\end{lemma}

\begin{proof}
    For each item, we apply induction on the structure of the ND-contexts and detail the base cases.
    For simplicity, we assume no names in $\Gamma$ and $\Delta$ were derived with \ttype{weaken}.
    \begin{enumerate}
        \item
            \begin{mathpar}
                \inferrule*{ }{
                    \pfwd{x}{y} \vdash x{:}A, y{:}\ol{A}
                }
                \and
                \inferrule{}{
                    Q \vdash \Delta, x{:}\ol{A}
                }
                \and
                \implies
                \and
                \inferrule{}{
                    Q\{y/x\} \vdash \Delta, y{:}\ol{A}
                }
            \end{mathpar}

        \item
            \begin{mathpar}
                \inferrule*{ }{
                    \pclose{x} \vdash x{:}\1
                }
                \and\implies\and
                \inferrule*{ }{
                    \0 \vdash \emptyset
                }
            \end{mathpar}

        \item
            \begin{mathpar}
                \inferrule{
                    Q \vdash \Gamma
                }{
                    \gclose{x};Q \vdash \Gamma, x{:}\bot
                }
                \and\implies\and
                \inferrule{}{
                    Q \vdash \Gamma
                }
            \end{mathpar}

        \item
            \begin{mathpar}
                \inferrule{
                    P \vdash \Gamma, y{:}A
                    \\
                    Q \vdash \Gamma', x{:}B
                }{
                    \pname{x}{y};(P \| Q) \vdash \Gamma, \Gamma', x{:}A \tensor B
                }
                \and
                \inferrule{
                    R \vdash \Delta, z{:}\ol{A}, x{:}\ol{B}
                }{
                    \gname{x}{z};R \vdash \Delta, x{:}\ol{A} \parr \ol{B}
                }
                \and\implies\and
                \inferrule{
                    Q \vdash \Gamma', x{:}B
                    \\
                    \inferrule*{
                        P \vdash \Gamma, y{:}A
                        \\
                        R\{y/z\} \vdash \Delta, y{:}\ol{A}, x{:}\ol{B}
                    }{
                        \res{y}(P \| R\{y/z\}) \vdash \Gamma, \Delta, x{:}\ol{B}
                    }
                }{
                    \res{x}(Q \| \res{y}(P \| R\{y/z\})) \vdash \Gamma, \Gamma', \Delta
                }
            \end{mathpar}

        \item
            \begin{mathpar}
                \inferrule{
                    P \vdash \Gamma, x{:}A_j
                    \\
                    j \in I
                }{
                    \psel{x}{j};P \vdash \Gamma, x{:}{\oplus}\{i:A_i\}_{i \in I}
                }
                \and\implies\and
                \inferrule{}{
                    P \vdash \Gamma, x{:}A_j
                }
            \end{mathpar}

        \item
            \begin{mathpar}
                \inferrule{
                    \forall i \in I.~ P_i \vdash \Gamma, x{:}A_i
                }{
                    \gsel{x}\{i:P_i\}_{i \in I} \vdash \Gamma, x{:}{\with}\{i:A_i\}
                }
                \and\implies\and
                \forall i \in I.~ \inferrule*{}{
                    P_i \vdash \Gamma, x{:}A_i
                }
            \end{mathpar}

        \item
            \begin{mathpar}
                \inferrule{
                    P \vdash \Gamma, x{:}A
                }{
                    \psome{x};P \vdash \Gamma, x{:}{\with}A
                }
                \and\implies\and
                \inferrule{}{
                    P \vdash \Gamma, x{:}A
                }
            \end{mathpar}

        \item
            \begin{mathpar}
                \inferrule*{ }{
                    \pnone{x} \vdash x{:}{\with}A
                }
                \and\implies\and
                \inferrule*{ }{
                    \0 \vdash \emptyset
                }
            \end{mathpar}

        \item
            \begin{mathpar}
                \inferrule{
                    P \vdash w_1{:}{\with}B_1, \ldots, w_n{:}{\with}B_n, x{:}A
                }{
                    \gsome{x}{w_1,\ldots,w_n};P \vdash w_1{:}{\with}B_1, \ldots, w_n{:}{\with}B_n, x{:}{\oplus}A
                }
                \and\implies\and
                \inferrule{}{
                    P \vdash w_1{:}{\with}B_1, \ldots, w_n{:}{\with}B_n, x{:}A
                }
                \and
                \inferruleDbl{
                    \inferrule*[fraction={---}]{ }{
                        \pnone{w_1} \vdash w_1{:}{\with}B_1
                    }
                    \\
                    \ldots
                    \\
                    \inferrule*[fraction={---}]{ }{
                        \pnone{w_n} \vdash w_n{:}{\with}B_n
                    }
                }{
                    \pnone{w_1} \| \ldots \| \pnone{w_n} \vdash w_1{:}{\with}B_1, \ldots, w_n{:}{\with}B_n
                }
            \end{mathpar}

        \item
            This item depends on whether $x \in \fn{P}$.
            \begin{itemize}
                \item
                    $x \in \fn{P}$:
                    \begin{mathpar}
                        \and
                        \inferrule*{
                            \inferrule*{
                                P\{x'/x\} \vdash \Gamma, y{:}A, x'{:}{?}A
                            }{
                                \puname{x}{y};(P\{x'/x\}) \vdash \Gamma, x{:}{?}A, x'{:}{?}A
                            }
                        }{
                            \puname{x}{y};P \vdash \Gamma, x{:}{?}A
                        }
                        \and
                        \inferrule*{
                            Q \vdash {?}\Delta, z{:}\ol{A}
                        }{
                            \guname{x}{z};Q \vdash {?}\Delta, x{:}{!}\ol{A}
                        }
                        \and\implies\and
                        \inferruleDbl{
                            \inferrule*[fraction={---}]{
                                \inferrule*{
                                    P \vdash \Gamma, y{:}A, x{:}{?}A
                                    \\
                                    Q\{y/z\}\{w'/w\}_{w \in {?}\Delta} \vdash {?}\Delta', y{:}\ol{A}
                                }{
                                    \res{y}(P \| Q\{y/z\}\{w'/w\}_{w \in {?}\Delta}) \vdash \Gamma, {?}\Delta', x{:}{?}A
                                }
                                \\
                                \guname{x}{z};Q \vdash {?}\Delta, x{:}{!}\ol{A}
                            }{
                                \res{x}(\res{y}(P \| Q\{y/z\}\{w'/w\}_{w \in {?}\Delta}) \| \guname{x}{z};Q) \vdash \Gamma, {?}\Delta, {?}\Delta'
                            }
                        }{
                            \res{x}(\res{y}(P \| Q\{y/z\}) \| \guname{x}{z};Q) \vdash \Gamma, {?}\Delta
                        }
                    \end{mathpar}

                \item
                    $x \notin \fn{P}$:
                    \begin{mathpar}
                        \inferrule*{
                            P \vdash \Gamma, y{:}A
                        }{
                            \puname{x}{y};P \vdash \Gamma, x{:}{?}A
                        }
                        \and
                        \inferrule*{
                            Q \vdash {?}\Delta, z{:}\ol{A}
                        }{
                            \guname{x}{z};Q \vdash {?}\Delta, x{:}{!}\ol{A}
                        }
                        \and\implies\and
                        \inferruleDbl{
                            \inferrule*[fraction={---}]{
                                \inferrule*{
                                    \inferrule*{
                                        P \vdash \Gamma, y{:}A
                                        \\
                                        Q\{y/z\}\{w'/w\}_{w \in {?}\Delta} \vdash {?}\Delta', y{:}\ol{A}
                                    }{
                                        \res{y}(P \| Q\{y/z\}\{w'/w\}_{w \in {?}\Delta}) \vdash \Gamma, {?}\Delta'
                                    }
                                }{
                                    \res{y}(P \| Q\{y/z\}\{w'/w\}_{w \in {?}\Delta}) \vdash \Gamma, {?}\Delta', x{:}{?}A
                                }
                                \\
                                \guname{x}{z};Q \vdash {?}\Delta, x{:}{!}\ol{A}
                            }{
                                \res{x}(\res{y}(P \| Q\{y/z\}\{w'/w\}_{w \in {?}\Delta}) \| \guname{x}{z};Q) \vdash \Gamma, {?}\Delta, {?}\Delta'
                            }
                        }{
                            \res{x}(\res{y}(P \| Q\{y/z\}) \| \guname{x}{z};Q) \vdash \Gamma, {?}\Delta
                        }
                    \end{mathpar}
            \end{itemize}
    \end{enumerate}
    The inductive cases follow straightforwardly.
    Notice that the conditions on the bound and free names of the ND-contexts in \cref{i:ctxRedOneName,i:ctxRedOneUname} make sure that no names are captured when embedding one context in the other.
\end{proof}

\begin{theorem}[SR for the Eager Semantics]\label{t:srOne}
    If $P \vdash \Gamma$ and $P \redone Q$, then $Q \vdash \Gamma$.
\end{theorem}

\begin{proof}
    By induction on the derivation of the reduction.
    \begin{itemize}
        \item
            Rule $\rredone{\scc{Id}}$.
            \begin{mathpar}
                \inferrule{
                    \pctx[\big]{N}[\pfwd{x}{y}] \vdash \Gamma, x{:}A
                    \\
                    Q \vdash \Delta, x{:}\ol{A}
                }{
                    \res{x}(\pctx[\big]{N}[\pfwd{x}{y}] \| Q) \vdash \Gamma, \Delta
                }
                \and\implies\and
                \inferrule{}{
                    \pctx*{N}[Q\{y/x\}] \vdash \Gamma, \Delta
                    ~~\text{(\refitem{l}{ctxRedOne}{Fwd})}
                }
            \end{mathpar}

        \item
            Rule $\rredone{\1\bot}$.
            \begin{mathpar}
                \inferrule{
                    \pctx{N}[\pclose{x}] \vdash \Gamma, x{:}\1
                    ~~\text{(\refitem{l}{ctxType}{Pclose})}
                    \\
                    \pctx{N'}[\gclose{x};Q] \vdash \Delta, x{:}\bot
                    ~~\text{(\refitem{l}{ctxType}{Gclose})}
                }{
                    \nu{x}(\pctx{N}[\pclose{x}] \| \pctx{N'}[\gclose{x};Q]) \vdash \Gamma, \Delta
                }
                \and\implies\and
                \inferrule{
                    \pctx*{N}[\0] \vdash \Gamma
                    ~~\text{(\refitem{l}{ctxRedOne}{Pclose})}
                    \\
                    \pctx*{N'}[Q] \vdash \Delta
                    ~~\text{(\refitem{l}{ctxRedOne}{Gclose})}
                }{
                    \pctx*{N}[\0] \| \pctx*{N'}[Q] \vdash \Gamma, \Delta
                }
            \end{mathpar}

        \item
            Rule $\rredone{\tensor\parr}$.
            \begin{mathpar}
                \inferrule{
                    \pctx{N}[\pname{x}{y};(P \| Q)] \vdash \Gamma, x{:}A \tensor B
                    ~~\text{(\refitem{l}{ctxType}{Pname})}
                    \\
                    \pctx{N'}[\gname{x}{z};R] \vdash \Delta, x{:}\ol{A} \parr \ol{B}
                    ~~\text{(\refitem{l}{ctxType}{Gname})}
                }{
                    \res{x}(\pctx{N}[\pname{x}{y};(P \| Q)] \| \pctx{N'}[\gname{x}{z};R]) \vdash \Gamma, \Delta
                }
                \and\implies\and
                \inferrule{}{
                    \pctx*{N}[\res{x}(Q \| \nu{y}(P \| \pctx*{N'}[R\{y/z\}]))] \vdash \Gamma, \Delta
                    ~~\text{(\refitem{l}{ctxRedOne}{Name})}
                }
            \end{mathpar}

        \item
            Rule $\rredone{{\oplus}{\with}}$.
            \begin{mathpar}
                \inferrule{
                    \pctx{N}[\psel{x}{j};P] \vdash \Gamma, x{:}{\oplus}\{i:A_i\}_{i \in I}
                    ~~ j \in I
                    ~~\text{(\refitem{l}{ctxType}{Psel})}
                    \\
                    \pctx{N'}[\gsel{x}\{i:Q_i\}_{i \in I}] \vdash \Delta, x{:}{\with}\{i:\ol{A_i}\}_{i \in I}
                    ~~\text{(\refitem{l}{ctxType}{Gsel})}
                }{
                    \res{x}(\pctx{N}[\psel{x}{j};P] \| \pctx{N'}[\gsel{x}\{i:Q_i\}_{i \in I}]) \vdash \Gamma, \Delta
                }
                \and\implies\and
                \inferrule{
                    \pctx*{N}[P] \vdash \Gamma, x{:}A_j
                    ~~\text{(\refitem{l}{ctxRedOne}{Psel})}
                    \\
                    \pctx*{N'}[Q_j] \vdash \Delta, x{:}\ol{A_j}
                    ~~\text{(\refitem{l}{ctxRedOne}{Gsel})}
                }{
                    \res{x}(\pctx*{N}[P] \| \pctx*{N'}[Q_j]) \vdash \Gamma, \Delta
                }
            \end{mathpar}

        \item
            Rule $\rredone{\some}$.
            \begin{mathpar}
                \inferrule{
                    \pctx{N}[\psome{x};P] \vdash \Gamma, x{:}{\with}A
                    ~~\text{(\refitem{l}{ctxType}{Psome})}
                    \\
                    \pctx{N'}[\gclose{x}{(w_1,\ldots,w_n)};Q] \vdash \Delta, x{:}{\oplus}\ol{A}
                    ~~\text{(\refitem{l}{ctxType}{Gsome})}
                }{
                    \res{x}(\pctx{N}[\psome{x};P] \| \pctx{N'}[\gclose{x}{(w_1,\ldots,w_n)};Q]) \vdash \Gamma, \Delta
                }
                \and\implies\and
                \inferrule{
                    \pctx*{N}[P] \vdash \Gamma, x{:}A
                    ~~\text{(\refitem{l}{ctxRedOne}{Psome})}
                    \\
                    \pctx*{N'}[Q] \vdash \Delta, x{:}\ol{A}
                    ~~\text{(\refitem{l}{ctxRedOne}{Gsome})}
                }{
                    \res{x}(\pctx*{N}[P] \| \pctx*{N'}[Q]) \vdash \Gamma, \Delta
                }
            \end{mathpar}

        \item
            Rule $\rredone{\none}$.
            \begin{mathpar}
                \inferrule{
                    \pctx{N}[\pnone{x}] \vdash \Gamma, x{:}{\with}A
                    ~~\text{(\refitem{l}{ctxType}{Pnone})}
                    \\
                    \pctx{N'}[\gclose{x}{(w_1,\ldots,w_n)};Q] \vdash \Delta, x{:}{\oplus}\ol{A}
                    ~~\text{(\refitem{l}{ctxType}{Gsome})}
                }{
                    \res{x}(\pctx{N}[\pnone{x}] \| \pctx{N'}[\gclose{x}{(w_1,\ldots,w_n)};Q]) \vdash \Gamma, \Delta
                }
                \and\implies\and
                \inferrule{
                    \pctx*{N}[\0] \vdash \Gamma
                    ~~\text{(\refitem{l}{ctxRedOne}{Pnone})}
                    \\
                    \pctx*{N'}[\pnone{w_1} \| \ldots \| \pnone{w_n}] \vdash \Delta
                    ~~\text{(\refitem{l}{ctxRedOne}{Gsome})}
                }{
                    \pctx*{N}[\0] \| \pctx*{N'}[\pnone{w_1} \| \ldots \| \pnone{w_n}]) \vdash \Gamma, \Delta
                }
            \end{mathpar}

        \item
            Rule $\rredone{{?}{!}}$.
            \begin{mathpar}
                \inferrule{
                    \pctx{N}[\puname{x}{y};P] \vdash \Gamma, x{:}{?}A
                    ~~\text{(\refitem{l}{ctxType}{Puname})}
                    \\
                    \pctx{N'}[\guname{x}{z};Q] \vdash \Delta, x{:}{!}\ol{A}
                    ~~\text{(\refitem{l}{ctxType}{Guname})}
                }{
                    \res{x}(\pctx{N}[\puname{x}{y};P] \| \pctx{N'}[\guname{x}{y};Q]) \vdash \Gamma, \Delta
                }
                \and\implies\and
                \inferrule{}{
                    \pctx*{N'}\big[\res{x}(\res{y}(\pctx*{N}[P] \| Q\{y/z\}) \| \guname{x}{z};Q)\big] \vdash \Gamma, \Delta
                    ~~\text{(\refitem{l}{ctxRedOne}{Uname})}
                }
            \end{mathpar}

        \item
            Rule $\rredone{\equiv}$.
            Assume $P \equiv P'$ and $P' \redone Q'$ and $Q' \equiv Q$.
            By \Cref{t:subcong}, $P' \vdash \Gamma$.
            By the IH, $Q' \vdash \Gamma$.
            By \Cref{t:subcong}, $Q \vdash \Gamma$.

        \item
            Rule $\rredone{\nu}$.
            Assume $P \redone P'$.
            \begin{mathpar}
                \inferrule{
                    P \vdash \Gamma, x{:}A
                    \\
                    Q \vdash \Delta, x{:}\ol{A}
                }{
                    \res{x}(P \| Q) \vdash \Gamma, \Delta
                }
                \and\implies\and
                \inferrule{
                    P' \vdash \Gamma, x{:}A
                    ~~\text{(IH)}
                    \\
                    Q \vdash \Delta, x{:}\ol{A}
                }{
                    \res{x}(P' \| Q) \vdash \Gamma, \Delta
                }
            \end{mathpar}

        \item
            Rule $\rredone{\|}$.
            Assume $P \redone P'$.
            \begin{mathpar}
                \inferrule{
                    P \vdash \Gamma
                    \\
                    Q \vdash \Delta
                }{
                    P \| Q \vdash \Gamma, \Delta
                }
                \and\implies\and
                \inferrule{
                    P' \vdash \Gamma
                    ~~\text{(IH)}
                    \\
                    Q \vdash \Delta
                }{
                    P' \| Q \vdash \Gamma, \Delta
                }
            \end{mathpar}

        \item
            Rule $\rredone{\nd}$.
            Assume $P \redone P'$.
            \begin{mathpar}
                \inferrule{
                    P \vdash \Gamma
                    \\
                    Q \vdash \Gamma
                }{
                    P \nd Q \vdash \Gamma
                }
                \and\implies\and
                \inferrule{
                    P' \vdash \Gamma
                    ~~\text{(IH)}
                    \\
                    Q \vdash \Gamma
                }{
                    P' \nd Q \vdash \Gamma
                }
            \end{mathpar}
            \qedhere
    \end{itemize}
\end{proof}

\subsubsection{Deadlock-freedom}

The proof uses several definitions and lemmas, which we summarize:
\begin{itemize}
    \item
        \Cref{d:sctx} defines single-choice multi-hole contexts, where holes may only appear on one side of non-deterministic choices.
        \Cref{d:scoll} yields deterministic multi-hole contexts from single-choice multi-hole contexts by committing non-deterministic choices to the sides of holes.
        \Cref{l:scoll} ensures typing remains consistent when committing a single-choice multi-hole context.

    \item
        \Cref{l:sctxform} states that any typable process not equivalent to $\0$ can be written as an S-context with each hole replaced by a prefixed process.
        Let us refer to this as the \emph{S-context form}.

    \item
        \Cref{l:sformfwd} states that if a process in S-context form is typable under empty context and has a forwarder as one of its prefixes, that process contains a cut on one of the forwarder's subjects.

    \item
        \Cref{l:sctxcuts} states that the number of prefixed processes of a process in S-context form is at least the number of cuts in the S-context.
        This lemma is key to the proof of Deadlock Freedom, as it is necessary to show the next lemma.

    \item
        \Cref{l:sctxsubjs} states that if a process in S-context form is typable under empty context, then there must be two of its prefixed processes that share a subject.
\end{itemize}

\begin{definition}[Single-choice Multi-hole Contexts]\label{d:sctx}
    We define \emph{single-choice multi-hole contexts} (S-contexts, for short) as follows:
    \[
        \pctx{S} ::= {\hole}_i \sepr \res{x}(\pctx{S} \| \pctx{S}) \sepr \pctx{S} \| \pctx{S} \sepr \pctx{S} \nd P
    \]
    An S-context is $n$-ary if it has $n$ holes ${\hole}_1, \ldots, {\hole}_n$.
    We write $\pctx{S}[P_1, \ldots, P_n]$ to denote the process obtained from the $n$-ary multi-hole context $\pctx{S}$ by replacing each $i$-th hole in $\pctx{S}$ with $P_i$.
    Given an S-context $\pctx{S}$ with hole indices $I$ and a sequence of processes ${(P_i)}_{i \in I}$, we write $\pctx{S}[P_i]_{i \in I}$ to denote the process obtained from $\pctx{S}$ by replacing each hole with index $i$ in $\pctx{S}$ with $P_i$.
    We say an S-context is a \emph{deterministic multi-hole context} (DM-context, for short) if its holes do not appear inside any non-deterministic choices.
\end{definition}

\begin{definition}[Commitment of Single-choice Multi-hole Contexts]\label{d:scoll}
    We define the \emph{commitment} of S-context $\pctx{S}$, by abuse of notation denoted $\D{\pctx{S}}$ (cf.\ \Cref{s:disc}), as follows, yielding a deterministic multi-hole context:
    \begin{align*}
        \D{{\hole}_i} &:= {\hole}_i
        & \D{\pctx{S} \| \pctx{S}'} &:= \D{\pctx{S}} \| \D{\pctx{S}'}
        & \D{\res{x}(\pctx{S} \| \pctx{S}')} &:= \res{x}(\D{\pctx{S}} \| \D{\pctx{S}'})
        & \D{\pctx{S} \nd P} &:= \D{\pctx{S}}
    \end{align*}
\end{definition}

\begin{lemma}\label{l:scoll}
    If $\pctx{S}[P_i]_{i \in I} \vdash \Gamma$, then $\D{\pctx{S}}[P_i]_{i \in I} \vdash \Gamma$.
\end{lemma}

\begin{proof}
    Straightforward, by induction on the structure of $\pctx{S}$.
\end{proof}

\begin{lemma}\label{l:sctxform}
    If $P \vdash \Gamma$ and $P \not\equiv \0$, then there exist S-context $\pctx{S}$ with indices $I$ and sequence of prefixed processes ${(\alpha_i;P_i)}_{i \in I}$ such that $P \equiv \pctx{S}[\alpha_i;P_i]_{i \in I}$.
\end{lemma}

\begin{proof}
    Using structural congruence, we first remove all cuts with unused servers and parallel compositions with $\0$, obtaining $P' \equiv P$.
    Since $P \not\equiv \0$, $P' \not\equiv \0$.
    Then, we construct $\pctx{S}$ by induction on the typing derivation of $P'$.
    Rules \ttype{empty} and \ttype{weaken} do not occur, because of how we obtained $P'$ from $P$.
    The structural rules \ttype{mix}, \ttype{cut}, and \ttype{weaken} are simply copied.
    In case of rule \ttype{$\nd$}, we arbitrarily pick a branch to continue the construction of $\pctx{S}$ with, while copying the entire other branch.
    The other rules, which type prefixes, add a hole to $\pctx{S}$; we mark the hole with index $i$ and refer to the prefixed process typed by the rule as $\alpha_i;P_i$.
    Clearly, $P \equiv P' = \pctx{S}[\alpha_i;P_i]_{i \in I}$.
\end{proof}

\begin{lemma}\label{l:sformfwd}
    If $P = \pctx{S}[\alpha_i;P_i]_{i \in I} \vdash \emptyset$ and there is $j \in I$ s.t.\ $\alpha_j = \pfwd{x}{y}$, then there are $\pctx{N},\pctx{N'},Q$ such that $P = \pctx[\Big]{N}[\res{x}(\pctx[\big]{N'}[\pfwd{x}{y}] \| Q)]$.
\end{lemma}

\begin{proof}
    Note that there must be a restriction on $x$ in $P$, because $x$ appears free in $\pfwd{x}{y}$ but $P \vdash \emptyset$.
    First, we obtain $\pctx{N}$ from $P$ by replacing the restriction on $x$ in $P$ with a hole, referring the parallel component in which $\pfwd{x}{y}$ appears as $P'$ and the other parallel component as $Q$.
    Then, we obtain $\pctx{N'}$ from $P'$ by replacing $\pfwd{x}{y}$ with a hole.
    Clearly, $P = \pctx[\Big]{N}[\res{x}(\pctx[\big]{N'}[\pfwd{x}{y}] \| Q)]$.
\end{proof}

\begin{lemma}\label{l:sctxcuts}
    If the derivation of $P = \pctx{S}[\alpha_i;P_i]_{i \in I} \vdash \Gamma$ and $\pctx{S}$ is deterministic and contains $n$ cuts, then $|I| \geq n+1$.
\end{lemma}

\begin{proof}
    We apply strong induction on the number $n$ of cuts in $\pctx{S}$:
    \begin{itemize}
        \item
            Case $n = 0$.
            Any S-context must have at least one hole, so $\pctx{S}$ has at least one hole.
            Hence, $|I| \geq 1 = n + 1$.

        \item
            Case $n = n' + 1$.
            By abuse of notation, $P = P_1 \| \ldots \| P_k$, where for each $1 \leq k' \leq k$, $P_{k'}$ is not a parallel composition.
            By assumption, $m \geq 1$ of the $P_1,\ldots,P_k$ are cuts.
            W.l.o.g., assume $P_1,\ldots,P_m$ are cuts.

            For each $1 \leq j \leq m$, by inversion of rule \ttype{mix}, $P_j \vdash \Gamma_j$, and by construction, there are $\pctx{S_j},I_j$ s.t.\ $P_j = \pctx{S_j}[\alpha_i;P_i]_{i \in I_j}$ where $\pctx{S_j}$ is deterministic.
            We have for each $1 \leq j \leq m$ and $1 \leq j' \leq m$ where $j \neq j'$ that $I_j \disj I_{j'}$, and $\bigcup_{1 \leq j \leq m} I_j \subseteq I$.
            Then, for each $1 \leq j \leq m$, let $1 \leq n_j \leq n$ be the number of cuts in $\pctx{S_j}$.
            Since $P_{m+1},\ldots,P_k$ are not cuts, we have $\sum_{1 \leq j \leq m} n_j = n$.

            Take any $1 \leq j \leq m$.
            We have $P_j = \res{x}(P'_j \| P''_j)$, and by inversion of rule \ttype{cut}, $P'_j \vdash \Gamma'_j, x{:}A$ and $P''_j \vdash \Gamma''_j, x{:}\ol{A}$ where $\Gamma_j = \Gamma'_j, \Gamma''_j$.
            By construction, there are $\pctx{S'_j},\pctx{S''_j},I'_j,I''_j$ s.t.\ $P'_j = \pctx{S'_j}[\alpha_i;P_i]_{i \in I'_j}$ and $P''_j = \pctx{S''_j}[\alpha_i;P_i]_{i \in I''_j}$ and $\pctx{S'_j}$ and $\pctx{S''_j}$ are deterministic.
            We have $I'_j \disj I''_j$ and $I'_j \cup I''_j = I_j$.

            Let $n'_j$ and $n''_j$ be the number of cuts in $\pctx{S'_j}$ and $\pctx{S''_j}$, respectively.
            We have that $n'_j + n''_j + 1 = n_j$.
            Since $n_j \leq n = n' + 1$, then $n'_j,n''_j \leq n'$.
            Then, by the IH, $|I'_j| \geq n'_j + 1$ and $|I''_j| \geq n''_j + 1$.
            Therefore, $|I_j| = |I'_j \cup I''_j| = |I'_j| + |I''_j| \geq n'_j + n''_j + 1 + 1 = n_j + 1$.

            In conclusion,
            \begin{align*}
                |I| \geq |\bigcup_{1 \leq j \leq m} I_j| = \sum_{1 \leq j \leq m} |I_j| \geq \sum_{1 \leq j \leq m} (n_j + 1) = \sum_{1 \leq j \leq m} n_j + m = n + m \geq n + 1.
                \tag*{\qedhere}
            \end{align*}
    \end{itemize}
\end{proof}

\begin{lemma}\label{l:sctxsubjs}
    If $P = \pctx{S}[\alpha_i;P_i]_{i \in I} \vdash \emptyset$ where for each $i \in I$, $\alpha_i \neq \pfwd{x}{y}$ for any $x$ and $y$, then there are $j,k \in I$ where $j \neq k$ and $x = \subj(\alpha_j) = \subj(\alpha_k)$, and there are $\pctx{N},\pctx{N_j},\pctx{N_k}$ such that $P = \pctx[\big]{N}[\res{x}(\pctx{N_j}[\alpha_j] \| \pctx{N_k}[\alpha_k])]$.
\end{lemma}

\begin{proof}
    Let $Q = \pctx*{S}[{(\alpha_i)}_{i \in I}]$.
    Then $Q$ is deterministic and, by \Cref{l:scoll}, $Q \vdash \emptyset$.
    Let $n$ be the number of cuts in $\pctx{S}$.
    By \Cref{l:sctxcuts}, $|I| \geq n + 1$.

    Suppose, for contradiction, that for every $j,k \in I$ where $j \neq k$, we have $\subj(\alpha_j) \neq \subj(\alpha_k)$.
    Since $Q \vdash \emptyset$, for each $j \in I$, $\subj(\alpha_j)$ must be bound by a cut, so $\pctx{S}$ must contain $|I|$ cuts.
    This means $|I| = n$, contradicting the fact that $|I| \geq n + 1$.
    Therefore, there must be $j,k \in I$ where $j \neq k$ such that $\subj(\alpha_j) = \subj(\alpha_k)$.

    Hence, we can take $x = \subj(\alpha_j) = \subj(\alpha_k)$.
    Since $P \vdash \emptyset$ but $x$ appears free in $\alpha_j;P_j$ and $\alpha_k;P_k$, there must be a restriction on $x$ in $\pctx{S}$ containing the holes ${\hole}_j$ and ${\hole}_k$.
    We now obtain $\pctx{N}$ from $P$ by replacing the restriction on $x$ in $P$ with a hole, referring to the parallel component in which $\alpha_j;P_j$ appears as $P_j$ and the component in which $\alpha_k;P_k$ appears as $P_k$.
    Then, we obtain $\pctx{N_j}$ and $\pctx{N_k}$ from $P_j$ and $P_k$, respectively, by replacing $\alpha_j;P_j$ and $\alpha_k;P_k$ with a hole.
    Clearly, $P = \pctx[\big]{N}[\res{x}(\pctx{N_j}[\alpha_j;P_j] \| \pctx{N_k}[\alpha_k;P_k])]$.
\end{proof}

\thmDlfreeOne*

\begin{proof}
    By \Cref{l:sctxform}, there are S-context $\pctx{S}$ with hole indices $I$ and sequence of prefixed processes ${(\alpha_i;P_i)}_{i \in I}$ such that $P \equiv \pctx{S}[\alpha_i;P_i]_{i \in I}$.
    The next step depends on whether there is a forwarder process among the $\alpha_i$.
    \begin{itemize}
        \item
            If there exists $j \in I$ s.t.\ $\alpha_j = \pfwd{x}{y}$ for some $x$ and $y$, then by \Cref{l:sformfwd} there are $\pctx{N},\pctx{N'},Q$ s.t.\ $\pctx{S}[\alpha_i;P_i] = \pctx[\Big]{N}[\res{x}(\pctx[\big]{N'}[\pfwd{x}{y}] \| Q)]$.
            \begin{align*}
                \res{x}(\pctx[\big]{N'}[\pfwd{x}{y}] \| Q) &\redone \pctx*{N'}[Q\{y/x\}] = R'
                &&\text{(by rule $\rredone{\scc{Id}}$)}
                \\
                \pctx{S}[\alpha_i;P_i]_{i \in I} = \pctx[\Big]{N}[\res{x}(\pctx[\big]{N'}[\pfwd{x}{y}] \| Q)] &\redone \pctx{N}[R'] = R
                &&\text{(by rules $\rredone{\nu},\rredone{\|},\rredone{\nd}$)}
                \\
                P &\redone R
                &&\text{(by rule $\rredone{\equiv}$)}
            \end{align*}

        \item
            If for each $i \in I$, $\alpha_i \neq x \fwd y$ for any $x$ and $y$, then by \Cref{l:sctxsubjs} there are $j,k \in I$ where $j \neq k$ and $x = \subj(\alpha_j) = \subj(\alpha_k)$ for some $x$, and $\pctx{N},\pctx{N_j},\pctx{N_k}$ such that $\pctx{S}[\alpha_i;P_i]_{i \in I} = \pctx[\big]{N}[\res{x}(\pctx{N_j}[\alpha_j;P_j] \| \pctx{N_k}[\alpha_k;P_k])]$.

            We now show by cases on $\alpha_j$ that there is $R'$ such that $\res{x}(\pctx{N_j}[\alpha_j;P_j] \| \pctx{N_k}[\alpha_k;P_k]) \redone R'$.
            First, note that by typability, if the type for $x$ in $\pctx{N_j}[\alpha_j;P_j]$ is $A$, then the type for $x$ in $\pctx{N_k}[\alpha_k;P_k]$ is $\ol{A}$.
            In the following cases, we determine more precisely the form of $A$ by typing inversion on $\pctx{N_j}[\alpha_j;P_j]$, and then determine the form of $\alpha_k$ by typing inversion using the form of $\ol{A}$.
            Note that we can exclude any cases where $\alpha_j$ or $\alpha_k$ are forwarder processes, as we assume they are not.
            \begin{itemize}
                \item
                    If $\alpha_j;P_j = \pclose{x}$, then $A = \1$ and $\ol{A} = \bot$.
                    Hence, $\alpha_k;P_k = \gclose{x};P_k$.
                    By rule $\rredone{\1\bot}$, there is $R'$ such that
                    \begin{align*}
                        \res{x}(\pctx{N_j}[\pclose{x}] \| \pctx{N_k}[\gclose{x};P_k]) \redone R'.
                    \end{align*}
                \item
                    If $\alpha_j;P_j = \pname{x}{y};(P'_j \| P''_j)$ for some $y$, then $A = B \tensor C$ and $\ol{A} = \ol{B} \parr \ol{C}$ for some $B$ and $C$.
                    Hence, $\alpha_k;P_k = \gname{x}{z}; P_k$ for some $z$.
                    By rule $\rredone{\tensor\parr}$, there is $R'$ such that
                    \begin{align*}
                        \res{x}(\pctx{N_j}[\pname{x}{y};(P'_j \| P''_j)] \| \pctx{N_k}[\gname{x}{z};P_k]) \redone R'.
                    \end{align*}

                \item
                    If $\alpha_j;P_j = \psel{x}{l'}; P_j$, then $A = {\oplus}\{l:B_l\}_{l \in L}$ and $\ol{A} = {\with}\{l:\ol{B_l}\}_{l \in L}$ for some ${(B_l)}_{l \in L}$ where $l' \in L$.
                    Hence, $\alpha_k;P_k = \gsel{x}\{l:P_k^l\}_{l \in L}$.
                    By rule $\rredone{\oplus\with}$, there is $R'$ such that
                    \begin{align*}
                        \res{x}(\pctx{N_j}[\psel{x}{l'}; P_j] \| \pctx{N_k}[\gsel{x}\{l:P_k^l\}_{l \in L}]) \redone R'.
                    \end{align*}

                \item
                    If $\alpha_j;P_j = \psome{x}; P_j$, then $A = {\oplus}B$ and $\ol{A} = {\with}\ol{B}$ for some $B$.
                    Hence, $\alpha_k = \gsome{x}{w_1,\ldots,w_n}; P_k$ for some $w_1,\ldots,w_n$.
                    By rule $\rredone{\some}$, there is $R'$ such that
                    \begin{align*}
                        \res{x}(\pctx{N_j}[\psome{x}; P_j] \| \pctx{N_k}[\gsome{x}{w_1,\ldots,w_n}; P_k]) \redone R'.
                    \end{align*}

                \item
                    If $\alpha_j;P_j = \pnone{x}$, then $A = {\oplus}B$ and $\ol{A} = {\with}\ol{B}$ for some $B$.
                    Hence, $\alpha_k = \gsome{x}{w_1,\ldots,w_n}; P_k$ for some $w_1,\ldots,w_n$.
                    By rule $\rredone{\none}$, there is $R'$ such that
                    \begin{align*}
                        \res{x}(\pctx{N_j}[\pnone{x}] \| \pctx{N_k}[\gsome{x}{w_1,\ldots,w_n}; P_k]) \redone R'.
                    \end{align*}

                \item
                    If $\alpha_j;P_j = \puname{x}{y}; P_j$ for some $y$, then $A = {?}B$ and $\ol{A} = {!}\ol{B}$ for some $B$.
                    Hence, $\alpha_k;P_k = \guname{x}{z}; P_k$ for some $z$.
                    By rule $\rredone{{?}{!}}$, there is $R'$ such that
                    \begin{align*}
                        \res{x}(\pctx{N_j}[\puname{x}{y}; P_j] \| \pctx{N_k}[\guname{x}{z}; P_k]) \redone R'.
                    \end{align*}

                \item
                    Otherwise, $\alpha_j$ is a receiving prefix and $\alpha_k$ is thus a sending prefix.
                    By cases on $\alpha_k$, the proof is analogous to above.
            \end{itemize}
            In conclusion,
            \begin{align*}
                \pctx{S}[\alpha_i;P_i]_{i \in I} = \pctx[\big]{N}[\res{x}(\pctx{N_j}[\alpha_j;P_j] \| \pctx{N_k}[\alpha_k;P_k])] &\redone \pctx{N}[R'] = R
                &&\text{(by rules $\rredone{\nu},\rredone{\|},\rredone{\nd}$)}
                \\
                P &\redone R.
                &&\text{(by rule $\rredone{\equiv}$)}
                \tag*{\qedhere}
            \end{align*}
    \end{itemize}
\end{proof}

\subsection{Lazy Semantics}
\label{ss:proofsLazy}

\subsubsection{Subject Reduction}
\label{ss:TPLazy}

\begin{lemma}\label{l:ctxRedTwo}
    For both of the following items, assume $\Gamma \disj \Delta$.
    \begin{enumerate}
        \item\label{i:ctxRedTwoName}
            If $\forall i \in I.~ \forall j \in J.~ \bn{\pctx{C_i}} \disj \fn{\pctx{D_j}}$ and $\forall i \in I.~ \pctx{C_i}[\pname{x}{y_i};(P_i \| Q_i)] \vdash \Gamma, x{:}A \tensor B$ and $\forall j \in J.~ \pctx{D_j}[\gname{x}{z};R_j] \vdash \Delta, x{:}\ol{A} \parr \ol{B}$, then $\bignd_{i \in I} \pctx[\Big]{C_i}[\res{x}\Big(Q_i \| \res{w}\Big(P_i\{w/y_i\} \| \bignd_{j \in J} \pctx{D_j}[R_j\{w/z\}]\Big)\Big)] \vdash \Gamma, \Delta$.

        \item\label{i:ctxRedTwoUname}
            If $\forall i \in I.~ \forall j \in J.~ \bn{\pctx{D_j}} \disj \fn{\pctx{C_i}}$ and $\forall i \in I.~ \pctx{C_i}[\puname{x}{y_i};P_i] \vdash \Gamma, x{:}{?}A$ and $\forall j \in J.~ \pctx{D_j}[\guname{x}{z};Q_j] \vdash \Delta, x{:}{!}\ol{A}$, then $\bignd_{j \in J} \pctx[\Big]{D_j}[\res{x}\Big(\res{w}\Big(\bignd_{i \in I} \pctx{C_i}[P_i\{w/y_i\}] \| Q_j\{w/z\}\Big) \| \guname{x}{z};Q_j\Big)] \vdash \Gamma, \Delta$.
    \end{enumerate}
\end{lemma}

\begin{proof}
    Both items follow by induction on the structures of the D-contexts.
    For each item, we detail the base case, where $\forall i \in I.~ \pctx{C_i} = \hole$ and $\forall j \in J.~ \pctx{D_j} = \hole$.
    The inductive cases follow from the IH straightforwardly.
    \begin{enumerate}
        \item
            \begin{mathpar}
                \forall i \in I.~
                \inferrule{
                    P_i \vdash \Gamma, y_i{:}A
                    \\
                    Q_i \vdash \Delta, x{:}B
                }{
                    \pname{x}{y_i};(P_i \| Q_i) \vdash \Gamma, \Delta, x{:}A \tensor B
                }
                \and
                \forall j \in J.~
                \inferrule{
                    R_j \vdash \Lambda, z{:}\ol{A}, x{:}\ol{B}
                }{
                    \gname{x}{z};R_j \vdash \Lambda, x{:}\ol{A} \parr \ol{B}
                }
                \and\implies\and
                \inferruleDbl{
                    \inferrule*[fraction={---}]{
                        \forall i \in I.~
                        Q_i \vdash \Delta, x{:}B
                        \\
                        \inferrule*{
                            \forall i \in I.~
                            P_i\{w/y_i\} \vdash \Gamma, w{:}A
                            \\
                            \inferruleDbl{
                                \forall j \in J.~ R_j\{w/z\} \vdash \Lambda, w{:}\ol{A}, x{:}\ol{B}
                            }{
                                \bignd_{j \in J} R_j\{w/z\} \vdash \Lambda, w{:}\ol{A}, x{:}\ol{B}
                            }
                        }{
                            \forall i \in I.~
                            \res{w}\Big(P_i\{w/y_i\} \| \bignd_{j \in J} R_j\{w/z\}\Big) \vdash \Gamma, \Lambda, x{:}\ol{B}
                        }
                    }{
                        \forall i \in I.~
                        \res{x}\Big(Q_i \| \res{w}\Big(P_i\{w/y_i\} \| \bignd_{j \in J} R_j\{w/z\}\Big)\Big) \vdash \Gamma, \Delta, \Lambda
                    }
                }{
                    \bignd_{i \in I} \res{x}\Big(Q_i \| \res{w}\Big(P_i\{w/y_i\} \| \bignd_{j \in J} R_j\{w/z\}\Big)\Big) \vdash \Gamma, \Delta, \Lambda
                }
            \end{mathpar}

        \item
            This item depends on whether $x \in \fn{P_i}$ or not, for each $i \in I$.
            For simplicity, we only consider the cases where either $\forall i \in I.~ x \in \fn{P_i}$ or $\forall i \in I.~ x \notin \fn{P_i}$.
            \begin{itemize}
                \item
                    $\forall i \in I.~ x \in \fn{P_i}$.
                    \begin{mathpar}
                        \forall i \in I.~
                        \inferrule{
                            \inferrule{
                                P_i\{x'/x\} \vdash \Gamma, y_i{:}A, x'{:}{?}A
                            }{
                                \puname{x}{y_i};(P_i\{x'/x\}) \vdash \Gamma, x{:}{?}A, x'{:}{?}A
                            }
                        }{
                            \puname{x}{y_i};P_i \vdash \Gamma, x{:}{?}A
                        }
                        \and
                        \forall j \in J.~
                        \inferrule{
                            Q_j \vdash {?}\Delta, x{:}\ol{A}
                        }{
                            \guname{x}{z};Q_j \vdash {?}\Delta, x{:}{!}\ol{A}
                        }
                        \and\implies\and
                        \inferruleDbl{
                            \inferruleDbl{
                                \inferrule*[fraction={---}]{
                                    \inferrule*{
                                        \inferruleDbl{
                                            {\begin{tarr}[b]{l}
                                                    \forall i \in I.
                                                    \\
                                                    P_i\{w/y_i\} \vdash \Gamma, w{:}A, x{:}{?}A
                                            \end{tarr}}
                                        }{
                                            \bignd_{i \in I} P_i\{w/y_i\} \vdash \Gamma, w{:}A, x{:}{?}A
                                        }
                                        \\
                                        {\begin{tarr}[b]{l}
                                                \forall j \in J.
                                                \\
                                                Q_j\{w/z\}\{v'/v\}_{v \in {?}\Delta} \vdash {?}\Delta', w{:}\ol{A}
                                        \end{tarr}}
                                    }{
                                        {\begin{tarr}[t]{l}
                                                \forall j \in J.
                                                \\
                                                \res{w}\Big(\bignd_{i \in I} P_i\{w/y_i\} \| Q_j\{w/z\}\{v'/v\}_{v \in {?}\Delta}\Big) \vdash \Gamma, {?}\Delta', x{:}{?}A
                                        \end{tarr}}
                                    }
                                    \\
                                    {\begin{tarr}[b]{l}
                                            \forall j \in J.
                                            \\
                                            \guname{x}{z};Q_j \vdash {?}\Delta, x{:}{!}\ol{A}
                                    \end{tarr}}
                                }{
                                    \forall j \in J.~
                                    \res{x}\Big(\res{w}\Big(\bignd_{i \in I} P_i\{w/y_i\} \| Q_j\{w/z\}\{v'/v\}_{v \in {?}\Delta}\Big) \| \guname{x}{z};Q_j\Big) \vdash \Gamma, {?}\Delta, {?}\Delta'
                                }
                            }{
                                \bignd_{j \in J} \res{x}\Big(\res{w}\Big(\bignd_{i \in I} P_i\{w/y_i\} \| Q_j\{w/z\}\{v'/v\}_{v \in {?}\Delta}\Big) \| \guname{x}{z};Q_j\Big) \vdash \Gamma, {?}\Delta, {?}\Delta'
                            }
                        }{
                            \bignd_{j \in J} \res{x}\Big(\res{w}\Big(\bignd_{i \in I} P_i\{w/y_i\} \| Q_j\{w/z\}\Big) \| \guname{x}{z};Q_j\Big) \vdash \Gamma, {?}\Delta
                        }
                    \end{mathpar}

                \item
                    $\forall i \in I.~ x \notin \fn{P_i}$.
                    \begin{mathpar}
                        \forall i \in I.~
                        \inferrule{
                            P_i \vdash \Gamma, y_i{:}A
                        }{
                            \puname{x}{y_i};P_i \vdash \Gamma, x{:}{?}A
                        }
                        \and
                        \forall j \in J.~
                        \inferrule{
                            Q_j \vdash {?}\Delta, x{:}\ol{A}
                        }{
                            \guname{x}{z};Q_j \vdash {?}\Delta, x{:}{!}\ol{A}
                        }
                        \and\implies\and
                        \inferruleDbl{
                            \inferruleDbl{
                                \inferrule*[fraction={---}]{
                                    \inferrule*{
                                        \inferrule*{
                                            \inferruleDbl{
                                                {\begin{tarr}[b]{l}
                                                        \forall i \in I.
                                                        \\
                                                        P_i\{w/y_i\} \vdash \Gamma, w{:}A
                                                \end{tarr}}
                                            }{
                                                \bignd_{i \in I} P_i\{w/y_i\} \vdash \Gamma, w{:}A
                                            }
                                            \\
                                            {\begin{tarr}[b]{l}
                                                    \forall j \in J.
                                                    \\
                                                    Q_j\{w/z\}\{v'/v\}_{v \in {?}\Delta} \vdash {?}\Delta', w{:}\ol{A}
                                            \end{tarr}}
                                        }{
                                            {\begin{tarr}[t]{l}
                                                    \forall j \in J.
                                                    \\
                                                    \res{w}\Big(\bignd_{i \in I} P_i\{w/y_i\} \| Q_j\{w/z\}\{v'/v\}_{v \in {?}\Delta}\Big) \vdash \Gamma, {?}\Delta'
                                            \end{tarr}}
                                        }
                                    }{
                                        {\begin{tarr}[t]{l}
                                                \forall j \in J.
                                                \\
                                                \res{w}\Big(\bignd_{i \in I} P_i\{w/y_i\} \| Q_j\{w/z\}\{v'/v\}_{v \in {?}\Delta}\Big) \vdash \Gamma, {?}\Delta', x{:}{?}A
                                        \end{tarr}}
                                    }
                                    \\
                                    {\begin{tarr}[b]{l}
                                            \forall j \in J.
                                            \\
                                            \guname{x}{z};Q_j \vdash {?}\Delta, x{:}{!}\ol{A}
                                    \end{tarr}}
                                }{
                                    \forall j \in J.~
                                    \res{x}\Big(\res{w}\Big(\bignd_{i \in I} P_i\{w/y_i\} \| Q_j\{w/z\}\{v'/v\}_{v \in {?}\Delta}\Big) \| \guname{x}{z};Q_j\Big) \vdash \Gamma, {?}\Delta, {?}\Delta'
                                }
                            }{
                                \bignd_{j \in J} \res{x}\Big(\res{w}\Big(\bignd_{i \in I} P_i\{w/y_i\} \| Q_j\{w/z\}\{v'/v\}_{v \in {?}\Delta}\Big) \| \guname{x}{z};Q_j\Big) \vdash \Gamma, {?}\Delta, {?}\Delta'
                            }
                        }{
                            \bignd_{j \in J} \res{x}\Big(\res{w}\Big(\bignd_{i \in I} P_i\{w/y_i\} \| Q_j\{w/z\}\Big) \| \guname{x}{z};Q_j\Big) \vdash \Gamma, {?}\Delta
                        }
                    \end{mathpar}
                    \qedhere
            \end{itemize}
    \end{enumerate}
\end{proof}

\begin{theorem}[SR for the Lazy Semantics]\label{t:srTwo}
    If $P \vdash \Gamma$ and $P \redtwo_S Q$, then $Q \vdash \Gamma$.
\end{theorem}

\begin{proof}
    By induction on the derivation of the reduction.
    \begin{itemize}
        \item
            Rule $\rredtwo{\scc{Id}}$.
            \begin{mathpar}
                \inferrule{
                    \inferruleDbl{
                        \forall i \in I.~
                        \pctx[\big]{C_i}[\pfwd{x}{y}] \vdash \Gamma, x{:}A
                    }{
                        \bignd_{i \in I} \pctx[\big]{C_i}[\pfwd{x}{y}] \vdash \Gamma, x{:}A
                    }
                    \\
                    Q \vdash \Delta, x{:}\ol{A}
                }{
                    \res{x}\Big(\bignd_{i \in I} \pctx[\big]{C_i}[\pfwd{x}{y}] \| Q\Big) \vdash \Gamma, \Delta
                }
                \and\implies\and
                \inferruleDbl{
                    \forall i \in I.~
                    \pctx{C_i}[Q\{y/x\}] \vdash \Gamma, \Delta
                    ~~\text{(\refitem{l}{ctxRedOne}{Fwd})}
                }{
                    \bignd_{i \in I} \pctx{C_i}[Q\{y/x\}] \vdash \Gamma, \Delta
                }
            \end{mathpar}

        \item
            Rule $\rredtwo{\1\bot}$.
            \begin{mathpar}
                \inferrule{
                    \inferruleDbl{
                        \forall i \in I.~
                        \pctx{C_i}[\pclose{x}] \vdash \Gamma, x{:}\1
                        ~~\text{(\refitem{l}{ctxType}{Pclose})}
                    }{
                        \bignd_{i \in I} \pctx{C_i}[\pclose{x}] \vdash \Gamma, x{:}\1
                    }
                    \\
                    \inferruleDbl{
                        \forall j \in J.~
                        \pctx{D_j}[\gclose{x};Q_j] \vdash \Delta, x{:}\bot
                        ~~\text{(\refitem{l}{ctxType}{Gclose})}
                    }{
                        \bignd_{j \in J} \pctx{D_j}[\gclose{x};Q_j] \vdash \Delta, x{:}\bot
                    }
                }{
                    \res{x}\Big(\bignd_{i \in I} \pctx{C_i}[\pclose{x}] \| \bignd_{j \in J} \pctx{D_j}[\gclose{x};Q_j]\Big) \vdash \Gamma, \Delta
                }
                \and\implies\and
                \inferrule{
                    \inferruleDbl{
                        \forall i \in I.~
                        \pctx{C_i}[\0] \vdash \Gamma
                        ~~\text{(\refitem{l}{ctxRedOne}{Pclose})}
                    }{
                        \bignd_{i \in I} \pctx{C_i}[\0] \vdash \Gamma
                    }
                    \\
                    \inferruleDbl{
                        \forall j \in J.~
                        \pctx{D_j}[Q_j] \vdash \Delta
                        ~~\text{(\refitem{l}{ctxRedOne}{Gclose})}
                    }{
                        \bignd_{j \in J} \pctx{D_j}[Q_j] \vdash \Delta
                    }
                }{
                    \bignd_{i \in I} \pctx{C_i}[\0] \| \bignd_{j \in J} \pctx{D_j}[Q_j] \vdash \Gamma, \Delta
                }
            \end{mathpar}

        \item
            Rule $\rredtwo{\tensor\parr}$.
            \begin{mathpar}
                \mprset{sep=0.8em}
                \inferrule{
                    \inferruleDbl{
                        \forall i \in I.~
                        \pctx{C_i}[\pname{x}{y_i};(P_i \| Q_i)] \vdash \Gamma, x{:}A \tensor B
                        ~~\text{(\refitem{l}{ctxType}{Pname})}
                    }{
                        \bignd_{i \in I} \pctx{C_i}[\pname{x}{y_i};(P_i \| Q_i)] \vdash \Gamma, x{:}A \tensor B
                    }
                    \\
                    \inferruleDbl{
                        \forall j \in J.~
                        \pctx{D_j}[\gname{x}{z};R_j] \vdash \Delta, x{:}\ol{A} \parr \ol{B}
                        ~~\text{(\refitem{l}{ctxType}{Gname})}
                    }{
                        \bignd_{j \in J} \pctx{D_j}[\gname{x}{z};R_j] \vdash \Delta, x{:}\ol{A} \parr \ol{B}
                    }
                }{
                    \res{x}\Big(\bignd_{i \in I} \pctx{C_i}[\pname{x}{y_i};(P_i \| Q_i)] \| \bignd_{j \in J} \pctx{D_j}[\gname{x}{z};R_j]\Big) \vdash \Gamma, \Delta
                }
                \and\implies\and
                \inferrule{}{
                    \bignd_{i \in I} \pctx[\Big]{C_i}[\res{x}\Big(Q_i \| \res{w}\Big(P_i\{w/y_i\} \| \bignd_{j \in J} \pctx{D_j}[R_j\{w/z\}]\Big)\Big)] \vdash \Gamma, \Delta
                    ~~\text{(\refitem{l}{ctxRedTwo}{Name})}
                }
            \end{mathpar}

        \item
            Rule $\rredtwo{{\oplus}{\with}}$.
            Take any $k' \in K$.
            \begin{mathpar}
                \inferrule{
                    \inferruleDbl{
                        {\begin{tarr}[b]{l}
                                \forall i \in I.
                                \\
                                \pctx{C_i}[\psel{x}{k'};P_i] \vdash \Gamma, x{:}{\oplus}\{k:A_k\}_{k \in K}
                                \\
                                \text{(\refitem{l}{ctxType}{Psel})}
                        \end{tarr}}
                    }{
                        \bignd_{i \in I} \pctx{C_i}[\psel{x}{k'};P_i] \vdash \Gamma, x{:}{\oplus}\{k:A_k\}_{k \in K}
                    }
                    \\
                    \inferruleDbl{
                        {\begin{tarr}[b]{l}
                                \forall j \in J.
                                \\
                                \pctx{D_j}[\gsel{x}\{k:Q_j^k\}_{k \in K}] \vdash \Delta, x{:}{\with}\{k:\ol{A_k}\}_{k \in K}
                                \\
                                \text{(\refitem{l}{ctxType}{Gsel})}
                        \end{tarr}}
                    }{
                        \bignd_{j \in J} \pctx{D_j}[\gsel{x}\{k:Q_j^k\}_{k \in K}] \vdash \Delta, x{:}{\with}\{k:\ol{A_k}\}_{k \in K}
                    }
                }{
                    \res{x}\Big(\bignd_{i \in I} \pctx{C_i}[\psel{x}{k'};P_i] \| \bignd_{j \in J} \pctx{D_j}[\gsel{x}\{k:Q_j^k\}_{k \in K}]\Big) \vdash \Gamma, \Delta
                }
                \and\implies\and
                \inferrule{
                    \inferruleDbl{
                        \forall i \in I.~
                        \pctx{C_i}[P_i] \vdash \Gamma, x{:}A
                        ~~\text{(\refitem{l}{ctxRedOne}{Psel})}
                    }{
                        \bignd_{i \in I} \pctx{C_i}[P_i] \vdash \Gamma, x{:}A
                    }
                    \\
                    \inferruleDbl{
                        \forall j \in J.~
                        \pctx{D_j}[Q_j^k] \vdash \Delta, x{:}\ol{A}
                        ~~\text{(\refitem{l}{ctxRedOne}{Gsel})}
                    }{
                        \bignd_{j \in J} \pctx{D_j}[Q_j^k] \vdash \Delta, x{:}\ol{A}
                    }
                }{
                    \res{x}\Big(\bignd_{i \in I} \pctx{C_i}[P_i] \| \bignd_{j \in J} \pctx{D_j}[Q_j^k]\Big) \vdash \Gamma, \Delta
                }
            \end{mathpar}

        \item
            Rule $\rredtwo{\some}$.
            \begin{mathpar}
                \inferrule{
                    \inferruleDbl{
                        {\begin{tarr}[b]{l}
                                \forall i \in I.
                                \\
                                \pctx{C_i}[\psome{x};P_i] \vdash \Gamma, x{:}{\with}A
                                \\
                                \text{(\refitem{l}{ctxType}{Psome})}
                        \end{tarr}}
                    }{
                        \bignd_{i \in I} \pctx{C_i}[\psome{x};P_i] \vdash \Gamma, x{:}{\with}A
                    }
                    \\
                    \inferruleDbl{
                        {\begin{tarr}[b]{l}
                                \forall j \in J.
                                \\
                                \pctx{D_j}[\gsome{x}{w_1,\ldots,w_n};Q_j] \vdash \Delta, x{:}{\oplus}\ol{A}
                                \\
                                \text{(\refitem{l}{ctxType}{Gsome})}
                        \end{tarr}}
                    }{
                        \bignd_{j \in J} \pctx{D_j}[\gsome{x}{w_1,\ldots,w_n};Q_j] \vdash \Delta, x{:}{\oplus}\ol{A}
                    }
                }{
                    \res{x}\Big( \bignd_{i \in I} \pctx{C_i}[\psome{x};P_i] \| \bignd_{j \in J} \pctx{D_j}[\gsome{x}{w_1,\ldots,w_n};Q_j]\Big) \vdash \Gamma, \Delta
                }
                \and\implies\and
                \inferrule{
                    \inferruleDbl{
                        \forall i \in I.~
                        \pctx{C_i}[P_i] \vdash \Gamma, x{:}A
                        ~~\text{(\refitem{l}{ctxRedOne}{Psome})}
                    }{
                        \bignd_{i \in I} \pctx{C_i}[P_i] \vdash \Gamma, x{:}A
                    }
                    \\
                    \inferruleDbl{
                        \forall j \in J.~
                        \pctx{D_j}[Q_j] \vdash \Delta, x{:}\ol{A}
                        ~~\text{(\refitem{l}{ctxRedOne}{Gsome})}
                    }{
                        \bignd_{j \in J} \pctx{D_j}[Q_j] \vdash \Delta, x{:}\ol{A}
                    }
                }{
                    \res{x}\Big(\bignd_{i \in I} \pctx{C_i}[P_i] \| \bignd_{j \in J} \pctx{D_j}[Q_j]\Big) \vdash \Gamma, \Delta
                }
            \end{mathpar}

        \item
            Rule $\rredtwo{\none}$.
            \begin{mathpar}
                \inferrule{
                    \inferruleDbl{
                        {\begin{tarr}[b]{l}
                                \forall i \in I.
                                \\
                                \pctx{C_i}[\pnone{x}] \vdash \Gamma, x{:}{\with}A
                                \\
                                \text{(\refitem{l}{ctxType}{Pnone})}
                        \end{tarr}}
                    }{
                        \bignd_{i \in I} \pctx{C_i}[\pnone{x}] \vdash \Gamma, x{:}{\with}A
                    }
                    \\
                    \inferruleDbl{
                        {\begin{tarr}[b]{l}
                                \forall j \in J.
                                \\
                                \pctx{D_j}[\gsome{x}{w_1,\ldots,w_n};Q_j] \vdash \Delta, x{:}{\oplus}\ol{A}
                                \\
                                \text{(\refitem{l}{ctxType}{Gsome})}
                        \end{tarr}}
                    }{
                        \bignd_{j \in J} \pctx{D_j}[\gsome{x}{w_1,\ldots,w_n};Q_j] \vdash \Delta, x{:}{\oplus}\ol{A}
                    }
                }{
                    \res{x}\Big( \bignd_{i \in I} \pctx{C_i}[\pnone{x}] \| \bignd_{j \in J} \pctx{D_j}[\gsome{x}{w_1,\ldots,w_n};Q_j]\Big) \vdash \Gamma, \Delta
                }
                \and\implies\and
                \inferrule{
                    \inferruleDbl{
                        \forall i \in I.~
                        \pctx{C_i}[\0] \vdash \Gamma
                        ~~\text{(\refitem{l}{ctxRedOne}{Pnone})}
                    }{
                        \bignd_{i \in I} \pctx{C_i}[P_i] \vdash \Gamma
                    }
                    \\
                    \inferruleDbl{
                        \forall j \in J.~
                        \pctx{D_j}[\pnone{w_1} \| \ldots \| \pnone{w_n}] \vdash \Delta
                        ~~\text{(\refitem{l}{ctxRedOne}{Gsome})}
                    }{
                        \bignd_{j \in J} \pctx{D_j}[\pnone{w_1} \| \ldots \| \pnone{w_n}] \vdash \Delta
                    }
                }{
                    \bignd_{i \in I} \pctx{C_i}[\0] \| \bignd_{j \in J} \pctx{D_j}[\pnone{w_1} \| \ldots \| \pnone{w_n}] \vdash \Gamma, \Delta
                }
            \end{mathpar}

        \item
            Rule $\rredtwo{{?}{!}}$.
            \begin{mathpar}
                \inferrule{
                    \inferruleDbl{
                        \forall i \in I.~
                        \pctx{C_i}[\puname{x}{y_i};P_i] \vdash \Gamma, x{:}{?}A
                        ~~\text{(\refitem{l}{ctxType}{Puname})}
                    }{
                        \bignd_{i \in I} \pctx{C_i}[\puname{x}{y_i};P_i] \vdash \Gamma, x{:}{?}A
                    }
                    \\
                    \inferruleDbl{
                        \forall j \in J.~
                        \pctx{D_j}[\guname{x}{z};Q_j] \vdash \Delta, x{:}{!}\ol{A}
                        ~~\text{(\refitem{l}{ctxType}{Guname})}
                    }{
                        \bignd_{j \in J} \pctx{D_j}[\guname{x}{z};Q_j] \vdash \Delta, x{:}{!}\ol{A}
                    }
                }{
                    \res{x}\Big(\bignd_{i \in I} \pctx{C_i}[\puname{x}{y_i};P_i] \| \bignd_{j \in J} \pctx{D_j}[\guname{x}{z};Q_j]\Big) \vdash \Gamma, \Delta
                }
                \and\implies\and
                \inferrule{}{
                    \bignd_{j \in J} \pctx[\Big]{D_j}[\res{x}\Big(\res{w}\Big(\bignd_{i \in I} \pctx{C_i}[P_i\{w/z\}] \| Q_j\{w/z\}\Big) \| \guname{x}{z};Q_j\Big)] \vdash \Gamma, \Delta
                    ~~\text{(\refitem{l}{ctxRedTwo}{Uname})}
                }
            \end{mathpar}
            \qedhere
    \end{itemize}
\end{proof}

\subsubsection{Deadlock Freedom}
\label{ss:DFLazy}

\begin{definition}[Multi-hole Non-deterministic Reduction Contexts]
    \begin{align*}
        \pctx{M} ::= \hole \sepr \res{x}(P \| \pctx{M}) \sepr P \| \pctx{M} \sepr \pctx{M} \nd \pctx{M}
    \end{align*}
\end{definition}

\begin{lemma}\label{l:ndCtxMultihole}
    If $\pctx{N}[\alpha;P] \vdash \Gamma, x{:}A$ and $x = \subj(\alpha)$, then there are $\pctx{M}$ and ${(\alpha_i;P_i)}_{i \in I}$ such that $\pctx{N}[\alpha;P] = \pctx{M}[\alpha_i;P_i]_{i \in I}$ where $x \notin \fn{\pctx{M}}$ and $x \in \bigcap_{i \in I} \fn{\alpha_i;P_i}$ and there is $i' \in I$ such that $\alpha_{i'};P_{i'} = \alpha;P$.
\end{lemma}

\begin{lemma}\label{l:multiNdDet}
    For every multi-hole ND-context $\pctx{M}$ with indices $I$:
    \begin{itemize}
        \item
            If $\pctx{M}$ has two or more holes, there are $\pctx{C}$, $\pctx{M_1}$ with indices $I_1$, $\pctx{M_2}$ with indices $I_2$ such that $\pctx{M} = \pctx{C}[\pctx{M_1} \nd \pctx{M_2}]$ where $I_1 \disj I_2$ and $I = I_1 \cup I_2$.

        \item
            If $\pctx{M}$ has only one hole, there is $\pctx{C}$ such that $\pctx{M} = \pctx{C}$.
    \end{itemize}
\end{lemma}

\begin{definition}\label{d:ndFlat}
    \begin{align*}
        \flat{\pctx{C}[\pctx{M} \nd \pctx{M'}]}
        &:= \flat{\pctx{C}[\pctx{M}]} \nd \flat{\pctx{C}[\pctx{M'}]}
        &
        \flat{\pctx{C}}
        &:= \pctx{C}
    \end{align*}
\end{definition}

\begin{lemma}\label{l:multiShape}
    If $\pctx{M}[P_i]_{i \in I} \vdash \Gamma$ where $x \notin \fn{\pctx{M}}$ and $\forall i \in I.~ x \in \fn{P_i}$, then there are ${(\pctx{C_i})}_{i \in I}$ such that $\flat{\pctx{M}}[P_i]_{i \in I} = \bignd_{i \in I} \pctx{C_i}[P_i]$ where $\forall i \in I.~ x \notin \fn{\pctx{C_i}}$.
\end{lemma}

\begin{lemma}\label{l:flatRed}
    If
    \begin{align*}
        \res{x}(\pctx[\big]{N}[\flat{\pctx{M}}[\alpha_i;P_i]_{i \in I}] \| \pctx[\big]{N'}[\flat{\pctx{M'}}[\beta_j;Q_j]_{j \in J}]) \redtwo_S R,
    \end{align*}
    then
    \begin{align*}
    \res{x}(\pctx[\big]{N}[\pctx{M}[\alpha_i;P_i]_{i \in I}] \| \pctx[\big]{N'}[\pctx{M'}[\beta_j;Q_j]_{j \in J}]) \redtwo_S R.
    \end{align*}
\end{lemma}

\begin{proof}
    By induction on the structures of $\pctx{M}$ and $\pctx{M'}$.
    By \Cref{l:multiNdDet}, we only have to consider two cases for $\pctx{M}$ ($\pctx{M} = \pctx{C}[\pctx{M_1} \nd \pctx{M_2}]$ and $\pctx{M} = \pctx{C}$), and similarly for $\pctx{M'}$.
    We only detail the base case ($\pctx{M} = \pctx{C}$ and $\pctx{M'} = \pctx{C'}$) and a representative inductive case ($\pctx{M} = \pctx{C}[\pctx{M_1} \nd \pctx{M_2}]$ and $\pctx{M'} = \pctx{C'}$).
    \begin{itemize}
        \item
            $\pctx{M} = \pctx{C}$ and $\pctx{M'} = \pctx{C'}$.
            Note that $\pctx{M}$ and $\pctx{M'}$ have only one hole; w.l.o.g., assume $I = J = \{1\}$.
            \begin{align*}
                \flat{\pctx{M}}[\alpha_1;P_1] &= \flat{\pctx{C}}[\alpha_1;P_1] = \pctx{C}[\alpha_1;P_1] = \pctx{M}[\alpha_1;P_1]
                \\
                \flat{\pctx{M'}}[\beta_1;Q_1] &= \flat{\pctx{C'}}[\beta_1;Q_1] = \pctx{C'}[\beta_1;Q_1] = \pctx{M'}[\beta_1;Q_1]
            \end{align*}
            The thesis follows by assumption and equality.

        \item
            $\pctx{M} = \pctx{C}[\pctx{M_1} \nd \pctx{M_2}]$ and $\pctx{M'} = \pctx{C'}$.
            Note that $\pctx{M'}$ has only one hole; w.l.o.g., assume $J = \{1\}$.
            \begin{align}
                & \res{x}(\pctx[\big]{N}[\flat{\pctx{M}}[\alpha_i;P_i]_{i \in I}] \| \pctx[\big]{N'}[\flat{\pctx{M'}}[\beta_1;Q_1]])
                \nonumber
                \\
                &= \res{x}(\pctx[\big]{N}[\flat{\pctx{C}[\pctx{M_1} \nd \pctx{M_2}]}[\alpha_i;P_i]_{i \in I}] \| \pctx[\big]{N'}[\flat{\pctx{C'}}[\beta_1;Q_1]])
                \nonumber
                \\
                &= \res{x}(\pctx[\big]{N}[(\flat{\pctx{C}[\pctx{M_1}]} \nd \flat{\pctx{C}[\pctx{M_2}]})[\alpha_i;P_i]_{i \in I}] \| \pctx[\big]{N'}[\pctx{C'}[\beta_1;Q_1]])
                \label{eq:oneImplTwoBeforeSplit}
            \end{align}
            There are $I_1$ and $I_2$ such that $I_1 \disj I_2$ and $I = I_1 \cup I_2$ and
            \begin{align}
                & \res{x}(\pctx[\big]{N}[\flat{\pctx{M}}[\alpha_i;P_i]_{i \in I}] \| \pctx[\big]{N'}[\flat{\pctx{M'}}[\beta_1;Q_1]])
                \nonumber
                \\
                &= \res{x}(\pctx[\big]{N}[\flat{\pctx{C}[\pctx{M_1}]}[\alpha_i;P_i]_{i \in I_1} \nd \flat{\pctx{C}[\pctx{M_2}]}[\alpha_i;P_i]_{i \in I_2}] \| \pctx[\big]{N'}[\pctx{C'}[\beta_1;Q_1]]).
                &&\text{(by~\eqref{eq:oneImplTwoBeforeSplit})}
                \label{eq:oneImplTwoAfterSplit}
            \end{align}
            Let $\pctx{N_1} = \pctx[\big]{N}[\hole \nd \flat{\pctx{C}[\pctx{M_2}]}[\alpha_i;P_i]_{i \in I_2}]$.
            \begin{align}
                & \res{x}(\pctx[\big]{N}[\flat{\pctx{M}}[\alpha_i;P_i]_{i \in I}] \| \pctx[\big]{N'}[\flat{\pctx{M'}}[\beta_1;Q_1]])
                \nonumber
                \\
                &= \res{x}(\pctx[\big]{N_1}[\flat{\pctx{C}[\pctx{M_1}]}[\alpha_i;P_i]_{i \in I_1}] \| \pctx[\big]{N'}[\pctx{C'}[\beta_1;Q_1]])
                &&\text{(by~\eqref{eq:oneImplTwoAfterSplit})}
                \nonumber
                \\
                &\redtwo_S R
                &&\text{(by assumption)}
                \label{eq:oneImplTwoLeftRed}
                \\
                & \res{x}(\pctx[\big]{N_1}[\pctx{C}[\pctx{M_1}][\alpha_i;P_i]_{i \in I_1}] \| \pctx[\big]{N'}[\pctx{C'}[\beta_1;Q_1]])
                \nonumber
                \\
                &= \res{x}(\pctx[\big]{N}[\pctx{C}[\pctx{M_1}][\alpha_i;P_i]_{i \in I_1} \nd \flat{\pctx{C}[\pctx{M_2}]}[\alpha_i;P_i]_{i \in I_2}] \| \pctx[\big]{N'}[\pctx{C'}[\beta_1;Q_1]])
                \nonumber
                \\
                &\redtwo_S R
                &&\text{(by IH on~\eqref{eq:oneImplTwoLeftRed})}
                \label{eq:oneImplTwoBeforeRightRed}
            \end{align}
        Let $\pctx{N_2} = \pctx[\big]{N}[\pctx{C}[\pctx{M_1}][\alpha_i;P_i]_{i \in I_1} \nd \hole]$.
            \begin{align}
                & \res{x}(\pctx[\big]{N}[\pctx{C}[\pctx{M_1}][\alpha_i;P_i]_{i \in I_1} \nd \flat{\pctx{C}[\pctx{M_2}]}[\alpha_i;P_i]_{i \in I_2}] \| \pctx[\big]{N'}[\pctx{C'}[\beta_1;Q_1]])
                \nonumber
                \\
                &= \res{x}(\pctx[\big]{N_2}[\flat{\pctx{C}[\pctx{M_2}]}[\alpha_i;P_i]_{i \in I_2}] \| \pctx[\big]{N'}[\pctx{C'}[\beta_1;Q_1]])
                \nonumber
                \\
                &\redtwo_S R
                &&\text{(by~\eqref{eq:oneImplTwoBeforeRightRed})}
                \label{eq:oneImplTwoRightRed}
                \\
                & \res{x}(\pctx[\big]{N_2}[\pctx{C}[\pctx{M_2}][\alpha_i;P_i]_{i \in I_2}] \| \pctx[\big]{N'}[\pctx{C'}[\beta_1;Q_1]])
                \nonumber
                \\
                &= \res{x}(\pctx[\big]{N}[\pctx{C}[\pctx{M_1}][\alpha_i;P_i]_{i \in I_1} \nd \pctx{C}[\pctx{M_2}][\alpha_i;P_i]_{i \in I_2}] \| \pctx[\big]{N'}[\pctx{C'}[\beta_1;Q_1]])
                \nonumber
                \\
                &= \res{x}(\pctx[\Big]{N}[\pctx[\big]{C}[\pctx{M_1}[\alpha_i;P_i]_{i \in I_1}] \nd \pctx[\big]{C}[\pctx{M_2}[\alpha_i;P_i]_{i \in I_2}]] \| \pctx[\big]{N'}[\pctx{C'}[\beta_1;Q_1]])
                \nonumber
                \\
                &\redtwo_S R
                &&\text{(by IH on~\eqref{eq:oneImplTwoRightRed})}
                \label{eq:oneImplTwoIHRed}
                \\
                & \res{x}(\pctx[\Big]{N}[\pctx[\big]{C}[\pctx{M_1}[\alpha_i;P_i]_{i \in I_1} \nd \pctx{M_2}[\alpha_i;P_i]_{i \in I_2}]] \| \pctx[\big]{N'}[\pctx{C'}[\beta_1;Q_1]])
                \nonumber
                \\
                &= \res{x}(\pctx[\Big]{N}[\pctx[\big]{C}[(\pctx{M_1} \nd \pctx{M_2})[\alpha_i;P_i]_{i \in I}]] \| \pctx[\big]{N'}[\pctx{C'}[\beta_1;Q_1]])
                \nonumber
                \\
                &= \res{x}(\pctx[\big]{N}[\pctx{C}[\pctx{M_1} \nd \pctx{M_2}][\alpha_i;P_i]_{i \in I}] \| \pctx[\big]{N'}[\pctx{C'}[\beta_1;Q_1]])
                \nonumber
                \\
                &= \res{x}(\pctx[\big]{N}[\pctx{M}[\alpha_i;P_i]_{i \in I}] \| \pctx[\big]{N'}[\pctx{M'}[\beta_1;Q_1]])
                \nonumber
                \\
                &\redtwo_S R
                &&\text{(by rule $\rredtwo{\nu\nd}$ on~\eqref{eq:oneImplTwoIHRed})}
                \nonumber
                \tag*{\qedhere}
            \end{align}
    \end{itemize}
\end{proof}

\begin{theorem}\label{t:oneImpliesTwo}
    If $P \vdash \Gamma$ and $P \redone R$, then $P \redtwo_S R$.
\end{theorem}

\begin{proof}
    By induction on the derivation of the reduction.
    The inductive cases of rules $\rredone{\equiv}$, $\rredone{\nu}$, $\rredone{\|}$, and $\rredone{\nd}$ follow from the IH straightforwardly, using the corresponding closure rule for $\redtwo$.
    As representative base case, we consider rule $\rredone{\1\bot}$: $P = \res{x}(\pctx{N}[\pclose{x}] \| \pctx{N'}[\gclose{x};Q]) \redone R$.

    By inversion of typing, $\pctx{N}[\pclose{x}] \vdash \Gamma, x{:}A$, so by \Cref{l:ndCtxMultihole}, there are $\pctx{M}$ and ${(\alpha_i;P_i)}_{i \in I}$ such that $\pctx{N}[\pclose{x}] = \pctx{M}[\alpha_i;P_i]_{i \in I}$ where $x \notin \fn{\pctx{M}}$ and $x \in \bigcap_{i \in I} \fn{\alpha_i;P_i}$ and there is $i' \in I$ such that $\alpha_{i'};P_{i'} = \pclose{x}$.
    Similarly, there are $\pctx{M'}$ and ${(\beta_j;Q_j)}_{j \in J}$ such that $\pctx{N'}[\gclose{x};Q] = \pctx{M'}[\beta_j;Q_j]_{j \in J}$ where $x \notin \fn{\pctx{M'}}$ and $x \in \bigcap_{j \in J} \fn{\beta_j;Q_j}$ and there is $j' \in J$ such that $\beta_{j'};Q_{j'} = \gclose{x};Q$.

    By \Cref{l:multiShape}, $\flat{\pctx{M}}[\alpha_i;P_i]_{i \in I} = \bignd_{i \in I} \pctx{C_i}[\alpha_i;P_i]$ and $\flat{\pctx{M'}}[\beta_j;Q_j]_{j \in J} = \bignd_{j \in J} \pctx{C'_j}[\beta_j;Q_j]$.
    By typability, there is $I' \subseteq I$ such that $\forall i \in I'.~ \alpha_i \relalpha \pclose{x}$ and $\forall i \in I \setminus I'.~ \alpha_i \not\relalpha \pclose{x}$; hence, $i' \in I'$.
    Similarly, there is $J' \subseteq J$ such that $\forall j \in J'.~ \beta_j \relalpha \gclose{x}$ and $\forall j \in J \setminus J'.~ \beta_j \not\relalpha \gclose{x}$; hence, $j' \in J'$.
    Then, by \Cref{d:rpreone}, $\flat{\pctx{M}}[\alpha_i;P_i]_{i \in I} \piprecong{x} \bignd_{i \in I'} \pctx{C_i}[\pclose{x}]$ and $\flat{\pctx{M'}}[\beta_j;Q_j]_{j \in J} \piprecong{x} \bignd_{j \in J'} \pctx{C_j}[\gclose{x};Q_j]$.
    By rule $\rredtwo{\1\bot}$,
    \begin{align*}
        \res{x}(\bignd_{i \in I'} \pctx{C_i}[\pclose{x}] \| \bignd_{j \in J'} \pctx{C_j}[\gclose{x};Q_j]) &\redtwo_x R,
    \end{align*}
    so by rule $\rredtwo{\piprecong{x}}$,
    \begin{align*}
        \res{x}(\flat{\pctx{M}}[\alpha_i;P_i]_{i \in I} \| \flat{\pctx{M'}}[\beta_j;Q_j]_{j \in J}) \redtwo_x R.
    \end{align*}
    Then, by \Cref{l:flatRed},
    \begin{align*}
        \res{x}(\pctx{M}[\alpha_i;P_i]_{i \in I} \| \pctx{M'}[\beta_j;Q_j]_{j \in J}) \redtwo_x R.
    \end{align*}

    As second base case, we consider rule $\rredone{\scc{Id}}$: $P = \res{x}(\pctx[\big]{N}[\pfwd{x}{y}] \| Q) \redone R$.

    By inversion of typing, $ \pctx[\big]{N}[\pfwd{x}{y}]  \vdash \Gamma, x{:}A, y{:}\dual{A}$, so by \Cref{l:ndCtxMultihole}, there are $\pctx{M}$ and ${(\alpha_i;P_i)}_{i \in I}$ such that $\pctx[\big]{N}[\pfwd{x}{y}]  = \pctx{M}[\alpha_i;P_i]_{i \in I}$ where $x \notin \fn{\pctx{M}}$ and $x \in \bigcap_{i \in I} \fn{\alpha_i;P_i}$ and there is $i' \in I$ such that $\alpha_{i'};P_{i'} = \pfwd{x}{y} $.

    By \Cref{l:multiShape}, $\flat{\pctx{M}}[\alpha_i;P_i]_{i \in I} = \bignd_{i \in I} \pctx{C_i}[\alpha_i;P_i]$.
    By typability, there is $I' \subseteq I$ such that $\forall i \in I'.~ \alpha_i = \pfwd{x}{y}$ and $\forall i \in I \setminus I'.~ \alpha_i \not= \pfwd{x}{y}$; hence, $i' \in I'$.

    Then, by \Cref{d:rpreone}, $\flat{\pctx{M}}[\alpha_i;P_i]_{i \in I} \piprecong{x,y} \bignd_{i \in I'} \pctx{C_i}[\pfwd{x}{y}]$.
    By rule $\rredtwo{\scc{Id}}$,
    \begin{align*}
        \res{x} \Big( \bignd_{i \in I}\pctx[\big]{C_i}[\pfwd{x}{y}] \| Q\Big)
                \redtwo_{x,y}
                 R,
    \end{align*}
    so by rule $\rredtwo{\piprecong{x,y}}$,
    \begin{align*}
        \res{x}(\flat{\pctx{M}}[\alpha_i;P_i]_{i \in I} \| Q ) \redtwo_{x,y} R.
    \end{align*}
    Then, by \Cref{l:flatRed},
    \begin{align*}
        \res{x}(\pctx{M}[\alpha_i;P_i]_{i \in I} \| Q ) \redtwo_{x,y} R.
        \tag*{\qedhere}
    \end{align*}
\end{proof}

\begin{theorem}[DF: Lazy Semantics]\label{t:dlfreeTwo}
    If $P \vdash \emptyset$ and $P \not\equiv \0$, then $P \redtwo_S R$ for some $S$ and $R$.
\end{theorem}

\begin{proof}
    As a corollary of \Cref{t:dlfreeOne,t:oneImpliesTwo}.
\end{proof}

%% file: appendix/lambda-correct.tex
\section{Proofs of Subject Reduction and Subject Expansion for \texorpdfstring{\lamcoldetsh}{Lambda}}

Here we prove \Cref{t:lamSRShort,t:lamSEShort} (subject reduction and subject expansion, respectively) for \lamcoldetsh.

\subsection{Subject Reduction}\label{a:lamTypes}

\begin{lemma}[Substitution Lemma for $\lamcoldetsh$]\label{l:lamrsharfailsubsunres}
    \leavevmode
    \begin{enumerate}
        \item (Linear) If $\Theta ; \Gamma ,  {x}:\sigma \wfdash M: \tau$, $\headf{M} =  {x}$, and $\Theta ; \Delta \wfdash N : \sigma$
            then
            $\Theta ; \Gamma  , \Delta \wfdash M \headlin{ N /  {x} }:\tau$.
        \item (Unrestricted) If $\Theta, \unvar{x}: \eta ; \Gamma \wfdash M: \tau$, $\headf{M} = {x}[i]$, $\eta_i = \sigma $, and $\Theta ; \cdot \wfdash N : \sigma$
            then
            $\Theta, \unvar{x}: \eta ; \Gamma  \wfdash M \headlin{ N / {x}[i] }$.
    \end{enumerate}
\end{lemma}

\begin{proof}
    \leavevmode
    \begin{enumerate}
        \item By structural induction on $M$ with $\headf{M}=  {x}$.
            There are six cases to be analyzed:
            \begin{enumerate}
                \item $M= {x}$

                    In this case, $\Theta ;  {x}:\sigma \wfdash  {x}:\sigma$ and $\Gamma=\emptyset$.  Observe that $ {x}\headlin{N/ {x}}=N$, since $\Delta\wfdash N:\sigma$, by hypothesis, the result follows.

                \item $M = M'\ B$.

                    Then $\headf{M'\ B} = \headf{M'} =  {x}$, and the derivation is the following:
                    \begin{prooftree}
                        \AxiomC{$\Theta ; \Gamma_1 ,  {x}:\sigma \wfdash M': (\delta^{j} , \eta )  \rightarrow \tau$}\
                        \AxiomC{\quad $\Theta ; \Gamma_2 \wfdash B : (\delta^{k} , \epsilon ) $}
                        \AxiomC{\quad $ \eta \relunbag \epsilon $}
                        \LeftLabel{\redlab{FS{:}app}}
                        \TrinaryInfC{$\Theta ; \Gamma_1 , \Gamma_2 ,  {x}:\sigma \wfdash M'B:\tau $}
                    \end{prooftree}
                    where $\Gamma=\Gamma_1 , \Gamma_2$, and  $j,k$ are non-negative integers, possibly different.  Since $\Delta \vdash N : \sigma$, by IH, the result holds for $M'$, that is,
                    \[\Gamma_1 , \Delta \wfdash M'\headlin{ N /  {x} }: (\delta^{j} , \eta )  \rightarrow \tau\]
                    which gives the  derivation:
                    \begin{prooftree}
                        \AxiomC{$\Theta ; \Gamma_1 , \Delta \wfdash M'\headlin{ N /  {x} }: (\delta^{j} , \eta )  \rightarrow \tau$}\
                        \AxiomC{\quad $\Theta ; \Gamma_2 \wfdash B : (\delta^{k} , \epsilon ) $}
                        \AxiomC{\quad$ \eta \relunbag \epsilon $}
                        \LeftLabel{\redlab{FS{:}app}}
                        \TrinaryInfC{$\Theta ; \Gamma_1 , \Gamma_2 , \Delta \wfdash ( M'\headlin{ N /  {x} } ) B:\tau $}
                    \end{prooftree}
                    From \Cref{f:lambda_red},   $(M'B) \headlin{ N /  {x} } = ( M'\headlin{ N /  {x} } ) B$, and the result follows.
                \item $M = M'[ {\widetilde{y}} \leftarrow  {y}] $.

                    Then $ \headf{M'[ {\widetilde{y}} \leftarrow  {y}]} = \headf{M'}= {x}$, for  $y\neq x$. Therefore,
                    \begin{prooftree}
                        \AxiomC{$\Theta ; \Gamma_1 ,  {y}_1: \delta, \dots,  {y}_k: \delta ,  {x}: \sigma \wfdash M' : \tau \quad  {y}\notin \Gamma_1 \quad k \not = 0$}
                        \LeftLabel{ \redlab{FS{:}share}}
                        \UnaryInfC{$ \Theta ; \Gamma_1 ,  {y}: \delta^k,  {x}: \sigma \wfdash M'[ {y}_1 , \dots ,  {y}_k \leftarrow x] : \tau $}
                    \end{prooftree}
                    where $\Gamma=\Gamma_1 ,  {y}: \delta^k$.
                    By IH, the result follows for $M'$, that is,
                    \[\Theta ; \Gamma_1 ,  {y}_1: \delta, \dots,  {y}_k: \delta ,\Delta \wfdash M'\headlin{N/ {x}} : \tau \]
                    and we have the derivation:
                    \begin{prooftree}
                        \AxiomC{$ \Theta ; \Gamma_1 ,  {y}_1: \delta, \dots,  {y}_k: \delta , \Delta \wfdash  M' \headlin{ N /  {x}} : \tau \quad  {y}\notin \Gamma_1 \quad k \not = 0$}
                        \LeftLabel{ \redlab{FS{:}shar} }
                        \UnaryInfC{$ \Theta ; \Gamma_1 ,  {y}: \delta^k, \Delta \wfdash M' \headlin{ N /  {x}} [ {\widetilde{y}} \leftarrow  {y}] : \tau $}
                    \end{prooftree}
                    From \Cref{f:lambda_red},  $M'[ {\widetilde{y}} \leftarrow  {y}] \headlin{ N /  {x} } = M' \headlin{ N /  {x}} [ {\widetilde{y}} \leftarrow  {y}]$, 
                    and the result follows.

                \item $M = M'[ \leftarrow  {y}] $.

                    Then $ \headf{M'[ \leftarrow  {y}]} = \headf{M'}= {x}$ with  $x \not  = y $,
                    \begin{prooftree}
                        \AxiomC{$ \Theta ; \Gamma  ,  {x}: \sigma  \wfdash M : \tau$}
                        \LeftLabel{ \redlab{FS{:}weak} }
                        \UnaryInfC{$ \Theta ;  \Gamma  ,  {y}: \omega,  {x}: \sigma  \wfdash M[\leftarrow  {y}]: \tau $}
                    \end{prooftree}
                    and $M'[ \leftarrow  {y}] \headlin{ N /  {x} } = M' \headlin{ N /  {x}} [ \leftarrow  {y}]$. Then by the induction hypothesis:
                    \begin{prooftree}
                        \AxiomC{$ \Theta ;  \Gamma , \Delta  \wfdash M \headlin{ N /  {x}}: \tau$}
                        \LeftLabel{ \redlab{FS{:}weak}}
                        \UnaryInfC{$ \Theta ;  \Gamma  ,  {y}: \omega, \Delta \wfdash M\headlin{ N /  {x}}[\leftarrow  {y}]: \tau $}
                    \end{prooftree}

                \item If $M =  M' \linexsub{C /  y_1 , \dots , y_k} $.

                    Then $\headf{M' \linexsub{C /  y_1 , \dots , y_k}} = \headf{M'} = x \not = y_1 , \dots , y_k$,
                    \begin{prooftree}
                            \AxiomC{$ \Theta ; \Gamma_1  ,  y_1:\delta, \dots , y_k:\delta , x: \sigma  \wfdash M' : \tau $}
                        \AxiomC{$\quad \Theta ; \Gamma_2 \wfdash C : \delta^k $}
                            \LeftLabel{\redlab{FS{:}Esub^{\ell}}}
                        \BinaryInfC{$ \Theta ; \Gamma_1 , \Gamma_2, x: \sigma  \wfdash M' \linexsub{C /  y_1 , \dots , y_k} : \tau $}
                    \end{prooftree}
                     and $M' \linexsub{C /  y_1 , \dots , y_k}  \headlin{ N / x } = M' \headlin{ N / x } \linexsub{C /  y_1 , \dots , y_k}  $. Then by the induction hypothesis:
                    \begin{prooftree}
                        \AxiomC{$ \Theta ; \Gamma_1  , \Delta,  y_1:\delta, \dots , y_k:\delta  \wfdash M' \headlin{ N / x } : \tau $}
                        \AxiomC{$\quad \Theta ; \Gamma_2 \wfdash C : \delta^k $}
                        \LeftLabel{\redlab{FS{:}Esub^{\ell}}}
                        \BinaryInfC{$ \Theta ; \Gamma_1 , \Gamma_2 , \Delta  \wfdash M' \headlin{ N / x } \linexsub{C /  y_1 , \dots , y_k} : \tau $}
                    \end{prooftree}

                \item If $M =  M' \unexsub {U / {y}} $ then $\headf{M' \unexsub {U / {y}}} = \headf{M'} = x $, and the proofs is similar to the case above.
            \end{enumerate}

        \item  By structural induction on $M$ with $\headf{M}= {x}[i]$.
            There are three cases to be analyzed:
            \begin{enumerate}
                \item $M= {x}[i]$.

                    In this case,
                    \begin{prooftree}
                        \AxiomC{}
                        \LeftLabel{ \redlab{FS{:}var^{ \ell}}}
                        \UnaryInfC{$ \Theta , \unvar{x}: \eta;  {x}: \eta_i  \wfdash  {x} : \sigma$}
                        \LeftLabel{\redlab{FS{:}var^!}}
                        \UnaryInfC{$ \Theta, \unvar{x}: \eta ; \cdot \wfdash {x}[i] : \sigma$}
                    \end{prooftree}
                    and $\Gamma=\emptyset$.  Observe that ${x}[i]\headlin{N/{x}[i]}=N$, since $\Theta, \unvar{x}: \eta  ; \Gamma  \wfdash M \headlin{ N / {x}[i] }$, by hypothesis, the result follows.

                \item $M = M'\ B$.

                    In this case, $\headf{M'\ B} = \headf{M'} =  {x}[i]$, and one has the following derivation:
                    \begin{prooftree}
                        \AxiomC{$ \Theta, \unvar{x}: \eta ;\Gamma_1 \wfdash M : (\delta^{j} , \epsilon ) \rightarrow \tau \quad \Theta, \unvar{x}: \sigma ; \Gamma_2 \wfdash B : (\delta^{k} , \epsilon' )  $}
                        \AxiomC{$ \epsilon \relunbag \epsilon' $}
                        \LeftLabel{\redlab{FS{:}app}}
                        \BinaryInfC{$ \Theta, \unvar{x}: \eta ;\Gamma_1 , \Gamma_2 \wfdash M\ B : \tau$}
                    \end{prooftree}
                    where $\Gamma=\Gamma_1 , \Gamma_2$, $\delta$ is a strict type and $j,k$ are non-negative  integers, possibly different.

                    By the induction hypothesis, we get $\Theta, \unvar{x}: \eta ;\Gamma_1 \wfdash M'\headlin{N/ {x[i]}}:(\delta^{j} , \epsilon ) \rightarrow \tau $, which gives the derivation:
                    \begin{prooftree}
                        \AxiomC{$\Theta , \unvar{x}: \eta;\Gamma_1\wfdash M'\headlin{N/ {x}[i]}:(\delta^{j} , \epsilon ) \rightarrow \tau $}\
                        \AxiomC{$\Theta , \unvar{x}: \eta; \Gamma_2 \wfdash B : (\delta^{k} , \epsilon' ) $}
                        \AxiomC{$ \epsilon \relunbag \epsilon' $}
                        \LeftLabel{\redlab{FS{:}app}}
                        \TrinaryInfC{$\Theta , \unvar{x}: \eta;\Gamma_1 , \Gamma_2  \wfdash ( M'\headlin{ N / {x}[i] } ) B:\tau $}
                    \end{prooftree}
                    From \Cref{f:lambda_red}, $M' \esubst{ B }{ y} \headlin{ N / {x}[i] } = M' \headlin{ N / {x}[i] } \esubst{ B }{ y}$, and  the result follows.

                \item $M = M'[ {\widetilde{y}} \leftarrow  {y}] $.

                    Then $ \headf{M'[ {\widetilde{y}} \leftarrow  {y}]} = \headf{M'}= {x}[i]$, for  $y\neq x$. Therefore,
                    \begin{prooftree}
                        \AxiomC{$\Theta , \unvar{x}: \eta; \Gamma_1 ,  {y}_1: \delta, \dots,  {y}_k: \delta  \wfdash M' : \tau \quad  {y}\notin \Gamma_1 \quad k \not = 0$}
                        \LeftLabel{ \redlab{FS{:}shar}}
                        \UnaryInfC{$ \Theta , \unvar{x}: \eta; \Gamma_1 ,  {y}: \delta^k \wfdash M'[ {y}_1 , \dots ,  {y}_k \leftarrow y] : \tau $}
                    \end{prooftree}
                    where $\Gamma=\Gamma_1 ,  {y}: \delta^k$.
                    By the induction hypothesis, the result follows for $M'$, that is,
                    \[\Theta, \unvar{x}: \eta ; \Gamma_1 ,  {y}_1: \delta, \dots,  {y}_k: \delta  \wfdash M'\headlin{N/ {x}[i] } : \tau \] and we have the derivation:
                    \begin{prooftree}
                        \AxiomC{$ \Theta , \unvar{x}: \eta; \Gamma_1 ,  {y}_1: \delta, \dots,  {y}_k: \delta  \wfdash  M' \headlin{ N / {x}[i] } : \tau \quad  {y}\notin \Gamma_1 \quad k \not = 0$}
                        \LeftLabel{ \redlab{FS{:}shar} }
                        \UnaryInfC{$ \Theta , \unvar{x}: \eta; \Gamma_1 ,  {y}: \delta^k \wfdash M' \headlin{ N / {x}[i]} [ {\widetilde{y}} \leftarrow  {y}] : \tau $}
                    \end{prooftree}
                    From \Cref{f:lambda_red}  $M'[ {\widetilde{y}} \leftarrow  {y}] \headlin{ N / {x}[i] } = M' \headlin{ N / {x}[i]} [ {\widetilde{y}} \leftarrow  {y}]$, 
                    and the result follows.

                \item $M = M'[ \leftarrow  {y}] $.

                    Then $ \headf{M'[ \leftarrow  {y}]} = \headf{M'}= {x}[i]$ with  $x \not  = y $,
                    \begin{prooftree}
                        \AxiomC{$ \Theta , \unvar{x}: \eta; \Gamma   \wfdash M : \tau$}
                        \LeftLabel{ \redlab{FS{:}weak} }
                        \UnaryInfC{$ \Theta , \unvar{x}: \eta;  \Gamma  ,  {y}: \omega \wfdash M[\leftarrow  {y}]: \tau $}
                    \end{prooftree}
                    and $M'[ \leftarrow  {y}] \headlin{ N / {x}[i] } = M' \headlin{ N / {x}[i] } [ \leftarrow  {y}]$.
                    By the induction hypothesis:
                    \begin{prooftree}
                        \AxiomC{$ \Theta , \unvar{x}: \eta;  \Gamma   \wfdash M \headlin{ N / {x}[i]  }: \tau$}
                        \LeftLabel{ \redlab{FS{:}weak}}
                        \UnaryInfC{$ \Theta , \unvar{x}: \eta;  \Gamma  ,  {y}: \omega \wfdash M\headlin{ N / {x}[i] }[\leftarrow  {y}]: \tau $}
                    \end{prooftree}

                \item $M =   M' \linexsub{C /  y_1 , \dots , y_k} $.

                    Then $\headf{ M' \linexsub{C /  y_1 , \dots , y_k} } = \headf{M'} = {x}[i]$ with $x \not = y$,
                    \begin{prooftree}
                        \AxiomC{$ \Theta, \unvar{x}: \eta ; \Gamma  ,  y_1:\delta, \dots , y_k:\delta \wfdash M : \tau $}
                        \AxiomC{$ \Theta, \unvar{x}: \eta ; \Delta \wfdash C : \delta^k $}
                        \LeftLabel{\redlab{FS{:}Esub^{\ell}}}
                        \BinaryInfC{$ \Theta, \unvar{x}: \eta ; \Gamma , \Delta \wfdash M \linexsub{C /  y_1, \dots , y_k} : \tau $}
                     \end{prooftree}
                    and $M' \linexsub{C /  y_1 , \dots , y_k} \headlin{ N / {x}[i]  } = M' \headlin{ N / {x}[i]  }  \linexsub{C /  y_1 , \dots , y_k}  $.
                    By the induction hypothesis:
                    \begin{prooftree}
                        \AxiomC{$ \Theta, \unvar{x}: \eta ; \Gamma  ,  y_1:\delta, \dots , y_k:\delta \wfdash M' \headlin{ N / {x}[i]  }  : \tau $}
                        \AxiomC{$ \Theta, \unvar{x}: \eta ; \Delta \wfdash C : \delta^k $}
                        \LeftLabel{\redlab{FS{:}Esub^{\ell}}}
                        \BinaryInfC{$ \Theta, \unvar{x}: \eta ; \Gamma , \Delta \wfdash M' \headlin{ N / {x}[i]  }  \linexsub{C /  y_1 , \dots , y_k} : \tau $}
                    \end{prooftree}

                \item $M =  M' \unexsub {U /\unvar{y}}$.

                    Then $\headf{M' \unexsub {U /\unvar{y}}} = \headf{M'} = {x}[i] $,
                    \begin{prooftree}
                        \AxiomC{$ \Theta , \unvar{x}: \eta , {y} : \epsilon_1; \Gamma  \wfdash M' : \tau $}
                        \AxiomC{$ \Theta , \unvar{x}: \eta; \dash \wfdash U : \epsilon_2 $}
                        \AxiomC{$ \epsilon_1 \relunbag \epsilon_2 $}
                            \LeftLabel{\redlab{FS{:}Esub^!}}
                        \TrinaryInfC{$ \Theta ; \Gamma \wfdash M' \unexsub{U / \unvar{y}}  : \tau $}
                    \end{prooftree}
                    and $M' \unexsub {U /\unvar{y}} \headlin{ N /  {x}[i] } = M'  \headlin{N /  {x}[i] } \unexsub {U /\unvar{y}}$.
                    Then by the induction hypothesis:
                    \begin{prooftree}
                        \AxiomC{$ \Theta , \unvar{x}: \eta , {y} : \epsilon_1; \Gamma  \wfdash M'  \headlin{N /  {x}[i] } : \tau $}
                        \AxiomC{$ \Theta , \unvar{x}: \eta; \dash \wfdash U : \epsilon_2 $}
                        \AxiomC{$ \epsilon_1 \relunbag \epsilon_2 $}
                        \LeftLabel{\redlab{FS{:}Esub^!}}
                        \TrinaryInfC{$ \Theta ; \Gamma \wfdash M'  \headlin{N /  {x}[i] } \unexsub{U / \unvar{y}}  : \tau $}
                    \end{prooftree}
                    \qedhere
            \end{enumerate}
    \end{enumerate}
\end{proof}

\begin{theorem}[SR in $\lamcoldetsh$]\label{t:lamSR}
    If $\Theta ; \Gamma \wfdash M:\tau$ and $M \red M'$ then $\Theta ; \Gamma \wfdash M' :\tau$.
\end{theorem}

\begin{proof}
    By structural induction on the reduction rule from \figref{fig:reduc_interm} applied in $M \red M'$.
    \begin{enumerate}
        \item \textbf{ Rule $\redlab{RS{:}Beta}$.}

            Then $M = (\lambda x. N[ {\widetilde{x}} \leftarrow  {x}]) B $  and the reduction is:

            \begin{prooftree}
                \AxiomC{}
                \LeftLabel{\redlab{RS{:}Beta}}
                \UnaryInfC{$(\lambda x. N[ {\widetilde{x}} \leftarrow  {x}]) B \red N[ {\widetilde{x}} \leftarrow  {x}]\ \esubst{ B }{ x }$}
            \end{prooftree}
            where $ M'  =  N[ {\widetilde{x}} \leftarrow  {x}]\ \esubst{ B }{ x }$. Since $\Theta ; \Gamma\wfdash M:\tau$ we get the following derivation:
            \begin{prooftree}
                \AxiomC{$\Theta , \unvar{x} : \eta; \Gamma' ,  {x}_1:\sigma , \dots ,  {x}_j:\sigma  \wfdash  N: \tau $}
                \LeftLabel{ \redlab{FS{:}share} }
                \UnaryInfC{$\Theta , \unvar{x} : \eta;  \Gamma' ,   {x}:\sigma^{j}  \wfdash  N[ {\widetilde{x}} \leftarrow  {x}]: \tau $}
                \LeftLabel{ \redlab{FS{:}abs \dash sh} }
                \UnaryInfC{$\Theta ; \Gamma' \wfdash \lambda x. N[ {\widetilde{x}} \leftarrow  {x}]: (\sigma^{j} , \eta ) \rightarrow \tau $}
                \AxiomC{$\Theta ;\Delta \wfdash B: (\sigma^{k} , \epsilon ) $}
                \AxiomC{$ \eta \relunbag \epsilon $}
                \LeftLabel{ \redlab{FS{:}app} }
                \TrinaryInfC{$ \Theta ;\Gamma' , \Delta \wfdash (\lambda x. N[ {\widetilde{x}} \leftarrow  {x}]) B:\tau $}
            \end{prooftree}
            for $\Gamma = \Gamma' , \Delta $ and $x\notin \dom{\Gamma'}$.
            Notice that:
            \begin{prooftree}
                \AxiomC{$\Theta , \unvar{x} : \eta; \Gamma' ,  {x}_1:\sigma , \dots ,  {x}_j:\sigma  \wfdash  N: \tau $}
                \LeftLabel{ \redlab{FS{:}share} }
                \UnaryInfC{$\Theta , \unvar{x} : \eta;  \Gamma' ,   {x}:\sigma^{j}  \wfdash  N[ {\widetilde{x}} \leftarrow  {x}]: \tau $}
                \AxiomC{$\Theta ;\Delta \wfdash B:(\sigma^{k} , \epsilon )  $}
                \AxiomC{$ \eta \relunbag \epsilon $}
                \LeftLabel{ \redlab{FS{:}Esub} }
                \TrinaryInfC{$ \Theta ;\Gamma' , \Delta \wfdash N[ {\widetilde{x}} \leftarrow  {x}]\ \esubst{ B }{ x }:\tau $}
            \end{prooftree}
            Therefore $ \Theta ; \Gamma' ,\Delta\wfdash M' :\tau$ and the result follows.

        \item \textbf{ Rule $ \redlab{RS{:}Ex \dash Sub}$}.

            Then $ M =  N[ {x}_1, \dots ,  {x}_k \leftarrow  {x}]\ \esubst{ C \bagsep U }{ x }$ where $C=  \bag{N_1}\cdot \cdots \cdot \bag{N_k} $.
            The reduction is:
            \begin{prooftree}
                \AxiomC{$ C = \bag{N_1}
                \dots  \bag{N_k} \qquad  N \not= \fail^{\widetilde{y}} $}
                \LeftLabel{\redlab{RS{:}Ex \dash Sub}}
                \UnaryInfC{$ N[ {x}_1, \!\dots\! ,  {x}_k \leftarrow  {x}]\esubst{ C \bagsep U }{ x } \red N\linexsub{C  /  x_1 , \dots , x_k} \unexsub{U / \unvar{x} }$}
            \end{prooftree}
            and $M' = N\linexsub{C  /  x_1 , \dots , x_k} \unexsub{U / \unvar{x} }$.
            To simplify the proof we take $k=2$, as the case $k>2$ is similar.
            Therefore $C=\bag{N_1}\cdot \bag{N_2}$.

            Since $\Theta ; \Gamma\wfdash M:\tau$ we get a derivation: (we omit the labels \redlab{FS:Esub} and \redlab{FS{:}share})
            {\small
                \begin{prooftree}
                    \AxiomC{$ \Theta, \unvar{x} : \eta  ;  \Gamma' ,  {x}_1: \sigma,  {x}_2: \sigma \wfdash N : \tau \quad  {x}\notin \dom{\Gamma} \quad k \not = 0$}
                    \UnaryInfC{$  \Theta , \unvar{x} : \eta ; \Gamma' ,  {x}: \sigma^{2} \wfdash N[ {x}_1,  {x}_2 \leftarrow  {x}] : \tau  $}

                    \AxiomC{$ \Theta ; \Delta \wfdash C \bagsep U : (\sigma^{2} , \epsilon ) $}
                    \AxiomC{$ \eta \relunbag \epsilon $}
                    \TrinaryInfC{$ \Theta ; \Gamma' , \Delta \wfdash N[ {x}_1,  {x}_2 \leftarrow  {x}]\esubst{ C \bagsep U }{ x }  : \tau $}
                \end{prooftree}
            }
            where $\Gamma = \Gamma' , \Delta $. Consider the wf derivation for $\Pi_{1,2}$: (we omit the labels \redlab{FS:Esub^!} and \redlab{FS:Esub^{\ell}})
            {\small
                \begin{prooftree}
                    \AxiomC{$ \Theta, \unvar{x} : \eta  ;  \Gamma' ,  {x}_1: \sigma,  {x}_2: \sigma \wfdash N : \tau$}
                    \AxiomC{$ \Theta ; \Delta \wfdash C : \sigma^k $}
                    \BinaryInfC{$ \Theta, \unvar{x} : \eta  ;  \Gamma' , \Delta \wfdash N \linexsub{C  / {x}_1,  {x}_2} : \tau $}
                    \AxiomC{$ \Theta ; \dash \wfdash U : \epsilon $}
                    \AxiomC{$ \eta \relunbag \epsilon $}
                    \TrinaryInfC{$ \Theta ; \Gamma' , \Delta \wfdash N\linexsub{C  / {x}_1,  {x}_2} \unexsub{U / \unvar{x} }  : \tau $}
                \end{prooftree}
            }
            and the result follows.

         \item Rule $ \redlab{RS{:}Fetch^{\ell}}$.

            Then $ M =  N \linexsub{C /  \widetilde{x}, x_j}  $ where  $\headf{N} =  {x}_j$.
            The reduction is:
            \begin{prooftree}
                \AxiomC{$ \headf{N} =  {x}_j$}
                 \LeftLabel{\redlab{RS{:}Fetch^{\ell}_{{i}}}}
                 \UnaryInfC{$  N \linexsub{C /  \widetilde{x}, x_j} \red  N \headlin{ C_i / x_j }  \linexsub{(C \setminus C_i ) /  \widetilde{x}  } $}
            \end{prooftree}
            and $M' =  N \headlin{ C_i / x_j }  \linexsub{(C \setminus C_i ) /  \widetilde{x}  } $.
            Since $\Theta ; \Gamma\wfdash M:\tau$ we get the following derivation:
            \begin{prooftree}
                \AxiomC{$ \Theta ; \Gamma'  ,  \widetilde{x}:\sigma^{k-1},  x_j:\sigma \wfdash N : \tau $}
                \AxiomC{$ \Theta ; \Delta \wfdash C : \sigma^k $}
                \LeftLabel{\redlab{FS{:}Esub^{\ell}}}
                \BinaryInfC{$ \Theta ; \Gamma' , \Delta \wfdash N \linexsub{C /  \widetilde{x}, x_j}  : \tau $}
            \end{prooftree}
            where $\Gamma = \Gamma' , \Delta $ and $\Delta = \Delta_i , \Delta_i'$ with $ \Theta ;  \Delta_i \wfdash  C_i : \sigma $.
            By \Cref{l:lamrsharfailsubsunres}, we obtain the derivation $ \Theta ; \Gamma' , \Delta \wfdash   N \headlin{ C_i / x_j }  \linexsub{(C \setminus C_i ) /  \widetilde{x}  } : \tau $ via:
            \begin{prooftree}
                \AxiomC{$ \Theta ; \Gamma' , \Delta_i ,  \widetilde{x}:\sigma^{k-1} \wfdash N \headlin{ C_i / x_j }: \tau $}
                \AxiomC{$ \Theta ; \Delta'_i \wfdash C \setminus C_i : \sigma^{k-1} $}
                \LeftLabel{\redlab{FS{:}Esub^{\ell}}}
                \BinaryInfC{$ \Theta ; \Gamma', \Delta \wfdash N \headlin{ C_i / x_j }  \linexsub{(C \setminus C_i ) /  \widetilde{x}  }  : \tau $}
            \end{prooftree}

        \item Rule $ \redlab{RS{:} Fetch^!}$.

            Then $ M =  N \unexsub{U /  \unvar{x}}  $ where  $\headf{M} = {x}[i]$. The reduction is:
            \begin{prooftree}
                \AxiomC{$ \headf{N} = {x}[i]$}
                \AxiomC{$ U_i = \unvar{\bag{N_i}}$}
                \LeftLabel{\redlab{RS{:} Fetch^!}}
                \BinaryInfC{$  N \unexsub{U / \unvar{x}} \red  N \headlin{ N_i /{x}[i] }\unexsub{U / \unvar{x}} $}
            \end{prooftree}
            and $M' =  N \headlin{ N_i /{x}[i] }\unexsub{U / \unvar{x}}  $.     Since $\Theta ; \Gamma \wfdash M:\tau$ we get the following derivation:
            \begin{prooftree}
                \AxiomC{$ \Theta , {x} : \eta; \Gamma  \wfdash N : \tau $}
                \AxiomC{$ \Theta ; \dash \wfdash U : \epsilon $}
                \AxiomC{$ \eta \relunbag \epsilon $}
                \LeftLabel{\redlab{FS{:}Esub^!}}
                \TrinaryInfC{$ \Theta ; \Gamma \wfdash N \unexsub{U / \unvar{x}}  : \tau $}
            \end{prooftree}
            By \Cref{l:lamrsharfailsubsunres}, we obtain the derivation $  \Theta , \unvar{x} : \eta; \Gamma  \wfdash  N \headlin{ N_i /{x}[i] } $, and the result follows from:
            \begin{prooftree}
                \AxiomC{$ \Theta , \unvar{x} : \eta; \Gamma  \wfdash N \headlin{ N_i /{x}[i] } : \tau $}
                \AxiomC{$ \Theta ; \dash \wfdash U : \epsilon $}
                \AxiomC{$ \eta \relunbag \epsilon $}
                \LeftLabel{\redlab{FS{:}Esub^!}}
                \TrinaryInfC{$ \Theta ; \Gamma \wfdash N \headlin{ N_i /{x}[i] }\unexsub{U / \unvar{x}}  : \tau $}
            \end{prooftree}

        \item Rule $\redlab{RS{:}TCont}$.

            Then $M = C[N]$ and the reduction is as follows:
            \begin{prooftree}
                    \AxiomC{$   N \red N' $}
                    \LeftLabel{\redlab{RS{:}TCont}}
                    \UnaryInfC{$ C[N] \red  C[N'] $}
            \end{prooftree}
            with $M'=  C[N'] $.
            The proof proceeds by analysing the context $C$.
            There are three cases:
            \begin{enumerate}
                \item $C=[\cdot]\ B$.

                    In this case $ M = N \ B$, for some $B$.
                    Since $\Gamma\vdash M:\tau$ one has a derivation:
                    \begin{prooftree}
                        \AxiomC{$ \Theta ;\Gamma' \wfdash N : (\sigma^{j} , \eta ) \rightarrow \tau $}
                        \AxiomC{$  \Theta ;\Delta \wfdash B : (\sigma^{k} , \epsilon )  $}
                        \AxiomC{$ \eta \relunbag \epsilon $}
                        \LeftLabel{\redlab{FS{:}app}}
                        \TrinaryInfC{$ \Theta ; \Gamma' , \Delta \wfdash N\ B : \tau$}
                    \end{prooftree}
                    where $\Gamma = \Gamma' , \Delta $.
                    From  $\Gamma'\wfdash N:\sigma^j\rightarrow\tau$ and the reduction $N \red N' $, one has by IH that  $\Gamma'\wfdash N':\sigma^j\rightarrow\tau$.
                    Finally, we may type the following:
                    \begin{prooftree}
                        \AxiomC{$ \Theta ;\Gamma' \wfdash N' : (\sigma^{j} , \eta ) \rightarrow \tau $}
                        \AxiomC{$  \Theta ;\Delta \wfdash B : (\sigma^{k} , \epsilon )  $}
                        \AxiomC{$ \eta \relunbag \epsilon $}
                        \LeftLabel{\redlab{FS{:}app}}
                        \TrinaryInfC{$ \Theta ; \Gamma' , \Delta \wfdash N' \ B : \tau$}
                    \end{prooftree}
                    Since $ M'  =   (C[N']) = N'B $, the result follows.

                \item  Cases $C=[\cdot]\linexsub{N/x} $ and $C=[\cdot][\widetilde{x} \leftarrow x]$ are similar to the previous.
            \end{enumerate}

        \item Rule $ \redlab{RS{:}Fail^{\ell}}$.

            Then $M =    N[x_1 , \dots , x_k \leftarrow  {x}]\ \esubst{C \bagsep U}{ x } $ where $ k \neq \size{C} $, $  \widetilde{y} = (\llfv{N} \setminus \{  \widetilde{x}\} ) \cup \llfv{C} $ and  the reduction is:
            \begin{prooftree}
                \AxiomC{$ k \neq \size{C} $}
                \AxiomC{$  \widetilde{y} = (\llfv{N} \setminus \{  \widetilde{x}\} ) \cup \llfv{C} $}
                \LeftLabel{\redlab{RS{:}Fail^{\ell}}}
                \BinaryInfC{$  N[x_1 , \dots , x_k \leftarrow  {x}]\ \esubst{C \bagsep U}{ x }  \red  \fail^{\widetilde{y}} $}
            \end{prooftree}
            where $M' =   \fail^{\widetilde{y}}$.
            Since $\Theta, x: \eta ; \Gamma' , x_1:\sigma,\ldots, x_k:\sigma \wfdash M$, one has a derivation:
            \begin{prooftree}
                \AxiomC{$ \Theta, x: \eta ; \Gamma' , x_1:\sigma,\ldots, x_k:\sigma \wfdash N: \tau $}
                \LeftLabel{ \redlab{FS{:}Esub} }
                \UnaryInfC{$ \Theta, x: \eta ;\Gamma' , x:\sigma^{k} \wfdash N[x_1, \dots , x_k \leftarrow x] : \tau $}
                \AxiomC{$\Theta ; \Delta \wfdash C \bagsep U : (\sigma^{j} , \epsilon ) $}
                \AxiomC{$ \eta \relunbag \epsilon $}
                \LeftLabel{ \redlab{FS{:}Esub} }
                \TrinaryInfC{$\Theta ; \Gamma' , \Delta \wfdash N[x_1 , \dots , x_k \leftarrow  {x}]\ \esubst{C \bagsep U}{ x }  : \tau $}
            \end{prooftree}
            where $\Gamma = \Gamma' , \Delta $.
            We may type the following:
            \begin{prooftree}
                \AxiomC{$ \dom{\Gamma' , \Delta} = \widetilde{x} $}
                \LeftLabel{ \redlab{FS{:}fail}}
                \UnaryInfC{$\Theta ; \Gamma' , \Delta \wfdash  \fail^{\widetilde{y}} : \tau  $}
            \end{prooftree}
            since $\Gamma' , \Delta$ contain assignments on the free variables in $M$ and $B$.
            Therefore, $\Theta ;\Gamma\wfdash \fail^{\widetilde{y}}:\tau$, by applying \redlab{FS{:}sum} as required.

        \item Rule $ \redlab{RS{:}Fail^!}$.

            Then $M = N \unexsub{U / \unvar{x} } $ where $\headf{M} = {x}[i]$ and $U_i = \unvar{\oneb} $ and  the reduction is:
            \begin{prooftree}
                \AxiomC{$\headf{N} = {x}[i]$}
                \AxiomC{$ U_i = \unvar{\oneb} $}
                \AxiomC{$  $}
                \LeftLabel{\redlab{RS{:}Fail^!}}
                \TrinaryInfC{$  N \unexsub{U / \unvar{x} } \red N \headlin{ \fail^{\emptyset} /{x}[i] } \unexsub{U / \unvar{x} }  $}
            \end{prooftree}
            with $M'=N \headlin{ \fail^{\emptyset} /{x}[i] } \unexsub{U / \unvar{x} }$.
            By the induction hypothesis, one has the derivation:
            \begin{prooftree}
                \AxiomC{$ \Theta , {x} : \eta; \Gamma  \wfdash N : \tau $}
                \AxiomC{$ \Theta ; \dash \wfdash U : \epsilon $}
                \AxiomC{$ \eta \relunbag \epsilon $}
                \LeftLabel{\redlab{FS{:}Esub^!}}
                \TrinaryInfC{$ \Theta ; \Gamma \wfdash N \unexsub{U / \unvar{x}}  : \tau $}
            \end{prooftree}
            By \Cref{l:lamrsharfailsubsunres}, there exists a derivation $\Pi_1$ of  $\Theta , \unvar{x} : \eta ; \Gamma'  \wfdash   N \headlin{ \fail^{\emptyset} /{x}[i] }  : \tau $.
            Thus,
            \begin{prooftree}
                \AxiomC{$ \Theta , \unvar{x} : \eta ; \Gamma \wfdash   N \headlin{ \fail^{\emptyset} /{x}[i] }  : \tau  $}
                \AxiomC{$ \Theta ; \dash \wfdash U : \epsilon $}
                \AxiomC{$ \eta \relunbag \epsilon $}
                \LeftLabel{\redlab{FS{:}Esub^!}}
                \TrinaryInfC{$ \Theta ; \Gamma \wfdash  N \headlin{ \fail^{\emptyset} /{x}[i] } \unexsub{U / \unvar{x}}  : \tau $}
            \end{prooftree}

        \item Rule $\redlab{RS{:}Cons_1}$.

            Then $M =   \fail^{\widetilde{x}}\ B $  and  the  reduction is:
            \begin{prooftree}
                \AxiomC{$ \widetilde{y} = \llfv{C} $}
                \LeftLabel{$\redlab{RS{:}Cons_1}$}
                \UnaryInfC{$\fail^{\widetilde{x}}\ C \bagsep U \red \displaystyle \fail^{\widetilde{x} \cup \widetilde{y}}  $}
            \end{prooftree}
            and $ M'  =   \fail^{\widetilde{x} \cup \widetilde{y}} $.
            Since $\Theta ; \Gamma\wfdash M:\tau$, one has the derivation:

            \begin{prooftree}
                \AxiomC{$ $}
                \LeftLabel{\redlab{F{:}fail}}
                \UnaryInfC{$ \Theta ;\Gamma' \wfdash \fail^{\widetilde{x}}: (\sigma^{j} , \eta ) \rightarrow \tau $}
                \AxiomC{$\Theta ; \Delta \wfdash C : \sigma^k $}
                \AxiomC{$\Theta ;\dash \wfdash  U : \epsilon $}
                \LeftLabel{$\redlab{FS{:}bag}$}
                \BinaryInfC{$ \Theta ;\Delta \wfdash C \bagsep U : (\sigma^{k} , \epsilon ) \quad \eta \relunbag \epsilon $}
                \LeftLabel{\redlab{FS{:}app}}
                \BinaryInfC{$\Theta ; \Gamma' , \Delta \wfdash \fail^{\widetilde{x}}:\ C \bagsep U : \tau$}
            \end{prooftree}

            Hence $\Gamma = \Gamma' , \Delta $ and we may type the following:
            \begin{prooftree}
                \AxiomC{$ $}
                \LeftLabel{\redlab{F{:}fail}}
                \UnaryInfC{$ \Theta ;\Gamma \wfdash \fail^{\widetilde{x} \cup \widetilde{y}} : \tau$}
            \end{prooftree}

        \item The proof for the cases of $\redlab{RS{:}Cons_2}$, $\redlab{RS{:}Cons_3}$ and $\redlab{RS{:}Cons_4}$ proceed similarly.
            \qedhere
    \end{enumerate}
\end{proof}

\subsection{Subject Expansion}\label{a:subexpan}

The full well-typed rules can be seen in \Cref{fig:wtsh_rulesunres}.

\begin{figure}[t]
\begin{mdframed}
\small
\begin{mathpar}
    \inferrule[$\redlab{TS{:}var^{\ell}}$]{ }{
        \Theta;  {x}: \sigma \wtdash  {x} : \sigma
    }
    \and
    \inferrule[$\redlab{TS{:}var^!}$]{
        \Theta , x^!: \eta;  {x}: \eta_i , \Delta \wtdash  {x} : \sigma
    }{
        \Theta ,  x^!: \eta; \Delta \wtdash {x}[i] : \sigma
    }
    \and
    \inferrule[$\redlab{TS{:}\oneb^{\ell}}$]{ }{
        \Theta ; \dash \wtdash \oneb : \omega
    }
    \and
    \inferrule[$\redlab{TS\!:\!weak}$]{
        \Theta ; \Gamma  \wtdash M : \tau
    }{
        \Theta ; \Gamma ,  {x}: \omega \wtdash M[\leftarrow  {x}]: \tau
    }
    \and
    \inferrule[$\redlab{TS{:}abs\dash sh}$]{
        \Theta , x^!:\eta ; \Gamma ,  {x}: \sigma^k \wtdash M[ {\widetilde{x}} \leftarrow  {x}] : \tau
        \\
        {x} \notin \dom{\Gamma}
    }{
        \Theta ; \Gamma \wtdash \lambda x . (M[ {\widetilde{x}} \leftarrow  {x}])  : (\sigma^k, \eta )  \rightarrow \tau
    }
    \and
    \inferrule[$\redlab{TS{:}bag^{\ell}}$]{
        \Theta ; \Gamma \wtdash M : \sigma
        \\
        \Theta ; \Delta \wtdash C : \sigma^k
    }{
        \Theta ; \Gamma , \Delta \wtdash \bag{M}\cdot C:\sigma^{k+1}
    }
    \and
    \inferrule[$\redlab{TS{:}app}$]{
        \Theta ;\Gamma \wtdash M : (\sigma^{j} , \eta ) \rightarrow \tau
        \\
        \Theta ;\Delta \wtdash B : (\sigma^{j} , \eta )
    }{
        \Theta ; \Gamma , \Delta \wtdash M\ B : \tau
    }
    \and
    \inferrule[$\redlab{TS{:} bag^{!}}$]{
        \Theta ; \dash \wtdash U : \epsilon
        \\
        \Theta ; \dash \wtdash V : \eta
    }{
        \Theta ; \dash  \wtdash U \concat V :\epsilon \concat \eta
    }
    \and
    \inferrule[$\redlab{TS{:}shar}$]{
        \Theta ;  \Gamma ,  {x}_1: \sigma, \dots,  {x}_k: \sigma \wtdash M : \tau
        \\
        {x}\notin \dom{\Gamma}
        \\
        k \not = 0
    }{
        \Theta ;  \Gamma ,  {x}: \sigma^{k} \wtdash M [  {x}_1 , \dots ,  {x}_k \leftarrow  {x} ]  : \tau
    }
    \and
    \inferrule[$\redlab{TS{:}bag}$]{
        \Theta ; \Gamma\wtdash C : \sigma^k
        \\
        \Theta ;\dash \wtdash  U : \eta
    }{
        \Theta ; \Gamma \wtdash C \bagsep U : (\sigma^{k} , \eta  )
    }
    \and
    \inferrule[$\redlab{TS{:}Esub^{\ell}}$]{
        \Theta ; \Gamma  ,  x_1:\sigma, \dots , x_k:\sigma \wtdash M : \tau ~~  \Theta ; \Delta \wtdash C : \sigma^k
    }{
        \Theta ; \Gamma , \Delta \wtdash M \linexsub{C /  x_1, \dots , x_k} : \tau
    }
    \and
    \inferrule[$\redlab{TS{:}Esub^!}$]{
        \Theta , x^! {:} \eta; \Gamma  \wtdash M : \tau
        \\
        \Theta ; \dash \wtdash U : \eta
    }{
        \Theta ; \Gamma \wtdash M \unexsub{U / \unvar{x}}  : \tau
    }
    \and
    \inferrule[$\redlab{TS{:}Esub}$]{
        \Theta , x^! : \eta ; \Gamma ,  {x}: \sigma^{j} \wtdash M[ {\widetilde{x}} \leftarrow  {x}] : \tau
        \\
        \Theta ; \Delta \wtdash B : (\sigma^{k} , \eta )
    }{
        \Theta ; \Gamma , \Delta \wtdash (M[ {\widetilde{x}} \leftarrow  {x}])\esubst{ B }{ x }  : \tau
    }
\end{mathpar}
\end{mdframed}
\caption{Well-Typed Rules for $\lamcoldetsh$.}
\label{fig:wtsh_rulesunres}
\end{figure}

\begin{lemma}[Anti-Substitution Lemma for $\lamcoldetsh$]\label{l:lamrsharfailantisub}
    \leavevmode
    \begin{enumerate}
        \item (Linear)
        If $\Theta ; \Gamma \wtdash M \headlin{ N /  {x} }:\tau$ then there exists $\Gamma' , \Delta$ and $ \sigma$ such that $\Theta ; \Gamma' ,  {x}:\sigma \wtdash M: \tau$ and $\Theta ; \Delta \wtdash N : \sigma$ with $\Gamma = \Gamma' , \Delta $.
        \item (Unrestricted)
        If $\Theta, \unvar{x}: \eta ; \Gamma  \wtdash M \headlin{ N / {x}[i] }$ with $\eta_i = \sigma $ then $\Theta, \unvar{x}: \eta ; \Gamma \wtdash M: \tau$ and $\Theta ; \cdot \wtdash N : \sigma$.
    \end{enumerate}
\end{lemma}

\begin{proof}
\leavevmode
\begin{enumerate}
\item By structural induction on $M$ with $\headf{M}=  {x}$.
There are six cases to be analyzed:
\begin{enumerate}
\item $M= {x}$.
\item
In this case, ${x}\headlin{N/ {x}}=N$ and so $\Gamma \wtdash N:\sigma$. Take $\Gamma' = \dash$ and $\Delta = \Gamma$ then $\Theta ;  {x}:\sigma \wtdash  {x}:\sigma$ and $\Delta \wtdash N:\sigma$ by hypothesis, the result follows.

\item $M = M'\ B$.

From \Cref{f:lambda_red}, $(M'B) \headlin{ N /  {x} } = ( M'\headlin{ N /  {x} } ) B$. Let $\Gamma = \Gamma_1, \Gamma_2$ for some $\Gamma_1,\Gamma_2$ and the derivation is the following:
\begin{prooftree}
\AxiomC{$\Theta ; \Gamma_1 \wtdash M'\headlin{ N /  {x} }: (\delta^{j} , \eta )  \rightarrow \tau$}\
\AxiomC{\quad $\Theta ; \Gamma_2 \wtdash B : (\delta^{j} , \eta ) $}
\LeftLabel{\redlab{TS{:}app}}
\BinaryInfC{$\Theta ; \Gamma_1 , \Gamma_2 \wtdash ( M'\headlin{ N /  {x} } ) B:\tau $}
\end{prooftree}
where $j \geq 0$. By IH there exists $\Gamma_1', \Delta, \sigma $ such that $\Gamma_1 = \Gamma_1' , \Delta$ with $\Theta ; \Gamma_1' ,  {x}:\sigma \wtdash M': (\delta^{j} , \eta )  \rightarrow \tau$ and $\Theta ; \Delta  \wtdash N: \sigma$. Which gives the derivation:
\begin{prooftree}
\AxiomC{$\Theta ; \Gamma_1' ,  {x}:\sigma \wtdash M': (\delta^{j} , \eta )  \rightarrow \tau$}\
\AxiomC{\quad $\Theta ; \Gamma_2 \wtdash B : (\delta^{j} , \eta ) $}
\LeftLabel{\redlab{TS{:}app}}
\BinaryInfC{$\Theta ; \Gamma_1' , \Gamma_2,  {x}:\sigma \wtdash M'B:\tau $}
\end{prooftree}
By taking $\Gamma' = \Gamma_1' , \Gamma_2$ the result follows.

\item $M = M'[ {\widetilde{y}} \leftarrow  {y}] $ with $y\neq x$.

Then from \Cref{f:lambda_red},  $M'[ {\widetilde{y}} \leftarrow  {y}] \headlin{ N /  {x} } = M' \headlin{ N /  {x}} [ {\widetilde{y}} \leftarrow  {y}]$. Let $\Gamma =  \Gamma_1 ,  {y}: \delta^k$ with $k \not = 0$. Therefore,
\begin{prooftree}
\AxiomC{$ \Theta ; \Gamma_1 ,  {y}_1: \delta, \dots,  {y}_k: \delta  \wtdash  M' \headlin{ N /  {x}} : \tau \quad  {y}\notin \Gamma_1 \quad k \not = 0$}
\LeftLabel{ \redlab{TS{:}shar} }
\UnaryInfC{$ \Theta ; \Gamma_1 ,  {y}: \delta^k \wtdash M' \headlin{ N /  {x}} [ {\widetilde{y}} \leftarrow  {y}] : \tau $}
\end{prooftree}
By IH there exists $ \Gamma_1 ' , \Delta, \sigma$ such that $\Gamma_1 , {y}_1: \delta, \dots,  {y}_k: \delta = \Gamma_1' , {y}_1: \delta, \dots,  {y}_k: \delta , \Delta$ with $\Theta ; \Gamma_1' , {y}_1: \delta, \dots,  {y}_k: \delta ,  {x}:\sigma \wtdash M': \tau$ and $\Theta ; \Delta  \wtdash N: \sigma$. Which gives the derivation:
\begin{prooftree}
\AxiomC{$\Theta ; \Gamma_1' ,  {y}_1: \delta, \dots,  {y}_k: \delta ,  {x}: \sigma \wtdash M' : \tau \quad  {y}\notin \Gamma_1 \quad k \not = 0$}
\LeftLabel{ \redlab{TS{:}share}}
\UnaryInfC{$ \Theta ; \Gamma_1' ,  {y}: \delta^k,  {x}: \sigma \wtdash M'[ {y}_1 , \dots ,  {y}_k \leftarrow x] : \tau $}
\end{prooftree}
By taking $\Gamma' = \Gamma_1' , {y}: \delta^k$ the result follows.

\item $M = M'[ \leftarrow  {y}] $ with  $x \not  = y $.

Then $M'[ \leftarrow  {y}] \headlin{ N /  {x} } = M' \headlin{ N /  {x}} [ \leftarrow  {y}]$. Therefore,
\begin{prooftree}
\AxiomC{$ \Theta ;  \Gamma  \wtdash M \headlin{ N /  {x}}: \tau$}
\LeftLabel{ \redlab{TS{:}weak}}
\UnaryInfC{$ \Theta ;  \Gamma  ,  {y}: \omega \wtdash M\headlin{ N /  {x}}[\leftarrow  {y}]: \tau $}
\end{prooftree}
By IH there exists $\Gamma' , \Delta, \sigma$ such that $\Gamma = \Gamma' , \Delta$ with $\Theta ; \Gamma' ,  {x}:\sigma \wtdash M': \tau$ and $\Theta ; \Delta  \wtdash N: \sigma$. Which gives the derivation:
\begin{prooftree}
\AxiomC{$ \Theta ; \Gamma'  ,  {x}: \sigma  \wtdash M : \tau$}
\LeftLabel{ \redlab{TS{:}weak} }
\UnaryInfC{$ \Theta ;  \Gamma'  ,  {y}: \omega,  {x}: \sigma  \wtdash M[\leftarrow  {y}]: \tau $}
\end{prooftree}
By taking $\Gamma' = \Gamma'  ,  {y}: \omega$ the result follows.

\item If $M =  M' \linexsub{C /  y_1 , \dots , y_k} $  with $ x \not = y_1 , \dots , y_k$.

Then $M' \linexsub{C /  y_1 , \dots , y_k}  \headlin{ N / x } = M' \headlin{ N / x } \linexsub{C /  y_1 , \dots , y_k}  $ and
\begin{prooftree}
\AxiomC{$ \Theta ; \Gamma_1 ,  y_1:\delta, \dots , y_k:\delta  \wtdash M' \headlin{ N / x } : \tau $}
\AxiomC{$\quad \Theta ; \Gamma_2 \wtdash C : \delta^k $}
\LeftLabel{\redlab{TS{:}Esub^{\ell}}}
\BinaryInfC{$ \Theta ; \Gamma_1 , \Gamma_2  \wtdash M' \headlin{ N / x } \linexsub{C /  y_1 , \dots , y_k} : \tau $}
\end{prooftree}
By IH there exists $ \Gamma_1 ' , \Delta, \sigma$ such that $\Gamma_1 , {y}_1: \delta, \dots,  {y}_k: \delta = \Gamma_1' , {y}_1: \delta, \dots,  {y}_k: \delta , \Delta$ with $\Theta ; \Gamma_1' , {y}_1: \delta, \dots,  {y}_k: \delta ,  {x}:\sigma \wtdash M': \tau$ and $\Theta ; \Delta  \wtdash N: \sigma$. Which gives the derivation:
\begin{prooftree}
    \AxiomC{$ \Theta ; \Gamma_1'  ,  y_1:\delta, \dots , y_k:\delta , x: \sigma  \wtdash M' : \tau $}
\AxiomC{$\quad \Theta ; \Gamma_2 \wtdash C : \delta^k $}
    \LeftLabel{\redlab{TS{:}Esub^{\ell}}}
\BinaryInfC{$ \Theta ; \Gamma_1' , \Gamma_2, x: \sigma  \wtdash M' \linexsub{C /  y_1 , \dots , y_k} : \tau $}
\end{prooftree}
By taking $\Gamma' = \Gamma_1' , \Gamma_2 $ the result follows.

\item If $M =  M' \unexsub {U / {y}} $ then $\headf{M' \unexsub {U / {y}}} = \headf{M'} = x $, and the proofs is similar to the case above.
\end{enumerate}

\item  By structural induction on $M$ with $\headf{M}= {x}[i]$.
There are three cases to be analyzed:
\begin{enumerate}
\item $M= {x}[i]$.

Observe that ${x}[i]\headlin{N/{x}[i]}=N$ and let $\Theta, \unvar{x}: \eta  ; \Gamma  \wtdash {x}[i]\headlin{N/{x}[i]}:\sigma$ with $\eta_i = \sigma$
In this case $\Gamma = \dash$, and we have both
\begin{prooftree}
\AxiomC{}
\LeftLabel{ \redlab{TS{:}var^{ \ell}}}
\UnaryInfC{$ \Theta , \unvar{x}: \eta;  {x}: \eta_i  \wtdash  {x} : \sigma$}
\LeftLabel{\redlab{TS{:}var^!}}
\UnaryInfC{$ \Theta, \unvar{x}: \eta ; \cdot \wtdash {x}[i] : \sigma$}
\end{prooftree}
and $\Theta, \unvar{x}: \eta  ; \Gamma  \wtdash N:\sigma$ and the result follows.

\item $M = M'\ B$.

From \Cref{f:lambda_red}, $M' \esubst{ B }{ y} \headlin{ N / {x}[i] } = M' \headlin{ N / {x}[i] } \esubst{ B }{ y}$, let $\Theta , \unvar{x}: \eta;\Gamma_1 , \Gamma_2  \wtdash ( M'\headlin{ N / {x}[i] } ) B:\tau$ where $\Gamma = \Gamma_1 , \Gamma_2 $ and $\eta_i = \sigma$ we derrive:
\begin{prooftree}
\AxiomC{$\Theta , \unvar{x}: \eta;\Gamma_1\wtdash M'\headlin{N/ {x}[i]}:(\delta^{j} , \epsilon ) \rightarrow \tau $}\
\AxiomC{$\Theta , \unvar{x}: \eta; \Gamma_2 \wtdash B : (\delta^{j} , \epsilon ) $}
\AxiomC{}
\LeftLabel{\redlab{TS{:}app}}
\TrinaryInfC{$\Theta , \unvar{x}: \eta;\Gamma_1 , \Gamma_2  \wtdash ( M'\headlin{ N / {x}[i] } ) B:\tau $}
\end{prooftree}
for some strict type $\delta$. By IH $\Theta, \unvar{x}: \eta ; \Gamma_1 \wtdash M': (\delta^{j} , \epsilon ) \rightarrow \tau $ and $\Theta ; \cdot \wtdash N : \sigma$ and we derrive:
\begin{prooftree}
\AxiomC{$ \Theta, \unvar{x}: \eta ;\Gamma_1 \wtdash M' : (\delta^{j} , \epsilon ) \rightarrow \tau \quad \Theta, \unvar{x}: \sigma ; \Gamma_2 \wtdash B : (\delta^{j} , \epsilon' )  $}
\AxiomC{}
\LeftLabel{\redlab{TS{:}app}}
\BinaryInfC{$ \Theta, \unvar{x}: \eta ;\Gamma_1 , \Gamma_2 \wtdash M'\ B : \tau$}
\end{prooftree}
and the result follows.

\item $M = M'[ {\widetilde{y}} \leftarrow  {y}] $.

From \Cref{f:lambda_red}  $M'[ {\widetilde{y}} \leftarrow  {y}] \headlin{ N / {x}[i] } = M' \headlin{ N / {x}[i]} [ {\widetilde{y}} \leftarrow  {y}]$, let $\Theta , \unvar{x}: \eta; \Gamma_1 ,  {y}: \delta^k \wtdash M' \headlin{ N / {x}[i]} [ {\widetilde{y}} \leftarrow  {y}] : \tau$ with $\Gamma = \Gamma_1 ,  {y}: \delta^k$ and $\eta_i = \sigma$, then we derive:
\begin{prooftree}
\AxiomC{$ \Theta , \unvar{x}: \eta; \Gamma_1 ,  {y}_1: \delta, \dots,  {y}_k: \delta  \wtdash  M' \headlin{ N / {x}[i] } : \tau \quad  {y}\notin \Gamma_1 \quad k \not = 0$}
\LeftLabel{ \redlab{TS{:}shar} }
\UnaryInfC{$ \Theta , \unvar{x}: \eta; \Gamma_1 ,  {y}: \delta^k \wtdash M' \headlin{ N / {x}[i]} [ {\widetilde{y}} \leftarrow  {y}] : \tau $}
\end{prooftree}

By IH $\Theta, \unvar{x}: \eta ; \Gamma_1 ,  {y}_1: \delta, \dots,  {y}_k: \delta \wtdash M': \tau $ and $\Theta ; \cdot \wtdash N : \sigma$ and we derrive:

\begin{prooftree}
\AxiomC{$\Theta , \unvar{x}: \eta; \Gamma_1 ,  {y}_1: \delta, \dots,  {y}_k: \delta  \wtdash M' : \tau \quad  {y}\notin \Gamma_1 \quad k \not = 0$}
\LeftLabel{ \redlab{TS{:}shar}}
\UnaryInfC{$ \Theta , \unvar{x}: \eta; \Gamma_1 ,  {y}: \delta^k \wtdash M'[ {y}_1 , \dots ,  {y}_k \leftarrow y] : \tau $}
\end{prooftree}
and the result follows.

\item $M = M'[ \leftarrow  {y}] $.

From \Cref{f:lambda_red}  $M'[ \leftarrow  {y}] \headlin{ N / {x}[i] } = M' \headlin{ N / {x}[i] } [ \leftarrow  {y}]$, let $\Theta , \unvar{x}: \eta;  \Gamma  ,  {y}: \omega \wtdash M\headlin{ N / {x}[i] }[\leftarrow  {y}]: \tau$ with $\Gamma = \Gamma  ,  {y}: \omega $ and $\eta_i = \sigma$, then we derive:
\begin{prooftree}
\AxiomC{$ \Theta , \unvar{x}: \eta;  \Gamma   \wtdash M \headlin{ N / {x}[i]  }: \tau$}
\LeftLabel{ \redlab{TS{:}weak}}
\UnaryInfC{$ \Theta , \unvar{x}: \eta;  \Gamma  ,  {y}: \omega \wtdash M\headlin{ N / {x}[i] }[\leftarrow  {y}]: \tau $}
\end{prooftree}

By IH $\Theta , \unvar{x}: \eta;  \Gamma \wtdash M': \tau $ and $\Theta ; \cdot \wtdash N : \sigma$ and we derrive:
\begin{prooftree}
\AxiomC{$ \Theta , \unvar{x}: \eta; \Gamma   \wtdash M : \tau$}
\LeftLabel{ \redlab{TS{:}weak} }
\UnaryInfC{$ \Theta , \unvar{x}: \eta;  \Gamma  ,  {y}: \omega \wtdash M[\leftarrow  {y}]: \tau $}
\end{prooftree}
and the result follows.

\item $M =   M' \linexsub{C /  y_1 , \dots , y_k} $.

From \Cref{f:lambda_red}  $M' \linexsub{C /  y_1 , \dots , y_k} \headlin{ N / {x}[i]  } = M' \headlin{ N / {x}[i]  }  \linexsub{C /  y_1 , \dots , y_k}  $, let $ \Theta, \unvar{x}: \eta ; \Gamma' , \Delta \wtdash M' \headlin{ N / {x}[i]  }  \linexsub{C /  y_1 , \dots , y_k} : \tau $ with $\Gamma =  \Gamma' , \Delta $ and $\eta_i = \sigma$, then we derive:
\begin{prooftree}
\AxiomC{$ \Theta, \unvar{x}: \eta ; \Gamma'  ,  y_1:\delta, \dots , y_k:\delta \wtdash M' \headlin{ N / {x}[i]  }  : \tau $}
\AxiomC{$ \Theta, \unvar{x}: \eta ; \Delta \wtdash C : \delta^k $}
\LeftLabel{\redlab{TS{:}Esub^{\ell}}}
\BinaryInfC{$ \Theta, \unvar{x}: \eta ; \Gamma' , \Delta \wtdash M' \headlin{ N / {x}[i]  }  \linexsub{C /  y_1 , \dots , y_k} : \tau $}
\end{prooftree}
By IH $\Theta , \unvar{x}: \eta;  \Gamma \wtdash M': \tau $ and $\Theta ; \cdot \wtdash N : \sigma$ and we derrive:
\begin{prooftree}
\AxiomC{$ \Theta, \unvar{x}: \eta ; \Gamma'  ,  y_1:\delta, \dots , y_k:\delta \wtdash M : \tau $}
\AxiomC{$ \Theta, \unvar{x}: \eta ; \Delta \wtdash C : \delta^k $}
\LeftLabel{\redlab{TS{:}Esub^{\ell}}}
\BinaryInfC{$ \Theta, \unvar{x}: \eta ; \Gamma' , \Delta \wtdash M \linexsub{C /  y_1, \dots , y_k} : \tau $}
\end{prooftree}
and the result follows.

\item $M =  M' \unexsub {U /\unvar{y}}$.

From \Cref{f:lambda_red}  $ M' \unexsub {U /\unvar{y}} \headlin{ N /  {x}[i] } = M'  \headlin{N /  {x}[i] } \unexsub {U /\unvar{y}} $, let $  \Theta ; \Gamma \wtdash M'  \headlin{N /  {x}[i] } \unexsub{U / \unvar{y}}  : \tau  $ and $\eta_i = \sigma$, then we derive:
\begin{prooftree}
\AxiomC{$ \Theta , \unvar{x}: \eta , {y} : \epsilon; \Gamma  \wtdash M'  \headlin{N /  {x}[i] } : \tau $}
\AxiomC{$ \Theta , \unvar{x}: \eta; \dash \wtdash U : \epsilon $}
\AxiomC{$ $}
\LeftLabel{\redlab{TS{:}Esub^!}}
\TrinaryInfC{$ \Theta ; \Gamma \wtdash M'  \headlin{N /  {x}[i] } \unexsub{U / \unvar{y}}  : \tau $}
\end{prooftree}
By IH $ \Theta , \unvar{x}: \eta , {y} : \epsilon; \Gamma  \wtdash M' : \tau  $ and $\Theta ; \cdot \wtdash N : \sigma$ and we derive:
\begin{prooftree}
\AxiomC{$ \Theta , \unvar{x}: \eta , {y} : \epsilon; \Gamma  \wtdash M' : \tau $}
\AxiomC{$ \Theta , \unvar{x}: \eta; \dash \wtdash U : \epsilon $}
\AxiomC{$  $}
    \LeftLabel{\redlab{TS{:}Esub^!}}
\TrinaryInfC{$ \Theta ; \Gamma \wtdash M' \unexsub{U / \unvar{y}}  : \tau $}
\end{prooftree}
and the result follows.
\qedhere
\end{enumerate}
\end{enumerate}
\end{proof}

\begin{theorem}[Subject Expansion in $\lamcoldetsh$]\label{t:lamSE}
If $\Theta ; \Gamma \wtdash M':\tau$ and $M \red M'$ then $\Theta ; \Gamma \wtdash M :\tau$.
\end{theorem}

\begin{proof}
By structural induction on the reduction rule from \figref{fig:reduc_interm} applied in $M \red M'$.
\begin{enumerate}
\item \textbf{ Rule $\redlab{RS{:}Beta}$.}

Then $M' = N[ {\widetilde{x}} \leftarrow  {x}]\ \esubst{ B }{ x } $  and the reduction is:

\begin{prooftree}
\AxiomC{}
\LeftLabel{\redlab{RS{:}Beta}}
\UnaryInfC{$(\lambda x. N[ {\widetilde{x}} \leftarrow  {x}]) B \red N[ {\widetilde{x}} \leftarrow  {x}]\ \esubst{ B }{ x }$}
\end{prooftree}

where $ M  =  (\lambda x. N[ {\widetilde{x}} \leftarrow  {x}]) B$. Since $\Theta ; \Gamma\wtdash M':\tau$ we get the following derivation:

\begin{prooftree}
\AxiomC{$\Theta , \unvar{x} : \eta; \Gamma' ,  {x}_1:\sigma , \dots ,  {x}_j:\sigma  \wtdash  N: \tau $}
\LeftLabel{ \redlab{TS{:}share} }
\UnaryInfC{$\Theta , \unvar{x} : \eta;  \Gamma' ,   {x}:\sigma^{j}  \wtdash  N[ {\widetilde{x}} \leftarrow  {x}]: \tau $}
\AxiomC{$\Theta ;\Delta \wtdash B:(\sigma^{j} , \eta )  $}
\AxiomC{$  $}
\LeftLabel{ \redlab{TS{:}Esub} }
\TrinaryInfC{$ \Theta ;\Gamma' , \Delta \wtdash N[ {\widetilde{x}} \leftarrow  {x}]\ \esubst{ B }{ x }:\tau $}
\end{prooftree}
for $\Gamma = \Gamma' , \Delta $. Notice that:
\begin{prooftree}
\AxiomC{$\Theta , \unvar{x} : \eta; \Gamma' ,  {x}_1:\sigma , \dots ,  {x}_j:\sigma  \wtdash  N: \tau $}
\LeftLabel{ \redlab{TS{:}share} }
\UnaryInfC{$\Theta , \unvar{x} : \eta;  \Gamma' ,   {x}:\sigma^{j}  \wtdash  N[ {\widetilde{x}} \leftarrow  {x}]: \tau $}
\LeftLabel{ \redlab{TS{:}abs \dash sh} }
\UnaryInfC{$\Theta ; \Gamma' \wtdash \lambda x. N[ {\widetilde{x}} \leftarrow  {x}]: (\sigma^{j} , \eta ) \rightarrow \tau $}
\AxiomC{$\Theta ;\Delta \wtdash B: (\sigma^{j} , \eta ) $}
\AxiomC{$  $}
\LeftLabel{ \redlab{TS{:}app} }
\TrinaryInfC{$ \Theta ;\Gamma' , \Delta \wtdash (\lambda x. N[ {\widetilde{x}} \leftarrow  {x}]) B:\tau $}
\end{prooftree}
Therefore $\Theta ;\Gamma' , \Delta \wtdash (\lambda x. N[ {\widetilde{x}} \leftarrow  {x}]) B:\tau$ and the result follows.

Then $M = (\lambda x. N[ {\widetilde{x}} \leftarrow  {x}]) B $  and the reduction is:

\item \textbf{ Rule $ \redlab{RS{:}Ex \dash Sub}.$}

Then $ M' =  N\linexsub{C  /  x_1 , \dots , x_k} \unexsub{U / \unvar{x} }$ where $C=  \bag{N_1}\cdot \dots \cdot \bag{N_k} $.
The reduction is:
\begin{prooftree}
\AxiomC{$ C = \bag{N_1}
\dots  \bag{N_k} \qquad  N \not= \fail^{\widetilde{y}} $}
\LeftLabel{\redlab{RS{:}Ex \dash Sub}}
\UnaryInfC{$ N[ {x}_1, \!\dots\! ,  {x}_k \leftarrow  {x}]\esubst{ C \bagsep U }{ x } \red N\linexsub{C  /  x_1 , \dots , x_k} \unexsub{U / \unvar{x} }$}
\end{prooftree}
and $M = N[ {x}_1, \!\dots\! ,  {x}_k \leftarrow  {x}]\esubst{ C \bagsep U }{ x }$.
To simplify the proof we take $k=2$, as the case $k>2$ is similar.
Therefore $C=\bag{N_1}\cdot \bag{N_2}$.

Since $\Theta ; \Gamma\wtdash M':\tau$ we get a derivation(we omit the labels \redlab{TS:Esub^!} and \redlab{TS:Esub^{\ell}}):
{\small
\begin{prooftree}
\AxiomC{$ \Theta, \unvar{x} : \eta  ;  \Gamma' ,  {x}_1: \sigma,  {x}_2: \sigma \wtdash N : \tau$}
\AxiomC{$ \Theta ; \Delta \wtdash C : \sigma^2 $}
\BinaryInfC{$ \Theta, \unvar{x} : \eta  ;  \Gamma' , \Delta \wtdash N \linexsub{C  / {x}_1,  {x}_2} : \tau $}
\AxiomC{$ \Theta ; \dash \wtdash U : \eta $}
\AxiomC{$  $}
\TrinaryInfC{$ \Theta ; \Gamma' , \Delta \wtdash N\linexsub{C  / {x}_1,  {x}_2} \unexsub{U / \unvar{x} }  : \tau $}
\end{prooftree}
}
where $\Gamma = \Gamma' , \Delta $. Consider the typing derivation:(we omit the labels \redlab{TS:Esub} and \redlab{TS{:}share})
{\small
\begin{prooftree}
\AxiomC{$ \Theta, \unvar{x} : \eta  ;  \Gamma' ,  {x}_1: \sigma,  {x}_2: \sigma \wtdash N : \tau $}
\UnaryInfC{$  \Theta , \unvar{x} : \eta ; \Gamma' ,  {x}: \sigma^{2} \wtdash N[ {x}_1,  {x}_2 \leftarrow  {x}] : \tau  $}

\AxiomC{$ \Theta ; \Delta \wtdash C \bagsep U : (\sigma^{2} , \eta ) $}
\AxiomC{$ $}
\TrinaryInfC{$ \Theta ; \Gamma' , \Delta \wtdash N[ {x}_1,  {x}_2 \leftarrow  {x}]\esubst{ C \bagsep U }{ x }  : \tau $}
\end{prooftree}
}
and the result follows.

\item Rule $ \redlab{RS{:}Fetch^{\ell}}$.

Then $ M' =   N \headlin{ C_i / x_j }  \linexsub{(C \setminus C_i ) /  \widetilde{x}  }  $ where  $\headf{N} =  {x}_j$.
The reduction is:
\begin{prooftree}
\AxiomC{$ \headf{N} =  {x}_j$}
\LeftLabel{\redlab{RS{:}Fetch^{\ell}_{{i}}}}
\UnaryInfC{$  N \linexsub{C /  \widetilde{x}, x_j} \red  N \headlin{ C_i / x_j }  \linexsub{(C \setminus C_i ) /  \widetilde{x}  } $}
\end{prooftree}
and $M =  N \linexsub{C /  \widetilde{x}, x_j} $.
Since $\Theta ; \Gamma\wtdash M':\tau$ we get the following derivation:
\begin{prooftree}
\AxiomC{$ \Theta ; \Gamma' , \Delta_i ,  \widetilde{x}:\sigma^{k-1} \wtdash N \headlin{ C_i / x_j }: \tau $}
\AxiomC{$ \Theta ; \Delta'_i \wtdash C \setminus C_i : \sigma^{k-1} $}
\LeftLabel{\redlab{TS{:}Esub^{\ell}}}
\BinaryInfC{$ \Theta ; \Gamma', \Delta \wtdash N \headlin{ C_i / x_j }  \linexsub{(C \setminus C_i ) /  \widetilde{x}  }  : \tau $}
\end{prooftree}
where $\Gamma = \Gamma' , \Delta $ and $\Delta = \Delta_i , \Delta_i'$.
By \Cref{l:lamrsharfailantisub}, we obtain the derivation $ \Theta ; \Gamma'  ,  \widetilde{x}:\sigma^{k-1},  x_j:\sigma \wtdash N : \tau $ and $ \Theta ;  \Delta_i \wtdash  C_i : \sigma $ via:
\begin{prooftree}
\AxiomC{$ \Theta ; \Gamma'  ,  \widetilde{x}:\sigma^{k-1},  x_j:\sigma \wtdash N : \tau $}
\AxiomC{$ \Theta ; \Delta \wtdash C : \sigma^k $}
\LeftLabel{\redlab{TS{:}Esub^{\ell}}}
\BinaryInfC{$ \Theta ; \Gamma' , \Delta \wtdash N \linexsub{C /  \widetilde{x}, x_j}  : \tau $}
\end{prooftree}

\item Rule $ \redlab{RS{:} Fetch^!}$.

Then $ M' = N \headlin{ N_i /{x}[i] }\unexsub{U / \unvar{x}}  $ where  $\headf{M} = {x}[i]$. The reduction is:
\begin{prooftree}
\AxiomC{$ \headf{N} = {x}[i]$}
\AxiomC{$ U_i = \unvar{\bag{N_i}}$}
\LeftLabel{\redlab{RS{:} Fetch^!}}
\BinaryInfC{$  N \unexsub{U / \unvar{x}} \red  N \headlin{ N_i /{x}[i] }\unexsub{U / \unvar{x}} $}
\end{prooftree}
and $M =  N \unexsub{U / \unvar{x}}  $.     Since $\Theta ; \Gamma \wtdash M':\tau$ we get the following derivation:
\begin{prooftree}
\AxiomC{$ \Theta , \unvar{x} : \eta; \Gamma  \wtdash N \headlin{ N_i /{x}[i] } : \tau $}
\AxiomC{$ \Theta ; \dash \wtdash U : \eta $}
\AxiomC{$  $}
\LeftLabel{\redlab{TS{:}Esub^!}}
\TrinaryInfC{$ \Theta ; \Gamma \wtdash N \headlin{ N_i /{x}[i] }\unexsub{U / \unvar{x}}  : \tau $}
\end{prooftree}
By \Cref{l:lamrsharfailantisub}, we obtain the derivation $ \Theta , {x} : \eta; \Gamma  \wtdash N : \tau  $, and the result follows from:
\begin{prooftree}
\AxiomC{$ \Theta , {x} : \eta; \Gamma  \wtdash N : \tau $}
\AxiomC{$ \Theta ; \dash \wtdash U : \eta $}
\AxiomC{$ $}
\LeftLabel{\redlab{TS{:}Esub^!}}
\TrinaryInfC{$ \Theta ; \Gamma \wtdash N \unexsub{U / \unvar{x}}  : \tau $}
\end{prooftree}

\item Rule $\redlab{RS{:}TCont}$.

Then $M' = C[N']$ and the reduction is as follows:
\begin{prooftree}
\AxiomC{$   N \red N' $}
\LeftLabel{\redlab{RS{:}TCont}}
\UnaryInfC{$ C[N] \red  C[N'] $}
\end{prooftree}
with $M=  C[N] $.
The proof proceeds by analysing the context $C$.
There are three cases:
\begin{enumerate}
\item $C=[\cdot]\ B$.

In this case $ M' = N' \ B$, for some $B$.
Since $\Gamma\vdash M':\tau$ one has a derivation:
\begin{prooftree}
\AxiomC{$ \Theta ;\Gamma' \wtdash N' : (\sigma^{j} , \eta ) \rightarrow \tau $}
\AxiomC{$  \Theta ;\Delta \wtdash B : (\sigma^{j} , \eta )  $}
\AxiomC{$ $}
\LeftLabel{\redlab{TS{:}app}}
\TrinaryInfC{$ \Theta ; \Gamma' , \Delta \wtdash N' \ B : \tau$}
\end{prooftree}
where $\Gamma = \Gamma' , \Delta $.
From  $\Gamma'\wtdash N':(\sigma^{j} , \eta ) \rightarrow \tau$ and the reduction $N \red N' $, one has by IH that  $\Gamma'\wtdash N:(\sigma^{j} , \eta ) \rightarrow \tau$.
Finally, we may type the following:
\begin{prooftree}
\AxiomC{$ \Theta ;\Gamma' \wtdash N : (\sigma^{j} , \eta ) \rightarrow \tau $}
\AxiomC{$  \Theta ;\Delta \wtdash B : (\sigma^{j} , \eta )  $}
\AxiomC{$  $}
\LeftLabel{\redlab{TS{:}app}}
\TrinaryInfC{$ \Theta ; \Gamma' , \Delta \wtdash N\ B : \tau$}
\end{prooftree}
Since $ M  =   (C[N]) = NB $, the result follows.

\item  Cases $C=[\cdot]\linexsub{N/x} $ and $C=[\cdot][\widetilde{x} \leftarrow x]$ are similar to the previous.
\end{enumerate}

\item Rules $ \redlab{RS{:}Fail^{\ell}}$, $ \redlab{RS{:}Fail^!}$ and $\redlab{RS{:}Cons_1}$.
These cases are trivial, since $\oneb$, $\oneb^!$ and $\fail^{\widetilde{x} \cup \widetilde{y}}$ are not well-typed.
    \qedhere

\end{enumerate}
\end{proof}

%% file: appendix/lazy-encod.tex
\section{Proof of Tight Correctness of the Translation under the Lazy Semantics}\label{a:tight}

Here we prove \Cref{t:correncLazy}.

\begin{definition}[Success]\label{def:app_Suc3unres}
    We define define $\headf{\sucs{\lambda}} = \sucs{\lambda}$ and $\piencodf{\sucs{\lambda}}_u = \sucs{\pi}$.
    We define {\em success} for terms  $M\in  \lamcoldet$ and $P\in \clpi$.
    \begin{itemize}
        \item \succp{{M}}{\sucs{\lambda}} if and only if, there exist  $M'_1 , \cdots , M_k'\in \lamcoldetsh$ such that ${M} \red^*  M'$ and $\headf{M'} = \sucs{\lambda}$.

        \item  ${P}\succone{\sucs{\pi}}$ if and only if, there exist  $Q_1 , Q_2\in \clpi$ such that $P \redone^*   (Q_1   \| \sucs{\pi}) \nd Q_2 $.

        \item  ${P}\succtwo{\sucs{\pi}}$ if and only if, there exist  $S$ and $ Q_1 , Q_2\in \clpi$ such that $P \redtwo_S^*   (Q_1   \| \sucs{\pi}) \nd Q_2 $.
    \end{itemize}
\end{definition}

\subsection{Type Preservation}\label{a:typepres}

\begin{lemma}\label{prop:app_auxunres}
 $ \piencodf{\sigma^{j}}_{(\tau_1, m)} = \piencodf{\sigma^{k}}_{(\tau_2, n)}$ and $ \piencodf{(\sigma^{j} , \eta)}_{(\tau_1, m)} = \piencodf{(\sigma^{k}, \eta)}_{(\tau_2, n)}$ hold, provided that  $\tau_1,\tau_2,n$ and $m$ are as follows:

        \begin{enumerate}
        \item If $j > k$ then take $\tau_1 $ to be an arbitrary type, $m = 0$,  take $\tau_2 $ to be $\sigma$ and $n = j-k$.

        \item If $j < k$ then take $\tau_1 $ to be $\sigma$, $m = k-j$,  take $\tau_2 $ to be an arbitrary type and $n = 0$.

        \item Otherwise, if $j = k$ then take $m = n = 0$. In this case, $\tau_1. \tau_2 $ are unimportant.
    \end{enumerate}

\end{lemma}

\begin{proof}
This proof  proceeds by analyzing the conditions on types, following Paulus et al.~\cite{DBLP:journals/corr/abs-2111}.
\end{proof}

\begin{lemma}\label{lem:relunbag-typeunres}
If
$ \eta \relunbag \epsilon $ then the following hold:
\begin{enumerate}
    \item If $\piencodf{M}_u\vdash \piencodf{\Gamma} , \piencodf{\Theta} , \unvar{x} : \dual{\piencodf{\eta}}$
    then
    $\piencodf{M}_u\vdash \piencodf{\Gamma} , \piencodf{\Theta}, \unvar{x} : \dual{\piencodf{\epsilon}}$.

    \item If $\piencodf{M}_u\vdash \piencodf{\Gamma}, u:\piencodf{(\sigma^{j} , \eta ) \rightarrow \tau} , \piencodf{\Theta}$
    then
    $\piencodf{M}_u\vdash \piencodf{\Gamma}, u:\piencodf{(\sigma^{j} , \epsilon ) \rightarrow \tau} , \piencodf{\Theta}$.

\end{enumerate}

\end{lemma}

\begin{proof} The proof is  by mutual induction on the the derivations of If $\piencodf{M}_u\vdash \piencodf{\Gamma} , \piencodf{\Theta} , \unvar{x} : \dual{\piencodf{\eta}}$ and  $\piencodf{M}_u\vdash \piencodf{\Gamma}, u:\piencodf{(\sigma^{j} , \eta ) \rightarrow \tau} , \piencodf{\Theta}$ and on the structure of $M$.

We use \defref{def:enc_sestypfailunres} that establishes   \(  \piencodf{ \eta } = ! \with_{\eta_i \in \eta} \{ i : \piencodf{\eta_i} \}\) and by duality $\dual{\piencodf{ \eta }} = ? \oplus_{\eta_i \in \eta} \{ i : \dual{\piencodf{\eta_i}} \}$.
\begin{enumerate}

\item $M =   {x}$.
\label{proof:relunbag-no}

By the translation in \figref{fig:encoding}:  $\piencodf{ {x}}_u =  \psome{x} ; \pfwd{x}{u} $. We have the following derivation,  for some type $A$:

\begin{prooftree}
    \AxiomC{}
    \LeftLabel{\ttype{id}}
    \UnaryInfC{$ \pfwd{x}{u} \vdash  {x}:  \overline{A}  , u :  A  $}
    \LeftLabel{\ttype{weaken}}
    \UnaryInfC{$ \pfwd{x}{u} \vdash  {x}:  \overline{A}  , u :  A, \unvar{x} : \dual{\piencodf{\eta}}$}
    \LeftLabel{\ttype{$\with$}}
    \UnaryInfC{$  \psome{x}; \pfwd{x}{u} \vdash  {x}: \with  \overline{A} , u :  A , \unvar{x} : \dual{\piencodf{\eta}} $}
\end{prooftree}

The derivation is independent of $\unvar{x} : \dual{\piencodf{\eta}}$, hence the result  trivially holds for $ \piencodf{\Gamma} , \piencodf{\Theta}={x}: \with  \overline{A} , u :  A$.
\item $ M =  {x}[k]$.
\label{proof:relunbag-yes}

By the translation in \figref{fig:encoding}:  $\piencodf{ {x}[k]}_u = \puname{ \unvar{x} }{ {x_i} }; \psel{ {x}_i }{k };  \pfwd{x_i}{u} $. We have the following derivation:

\begin{prooftree}
\AxiomC{}
\LeftLabel{\ttype{id}}
\UnaryInfC{$ \pfwd{x_i}{u}  \vdash  u :  \piencodf{ \tau },  x_i:  \overline{\piencodf{\eta_{k} }}  $}`'
\LeftLabel{ \ttype{$\oplus$}}
\UnaryInfC{$  \psel{ {x}_i }{ k }; \pfwd{x_i}{u} \vdash  u :  \piencodf{ \tau }, {x}_i :  \oplus_{\eta_i \in \eta} \{ i : \dual{\piencodf{\eta_i}}  \} $}
\LeftLabel{\ttype{$?$}}
\UnaryInfC{$ \puname{ \unvar{x} }{ x_i }; \psel{ {x}_i }{ k }; \pfwd{x_i}{u} \vdash  u :  \piencodf{ \tau }, \unvar{x}:? \oplus_{\eta_i \in \eta} \{ i : \dual{\piencodf{\eta_i}} \}  $}
\end{prooftree}

  Since $ \eta \relunbag \epsilon $ we have that $ \epsilon_{k} = \eta_{k} $ for each $k=1,\ldots |\eta|$. Thus, the  same derivation above, replacing $\eta_i$'s for $\epsilon_i$'s entails $ \puname{ \unvar{x} }{ x_i }; \psel{ {x}_i }{ k}; \pfwd{x_i}{u} \vdash  u :  \piencodf{ \tau }; \unvar{x}:\dual{\piencod{\epsilon}}$, and the result follows. For the case of $ M =  {y}[k]$ with $y \not =  x$ we use the argument that the typing of $y$ is independent on $x$.

\item $ M =  \lambda y . (M'[ {\widetilde{y}} \leftarrow  {y}])$.

From the translation in
\figref{fig:encoding}:

$ \piencodf{\lambda y.M'[ {\widetilde{y}} \leftarrow y]}_u = \psome{u}; \gname{u}{y}; \psome{y};\underbrace{ \gname{ y }{ \linvar{y} }; \gname{ y }{ \unvar{y} }; \gclose{ y } ;  \piencodf{M'[ {\widetilde{y}}\leftarrow  {y}]}_u}_{P}$.

    \begin{prooftree}
    \AxiomC{$ \Pi_1 $}
        \noLine
        \UnaryInfC{$ \vdots $}
        \noLine
        \UnaryInfC{$\piencodf{M'[ {\widetilde{y}} \leftarrow  {y}]}_u \vdash  u:\piencodf{\tau} , \piencodf{\Gamma'} , \linvar{y}: \overline{\piencodf{\sigma^k}_{(\sigma, i)}}  , \piencodf{\Theta} , \unvar{y}: \overline{\piencodf{\eta}}$}
        \LeftLabel{\ttype{$\bot$}}
        \UnaryInfC{$ \gclose{ y } ; \piencodf{M'[ {\widetilde{y}} \leftarrow  {y}]}_u \vdash y{:}\bot, u:\piencodf{\tau} , \piencodf{\Gamma'} , \linvar{y}: \overline{\piencodf{\sigma^k}_{(\sigma, i)}}  , \piencodf{\Theta} , \unvar{y}: \overline{\piencodf{\eta}}$}
        \LeftLabel{\ttype{$\parr$}}
        \UnaryInfC{$\gname{ y }{ \unvar{y} }; \gclose{ y } ; \piencodf{M'[ {\widetilde{y}} \leftarrow  {y}]}_u \vdash y: ( \overline{\piencodf{\eta}}) \ampy (\bot) , u:\piencodf{\tau} , \piencodf{\Gamma'} , \linvar{y}: \overline{\piencodf{\sigma^k}_{(\sigma, i)}} , \piencodf{\Theta} $}
        \LeftLabel{\ttype{$\parr$}}
        \UnaryInfC{$ \gname{ y }{ \linvar{y} }; \gname{ y }{ \unvar{y} }; \gclose{ y } ; \piencodf{M'[ {\widetilde{y}} \leftarrow  {y}]}_u \vdash  y: \overline{\piencodf{\sigma^k}_{(\sigma, i)}} \ampy (( \overline{\piencodf{\eta}}) \ampy (\bot)) , u:\piencodf{\tau} , \piencodf{\Gamma'} , \piencodf{\Theta} $}
        \LeftLabel{\ttype{${\with}\some$}}
        \UnaryInfC{$ \psome{y}; P \vdash y :\with(  \overline{\piencodf{\sigma^k}_{(\sigma, i)}} \ampy (( \overline{\piencodf{\eta}}) \ampy (\bot))) , u:\piencodf{\tau} , \piencodf{\Gamma'} , \piencodf{\Theta} $}
        \LeftLabel{\ttype{$\parr$}}
        \UnaryInfC{$\gname{ u }{ y }; \psome{y}; P \vdash u: \with(  \overline{\piencodf{\sigma^k}_{(\sigma, i)}} \ampy (( \overline{\piencodf{\eta}}) \ampy (\bot))) \ampy \piencodf{\tau} , \piencodf{\Gamma'} , \piencodf{\Theta}  $}
        \LeftLabel{\ttype{${\with}\some$}}
        \UnaryInfC{$\psome{u}; \gname{ u }{ y }; \psome{y}; P \vdash u :\underbrace{\with (\with(  \overline{\piencodf{\sigma^k}_{(\sigma, i)}} \ampy (( \overline{\piencodf{\eta}}) \ampy (\bot))) \ampy \piencodf{\tau})}_{\piencodf{(\sigma^{k} , \eta )   \rightarrow \tau} \qquad \defref{def:enc_sestypfailunres}} , \piencodf{\Gamma'} , \piencodf{\Theta} $}
    \end{prooftree}

    Let us consider the following two cases:
    \begin{itemize}
        \item $y = x$

        By the IH  there exists a derivation $\Pi_1'$ of \(\piencodf{M'[ {\widetilde{y}} \leftarrow  {y}]}_u \vdash  u:\piencodf{\tau} , \piencodf{\Gamma'} , \linvar{y}: \overline{\piencodf{\sigma^k}_{(\sigma, i)}}  , \piencodf{\Theta} , \unvar{y}: \overline{\piencodf{\epsilon}}\). Following the steps above we obtain \(\psome{u}; \gname{ u }{ y }; \psome{y}; P \vdash u:\piencod{(\sigma^{k} , \epsilon )   \rightarrow \tau}, \piencodf{\Gamma'} , \piencodf{\Theta}\).

        \item $ y \not = x$

        Then we have that $\piencodf{\Theta} = \piencodf{\Theta'}, \unvar{x} : \dual{\piencodf{\epsilon}}$ By the IH  there exists a derivation $\Pi_1'$ of \(\piencodf{M'[ {\widetilde{y}} \leftarrow  {y}]}_u \vdash  u:\piencodf{\tau} , \piencodf{\Gamma'} , \linvar{y}: \overline{\piencodf{\sigma^k}_{(\sigma, i)}}  , \piencodf{\Theta'} , \unvar{y}: \overline{\piencodf{\eta}}, \unvar{x} : \dual{\piencodf{\epsilon}} \) and then redo the steps above and obtain \(\psome{u}; \gname{ u }{ y }; \psome{y}; P \vdash u:\piencod{(\sigma^{k} , \eta )   \rightarrow \tau}, \piencodf{\Gamma'} , \piencodf{\Theta'}, \unvar{x} : \dual{\piencodf{\epsilon}} \).

    \end{itemize}

    \item The analysis of the other cases proceeds similarly. \qedhere

    \end{enumerate}

\end{proof}

\begin{restatable}[Type Preservation]{theorem}{thmEncTypePres}\label{t:preservationencode}
    Let $B$ and ${M}$ be a bag and an term in \lamcoldetsh, respectively.
    \begin{enumerate}
        \item If $\Theta ; \Gamma \wfdash B : (\sigma^{k} , \eta )$
        then
        $\piencodf{B}_u \wfdash  \piencodf{\Gamma}, u : \piencodf{(\sigma^{k} , \eta )}_{(\sigma, i)} , \piencodf{\Theta}$.

        \item If $\Theta ; \Gamma \wfdash M : \tau$
        then
        $\piencodf{{M}}_u \wfdash  \piencodf{\Gamma}, u :\piencodf{\tau} , \piencodf{\Theta}$.
    \end{enumerate}
\end{restatable}

\begin{proof}

The proof is by mutual induction on the typing derivation of $B$ and ${M'}$, with an analysis for the last rule applied.
Recall that the translation of types ($\piencodf{-}$) has been given in
\defref{def:enc_sestypfailunres}. We will be silently performing \ttype{contract} to split the translation of unrestricted context in derivation trees of \clpi processes as well as combining multiple \ttype{weaken} when convenient.

\begin{enumerate}
\item For $ B =  C \bagsep U$:

\begin{prooftree}
    \AxiomC{\( \Theta ; \Gamma\wfdash C : \sigma^k\)}
    \AxiomC{\(  \Theta ;\cdot \wfdash  U : \eta \)}
\LeftLabel{\redlab{FS{:}bag}}
\BinaryInfC{\( \Theta ; \Gamma \wfdash C \bagsep U : (\sigma^{k} , \eta  ) \)}
\end{prooftree}

        Our translation gives:
$ \piencodf{C \bagsep U}_u = \gsome{ x }{ \llfv{C} }; \pname{x}{\linvar{x}} .( \piencodf{ C }_{\linvar{x}}   \| \pname{x}{\unvar{x}} .(  \guname{ \unvar{x} }{ x_i };  \piencodf{ U }_{x_i}   \|  \pclose{ x } ) )$.
        In addition, the translation of $(\sigma^{k} , \eta  )$ is:
        $$\piencodf{ (\sigma^{k} , \eta  )  }_{(\sigma, i)} = \oplus( (\piencodf{\sigma^{k} }_{(\sigma, i)}) \otimes ((\piencodf{\eta}) \otimes (\onef))  )\quad \text{(for some  $i \geq 0$ and  strict type $\sigma$)}$$

        And one can build the following type derivation (rules from \figref{fig:trulespi}):
{
\small
\begin{prooftree}
\AxiomC{$ \piencodf{ C }_{\linvar{x}} \vdash \piencodf{\Gamma}, \linvar{x}:\piencodf{\sigma^{k} }_{(\sigma, i)} , \piencodf{\Theta}$}

\AxiomC{$ \piencodf{ U }_{x_i} \vdash x_i: \with_{\eta_i \in \eta} \{ i ; \piencodf{\eta_i} \}, \piencodf{\Theta}$}
\LeftLabel{\ttype{$!$}}
\UnaryInfC{$ \guname{ \unvar{x} }{ x_i } ;  \piencodf{ U }_{x_i} \vdash \unvar{x}: \piencodf{\eta} , \piencodf{\Theta}$}

\AxiomC{\mbox{\ }}
\LeftLabel{\ttype{$\1$}}
\UnaryInfC{$ \pclose{ x } \vdash x: \onef $}
\UnaryInfC{$ \pclose{ x } \vdash x: \onef , \piencodf{\Theta}$}
\LeftLabel{\ttype{$\tensor$}}
\BinaryInfC{$ \pname{x}{\unvar{x}} .(  \guname{ \unvar{x} }{ x_i } ;  \piencodf{ U }_{x_i}   \|  \pclose{ x } ) \vdash x: (\piencodf{\eta}) \otimes (\onef) , \piencodf{\Theta}$}
\LeftLabel{\ttype{$\tensor$}}
\BinaryInfC{$\pname{x}{\linvar{x}} .( \piencodf{ C }_{\linvar{x}}   \| \pname{x}{\unvar{x}} .(  \guname{ \unvar{x} }{ x_i } ;  \piencodf{ U }_{x_i}   \|  \pclose{ x } ) ) \vdash  \piencodf{\Gamma}, x:(\piencodf{\sigma^{k} }_{(\sigma, i)}) \otimes ((\piencodf{\eta}) \otimes (\onef)) , \piencodf{\Theta} $}
\LeftLabel{\ttype{${\oplus}\some$}}
\UnaryInfC{$\gsome{ x }{ \llfv{C} };  \pname{x}{\linvar{x}} .( \piencodf{ C }_{\linvar{x}}   \| \pname{x}{\unvar{x}} .(  \guname{ \unvar{x} }{ x_i } ;  \piencodf{ U }_{x_i}   \|  \pclose{ x } ) ) \vdash \piencodf{\Gamma}, x{:}\piencodf{ (\sigma^{k} , \eta  )  }_{(\sigma, i)}, \piencodf{\Theta}$}
\end{prooftree}
}

The result follows  provided both $ \piencodf{ C }_{\linvar{x}} \vdash \piencodf{\Gamma}, \linvar{x}:\piencodf{\sigma^{k} }_{(\sigma, i)} , \piencodf{\Theta}$ and $ \piencodf{ U }_{x_i} \vdash  x_i: \with_{\eta_i \in \eta} \{ i ; \piencodf{\eta_i} \}, \piencodf{\Theta}$ hold.

\begin{enumerate}

\item For $ \piencodf{ C }_{\linvar{x}} \vdash \piencodf{\Gamma}, \linvar{x}:\piencodf{\sigma^{k} }_{(\sigma, i)} , \piencodf{\Theta}$ to hold analyze the shape of $C$:

\noindent{\bf For $C = \oneb$:}
\begin{prooftree}
    \AxiomC{\(  \)}
    \LeftLabel{\redlab{FS{:}\oneb^{\ell}}}
    \UnaryInfC{\( \Theta ; \dash \wfdash \oneb : \omega \)}
\end{prooftree}

Our translation gives:         $ \piencodf{\oneb}_{\linvar{x}} = \gsome{ \linvar{x} }{ \emptyset }; \gname{ \linvar{x} }{ y_n }; ( \psome{y_n}; \pclose{ y_n }    \| \gsome{ \linvar{x} }{ \emptyset };   \pnone{ \linvar{x} }  ). $

and  the translation of $\omega$ can be either:
\begin{enumerate}
\item  $\piencodf{\omega}_{(\sigma,0)} =  \overline{\with(( \oplus \bot )\otimes ( \with \oplus \bot ))}$; or
\item $\piencodf{\omega}_{(\sigma, i)} =  \overline{   \with(( \oplus \bot) \otimes ( \with  \oplus (( \with  \overline{\piencodf{ \sigma }} )  \ampy (\overline{\piencodf{\omega}_{(\sigma, i - 1)}})))) }$
\end{enumerate}

And one can build the following type derivation (rules from \figref{fig:trulespi}):

\begin{prooftree}
\AxiomC{\mbox{\ }}
\UnaryInfC{$ \pclose{ y_n } \vdash y_n: \onef$}
\UnaryInfC{$ \pclose{ y_n } \vdash y_n: \onef, \piencodf{\Theta}$}
\UnaryInfC{$\psome{y_n}; \pclose{ y_n }  \vdash  y_n :\with \onef , \piencodf{\Theta}$}
\AxiomC{}
\UnaryInfC{$  \pnone{ \linvar{x} } \vdash \linvar{x} :\with A $}
\UnaryInfC{$  \pnone{ \linvar{x} } \vdash \linvar{x} :\with A , \piencodf{\Theta}$}
\UnaryInfC{$\gsome{ \linvar{x} }{ \emptyset };   \pnone{ \linvar{x} }\vdash  \linvar{x}{:}\oplus \with A, \piencodf{\Theta}$}
\BinaryInfC{$(\psome{y_n}; \pclose{ y_n }   \| \gsome{ \linvar{x} }{ \emptyset };   \pnone{ \linvar{x} } ) \vdash y_n :\with \onef, \linvar{x}{:}\oplus \with A , \piencodf{\Theta}$}
\UnaryInfC{$\gname{\linvar{x}}{y_n}; ( \psome{y_n}; \pclose{ y_n }    \| \gsome{ \linvar{x} }{ \emptyset };   \pnone{ \linvar{x} } ) \vdash  \linvar{x}: (\with \onef) \ampy (\oplus \with A) , \piencodf{\Theta}$}
\UnaryInfC{$\gsome{ \linvar{x} }{ \emptyset };  \gname{ \linvar{x} }{y_n  };  ( \psome{y_n} ; \pclose{ y_n }    \| \gsome{ \linvar{x} }{ \emptyset };    \pnone{ \linvar{x} }) \vdash  \linvar{x}{:}\oplus ((\with \onef) \ampy (\oplus \with A)), \piencodf{\Theta}$}
\end{prooftree}

Since $A$ is arbitrary,  we can take $A=\oneb$ for $\piencodf{\omega}_{(\sigma,0)} $ and  $A= \overline{(( \with  \overline{\piencodf{ \sigma }} )  \ampy (\overline{\piencodf{\omega}_{(\sigma, i - 1)}}))}$  for $\piencodf{\omega}_{(\sigma,i)} $, in both cases, the result follows.

\item
\noindent{\bf For  $C = \bag{M'}  \cdot C'$:}

\begin{prooftree}
\AxiomC{\( \Theta ; \Gamma' \wfdash M' : \sigma\)}
\AxiomC{\( \Theta ; \Delta \wfdash C' : \sigma^k\)}
\LeftLabel{\redlab{FS{:}bag^{\ell}}}
\BinaryInfC{\( \Theta ; \Gamma' \contexcat \Delta \wfdash \bag{M'}  \cdot C':\sigma^{k}\)}
\end{prooftree}

Where $ \Gamma = \Gamma' \contexcat \Delta$.
To simplify the proof, we will consider $k=3$.

By IH we have \(
\piencodf{M'}_{x_i}  \vdash \piencodf{\Gamma'}, x_i: \piencodf{\sigma}; \piencodf{\Theta} \) and
\( \piencodf{C'}_{\linvar{x}}  \vdash \piencodf{\Delta}, \linvar{x}: \piencodf{\sigma\wedge \sigma}_{(\tau, j)}; \piencodf{\Theta}\).

Our translation  \figref{fig:encoding} gives:
\begin{equation}
\begin{aligned}
\piencodf{C}_{\linvar{x}} & =
        \gsome{ \linvar{x} }{ \llfv{C} }; \gname{\linvar{x}}{y_i}; \gsome{ \linvar{x} }{ y_i, \llfv{C} }; \psome{\linvar{x}} ;  \pname{\linvar{x}}{x_i}. \\
        & \qquad  (\gsome{ x_i }{ \llfv{M'} };  \piencodf{M'}_{x_i}   \| \piencodf{(C' )}_{\linvar{x}}   \|   \pnone{ y_i })
\end{aligned}
\end{equation}

Let $\Pi_{1}$ be the derivation:

\begin{prooftree}
\AxiomC{$\piencodf{M'}_{x_i} \;{ \vdash} \piencodf{\Gamma'}, x_i: \piencodf{\sigma}, \piencodf{\Theta} $}
\UnaryInfC{$\gsome{ x_i }{ \llfv{M'} };  \piencodf{M'}_{x_i} \vdash \piencodf{\Gamma'} ,x_i: \oplus \piencodf{\sigma} , \piencodf{\Theta}$}
\AxiomC{}
\UnaryInfC{$   \pnone{ y_i }  \vdash y_i :\with \onef $}
\UnaryInfC{$   \pnone{ y_i }  \vdash y_i :\with \onef, \piencodf{\Theta}$}
\BinaryInfC{$\gsome{  x_i }{ \llfv{M'} }; \piencodf{M'}_{x_i}   \|   \pnone{ y_i } \vdash \piencodf{\Gamma'} ,x_i: \oplus \piencodf{\sigma}, y_i :\with \onef , \piencodf{\Theta}$}
\end{prooftree}

Let $ P =   \piencodf{M'}_{x_i}   \| \piencodf{C'}_{\linvar{x}}   \|   \pnone{ y_i }$, in the derivation $\Pi_{2}$ below:

\begin{prooftree}
\AxiomC{$ \Pi_{1}$}

\AxiomC{$ \piencodf{C' }_{\linvar{x}}  \vdash  \piencodf{\Delta}, \linvar{x}: \piencodf{\sigma\wedge \sigma}_{(\tau, j)}, \piencodf{\Theta} $}

\LeftLabel{\ttype{$\tensor$}}
\BinaryInfC{$ \pname{\linvar{x}}{x_i}. P \vdash  \piencodf{\Gamma'}  ,  \piencodf{\Delta}, y_i :\with \onef, \linvar{x}: (\oplus \piencodf{\sigma})  \otimes (\piencodf{\sigma\wedge \sigma}_{(\tau, j)}) , \piencodf{\Theta}$}
\LeftLabel{\ttype{${\with}\some$}}
\UnaryInfC{$\underbrace{ \psome{\linvar{x}} ;   \pname{\linvar{x}}{x_i}. P }_{P_{2}} \vdash \piencodf{\Gamma'}  ,  \piencodf{\Delta}, y_i :\with \onef, \linvar{x}: \with (( \oplus \piencodf{\sigma} ) \otimes (\piencodf{\sigma\wedge \sigma}_{(\tau, j)})) , \piencodf{\Theta} $}
\end{prooftree}

Let $P_2 =   \psome{\linvar{x}} ;  \pname{\linvar{x}}{x_i}. P$ in the derivation below:

\begin{prooftree}
\AxiomC{$ \Pi_{2}$}
\noLine
\UnaryInfC{$\vdots$}
\noLine
\UnaryInfC{$P_2\vdash \piencodf{\Gamma}, y_i :\with \onef, \linvar{x}: \with (( \oplus \piencodf{\sigma} ) \otimes (\piencodf{\sigma\wedge \sigma}_{(\tau, j)})) , \piencodf{\Theta} $}
\LeftLabel{\ttype{${\oplus}\some$}}
\UnaryInfC{$\gsome{ \linvar{x} }{ y_i, \llfv{C} };P_2  \vdash \piencodf{\Gamma}, y_i :\with \onef, \linvar{x}:\oplus  \with (( \oplus \piencodf{\sigma} ) \otimes (\piencodf{\sigma\wedge \sigma}_{(\tau, j)})), \piencodf{\Theta}$}
\LeftLabel{\ttype{$\parr$}}
\UnaryInfC{$
\gname{\linvar{x}}{y_i}; \gsome{ \linvar{x} }{ y_i, \llfv{C} }; P_2
\vdash \piencodf{\Gamma}, \linvar{x}: ( \with \onef) \ampy ( \oplus  \with (( \oplus \piencodf{\sigma} ) \otimes (\piencodf{\sigma\wedge \sigma}_{(\tau, j)}))), \piencodf{\Theta} $}
\LeftLabel{\ttype{${\oplus}\some$}}
\UnaryInfC{$\piencodf{C}_{\linvar{x}} \vdash   \piencodf{\Gamma}, \linvar{x}: \underbrace{\oplus(( \with \onef) \ampy ( \oplus  \with (( \oplus \piencodf{\sigma} ) \otimes (\piencodf{\sigma\wedge \sigma}_{(\tau, j)}))))}_{\piencodf{\sigma\wedge \sigma \wedge \sigma}_{(\tau, j)} \qquad \defref{def:enc_sestypfailunres}}, \piencodf{\Theta} $}
\end{prooftree}

Therefore, $ \piencodf{C}_{\linvar{x}} \vdash \piencodf{\Gamma}, \linvar{x}: \piencodf{\sigma\wedge \sigma \wedge \sigma}_{(\tau, j)} , \piencodf{\Theta} $ and the result follows.

             \item For $ \piencodf{ U }_{x_i} \vdash x_i: \with_{\eta_i \in \eta} \{ i : \piencodf{\eta_i} \}, \piencodf{\Theta}$.

To simplify the proof, we  consider $U= \unvar{\oneb} \concat \unvar{\bag{M'}}$ with $\eta =  \sigma_1 \concat \sigma_2 $ and $\with_{\eta_i \in \eta} = \with \{ 1 : \piencodf{ \sigma_1} , 2 : \piencodf{\sigma_2} \}$.

Our translation gives $ \piencodf{ U }_{x_i} = \gsel{ x_i }\{ {1} :   \pnone{ x_i }  , {2} : \piencodf{M'}_{x_i}  \}_{  }  $. Hence, we have:

\begin{prooftree}
\AxiomC{\(  \)}
\LeftLabel{\redlab{FS{:}bag^!}}
\UnaryInfC{\( \Theta ;  \dash  \wfdash \unvar{\oneb} : \sigma_1 \)}

\AxiomC{\( \Theta ; \cdot \wfdash M' : \sigma_2\)}
\LeftLabel{\redlab{FS{:}bag^!}}
\UnaryInfC{\( \Theta ; \cdot  \wfdash \unvar{\bag{M'}}:\sigma_2 \)}
\LeftLabel{\redlab{FS{:}\concat-bag^{!}}}
\BinaryInfC{\( \Theta ; \cdot  \wfdash \unvar{\oneb} \concat \unvar{\bag{M'}} : \sigma_1 \concat \sigma_2 \)}
\end{prooftree}

By the induction hypothesis we have that $ \Theta ; \cdot \wfdash M' : \sigma $ implies $ \piencodf{M'}_{x_i} \wfdash x_i : \piencodf{\sigma} ,  \piencodf{\Theta}$. Thus,

\begin{prooftree}
\AxiomC{}
\LeftLabel{\ttype{${\with}\none$}}
\UnaryInfC{$   \pnone{ x_i } \vdash x_i:  \piencodf{ \sigma_1}  $}
\LeftLabel{\ttype{weaken}}
\UnaryInfC{$   \pnone{ x_i } \vdash x_i:  \piencodf{ \sigma_1}  , \piencodf{\Theta} $}

\AxiomC{$ \piencodf{M'}_{x_i} \vdash  x_i:  \piencodf{\sigma_2} , \piencodf{\Theta} $}
\LeftLabel{\ttype{$\with$}}
\BinaryInfC{$ \gsel{ x_i }\{ {1} :   \pnone{ x_i }  , {2} : \piencodf{M'}_{x_i} \}_{  } \vdash  x_i: \with \{ 1 : \piencodf{ \sigma_1} , 2 : \piencodf{\sigma_2} \} , \piencodf{\Theta} $}
\end{prooftree}

Therefore, $\gsel{ x_i }\{ {1} :   \pnone{  x_i } , {2} : \piencodf{M'}_{x_i}  \}_{  } \vdash  x_i: \with \{ 1 : \piencodf{ \sigma_1} , 2 : \piencodf{\sigma_2} \} , \piencodf{\Theta}  $ and the result follows.

\end{enumerate}

\item  The proof of type preservation for terms, relies on the analysis of ten cases:

\begin{enumerate}

\item {\bf Rule \redlab{FS{:}var^{\ell}}:}
Then we have the following derivation:

\begin{prooftree}
\AxiomC{}
\LeftLabel{\redlab{FS{:}var^{\ell}}}
\UnaryInfC{\( \Theta;  {x}: \tau \wfdash  {x} : \tau\)}
\end{prooftree}

By \defref{def:enc_sestypfailunres},  $\piencodf{ {x}:\tau}=  {x}:\with \overline{\piencodf{\tau }}$, and by \figref{fig:encoding},  $\piencodf{ {x}}_u = \psome{x} ;   \pfwd{x}{u} $. The result follows from the derivation:

\begin{prooftree}
    \AxiomC{}
    \LeftLabel{\ttype{id}}
    \UnaryInfC{$ \pfwd{x}{u} \vdash  {x}:  \overline{\piencodf{\tau }}  , u :  \piencodf{ \tau } , \piencodf{\Theta} $}
    \LeftLabel{\ttype{${\with}\some$}}
    \UnaryInfC{$  \psome{x} ; \pfwd{x}{u} \vdash  {x}: \with  \overline{\piencodf{ \tau }} , u :  \piencodf{ \tau } , \piencodf{\Theta} $}
\end{prooftree}

\item {\bf Rule \redlab{FS{:}var^!}:}
Then we have the following derivation provided $\eta_{k} = \tau $:

\begin{prooftree}
\AxiomC{}
\LeftLabel{\redlab{FS{:}var^{\ell}}}
\UnaryInfC{\( \Theta , \unvar{x}: \eta;  {x}: \eta_{k}  \wfdash  {x} : \tau\)}
\LeftLabel{\redlab{FS{:}var^!}}
\UnaryInfC{\( \Theta , \unvar{x}: \eta; \dash \wfdash  {x}[k] : \tau\)}
\end{prooftree}

By \defref{def:enc_sestypfailunres},  $\piencodf{\Theta , \unvar{x}: \eta}= \piencodf{\Theta} , \unvar{x}: \dual{!  \with_{\eta_i \in \eta} \{ i ; \piencodf{\eta_i} \} }$, and by our translation (in \figref{fig:encoding}),  $\piencodf{ {x}[k]}_u = \puname{ \unvar{x} }{ x_i };\psel{ {x}_i }{ k }; \pfwd{x_i}{u} $. The result follows from the derivation:

\begin{prooftree}
    \AxiomC{}
    \LeftLabel{\ttype{id}}
    \UnaryInfC{$ \pfwd{x_i}{u}  \vdash  u :  \piencodf{ \tau },  x_i:  \overline{\piencodf{\eta_{k} }}   , \piencodf{\Theta} $}
\LeftLabel{ \ttype{$\oplus$}}
\UnaryInfC{$  \psel{ {x}_i }{ k }; \pfwd{x_i}{u} \vdash  u :  \piencodf{ \tau }, {x}_i :  \oplus_{\eta_i \in \eta} \{  i ; \dual{\piencodf{\eta_i}}  \} ,  \piencodf{\Theta}  $}
\LeftLabel{\ttype{$?$}}
\UnaryInfC{$ \puname{ \unvar{x} }{ x_i }; \psel{ {x}_i }{ k }; \pfwd{x_i}{u} \vdash  u :  \piencodf{ \tau } , \unvar{x}: ? \oplus_{\eta_i \in \eta} \{  i ; \dual{\piencodf{\eta_i}} \} , \piencodf{\Theta} $}
\end{prooftree}

\item {\bf Rule \redlab{FS\!:\!weak}:}
        Then we have the following derivation:

\begin{prooftree}
\AxiomC{\( \Theta ; \Gamma  \wfdash M' : \tau\)}
\LeftLabel{ \redlab{FS\!:\!weak}}
\UnaryInfC{\( \Theta ; \Gamma ,  {x}: \omega \wfdash M'[\leftarrow  {x}]: \tau \)}
\end{prooftree}

By \defref{def:enc_sestypfailunres},  $\piencodf{\Gamma ,  {x}: \omega}= \piencodf{\Gamma}, \linvar{x}: \overline{\piencodf{\omega }_{(\sigma, i_1)}}$, and by our translation \figref{fig:encoding},

$\piencodf{M'[  \leftarrow  {x}]}_u = \psome{\linvar{x}} ;  \pname{\linvar{x}}{y_i} . (\gsome{  y_i }{ u,\llfv{M'} }; \gclose{ y_{i} } ; \piencodf{M'}_u   \|   \pnone{ \linvar{x} } ) $.

By IH, we have $\piencodf{M'}_u\vdash  \piencodf{\Gamma }, u:\piencodf{\tau} , \piencodf{\Theta }$.
The result follows from the derivation, omitting labels:

        \begin{prooftree}
    \AxiomC{$\piencodf{M'}_u\vdash  \piencodf{\Gamma }, u:\piencodf{\tau} , \piencodf{\Theta} $}
    \UnaryInfC{$\gclose{ y_{i} } ; \piencodf{M'}_u \vdash y_i{:}\bot, \piencodf{\Gamma }, u:\piencodf{\tau} , \piencodf{\Theta}$}
    \UnaryInfC{$\gsome{ y_i }{ u,\llfv{M'} };  \gclose{ y_{i} } ; \piencodf{M'}_u \vdash  y_i{:}\oplus \bot , \piencodf{\Gamma }, u:\piencodf{\tau} , \piencodf{\Theta}$}

    \AxiomC{}
    \UnaryInfC{$  \pnone{ \linvar{x} }  \vdash \linvar{x} :\with A $}
    \UnaryInfC{$  \pnone{ \linvar{x} }  \vdash \linvar{x} :\with A , \piencodf{\Theta}$}
\BinaryInfC{$\pname{\linvar{x}}{y_i} . ( \gsome{ y_i }{ u,\llfv{M'} }; \gclose{ y_{i} } ; \piencodf{M'}_u   \|   \pnone{ \linvar{x} } ) \vdash  \linvar{x}: (\oplus \bot) \otimes (\with A) , \piencodf{\Gamma }, u:\piencodf{\tau} , \piencodf{\Theta}$}
\UnaryInfC{$\piencodf{M'[  \leftarrow  {x}]}_u \vdash  \linvar{x} :\with ((\oplus \bot) \otimes (\with A))  , \piencodf{\Gamma }, u:\piencodf{\tau} , \piencodf{\Theta} $}
\end{prooftree}

Since $A$ is arbitrary,  we can take $A=\oneb$ for $\piencodf{\omega}_{(\sigma,0)} $ and  $A= \overline{(( \with  \overline{\piencodf{ \sigma }} )  \ampy (\overline{\piencodf{\omega}_{(\sigma, i - 1)}}))}$  for $\piencodf{\omega}_{(\sigma,i)} $ where $i > 0$, in both cases, the result follows.

\item {\bf Rule $\redlab{FS:abs \dash sh}$:}

Then $M= \lambda x . (M'[ {\widetilde{x}} \leftarrow  {x}])$, and the derivation is:

\begin{prooftree}
\AxiomC{\( \Theta , \unvar{x}:\eta ; \Gamma ,  {x}: \sigma^k \wfdash M'[ {\widetilde{x}} \leftarrow  {x}] : \tau \quad  {x} \notin \dom{\Gamma} \)}
\LeftLabel{\redlab{FS{:}abs\dash sh}}
\UnaryInfC{\( \Theta ; \Gamma \wfdash \lambda x . (M'[ {\widetilde{x}} \leftarrow  {x}])  : (\sigma^k, \eta )  \rightarrow \tau \)}
\end{prooftree}

By IH, we have $\piencodf{M'[ {\widetilde{x}} \leftarrow  {x}]}_u \vdash  u:\piencodf{\tau} , \piencodf{\Gamma} , \linvar{x}: \overline{\piencodf{\sigma^k}_{(\sigma, i)}}  , \piencodf{\Theta} , \unvar{x}: \overline{\piencodf{\eta}} $ and our translation (in
\figref{fig:encoding}) gives
$ \piencodf{\lambda x.M'[ {\widetilde{x}} \leftarrow x]}_u = \psome{u}; \gname{u}{x}; \psome{x}; \gname{x}{\linvar{x}}; \gname{x}{\unvar{x}}; \gclose{ x } ; \piencodf{M'[ {\widetilde{x}} \leftarrow  {x}]}_u $, whose type derivation $\piencodf{M'[ {\widetilde{x}} \leftarrow  {x}]}_u \vdash u:\piencodf{(\sigma^{k} , \eta )   \rightarrow \tau}  , \piencodf{\Gamma} , \piencodf{\Theta} $,  was given in the proof of Lemma~\ref{lem:relunbag-typeunres}, item 3.

\item {\bf Rule $\redlab{FS:app}$:}
Then $M = M'\ B$, where $ B = C \bagsep U $ and the derivation is:

\begin{prooftree}
\AxiomC{\( \Theta ;\Gamma \wfdash M' : (\sigma^{j} , \eta ) \rightarrow \tau \)}
\AxiomC{\(  \Theta ;\Delta \wfdash B : (\sigma^{k} , \epsilon )  \)}
\AxiomC{\( \eta \relunbag \epsilon \)}
\LeftLabel{\redlab{FS{:}app}}
\TrinaryInfC{\( \Theta ; \Gamma, \Delta \wfdash M'\ B : \tau\)}
\end{prooftree}

By IH, we have both

\begin{itemize}
\item  $\piencodf{M'}_u\vdash \piencodf{\Gamma}, u:\piencodf{(\sigma^{j} , \eta ) \rightarrow \tau} , \piencodf{\Theta}$;
\item $\piencodf{M'}_u\vdash \piencodf{\Gamma}, u:\piencodf{(\sigma^{j} , \epsilon ) \rightarrow \tau} , \piencodf{\Theta}$,   by Lemma \ref{lem:relunbag-typeunres};
\item $\piencodf{B}_u\vdash \piencodf{\Delta}, u:\overline{\piencodf{(\sigma^{k} , \epsilon ) }_{(\tau_2, n)}}  , \piencodf{\Theta} $, for some $\tau_2$ and some $n$.
\end{itemize}

Therefore, from the fact that $M$ is well-formed and \figref{fig:encoding} and \ref{def:enc_sestypfailunres}, we  have:

\begin{itemize}
\item $\displaystyle{\piencodf{M' (C \bagsep U)}_u =   \res{ v } (\piencodf{M'}_v   \| \gsome{ v }{ u , \llfv{C} };  \pname{v}{x} . (\pfwd{v}{u}   \| \piencodf{C \bagsep U}_x ) )} $;
\item $\piencodf{(\sigma^{j} , \eta ) \rightarrow \tau}= \oplus( (\piencodf{\sigma^{k} }_{(\tau_1, m)}) \otimes ((!\piencodf{\eta})  $, for some $\tau_1$ and some $m$.
\end{itemize}

Also, since $\piencodf{B}_u\vdash \piencodf{\Delta}, u:\piencodf{(\sigma^{k} , \epsilon )}_{(\tau_2, n)}, \piencodf{\Theta}$, we have the following derivation $\Pi$:

\begin{prooftree}
\AxiomC{$\piencodf{C \bagsep U}_x\vdash \piencodf{\Delta}, x:\piencodf{(\sigma^{k} , \epsilon )}_{(\tau_2, n)}  , \piencodf{\Theta} $ }
\AxiomC{\(\)}
\UnaryInfC{$ \pfwd{v}{u}  \vdash v:  \overline{\piencodf{ \tau }} , u: \piencodf{ \tau }  $}
\UnaryInfC{$ \pfwd{v}{u}  \vdash v:  \overline{\piencodf{ \tau }} , u: \piencodf{ \tau }  , \piencodf{\Theta}$}
\BinaryInfC{$\pname{v}{x} . (\pfwd{v}{u}   \| \piencodf{C \bagsep U}_x ) \vdash \piencodf{ \Delta }, v:\piencodf{(\sigma^{k} , \epsilon )}_{(\tau_2, n)} \otimes \overline{ \piencodf{ \tau }} , u:\piencodf{ \tau }  , \piencodf{\Theta} $}
\UnaryInfC{$ \gsome{ v }{ u , \llfv{C} };   \pname{v}{x} . (\pfwd{v}{u}   \| \piencodf{C \bagsep U}_x )  \vdash\piencodf{ \Delta }, v:\underbrace{\oplus (\piencodf{(\sigma^{k} , \epsilon )}_{(\tau_2, n)} \otimes  \overline{\piencodf{ \tau }})}_{\overline{\piencodf{(\sigma^{k} , \epsilon ) \rightarrow \tau}}}, u:\piencodf{ \tau } , \piencodf{\Theta} $}
\end{prooftree}
  In order to apply \ttype{cut}, we must have that $\piencodf{\sigma^{j}}_{(\tau_1, m)} = \piencodf{\sigma^{k}}_{(\tau_2, n)}$, therefore, the choice of $\tau_1,\tau_2,n$ and $m$, will consider the different possibilities for $j$ and $k$, as in Proposition~\ref{prop:app_auxunres}.
\begin{prooftree}
\AxiomC{\( \piencodf{M'}_v\vdash \piencodf{\Gamma}, v:\piencodf{(\sigma^{j} , \epsilon ) \rightarrow \tau} ; \piencodf{\Theta} \)}
\AxiomC{$\Pi$}
\LeftLabel{\ttype{cut}}
\BinaryInfC{$  \res{ v }  ( \piencodf{ M'}_v   \| \gsome{ v }{ u , \llfv{C} };  \pname{v}{x} . (\pfwd{v}{u}   \| \piencodf{B}_x ) ) \vdash \piencodf{ \Gamma } ,\piencodf{ \Delta } , u: \piencodf{ \tau }  ; \piencodf{\Theta} $}
\end{prooftree}

We can then conclude that $\piencodf{M' B}_u \vdash \piencodf{ \Gamma}, \piencodf{ \Delta }, u:\piencodf{ \tau } , \piencodf{\Theta} $ and the result follows.

\item {\bf Rule $\redlab{FS:share}$:}
Then $M = M' [  {x}_1, \dots  {x}_k \leftarrow x ]$ and the derivation is:
\begin{prooftree}
    \AxiomC{\( \Theta ; \Delta ,  {x}_1: \sigma, \cdots,  {x}_k: \sigma \wfdash M' : \tau \quad  {x} \notin \Delta \quad k \not = 0\)}
    \LeftLabel{ \redlab{FS:share}}
    \UnaryInfC{\( \Theta ; \Delta ,  {x}: \sigma_{k} \wfdash M'[ {x}_1 , \cdots ,  {x}_k \leftarrow  {x}] : \tau \)}
\end{prooftree}

To simplify the proof we will consider $k=1$ (the case in which $k>1$ follows similarly).

By IH, we have $\piencodf{M'}_u\vdash  \piencodf{\Delta ,  {x}_1:\sigma }, u:\piencodf{\tau} ; \piencodf{\Theta}$.
From
\figref{fig:encoding} and \ref{def:enc_sestypfailunres}, it follows

\begin{itemize}
    \item $\piencodf{ \Delta ,  {x}_1: \sigma} = \piencodf{\Delta}, \linvar{x}_1:\with\overline{\piencodf{\sigma}} $.
\item
$
\piencodf{M'[ {x}_1, \leftarrow  {x}]}_u =
    \begin{array}[t]{l}
        \psome{\linvar{x}}; \pname{\linvar{x}}{y_1}. (\gsome{ y_1 }{ \emptyset };   \gclose{ y_{1} } ;\0
            \| \psome{\linvar{x}};\gsome{ \linvar{x} }{ u, \llfv{M'} \setminus  {x}_1  };
        \\
        \bignd_{x_i \in x_1}\gname{x}{{x}_i}; \psome{\linvar{x}}; \pname{\linvar{x}}{y_2} . (\gsome{ y_2 }{ u,\llfv{M'} };  \gclose{ y_{2} } ; \piencodf{M'}_u   \|   \pnone{ \linvar{x} } ) )
    \end{array}
    $

\end{itemize}

We shall split the expression into two parts:
\[
\begin{aligned}
N_1 &= \psome{\linvar{x}}; \pname{\linvar{x}}{y_2} . ( \gsome{ y_2 }{ u,\llfv{M'} }; \gclose{ y_{2} } ; \piencodf{M'}_u   \|   \pnone{ \linvar{x} } ) \\
N_2 &= \psome{\linvar{x}}; \pname{\linvar{x}}{y_1}. (\gsome{ y_1 }{ \emptyset }; \gclose{ y_{1} } ;\0   \| \psome{\linvar{x}}; \gsome{ \linvar{x} }{ u, \llfv{M'} \setminus  {x}_1  }; \gname{x}{{x}_1};N_1)
\end{aligned}
\]

and we obtain the  derivation for term $N_1$ as follows where we omit $, \piencodf{\Theta}$ and derivation labels:
\begin{prooftree}
\AxiomC{$\piencodf{M'}_u \vdash  \piencodf{\Delta ,  {x}_1:\sigma }, u:\piencodf{\tau}  $}
\UnaryInfC{$\gclose{ y_{2} } ; \piencodf{M'}_u \vdash \piencodf{\Delta ,  {x}_1:\sigma }, u:\piencodf{\tau}, y_{2}{:}\bot   $}
\UnaryInfC{$\gsome{ y_2 }{ u,\llfv{M'} };   \gclose{ y_{2} } ; \piencodf{M'}_u \vdash \piencodf{\Delta ,  {x}_1:\sigma }, u:\piencodf{\tau}, y_{2}{:}\oplus \bot    $}

\AxiomC{}
\UnaryInfC{$  \pnone{ \linvar{x} }  \vdash \linvar{x} :\with A  $}
\BinaryInfC{$  \pname{\linvar{x}}{y_2} . (  \gsome{ y_2 }{ u,\llfv{M'} }; \gclose{ y_{2} } ; \piencodf{M'}_u   \|   \pnone{ \linvar{x} } ) \vdash \piencodf{\Delta ,  {x}_1:\sigma }, u:\piencodf{\tau} , \linvar{x}: ( \oplus \bot )\otimes ( \with A )    $}
\UnaryInfC{$\underbrace{ \psome{\linvar{x}}; \pname{\linvar{x}}{y_2} . (  \gsome{ y_2 }{ u,\llfv{M'} };  \gclose{ y_{2} } ; \piencodf{M'}_u   \|   \pnone{ \linvar{x} } ) }_{N_1} \vdash \piencodf{\Delta ,  {x}_1:\sigma } , u:\piencodf{\tau} , \linvar{x}: \overline{\piencodf{\omega}_{(\sigma, i)}}   $}
\end{prooftree}

Notice that the last rule applied \ttype{${\with}\some$} assigns $x: \with ((\oplus \bot) \otimes (\with A))$. Again, since $A$ is arbitrary, we can take $A= \oplus (( \with  \overline{\piencodf{ \sigma }} )  \ampy (\overline{\piencodf{\omega}_{(\sigma, i - 1)}}))$, obtaining $x:\overline{\piencodf{\omega}_{(\sigma,i)}}$.

In order to obtain a type derivation for $N_2$, consider the derivation $\Pi_1$:
{
\small
\begin{prooftree}
\AxiomC{$N_1 \vdash \piencodf{\Delta},  {x}_1:\with\overline{\piencodf{\sigma}} , u:\piencodf{\tau}, \linvar{x}: \overline{\piencodf{\omega}_{(\sigma, i)}}  , \piencodf{\Theta} $}
\LeftLabel{\ttype{$\parr$}}
\UnaryInfC{$\gname{x}{{x}_i};N_1   \vdash \piencodf{\Delta} ,  u:\piencodf{\tau}, \linvar{x}: ( \with\overline{\piencodf{\sigma}} ) \ampy (\overline{\piencodf{\omega}_{(\sigma, i)}})  , \piencodf{\Theta} $}
\LeftLabel{\ttype{$\nd$}}
\UnaryInfC{$\bignd_{x_i \in x_1} \gname{x}{{x}_i};N_1   \vdash \piencodf{\Delta} ,  u:\piencodf{\tau}, \linvar{x}: ( \with\overline{\piencodf{\sigma}} ) \ampy (\overline{\piencodf{\omega}_{(\sigma, i)}})  , \piencodf{\Theta} $}
\UnaryInfC{$  \gsome{ \linvar{x} }{ u, \llfv{M'} \setminus  {x}_1  };  \bignd_{x_i \in x_1} \gname{x}{{x}_i};N_1  \vdash \piencodf{\Delta} ,  u:\piencodf{\tau}, \linvar{x} {:}\oplus (( \with \overline{\piencodf{\sigma}} ) \ampy (\overline{\piencodf{\omega}_{(\sigma, i)}})) , \piencodf{\Theta} $}
\UnaryInfC{$ \psome{\linvar{x}}; \gsome{ \linvar{x} }{ u, \llfv{M'} \setminus  {x}_1  };  \bignd_{x_i \in x_1} \gname{x}{{x}_i};N_1  \vdash \piencodf{\Delta},  u:\piencodf{\tau} , \linvar{x} :\with \oplus (( \with\overline{\piencodf{\sigma}} ) \ampy ( \overline{\piencodf{\omega}_{(\sigma, i)}} ))  , \piencodf{\Theta} $}
\end{prooftree}
}

We take $ P_1 = \psome{\linvar{x}};  \gsome{ \linvar{x} }{ u, \llfv{M'} \setminus  {x}_1  }; \bignd_{x_i \in x_1} \gname{x}{{x}_i};N_1 $ and $\Gamma_1 =   \piencodf{\Delta},  u:\piencodf{\tau} $ and continue the derivation of $ N_2 $

\begin{prooftree}
\AxiomC{}
\LeftLabel{\redlab{T\cdot}}
\UnaryInfC{$\0\vdash \dash , \piencodf{\Theta} $}
\LeftLabel{\ttype{$\bot$}}
\UnaryInfC{$ \gclose{ y_{1} } ;\0 \vdash y_{1} : \bot , \piencodf{\Theta} $}
\UnaryInfC{$  \gsome{ y_1 }{ \emptyset };  \gclose{ y_{1} };\0 \vdash  y_1{:}\oplus \bot , \piencodf{\Theta} $}

\AxiomC{$\Pi_1 $}
\noLine
\UnaryInfC{$\vdots$}
\noLine
\UnaryInfC{$P_1\vdash \Gamma_1, \linvar{x} :\with \oplus (( \with\overline{ \piencodf{\sigma}} ) \ampy ( \overline{\piencodf{\omega}_{(\sigma, i)}} )) , \piencodf{\Theta} $}
\LeftLabel{\ttype{$\tensor$}}
\BinaryInfC{$\pname{\linvar{x}}{y_1}. ( \gsome{ y_1 }{ \emptyset };  \gclose{ y_{1} } ;\0   \| P_1) \vdash \Gamma_1 ,  \linvar{x} : (\oplus \bot)\otimes (\with \oplus (( \with\overline{\piencodf{\sigma}} ) \ampy ( \overline{\piencodf{\omega}_{(\sigma, i)}} )) ) , \piencodf{\Theta} $}
\UnaryInfC{$\underbrace{ \psome{\linvar{x}}; \pname{\linvar{x}}{y_1}. ( \gsome{ y_1 }{ \emptyset }; \gclose{ y_{1} } ;\0   \| P_1)}_{N_2} \vdash \Gamma_1 , \linvar{x} : \overline{ \piencodf{\sigma \wedge \omega}_{(\sigma, i)}}  , \piencodf{\Theta}$}
\end{prooftree}

Hence the theorem holds for this case.

\item {\bf Rule $\redlab{FS:Esub}$:}
Then $M = (M'[ {\widetilde{x}} \leftarrow  {x}])\esubst{ B }{ x }$ and

\begin{prooftree}
\AxiomC{\( \Theta , \unvar{x} : \eta ; \Gamma ,  {x}: \sigma^{j} \wfdash M'[ {\widetilde{x}} \leftarrow  {x}] : \tau  \)}
\AxiomC{\( \Theta ; \Delta \wfdash B : (\sigma^{k} , \epsilon ) \)}
\AxiomC{\( \eta \relunbag \epsilon \)}
\LeftLabel{\redlab{FS{:}Esub}}
\TrinaryInfC{\( \Theta ; \Gamma, \Delta \wfdash (M'[ {\widetilde{x}} \leftarrow  {x}])\esubst{ B }{ x }  : \tau \)}
\end{prooftree}

By Proposition~\ref{prop:app_auxunres} and IH we have:
$$
\begin{array}{rl}
\piencodf{ M'[x_1, \cdots , x_k \leftarrow x]}_u&\vdash \piencodf{\Gamma}, \linvar{x}: \overline{ \piencodf{ \sigma^j }_{(\tau, n)}} , u:\piencodf{\tau} , \piencodf{\Theta} , \unvar{x} : \dual{\piencodf{\eta}} \\
\piencodf{ M'[x_1, \cdots , x_k \leftarrow x]}_u&\vdash \piencodf{\Gamma}, \linvar{x}: \overline{ \piencodf{ \sigma^j }_{(\tau, n)}} , u:\piencodf{\tau} , \piencodf{\Theta} , \unvar{x} : \dual{\piencodf{\epsilon}},\text{ by Lemma \ref{lem:relunbag-typeunres} }\\
\piencodf{B}_x&\vdash \piencodf{\Delta}, x:\piencodf{ (\sigma^{k} , \epsilon ) }_{(\tau, m)}  , \piencodf{\Theta}
\end{array}
$$

From \figref{fig:encoding}, we have
\begin{equation*}
\piencodf{ M'[ {\widetilde{x}} \leftarrow  {x}]\ \esubst{ B }{ x }}_u =  \res{ x } (\psome{x}; \gname{x}{\linvar{x}}; \gname{x}{\unvar{x}}; \gclose{ x } ;\piencodf{ M'[ {\widetilde{x}} \leftarrow  {x}]}_u   \| \piencodf{ C \bagsep U}_x )
\end{equation*}

Therefore we obtain the following derivation $\Pi$:
\begin{prooftree}
\AxiomC{$ \piencodf{ M'[ {\widetilde{x}} \leftarrow  {x}]}_u \vdash \piencodf{\Gamma}, \linvar{x}: \overline{ \piencodf{ \sigma^j }_{(\tau, n)}} , u:\piencodf{\tau} , \piencodf{\Theta} , \unvar{x} : \dual{\piencodf{\epsilon}} $}
\LeftLabel{\ttype{$\bot$}}
\UnaryInfC{$ \gclose{ x } ;\piencodf{ M'[ {\widetilde{x}} \leftarrow  {x}]}_u \vdash x{:}\bot, \piencodf{\Gamma}, \linvar{x}: \overline{ \piencodf{ \sigma^j }_{(\tau, n)}} , u:\piencodf{\tau} , \piencodf{\Theta} , \unvar{x} : \dual{\piencodf{\epsilon}}$}
\LeftLabel{\ttype{$\parr$}}
\UnaryInfC{$\gname{x}{\unvar{x}}; \gclose{ x } ;\piencodf{ M'[ {\widetilde{x}} \leftarrow  {x}]}_u \vdash x:  (  \dual{\piencodf{\epsilon}} ) \ampy \bot , \piencodf{\Gamma}, \linvar{x}: \overline{ \piencodf{ \sigma^j }_{(\tau, n)}} , u:\piencodf{\tau} , \piencodf{\Theta} $}
\LeftLabel{\ttype{$\parr$}}
\UnaryInfC{$\gname{x}{\linvar{x}}; \gname{x}{\unvar{x}}; \gclose{ x } ;\piencodf{ M'[ {\widetilde{x}} \leftarrow  {x}]}_u \vdash x: \overline{ \piencodf{ \sigma^j }_{(\tau, n)}} \ampy ( (  \dual{\piencodf{\epsilon}} ) \ampy \bot) , \piencodf{\Gamma}, u:\piencodf{\tau} , \piencodf{\Theta} $}
\LeftLabel{\ttype{${\with}\some$}}
\UnaryInfC{$\psome{x}; \gname{x}{\linvar{x}}; \gname{x}{\unvar{x}};\gclose{ x } ; \piencodf{ M'[ {\widetilde{x}} \leftarrow  {x}]}_u \vdash x : \overline{\piencodf{ (\sigma^{j} , \epsilon  )  }_{(\tau, n)}}  , \piencodf{\Gamma}, u:\piencodf{\tau} , \piencodf{\Theta}  $}
\end{prooftree}

We take $ P_1 = \psome{x}; \gname{x}{\linvar{x}}; \gname{x}{\unvar{x}}; \gclose{ x } ;\piencodf{ M'[ {\widetilde{x}} \leftarrow  {x}]}_u$ and continue the derivation of $ \Pi $
\begin{prooftree}
\AxiomC{$ P_1 \vdash  x : \overline{\piencodf{ (\sigma^{j} , \epsilon  )  }_{(\tau, n)}}  , \piencodf{\Gamma}, u:\piencodf{\tau} ; \piencodf{\Theta}  $}
\AxiomC{$  \piencodf{ C \bagsep U}_x \vdash \piencodf{\Delta}, x:\piencodf{ (\sigma^{k} , \epsilon ) }_{(\tau, m)}  , \piencodf{\Theta} $}
\LeftLabel{\ttype{cut}}
\BinaryInfC{$  \res{ x }  ( P_1   \| \piencodf{ C \bagsep U}_x )  \vdash \piencodf{ \Gamma} , \piencodf{ \Delta } , u: \piencodf{ \tau }  , \piencodf{\Theta}  $}
\end{prooftree}

We must have that $\piencodf{\sigma^{j}}_{(\tau, m)} = \piencodf{\sigma^{k}}_{(\tau, n)}$ which by our restrictions allows. It follows that $\piencodf{M'[  {x}_1 \leftarrow  {x}]\ \esubst{ B }{ x }} $
$ \vdash \piencodf{\Gamma, \Delta}, u:\piencodf{\tau}  , \piencodf{\Theta}$ and the result follows.

\item {\bf Rule $\redlab{FS{:}Esub^{\ell}}$: }
Then $M =  M' \linexsub{C /  {x}_1 , \cdots , x_k}$, with $C = \bag{N_1} \cdot \cdots \cdot \bag{N_k}$ and

\begin{prooftree}
\AxiomC{\( \Theta ; \Delta ,  x_1:\sigma, \cdots , x_k:\sigma \wfdash M' : \tau \)}
\AxiomC{\( \Theta ; \Gamma,  \wfdash C:\sigma^{k}\)}
\LeftLabel{\redlab{FS{:}Esub^{\ell}}}
\BinaryInfC{\( \Theta ; \Gamma, \Delta \wfdash M' \linexsub{C /  x_1, \cdots , x_k} : \tau \)}
\end{prooftree}

with $ \Gamma = \Gamma_1 , \cdots ,\Gamma_k $ and:

\begin{prooftree}
\AxiomC{\( \Theta ; \Gamma_1 \wfdash N_1 : \sigma\)}
\AxiomC{\( \Theta ; \Gamma_k \wfdash N_k : \sigma\)}
\AxiomC{\(  \)}
\LeftLabel{\redlab{FS{:}\oneb^!}}
\UnaryInfC{\( \Theta ;  \dash  \wfdash \unvar{\oneb} : \sigma \)}
\LeftLabel{\redlab{FS{:}bag^{\ell}}}
\BinaryInfC{\( \vdots \)}
\LeftLabel{\redlab{FS{:}bag^{\ell}}}
\UnaryInfC{\( \Theta ; \Gamma_2 , \cdots ,\Gamma_k \wfdash C:\sigma^{k-1}\)}
\LeftLabel{\redlab{FS{:}bag^{\ell}}}
\BinaryInfC{\( \Theta ; \Gamma_1 , \cdots ,\Gamma_k  \wfdash \bag{M'}\cdot C:\sigma^{k}\)}
\end{prooftree}

By IH we have both
$$\piencodf{ N_1 }_{{x}_1} \vdash \piencodf{\Gamma_1}, {x}_1:\piencodf{\sigma } , \piencodf{\Theta} \qquad \cdots \qquad
\piencodf{ N_k }_{{x}_k} \vdash \piencodf{\Gamma_k}, {x}_k:\piencodf{\sigma } , \piencodf{\Theta}$$
$$\piencodf{ C }_{{x}} \vdash \piencodf{\Gamma}, {x}:\piencodf{\sigma^{k} }_{(\sigma, i)} , \piencodf{\Theta}
\qquad \text{ and } \qquad \piencodf{M'}_u\vdash \piencodf{\Gamma},  {x}_1: \with\overline{\piencodf{\sigma}} , \cdots , {x}_k: \with\overline{\piencodf{\sigma}}, u:\piencodf{\tau} , \piencodf{\Theta}$$

From \figref{fig:encoding},

$$
\begin{aligned}
\piencodf{ M' \linexsub{\bag{N_1} \cdot \cdots \cdot \bag{N_k} /  x_1 , \cdots , x_k}  }_u    & =
\res{z_1}( \gsome{z_1}{\llfv{N_{1}}};\piencodf{ N_{1} }_{ {z_1}}  \|  \cdots \res{z_k} ( \gsome{z_k}{\llfv{N_{k}}};\piencodf{ N_{k} }_{ {z_k}} \\
& \qquad \qquad  \| \bignd_{x_{i_1} \in \{ x_1 ,\cdots , x_k  \}} \cdots \bignd_{x_{i_k} \in \{ x_1 ,\cdots , x_k \setminus x_{i_1} , \cdots , x_{i_{k-1}}  \}}\\
& \qquad \qquad \qquad  \piencodf{ M' }_u \{ z_1 / x_{i_1} \} \cdots \{ z_k / x_{i_k} \} ) \cdots )
\\
\end{aligned}
$$
Let us take $k = 1$ and for $k > 1$ cases follow similarly omitting labels:
{
\small
\begin{prooftree}

\AxiomC{\( \piencodf{ M' }_u \{ z_1 / x_{i} \}  \vdash   \piencodf{\Delta},  {z_1}: \with\overline{\piencodf{\sigma}}, u:\piencodf{\tau} , \piencodf{\Theta}\)}
\UnaryInfC{\( \bignd_{x_{i} \in \{ x_1 \}}\piencodf{ M' }_u \{ z_1 / x_{i} \}  \vdash   \piencodf{\Delta},  {z_1}: \with\overline{\piencodf{\sigma}}, u:\piencodf{\tau} , \piencodf{\Theta}\)}

\AxiomC{\( \piencodf{ N_1 }_x  \vdash  \piencodf{\Gamma},  {x}: \piencodf{\sigma} , \piencodf{\Theta} \)}
\UnaryInfC{\(   \gsome{ {z_1} }{ \llfv{N_1} }; \piencodf{ N_1 }_{z_1} \vdash  \piencodf{ \Gamma } ,  {z_1}: : \oplus \piencodf{\sigma}, \piencodf{\Theta}\)}
\BinaryInfC{\(   \res{ {z_1} }  ( \bignd_{x_{i} \in \{ x_1 \}}\piencodf{ M' }_u \{ z_1 / x_{i} \}  \|   \gsome{ {z_1} }{ \llfv{N_1} };  \piencodf{ N_1 }_{ {z_1}} ) \vdash  \piencodf{ \Gamma} , \piencodf{ \Delta } , u : \piencodf{ \tau } , \piencodf{\Theta} \)}
\end{prooftree}
}

Therefore, $\piencodf{M' \linexsub{N_1 /  {x_1}} }_u\vdash \piencodf{ \Gamma} , \piencodf{ \Delta } , u : \piencodf{ \tau }, \piencodf{\Theta} $ and the result follows.

\item {\bf Rule $\redlab{FS{:}Esub^!}$: }
Then $M =  M' \unexsub{U / \unvar{x}}$ and
\begin{prooftree}
\AxiomC{\( \Theta , \unvar{x} : \eta; \Gamma \wfdash M' : \tau \quad  \Theta ; \dash \wfdash U : \eta \)}
\LeftLabel{\redlab{FS{:}Esub^!}}
\UnaryInfC{\( \Theta ; \Gamma \wfdash M' \unexsub{U / \unvar{x}}  : \tau \)}
\end{prooftree}

By IH we have both
$$
\begin{array}{c}
\piencodf{U}_{x_i}\vdash x_i : \with_{\eta_i \in \eta} \{ i ; \piencodf{\eta_i} \}  , \piencodf{\Theta}\qquad \text{ and } \qquad
\piencodf{M'}_u\vdash \piencodf{\Gamma} , u:\piencodf{\tau} , \unvar{x}: \overline{\piencodf{\eta}} , \piencodf{\Theta}
\end{array}
$$

From Definition \figref{fig:encoding}, $ \piencodf{ M' \unexsub{U / \unvar{x}}  }_u   =    \res{ \unvar{x} }  ( \piencodf{ M' }_u   \|   \guname{ \unvar{x} }{ x_i } ; \piencodf{ U }_{x_i} ) $ and

\begin{prooftree}
\AxiomC{$\piencodf{M'}_u\vdash \piencodf{\Gamma} , u:\piencodf{\tau}, \unvar{x}: \overline{\piencodf{\eta}} , \piencodf{\Theta}$}

\AxiomC{$ \piencodf{ U }_{x_i} \vdash x_i : \with_{\eta_i \in \eta} \{ i ; \piencodf{\eta_i} \} , \piencodf{\Theta} $}
\LeftLabel{\ttype{$!$}}
\UnaryInfC{$ \guname{ \unvar{x} }{ x_i } ; \piencodf{ U }_{x_i} \vdash \unvar{x}: \piencodf{\eta} , \piencodf{\Theta}  $}
\LeftLabel{\ttype{cut}}
\BinaryInfC{\(    \res{ \unvar{x} } ( \piencodf{ M' }_u   \|    \guname{ \unvar{x} }{ x_i } ; \piencodf{ U }_{x_i} ) \vdash \piencodf{\Gamma} , u:\piencodf{\tau} , \piencodf{\Theta} \)}
\end{prooftree}

Therefore, $\piencodf{M' \unexsub{U / \unvar{x}} }_u \vdash \piencodf{\Gamma} , u:\piencodf{\tau} , \piencodf{\Theta} $ and the result follows.

\item {\bf Rule $\redlab{FS:fail}$:}
Then $M= \fail^{\widetilde{x}}$ where $ \widetilde{x} = x_1, \cdots , x_n$ and

\begin{prooftree}
\AxiomC{\( \dom{\Gamma} = \widetilde{x}\)}
\LeftLabel{\redlab{FS{:}fail}}
\UnaryInfC{\( \Theta ; \Gamma \wfdash  \fail^{\widetilde{x}} : \tau \)}
\end{prooftree}

From \figref{fig:encoding}, $\piencodf{\fail^{x_1, \cdots , x_n} }_u=   \pnone{ u }    \|   \pnone{ x_1 }   \| \cdots   \|   \pnone{ x_k }  $ and

\begin{prooftree}
\AxiomC{}
\UnaryInfC{$  \pnone{ u }  \vdash u : \piencodf{ \tau }  $}
\UnaryInfC{$  \pnone{ u }  \vdash u : \piencodf{ \tau } , \piencodf{\Theta} $}

\AxiomC{}
\UnaryInfC{$  \pnone{ x_1 }  \vdash_1 : \with \overline{\piencodf{\sigma_1}} $}
\UnaryInfC{$  \pnone{ x_1 }  \vdash_1 : \with \overline{\piencodf{\sigma_1}} , \piencodf{\Theta} $}

\AxiomC{}
\UnaryInfC{$  \pnone{ x_n }  \vdash x_n : \with \overline{\piencodf{\sigma_n}}  $}
\UnaryInfC{$  \pnone{ x_n }  \vdash x_n : \with \overline{\piencodf{\sigma_n}} , \piencodf{\Theta} $}
\UnaryInfC{$\vdots$}
\BinaryInfC{$  \pnone{ x_1 }    \| \cdots   \|   \pnone{ x_k }  \vdash  x_1 : \with \overline{\piencodf{\sigma_1}}, \cdots  ,x_n : \with \overline{\piencodf{\sigma_n}} , \piencodf{\Theta} $}
\LeftLabel{\ttype{mix}}
\BinaryInfC{$  \pnone{ u }    \|   \pnone{ x_1 }    \| \cdots   \|   \pnone{ x_k }  \vdash x_1 : \with \overline{\piencodf{\sigma_1}}, \cdots  ,x_n : \with \overline{\piencodf{\sigma_n}}, u : \piencodf{ \tau } , \piencodf{\Theta} $}
\end{prooftree}

Thus, $\piencodf{\fail^{x_1, \cdots , x_n} }_u\vdash  x_1 : \with \overline{\piencodf{\sigma_1}}, \cdots  ,x_n : \with \overline{\piencodf{\sigma_n}}, u : \piencodf{ \tau } , \piencodf{\Theta} $ and the result follows.
\qedhere

\end{enumerate}
\end{enumerate}
\end{proof}

\subsection{Completeness}\label{a:tcompletness}

\begin{definition}[Linearly Partially Open Terms]
    We say that a $\lamcoldetsh$-term $M$ is \emph{linearly partially open} if $\forall x \in \llfv{M}$ implies that $x$ is not a sharing variable.
\end{definition}

\begin{proposition}
    \label{prop:correctformfailunres}
    Suppose $N$ is a well-formed linearly partially open $\lamcoldetsh$-term with $\headf{N} = x$ ($x$ denoting either linear or unrestricted occurrence of $x$) .
    Then,
    \[
    \piencodf{ N }_{u} = \res{ y_1 }( \cdots \res{y_m}( \piencodf{ x }_{n} \| P_m ) \cdots \| P_1  )
    \]
    which we shall denote as:
    \(
    \piencodf{ N }_{u} =  \res{ \widetilde{y} } (\piencodf{ x }_{n}   \| P)
    \),
    for some names $\widetilde{y} =$ and $n$, and processes $P$.
    \end{proposition}

    \begin{proof}
    The proof is by induction on the structure of $N$.

    \begin{enumerate}

        \item $N =  {x}$:  then $\piencodf{ {x}}_u  =  \psome{x}; \pfwd{x}{u} $. Hence $ P = \zero$ and $ \widetilde{y} = \emptyset$.

        \item $N =  {x}[j]$:  then $\piencodf{ {x}[j]}_u = \puname{ \unvar{x} }{ x_i };\psel{ {x}_i }{j }; \pfwd{x_i}{u}$. Hence $ P = \zero$ and $ \widetilde{y} = \emptyset$.

        \item  $N = M\ (C \bagsep U)$: then $\headf{M\ (C \bagsep U)} = \headf{M} = x$ and
        \[ \piencodf{N}_u = \piencodf{M\ (C \bagsep U)}_u  =   \res{ v } (\piencodf{M}_v   \|  \gsome{ v }{ u, \llfv{C} };  \pname{v}{x} . (\pfwd{v}{u}   \| \piencodf{(C \bagsep U)}_x ) ) \]
        The result follows by  induction hypothesis applied on $\piencodf{M}_u$.

        \item $N = (M[\widetilde{y} \leftarrow y])\esubst{ C \bagsep U }{ y }$:  this case does not  apply since $\head{N}\neq x$.

        \item $N = M \linexsub{C /  x_1 , \cdots , x_k}$:  then $\headf{M \linexsub{C /  x_1 , \cdots , x_k}} = \headf{M} = x$. Let $C = \bag{M_1} \cdot \cdots \cdot \bag{M_k} $
        \[
        \begin{aligned}
            \piencodf{N}_u &= \res{z_1}( \gsome{z_1}{\llfv{M_{1}}};\piencodf{ M_{1} }_{ {z_1}}  \|      \cdots \res{z_k} ( \gsome{z_k}{\llfv{M_{k}}};\piencodf{ M_{k} }_{ {z_k}} \\
            & \qquad \qquad  \| \bignd_{x_{i_1} \in \{ x_1 ,\cdots , x_k  \}} \cdots \bignd_{x_{i_k} \in \{ x_1 ,\cdots , x_k \setminus x_{i_1} , \cdots , x_{i_{k-1}}  \}} \piencodf{ M }_u \{ z_1 / x_{i_1} \} \cdots \{ z_k / x_{i_k} \} ) \cdots )
        \end{aligned}
           \]
        The result follows by  induction hypothesis applied on $\piencodf{M}_u$.

        \item  $N = M \unexsub{U / \unvar{x}} $: then $\headf{M \unexsub{U / \unvar{x}} } = \headf{M} = x$ and
        \[ \piencodf{N}_u = \piencodf{M \unexsub{U / \unvar{x}} }_u  =   \res{ \unvar{x} } ( \piencodf{ M }_u   \|   ~ \guname{ \unvar{x} }{ x_i } ; \piencodf{ U }_{x_i} ) \]
        The result follows by  induction hypothesis applied on $\piencodf{M}_u$.

        \qedhere

    \end{enumerate}

    \end{proof}

\begin{restatable}[Completeness (Under $\redtwo$)]{theorem}{thmEncTWCompl}\label{l:app_completenesstwo}
    If $ {N}\red {M}$ for a well-formed closed $\lamcoldetsh$-term $N$, then $\piencodf{{N}}_u \redtwo^\ast \piencodf{{M}}_u$.
\end{restatable}

\begin{proof}
    By induction on the reduction rule applied to infer ${N}\red {M}$.
    We have five cases.

 \begin{enumerate}
        \item  Case $\redlab{RS:Beta}$:

        Then  $ N= (\lambda x . (M'[ {\widetilde{x}} \leftarrow  {x}])) B  \red (M' [ {\widetilde{x}} \leftarrow  {x}])\esubst{ B }{ x }  = M$ , where $B = C \bagsep U$.
    The result follows from

        \begin{equation*}\label{eq:comp_sh_beta}
        \begin{aligned}
        \piencodf{N}_u & = \res{v} (\piencodf{\lambda x . (M'[ {\widetilde{x}} \leftarrow  {x}])}_v \| \gsome{v}{u , \llfv{C}};\pname{v}{x}; ( \piencodf{C \bagsep U}_x   \| \pfwd{v}{u}  ) )\\
        & = \res{v} (\psome{v};\gname{v}{x};  \psome{x};\gname{x}{\linvar{x}}; \gname{x}{\unvar{x}};  \gclose{x} ; \piencodf{M'[ {\widetilde{x}} \leftarrow  {x}]}_v \| \\
        & \qquad \gsome{v}{u , \llfv{C}};\pname{v}{x}; ( \piencodf{C \bagsep U}_x   \| \pfwd{v}{u}  ) )\\
        & \redtwo \res{v} (\gname{v}{x};  \psome{x};\gname{x}{\linvar{x}}; \gname{x}{\unvar{x}};  \gclose{x} ; \piencodf{M'[ {\widetilde{x}} \leftarrow  {x}]}_v \| \pname{v}{x}; ( \piencodf{C \bagsep U}_x   \| \pfwd{v}{u}  ) )\\
        & \redtwo \res{v} ( \res{x}( \psome{x};\gname{x}{\linvar{x}}; \gname{x}{\unvar{x}};  \gclose{x} ; \piencodf{M'[ {\widetilde{x}} \leftarrow  {x}]}_v \|   \piencodf{C \bagsep U}_x )  \| \pfwd{v}{u}   )\\
        & \redtwo \res{x}( \psome{x};\gname{x}{\linvar{x}}; \gname{x}{\unvar{x}};  \gclose{x} ; \piencodf{M'[ {\widetilde{x}} \leftarrow  {x}]}_u \|   \piencodf{C \bagsep U}_x ) =   \piencodf{M}_u    \\
        \end{aligned}
        \end{equation*}

        \item Case $ \redlab{RS:Ex \dash Sub}$: Then $ N =M'[ {x}_1, \!\cdots\! ,  {x}_k \leftarrow  {x}]\esubst{ C \bagsep U }{ x }$, with $C = \bag{M_1}
            \cdots  \bag{M_k}$, $k\geq 0$ and $M' \not= \fail^{\widetilde{y}}$.

        The reduction is $N = M'[ {x}_1, \!\cdots\! ,  {x}_k \leftarrow  {x}]\esubst{ C \bagsep U }{ x } \red  M'\linexsub{C  /  x_1 , \cdots , x_k} \unexsub{U / \unvar{x} } = M.$

        We detail the translations of $\piencodf{N}_u$ and $\piencodf{M}_u$. To simplify the proof, we will consider $k=2$ (the case in which $k> 2$ is follows analogously. Similarly the case of $k < 2$ it contained within $k = 2$). The result follows from:

        \begin{equation*}\label{eq:comp_sh_esub}
        \begin{aligned}
        \piencodf{N}_u
        &= \piencodf{M'[ {x}_1 \leftarrow  {x}]\esubst{ C \bagsep U }{ x }}_u
        = \res{x}( \psome{x}; \gname{x}{\linvar{x}}; \gname{x}{\unvar{x}};  \gclose{x} ;\piencodf{ M'[ {\widetilde{x}} \leftarrow  {x}]}_u \| \piencodf{ C \bagsep U}_x ) \\
        &= \res{x}( \psome{x}; \gname{x}{\linvar{x}}; \gname{x}{\unvar{x}};  \gclose{x} ;\piencodf{ M'[ {\widetilde{x}} \leftarrow  {x}]}_u \| \gsome{x}{\llfv{C}};  \pname{x}{\linvar{x}}; \big( \piencodf{ C }_{\linvar{x}} \|  \pname{x}{\unvar{x}};\\
        & \qquad \qquad \qquad ( \guname{\unvar{x}}{x_i}; \piencodf{ U }_{x_i} \| \pclose{x} ) \big) ) (:= P_{\mathbb{N}})\\
            & \redtwo^* \res{\unvar{x}} ( \res{\linvar{x}}( \piencodf{ M'[ x_1, x_2 \leftarrow  {x}]}_u \|   \piencodf{ \bag{M_1} \cdot \bag{M_2} }_{\linvar{x}} ) \|   \guname{\unvar{x}}{x_i}; \piencodf{ U }_{x_i}   )  \\
            & = \res{\unvar{x}} ( \res{\linvar{x}}( \psome{\linvar{x}}; \pname{\linvar{x}}{y_1}; \big( \gsome{y_1}{ \emptyset }; \gclose{ y_{1} } ; \0   \\
            & \qquad \qquad \|\psome{\linvar{x}}; \gsome{\linvar{x}}{u, \llfv{M'} \setminus  \widetilde{x} }; \bignd_{x_{i_1} \in \widetilde{x}} \gname{x}{{x}_{i_1}};\piencodf{M'[ (\widetilde{x} \setminus x_{i_1} ) \leftarrow  {x}]}_u \big) \| \\
            & \qquad \qquad \gsome{\linvar{x}}{\llfv{C} }; \gname{x}{y_1}; \gsome{\linvar{x}}{y_i, \llfv{C}}; \psome{\linvar{x}}; \pname{\linvar{x}}{z_1}; \\
            & \qquad   ( \gsome{z_1}{\llfv{M_1}};  \piencodf{M_1}_{z_1} \| \piencodf{\bag{M_2}}_{\linvar{x}} \| \pnone{y_1} ) ) \|   \guname{\unvar{x}}{x_i}; \piencodf{ U }_{x_i}   )  \\
            & \redtwo^* \res{\unvar{x}} ( \res{\linvar{x}}( \piencodf{\bag{M_2}}_{\linvar{x}}  \|  \res{z_1}  \big( \bignd_{x_{i_1} \in \widetilde{x}} \piencodf{M'[ (\widetilde{x} \setminus x_{i_1} ) \leftarrow  {x}]}_u \{ z_1 / x_{i_1} \}  \| \\
            & \qquad \qquad    \gsome{z_1}{\llfv{M_1}};  \piencodf{M_1}_{z_1}  \big) ) \|   \guname{\unvar{x}}{x_i}; \piencodf{ U }_{x_i}   )  \\
            & = \res{\unvar{x}} ( \res{\linvar{x}}( \gsome{\linvar{x}}{\llfv{C} }; \gname{x}{y_2}; \gsome{\linvar{x}}{y_2, \llfv{C}}; \psome{\linvar{x}}; \pname{\linvar{x}}{z_2}; \\
            & \qquad   ( \gsome{z_2}{\llfv{M_2}};  \piencodf{M_2}_{z_2} \| \piencodf{\oneb}_{\linvar{x}} \| \pnone{y_2} )  \|\\
            &   \res{z_1}  \big( \bignd_{x_{i_1} \in \widetilde{x}} \psome{\linvar{x}}; \pname{\linvar{x}}{y_2}; \big( \gsome{y_2}{ \emptyset }; \gclose{ y_{2} } ; \0   \\
            & \qquad \qquad \|\psome{\linvar{x}}; \gsome{\linvar{x}}{u, \llfv{M'} \setminus  (\widetilde{x} \setminus x_{i_1}  )}; \bignd_{x_{i_2} \in (\widetilde{x} \setminus x_{i_1} )} \gname{x}{{x}_i};\piencodf{M'[  \leftarrow  {x}]}_u\{ z_1 / x_{i_1} \} \big)   \| \\
            & \qquad \qquad    \gsome{z_1}{\llfv{M_1}};  \piencodf{M_1}_{z_1}  \big) ) \|   \guname{\unvar{x}}{x_i}; \piencodf{ U }_{x_i}   )  \\
            & \red^* \res{\unvar{x}} ( \res{\linvar{x}}(  \piencodf{\oneb}_{\linvar{x}} \| \\
            & \qquad \res{z_2}  ( \res{z_1}  \big( \bignd_{x_{i_1} \in \widetilde{x}}  \bignd_{x_{i_2} \in (\widetilde{x} \setminus x_{i_1} )} \gname{x}{{x}_i};\piencodf{M'[  \leftarrow  {x}]}_u\{ z_1 / x_{i_1} \}   \| \\
            & \qquad \qquad    \gsome{z_1}{\llfv{M_1}};  \piencodf{M_1}_{z_1}  \big) \|
            \gsome{z_2}{\llfv{M_2}};  \piencodf{M_2}_{z_2} )      ) \|   \guname{\unvar{x}}{x_i}; \piencodf{ U }_{x_i}   )  \\
            & \red^* \res{\unvar{x}} (  \res{z_2}  ( \res{z_1}  \big( \bignd_{x_{i_1} \in \widetilde{x}}  \bignd_{x_{i_2} \in (\widetilde{x} \setminus x_{i_1} )} \gname{x}{{x}_i};\piencodf{M'}_u\{ z_1 / x_{i_1} \}   \| \\
            & \qquad \qquad    \gsome{z_1}{\llfv{M_1}};  \piencodf{M_1}_{z_1}  \big) \|
            \gsome{z_2}{\llfv{M_2}};  \piencodf{M_2}_{z_2} ) \|   \guname{\unvar{x}}{x_i}; \piencodf{ U }_{x_i}   ) =    \piencodf{M}_u \\
        \end{aligned}
        \end{equation*}

        \item Case $\redlab{RS{:}Fetch^{\ell}}$:

        Then we have
        $N = M' \linexsub{C /  x_1 , \cdots , x_k} $ with $\headf{M'} =  {x}_j$, $C = M_1 , \cdots , M_k$ and $N \red  M' \headlin{ M_i / x_j }  \linexsub{(C \setminus M_i ) /  x_1 , \cdots , x_k \setminus x_j }  = M$, for some $M_i \in C$.

        On the one hand, we have:
            \begin{equation}\label{eq:comp_sh_linfet1}
            \begin{aligned}
            \piencodf{N}_u = \piencodf{M' \linexsub{C /  x_1 , \cdots , x_k, x_j}}_u
            &= \res{z_1}( \gsome{z_1}{\llfv{M_{1}}};\piencodf{ M_{1} }_{ {z_1}}  \|  \cdots \res{z_k} ( \gsome{z_k}{\llfv{M_{k}}};\piencodf{ M_{k} }_{ {z_k}} \\
            &  \quad  \| \bignd_{x_{i_1} \in \{ x_1 ,\cdots , x_k  \}} \cdots \\
            &  \quad \qquad  \bignd_{x_{i_k} \in \{ x_1 ,\cdots , x_k \setminus x_{i_1} , \cdots , x_{i_{k-1}}  \}} \piencodf{ M' }_u \{ z_1 / x_{i_1} \} \cdots \{ z_k / x_{i_k} \} ) \cdots )\\
            &= \res{z_1}( \gsome{z_1}{\llfv{M_{1}}};\piencodf{ M_{1} }_{ {z_1}}  \|  \cdots \res{z_k} ( \gsome{z_k}{\llfv{M_{k}}};\piencodf{ M_{k} }_{ {z_k}} \\
            &  \quad  \| \bignd_{x_{i_1} \in \{ x_1 ,\cdots , x_k  \}} \cdots \\
            &  \quad \qquad \bignd_{x_{i_k} \in \{ x_1 ,\cdots , x_k \setminus x_{i_1} , \cdots , x_{i_{k-1}}  \}} \res{y}(\piencodf{  {x}_{j} }_y \| P) \{ z_1 / x_{i_1} \} \cdots \{ z_k / x_{i_k} \} )\\
            &  \quad \cdots ) \qquad (*)\\
            &= \res{z_1}( \gsome{z_1}{\llfv{M_{1}}};\piencodf{ M_{1} }_{ {z_1}}  \| \cdots \res{z_k} ( \gsome{z_k}{\llfv{M_{k}}};\piencodf{ M_{k} }_{ {z_k}} \\
            &  \quad  \| \bignd_{x_{i_1} \in \{ x_1 ,\cdots , x_k  \}} \cdots \\
            &  \quad \qquad \bignd_{x_{i_k} \in \{ x_1 ,\cdots , x_k \setminus x_{i_1} , \cdots , x_{i_{k-1}}  \}}\\
            &  \quad \qquad \qquad \res{y}(  \psome{x_j}; \pfwd{x_j}{y}  \| P) \{ z_1 / x_{i_1} \} \cdots \{ z_k / x_{i_k} \} )\cdots )\\
            \end{aligned}
            \end{equation}
        where $(*)$ is inferred via Proposition~\ref{prop:correctformfailunres}.

        Let us consider the case when $j = k$ and $ M_i = M_1$ the other cases proceed similarly. Then we have the following reduction:

        \begin{equation}\label{eq:comp_sh_linfet2}
            \begin{aligned}
                &= \res{z_1}( \gsome{z_1}{\llfv{M_{1}}};\piencodf{ M_{1} }_{ {z_1}}  \| \cdots \res{z_k} ( \gsome{z_k}{\llfv{M_{k}}};\piencodf{ M_{k} }_{ {z_k}} \\
                & \qquad \qquad  \| \bignd_{x_{i_1} \in \{ x_1 ,\cdots , x_k  \}} \cdots \\
                & \qquad \qquad \qquad \bignd_{x_{i_k} \in \{ x_1 ,\cdots , x_k \setminus x_{i_1} , \cdots , x_{i_{k-1}}  \}}\\
                & \qquad \qquad \qquad \qquad \res{y}(\psome{x_k}; \pfwd{x_k}{y} \| P) \{ z_1 / x_{i_1} \} \cdots \{ z_k / x_{i_k} \} ) \cdots )\\
                & \redtwo \res{z_1}( \piencodf{ M_{1} }_{ {z_1}}  \|  \cdots \res{z_k} ( \gsome{z_k}{\llfv{M_{k}}};\piencodf{ M_{k} }_{ {z_k}} \\
                & \qquad \qquad  \| \bignd_{x_{i_2} \in \{ x_1 ,\cdots , x_{k-1}  \}} \cdots \\
                & \qquad \qquad \qquad \bignd_{x_{i_k} \in \{ x_1 ,\cdots , x_{k-1} \setminus x_{i_1} , \cdots , x_{i_{k-1}}  \}}\\
                & \qquad \qquad \qquad \qquad \res{y}(  \pfwd{z_1}{y} \| P) \{ z_2 / x_{i_2} \} \cdots \{ z_k / x_{i_k} \} )\cdots )\\
                & \redtwo \res{z_2}( \gsome{z_2}{\llfv{M_{2}}};\piencodf{ M_{2} }_{ {z_2}}  \|  \cdots \res{z_k} ( \gsome{z_k}{\llfv{M_{k}}};\piencodf{ M_{k} }_{ {z_k}} \\
                & \qquad \qquad  \| \bignd_{x_{i_2} \in \{ x_1 ,\cdots , x_{k-1}  \}} \cdots \\
                & \qquad \qquad \qquad \bignd_{x_{i_k} \in \{ x_1 ,\cdots , x_{k-1} \setminus x_{i_1} , \cdots , x_{i_{k-1}}  \}}\\
                & \qquad \qquad \qquad \qquad \res{y}( \piencodf{ M_{1} }_{y}  \| P) \{ z_2 / x_{i_2} \} \cdots \{ z_k / x_{i_k} \} )\cdots )\\
            \end{aligned}
            \end{equation}

         On the other hand, we have:
        \begin{equation}\label{eq:comp_sh_linfet3}
        \begin{aligned}
        \piencodf{M}_u &= \piencodf{M' \headlin{ M_1 / x_k }  \linexsub{(C \setminus M_1 ) /  x_1 , \cdots , x_{k-1} } }_u \\
            & =   \res{z_1}( \gsome{z_1}{\llfv{M_{2}}};\piencodf{ M_{2} }_{ {z_1}}  \|  \cdots \res{z_{k-1}} ( \gsome{z_{k-1}}{\llfv{M_{k}}};\piencodf{ M_{k} }_{ {z_{k-1}}} \\
            & \quad \qquad  \| \bignd_{x_{i_1} \in \{ x_1 ,\cdots , x_{k-1}  \}}
            \cdots \bignd_{x_{i_{k-1}} \in \{ x_1 ,\cdots , x_{k-1} \setminus x_{i_1} , \cdots , x_{i_{k-2}}  \}}
            \piencodf{ M' }_u \{ z_1 / x_{i_1} \} \cdots \{ z_{k-1} / x_{i_{k-1}} \} ) \cdots ) \\
        \end{aligned}
        \end{equation}
                Therefore, by \eqref{eq:comp_sh_linfet2}, \eqref{eq:comp_sh_linfet3} and taking $M_{j_k} = M_i$ the result follows.

        \item Case $ \redlab{RS{:} Fetch^!}$:

        Then,
        $N = M' \unexsub{U / \unvar{x}}$ with $\headf{M' } = \unvar{x}[k]$, $U_k = \unvar{\bag{N}}$ and $N \red  M' \headlin{ N /\unvar{x} }\unexsub{U / \unvar{x}} = M$. The result follows from

            \begin{equation}\label{eq:compl_sh_fetchun1}
            \begin{aligned}
            \piencodf{N}_u &= \piencodf{M' \unexsub{U / \unvar{x}}}_u = \res{\unvar{x}} ( \piencodf{ M' }_u \|   ~ \guname{\unvar{x}}{x_i}; \piencodf{ U }_{x_i} ) \\
            & = \res{\unvar{x}} ( \res{y}(\piencodf{ \unvar{x}[k] }_{j} \| P) \|   ~ \guname{\unvar{x}}{x_i}; \piencodf{ U }_{x_i} ) \qquad (*)\\
            & = \res{\unvar{x}} ( \res{y}( \puname{\unvar{x}}{{x_i}}; \psel{x_i}{k}; \pfwd{x_i}{j} \| P) \|   ~ \guname{\unvar{x}}{x_i}; \piencodf{ U }_{x_i} )
            \\
            & \redtwo \res{\unvar{x}} ( \res{y}( \puname{\unvar{x}}{{x_i}}; \psel{x_i}{k}; \pfwd{x_i}{j} \| P) \|   ~ \guname{\unvar{x}}{x_i}; \piencodf{ U }_{x_i} )
            \\
             & \redtwo \res{\unvar{x}} ( \guname{\unvar{x}}{x_i}; \piencodf{ U }_{x_i} \| \res{x_i} ( \res{y}( \psel{x_i}{k}; \pfwd{x_i}{j} \| P) \|  \piencodf{ U }_{x_i}   )  ~  )
            \\
             & = \res{\unvar{x}} ( \guname{\unvar{x}}{x_i}; \piencodf{ U }_{x_i} \| \res{x_i} ( \res{y}( \psel{x_i}{k}; \pfwd{x_i}{j} \| P) \|  \gsel{x_i}\{i:\piencodf{U_i}_{x_i}\}_{U_i \in U}   )  ~  )
            \\
            & \redtwo \res{\unvar{x}} ( \guname{\unvar{x}}{x_i}; \piencodf{ U }_{x_i} \| \res{x_i} ( \res{y}( \pfwd{x_i}{j} \| P) \| \piencodf{U_i}_{x_i}   )  ~  )
            \\
             & \redtwo \res{\unvar{x}} ( \guname{\unvar{x}}{x_i}; \piencodf{ U }_{x_i} \| \res{y}( \piencodf{U_i}_{j} \| P)  ~  ) \\
             & \redtwo \res{\unvar{x}} ( \guname{\unvar{x}}{x_i}; \piencodf{ U }_{x_i} \| \res{y}( \piencodf{N}_{j} \| P)  ~  ) =\piencodf{M}_u
            \end{aligned}
            \end{equation}
       where the reductions denoted by $(*)$ are inferred via Proposition~\ref{prop:correctformfailunres}.

        \item Case $\redlab{RS:TCont}$:
            This case follows by IH.

        \item Case $\redlab{RS{:}Fail^{\ell}}$:

        Then we have
        $N = M' [ {x}_1, \!\cdots\! ,  {x}_k \leftarrow  {x}]\ \esubst{C \bagsep U}{ x } $ with $k \neq \size{C}$ and
        $N \red  \fail^{\widetilde{y}} = M$, where $\widetilde{y} = (\llfv{M' } \setminus \{   {x}_1, \cdots ,  {x}_k \} ) \cup \llfv{C}$. Let $ C = \bag{M_1} \cdot \cdots \cdot \bag{M_l}$ and we assume that $k > l$ and we proceed similarly for $k > l$. Hence $k = l + m$ for some $m \geq 1$

        \begin{equation*}\label{eq:compl_sh_faillin1}
        \begin{aligned}
            \piencodf{N}_u
            &= \piencodf{M' [ {x}_1, \!\cdots\! ,  {x}_k \leftarrow  {x}]\ \esubst{C \bagsep U}{ x } }_u= \res{x}( \psome{x}; \gname{x}{\linvar{x}}; \gname{x}{\unvar{x}};  \gclose{x} ;\piencodf{ M' [ {\widetilde{x}} \leftarrow  {x}]}_u \| \piencodf{ C \bagsep U}_x )\\
            &= \res{x}( \psome{x}; \gname{x}{\linvar{x}}; \gname{x}{\unvar{x}};  \gclose{x} ;\piencodf{ M' [ {\widetilde{x}} \leftarrow  {x}]}_u \| \\
            & \qquad \qquad \gsome{x}{\llfv{C}};  \pname{x}{\linvar{x}}; \big( \piencodf{ C }_{\linvar{x}} \|  \pname{x}{\unvar{x}}; ( \guname{\unvar{x}}{x_i}; \piencodf{ U }_{x_i} \| \pclose{x} ) \big) )\\
            & \redtwo^*  \res{\unvar{x}} ( \res{\linvar{x} } \big( \piencodf{ M' [ {\widetilde{x}} \leftarrow  {x}]}_u \|  \piencodf{ C }_{\linvar{x}}   \big) \|  \guname{\unvar{x}}{x_i}; \piencodf{ U }_{x_i}  )  \\
            & = \res{\unvar{x}} ( \res{\linvar{x} } \big( \psome{\linvar{x}}; \pname{\linvar{x}}{y_1}; \big( \gsome{y_1}{ \emptyset }; \gclose{ y_{1} } ; \0   \\
            & \qquad \qquad \|\psome{\linvar{x}}; \gsome{\linvar{x}}{u, \llfv{M' } \setminus  \widetilde{x} }; \bignd_{x_{i_1} \in \widetilde{x}} \gname{x}{{x}_{i_1}}; \cdots \\
            & \qquad \qquad \psome{\linvar{x}}; \pname{\linvar{x}}{y_k}; \big( \gsome{y_k}{ \emptyset }; \gclose{ y_{k} } ; \0   \\
            & \qquad \qquad \|\psome{\linvar{x}}; \gsome{\linvar{x}}{u, \llfv{M' } \setminus  (\widetilde{x} \setminus x_{i_1} , \cdots , x_{i_{k-1}}   )}; \bignd_{x_{i_k} \in (\widetilde{x} \setminus x_{i_1} , \cdots , x_{i_{k-1}}  )} \gname{x}{{x}_{i_k}};\piencodf{M' [  \leftarrow  {x}]}_u \big)
            \big) \| \\
            & \qquad \qquad  \gsome{\linvar{x}}{\llfv{C} }; \gname{x}{y_1}; \gsome{\linvar{x}}{y_1, \llfv{C}}; \psome{\linvar{x}}; \pname{\linvar{x}}{z_1}; \\
            & \qquad \qquad  ( \gsome{z_1}{\llfv{M_1}};  \piencodf{M_1}_{z_1} \| \pnone{y_1}  \| \cdots  \gsome{\linvar{x}}{\llfv{C} }; \gname{x}{y_l}; \gsome{\linvar{x}}{y_l, \llfv{M_l}}; \psome{\linvar{x}}; \pname{\linvar{x}}{z_l}; \\
            & \qquad \qquad   ( \gsome{z_l}{\llfv{M_l}};  \piencodf{M_l}_{z_l} \| \piencodf{ \oneb }_{\linvar{x}} \| \pnone{y_l} ) \cdots   )   \big) \|  \guname{\unvar{x}}{x_i}; \piencodf{ U }_{x_i}  ) \qquad
            (:= P_\mathbb{N}) \\
            \end{aligned}
            \end{equation*}

            we reduce $P_\mathbb{N}$ arbitrarily synchronising along channels $\linvar{x} , y_1, \cdots y_l$.

            \begin{equation*}
            \begin{aligned}
                P_\mathbb{N}
                & \redtwo^* \res{\unvar{x}} ( \res{\linvar{x} } \big( \piencodf{ \oneb }_{\linvar{x}} \|
                \res{z_1} (\gsome{z_1}{\llfv{M_1}};  \piencodf{M_1}_{z_1}   \| \cdots
                \res{z_l} (\gsome{z_l}{\llfv{M_l}};  \piencodf{M_l}_{z_l} \| \\
                &  \qquad \qquad \bignd_{x_{i_1} \in \widetilde{x}} \cdots \bignd_{x_{i_l} \in (\widetilde{x} \setminus x_{i_1} , \cdots , x_{i_{l-1}} )}  \psome{\linvar{x}}; \pname{\linvar{x}}{y_{l+1}}; \big( \gsome{y_{l+1}}{ \emptyset }; \gclose{ y_{{l+1}} } ; \0   \\
                & \qquad \qquad \|\psome{\linvar{x}}; \gsome{\linvar{x}}{u, \llfv{M' } \setminus  (\widetilde{x} \setminus x_{i_1} , \cdots , x_{i_{l}}  )}; \bignd_{x_{i_{l+1}} \in (\widetilde{x} \setminus x_{i_1} , \cdots , x_{i_{l}} )} \gname{x}{{x}_{i_{l+1}}}; \cdots   \\
                & \qquad \qquad \psome{\linvar{x}}; \pname{\linvar{x}}{y_k}; \big( \gsome{y_k}{ \emptyset }; \gclose{ y_{k} } ; \0   \\
                & \qquad \qquad \|\psome{\linvar{x}}; \gsome{\linvar{x}}{u, \llfv{M' } \setminus  (\widetilde{x} \setminus x_{i_1} , \cdots , x_{i_{k-1}}   )}; \bignd_{x_{i_k} \in (\widetilde{x} \setminus x_{i_1} , \cdots , x_{i_{k-1}}  )} \gname{x}{{x}_{i_k}};\\
                & \qquad \qquad \piencodf{M' [  \leftarrow  {x}]}_u \{ z_1 / x_{i_1} \} \cdots \{ z_l / {x}_{i_l} \}
                \big) \cdots \big)  ) \cdots ) \big) \|\guname{\unvar{x}}{x_i}; \piencodf{ U }_{x_i}  )\\
                & = \res{\unvar{x}} ( \res{\linvar{x} } \big(
               \gsome{\linvar{x}}{\emptyset}; \gname{x}{y_{l+1}};  ( \psome{ y_{l+1}}; \pclose{y_{l+1}}  \| \gsome{\linvar{x}}{\emptyset}; \pnone{\linvar{x}} ) \\
                & \qquad \qquad \|
                \res{z_1} (\gsome{z_1}{\llfv{M_1}};  \piencodf{M_1}_{z_1}   \| \cdots
                \res{z_l} (\gsome{z_l}{\llfv{M_l}};  \piencodf{M_l}_{z_l} \| \\
                &  \qquad \qquad \bignd_{x_{i_1} \in \widetilde{x}} \cdots \bignd_{x_{i_l} \in (\widetilde{x} \setminus x_{i_1} , \cdots , x_{i_{l-1}} )}\psome{\linvar{x}}; \pname{\linvar{x}}{y_{l+1}}; \big( \gsome{y_{l+1}}{ \emptyset }; \gclose{ y_{{l+1}} } ; \0   \\
                & \qquad \qquad \|\psome{\linvar{x}}; \gsome{\linvar{x}}{u, \llfv{M' } \setminus  (\widetilde{x} \setminus x_{i_1} , \cdots , x_{i_{l}}  )}; \bignd_{x_{i_{l+1}} \in (\widetilde{x} \setminus x_{i_1} , \cdots , x_{i_{l}} )} \gname{x}{{x}_{i_{l+1}}}; \cdots   \\
                & \qquad \qquad \psome{\linvar{x}}; \pname{\linvar{x}}{y_k}; \big( \gsome{y_k}{ \emptyset }; \gclose{ y_{k} } ; \0   \\
                & \qquad \qquad \|\psome{\linvar{x}}; \gsome{\linvar{x}}{u, \llfv{M' } \setminus  (\widetilde{x} \setminus x_{i_1} , \cdots , x_{i_{k-1}}   )}; \bignd_{x_{i_k} \in (\widetilde{x} \setminus x_{i_1} , \cdots , x_{i_{k-1}}  )} \gname{x}{{x}_{i_k}};\\
                & \qquad \qquad \piencodf{M' [  \leftarrow  {x}]}_u \{ z_1 / x_{i_1} \} \cdots \{ z_l / {x}_{i_l} \}
                \big) \cdots \big)  ) \cdots ) \big) \| \guname{\unvar{x}}{x_i}; \piencodf{ U }_{x_i}  )\\
                & \redtwo^* \res{\unvar{x}} ( \res{\linvar{x} } \big( \pnone{\linvar{x}}  \|
                \res{z_1} (\gsome{z_1}{\llfv{M_1}};  \piencodf{M_1}_{z_1}   \| \cdots
                \res{z_l} (\gsome{z_l}{\llfv{M_l}};  \piencodf{M_l}_{z_l} \| \\
                &  \qquad \qquad \bignd_{x_{i_1} \in \widetilde{x}} \cdots \bignd_{x_{i_l} \in (\widetilde{x} \setminus x_{i_1} , \cdots , x_{i_{l-1}} )} \gsome{\linvar{x}}{u, \llfv{M' } \setminus  (\widetilde{x} \setminus x_{i_1} , \cdots , x_{i_{l}}  )}; \bignd_{x_{i_{l+1}} \in (\widetilde{x} \setminus x_{i_1} , \cdots , x_{i_{l}} )} \gname{x}{{x}_{i_{l+1}}}; \cdots   \\
                & \qquad \qquad \psome{\linvar{x}}; \pname{\linvar{x}}{y_k}; \big( \gsome{y_k}{ \emptyset }; \gclose{ y_{k} } ; \0   \\
                & \qquad \qquad \|\psome{\linvar{x}}; \gsome{\linvar{x}}{u, \llfv{M' } \setminus  (\widetilde{x} \setminus x_{i_1} , \cdots , x_{i_{k-1}}   )}; \bignd_{x_{i_k} \in (\widetilde{x} \setminus x_{i_1} , \cdots , x_{i_{k-1}}  )} \gname{x}{{x}_{i_k}};\\
                & \qquad \qquad \piencodf{M' [  \leftarrow  {x}]}_u \{ z_1 / x_{i_1} \} \cdots \{ z_l / {x}_{i_l} \}
                \big) \cdots \big)  ) \cdots ) \big) \|  \guname{\unvar{x}}{x_i}; \piencodf{ U }_{x_i}  )\\
                & \redtwo \res{\unvar{x}} (
                \res{z_1} (\gsome{z_1}{\llfv{M_1}};  \piencodf{M_1}_{z_1}   \| \cdots
                \res{z_l} (\gsome{z_l}{\llfv{M_l}};  \piencodf{M_l}_{z_l} \| \\
                &  \qquad \qquad \bignd_{x_{i_1} \in \widetilde{x}} \cdots \bignd_{x_{i_l} \in (\widetilde{x} \setminus x_{i_1} , \cdots , x_{i_{l-1}} )} \pnone{u } \| \pnone{(\llfv{M' } \setminus  \widetilde{x}) } \|  \pnone{(z_1 , \cdots ,z_l ) }
                ) \cdots ) \| \\
                & \qquad \qquad  \guname{\unvar{x}}{x_i}; \piencodf{ U }_{x_i}  )\\
                & \redtwo^* \res{\unvar{x}} (
                 \bignd_{x_{i_1} \in \widetilde{x}} \cdots \bignd_{x_{i_l} \in (\widetilde{x} \setminus x_{i_1} , \cdots , x_{i_{l-1}} )} \pnone{u } \| \pnone{(\llfv{M' } \setminus  \widetilde{x}) }  \|\guname{\unvar{x}}{x_i}; \piencodf{ U }_{x_i}  )\\
                & \equiv \pnone{u } \| \pnone{(\llfv{M' } \setminus  \widetilde{x}) }   =\piencodf{M}_u\\
        \end{aligned}
        \end{equation*}

       \item Case $\redlab{RS{:}Fail^!}$:

        Then,
        $N = M' \unexsub{U /\unvar{x}}  $ with $\headf{M' } =  {x}[i]$, $U_i = \unvar{\oneb} $ and
        $N \red   M' \headlin{ \fail^{\emptyset} /\unvar{x} } \unexsub{U /\unvar{x} } $. The result follows from

        \begin{equation*}\label{eq:compl_sh_failun1}
        \begin{aligned}
            \piencodf{N}_u &= \piencodf{ M' \unexsub{U /\unvar{x}} }_u  =   \res{\unvar{x}} ( \piencodf{ M' }_u \|   ~ \guname{\unvar{x}}{x_i}; \piencodf{ U }_{x_i} )  \\
            & = \res{\unvar{x}} (  \res{\widetilde{y}}(\piencodf{  {x}[i] }_{j} \| P)   \|   ~ \guname{\unvar{x}}{x_i}; \piencodf{ U }_{x_i} ) \qquad (*)
            \\
            & = \res{\unvar{x}} (  \res{\widetilde{y}}(  \puname{\unvar{x}}{{x_i}}; \psel{x_i}{i}; \pfwd{x_i}{j} \| P)   \|   ~ \guname{\unvar{x}}{x_i}; \piencodf{ U }_{x_i} )
            \\
            & \redtwo \res{\unvar{x}} ( \res{x_i} ( \res{\widetilde{y}}(  \psel{x_i}{i}; \pfwd{x_i}{j} \| P) \|  \piencodf{ U }_{x_i} )   \|   ~ \guname{\unvar{x}}{x_i}; \piencodf{ U }_{x_i} )
            \\
            & = \res{\unvar{x}} ( \res{x_i} ( \res{\widetilde{y}}(  \psel{x_i}{i}; \pfwd{x_i}{j} \| P) \|  \gsel{x_i}\{i:\piencodf{ U_i }_{x_i} \}_{U_i \in U} )   \|   ~ \guname{\unvar{x}}{x_i}; \piencodf{ U }_{x_i} )
            \\
            & \redtwo* \res{\unvar{x}} (  \res{\widetilde{y}}( \piencodf{ U_i }_{j}  \| P)     \|   ~ \guname{\unvar{x}}{x_i}; \piencodf{ U }_{x_i} )
            \\
            & = \res{\unvar{x}} (  \res{\widetilde{y}}( \piencodf{ \unvar{\oneb} }_{j}  \| P)    \|   ~ \guname{\unvar{x}}{x_i}; \piencodf{ U }_{x_i} )
            \\
            & = \res{\unvar{x}} (  \res{\widetilde{y}}( \pnone{j}   \| P)    \|   ~ \guname{\unvar{x}}{x_i}; \piencodf{ U }_{x_i} ) =   \piencodf{M}_u
            \\
        \end{aligned}
        \end{equation*}

        \item Case $\redlab{RS:Cons_1}$:
        Then we have
        $N = \fail^{\widetilde{x}}\ C \bagsep U$ and $N \red \fail^{\widetilde{x} \cup \widetilde{y}}  = M$ where $ \widetilde{y} = \llfv{C}$. Also,

        \begin{equation*}\label{eq:compl_sh_con1}
        \begin{aligned}
            \piencodf{N}_u &= \piencodf{ \fail^{\widetilde{x}}\ C \bagsep U }_u = \res{v} (\piencodf{\fail^{\widetilde{x}}}_v \| \gsome{v}{u , \llfv{C}};\pname{v}{x}; ( \piencodf{C \bagsep U}_x  \| \pfwd{v}{u}  ) ) \\
            & = \res{v} ( \pnone{v } \| \pnone{\widetilde{x}}  \| \gsome{v}{u , \llfv{C}};\pname{v}{x}; ( \piencodf{C \bagsep U}_x  \| \pfwd{v}{u}  ) ) \\
            & \redtwo  \pnone{\widetilde{x}} \| \pnone{u} \| \pnone{\widetilde{y}} =  \piencodf{M}_u
        \end{aligned}
        \end{equation*}

        \item Cases $\redlab{RS:Cons_2}$ and $\redlab{RS:Cons_3}$: These cases follow by IH similarly to Case 7.

        \item Case $\redlab{RS{:}Cons_4}$:
        Then we have
        $N =  \fail^{\widetilde{y}} \unexsub{U / \unvar{x}} $ and $N \red \fail^{\widetilde{y}}  = M$, and

        \begin{align}\label{eq:compl_sh_con4_1}
            \piencodf{N}_u &= \piencodf{ \fail^{\widetilde{y}} \unexsub{U / \unvar{x}}}_u = \res{\unvar{x}} ( \piencodf{ \fail^{\widetilde{y}} }_u \|   ~ \guname{\unvar{x}}{x_i}; \piencodf{ U }_{x_i} ) \\
            &= \res{\unvar{x}} ( \pnone{u }  \| \pnone{\widetilde{x}} \|   ~ \guname{\unvar{x}}{x_i}; \piencodf{ U }_{x_i} )
            \equiv   \pnone{u }  \| \pnone{\widetilde{x}}=  \piencodf{M}_u
            \tag*{\qedhere}
        \end{align}

    \end{enumerate}

\end{proof}

\subsection{Soundness}\label{a:tsoundness}

\begin{restatable}[Weak Soundness (Under $\redtwo$)]{theorem}{thmEncTWSound}\label{t:soundnesstwounres}
    If $ \piencodf{N}_u \redtwo^* Q$ for a well-formed closed $\lamcoldetsh$-term $N$, then there exist $Q'$  and $N' $ such that $Q \redtwo^* Q'$, $N  \red^* N'$ and $\piencodf{N'}_u \equiv Q'$.
\end{restatable}

\begin{proof}
By induction on the structure of $N $ and then induction on the number of reductions of $\piencodf{N} \redtwo^* Q$.

\begin{enumerate}
    \item {\bf Base case:} $N =  {x}$, $N =  {x}[j]$, $N = \fail^{\emptyset}$ and $N = \lambda x . (M[ {\widetilde{x}} \leftarrow  {x}])$.
.

    No reductions can take place, and the result follows trivially.
    $Q =  \piencodf{N}_u \redtwo^0 \piencodf{N}_u = Q'$ and $ N \red^0  N = N'$.

    \item $N =  M (C \bagsep U) $.

        Then,
        $ \piencodf{M (C \bagsep U)}_u = \res{v} (\piencodf{M}_v \| \gsome{v}{u , \llfv{C}};\pname{v}{x}; ( \piencodf{C \bagsep U}_x  \| \pfwd{v}{u}  ) )$, and we are able to perform the  reductions from $\piencodf{M (C \bagsep U)}_u$.

        We now proceed by induction on $k$, with  $\piencodf{N}_u \redtwo^k Q$. There are two main cases:
        \begin{enumerate}
            \item When $k = 0$ the thesis follows easily:

            We have
    $Q =  \piencodf{M (C \bagsep U)}_u \redtwo^0 \piencodf{M (C \bagsep U)}_u = Q'$ and $M (C \bagsep U) \red^0 M (C \bagsep U) = N'$.

            \item The interesting case is when $k \geq 1$.

            Then, for some process $R$ and $n, m$ such that $k = n+m$, we have the following:
            \[
            \begin{aligned}
               \piencodf{{N}}_u & =  \res{v} (\piencodf{M}_v \| \gsome{v}{u , \llfv{C}};\pname{v}{x}; ( \piencodf{C \bagsep U}_x  \| \pfwd{v}{u}  ) )\\
               & \redtwo^m  \res{v} ( R \| \gsome{v}{u , \llfv{C}};\pname{v}{x}; ( \piencodf{C \bagsep U}_x  \| \pfwd{v}{u}  ) ))
               \redtwo^n  Q\\
            \end{aligned}
            \]
            Thus, the first $m \geq 0$ reduction steps are  internal to $\piencodf{ M}_v$; type preservation in \clpi ensures that, if they occur,  these reductions  do not discard the possibility of synchronizing with $\psome{v}$. Then, the first of the $n \geq 0$ reduction steps towards $Q$ is a synchronization between $R$ and $\gsome{v}{u , \llfv{C}}$.

            We consider two sub-cases, depending on the values of  $m$ and $n$:
            \begin{enumerate}
                \item $m = 0$ and $n \geq 1$:

            Then $R = \piencodf{M}_v$ as $\piencodf{M}_v \redtwo^0 \piencodf{M}_v$.
            Notice that there are two possibilities of having an unguarded:

            \begin{enumerate}
            \item $M =  (\lambda x . (M'[ {\widetilde{x}} \leftarrow  {x}])) \linexsub{C_1 / \widetilde{ y}_1} \cdots \linexsub{C_p / \widetilde{ y}_p} \unexsub{U_1 / \unvar{z}_1} \cdots \unexsub{U_q / \unvar{z}_q}   \quad (p, q \geq 0)$

            \[
            \begin{aligned}
            \piencodf{M}_v &= \piencodf{ (\lambda x . (M'[ {\widetilde{x}} \leftarrow  {x}])) \linexsub{C_1 / \widetilde{ y}_1} \cdots \linexsub{C_p / \widetilde{ y}_p} \unexsub{U_1 / \unvar{z}_1} \cdots \unexsub{U_q / \unvar{z}_q}  }_v \\
            &=  \res{\unvar{z}_q} ( \cdots \res{\unvar{z}_1} ( \piencodf{(\lambda x . (M'[ {\widetilde{x}} \leftarrow  {x}])) \linexsub{C_1 / \widetilde{ y}_1} \cdots \linexsub{C_p / \widetilde{ y}_p}}_v  \|   ~ \guname{\unvar{z}_1}{z_1}; \piencodf{ U_1 }_{z_1}  ) \\
            & \qquad \qquad \cdots   \|   ~ \guname{\unvar{z}_q}{z_q}; \piencodf{ U_q }_{z_q} ) \quad = Q
            \\
            \end{aligned}
            \]
            Which we shall write as:
            \[
            \begin{aligned}
            Q = \res{\unvar{z}_q,  \cdots , \unvar{z}_1} ( \piencodf{(\lambda x . (M'[ {\widetilde{x}} \leftarrow  {x}])) \linexsub{C_1 / \widetilde{ y}_1} \cdots \linexsub{C_p / \widetilde{ y}_p}}_v  \|   ~ \guname{\unvar{z}_1}{z_1}; \piencodf{ U_1 }_{z_1}  \cdots   \|   ~ \guname{\unvar{z}_q}{z_q}; \piencodf{ U_q }_{z_q} )
            \end{aligned}
                \]
                for simplicity to represent the process. We also use this to simplify the translation of linear explicit substitutions of bags from:
                \[
                \begin{aligned}
                \piencodf{ M \linexsub{\bag{M_1} \cdot \cdots \cdot \bag{M_k} /  x_1 , \cdots , x_k}  }_v    & =
                    \res{z_1}( \gsome{z_1}{\llfv{M_{1}}};\piencodf{ M_{1} }_{ {z_1}}  \| \cdots \res{z_k} ( \gsome{z_k}{\llfv{M_{k}}};\piencodf{ M_{k} }_{ {z_k}} \\
                    & \qquad \qquad  \| \bignd_{x_{i_1} \in \{ x_1 ,\cdots , x_k  \}} \cdots \bignd_{x_{i_k} \in \{ x_1 ,\cdots , x_k \setminus x_{i_1} , \cdots , x_{i_{k-1}}  \}} \\
                    & \qquad \qquad \qquad \piencodf{ M }_v \{ z_1 / x_{i_1} \} \cdots \{ z_k / x_{i_k} \} ) \cdots )
                \end{aligned}
                \]
                to be represented as:
                \(
                \begin{aligned}
                \piencodf{ M \linexsub{C /  \widetilde{x}}  }_v    & =
                    \res{\widetilde{z}}( \gsome{\widetilde{z}}{\llfv{C}};\piencodf{ C }_{\widetilde{z} }  \| \bignd_{\widetilde{x}_{i} \in \perm{\widetilde{x}}} \piencodf{ M }_v \{ \widetilde{z} / \widetilde{x}_{i} \}).
                \end{aligned}
                \)
                These representations are purely for simplicity and are not an alternative to the actual translation. We continue expanding the sub-process\\
                $\piencodf{(\lambda x . (M'[ {\widetilde{x}} \leftarrow  {x}])) \linexsub{C_1 / \widetilde{ y}_1} \cdots \linexsub{C_p / \widetilde{ y}_p}}_v$ using the above shortened and simplified notation:
                \[
            \begin{aligned}
            \piencodf{(\lambda x . (M'[ {\widetilde{x}} \leftarrow  {x}])) \linexsub{C_1 / \widetilde{ y}_1} \cdots \linexsub{C_p / \widetilde{ y}_p}}_v & =
                    \res{\widetilde{w}_p}( \gsome{\widetilde{w}_p}{\llfv{C_p}};\piencodf{ C_p }_{\widetilde{w}_p }  \| \\
                    & \qquad  \bignd_{\widetilde{y}_{p_i} \in \perm{\widetilde{y}_p}} \cdots   \res{\widetilde{w}_1}( \gsome{\widetilde{w}_1}{\llfv{C_1}};\piencodf{ C_1 }_{\widetilde{w}_1 }  \| \\
                    & \qquad  \bignd_{\widetilde{y}_{1_i} \in \perm{\widetilde{y}_1}}  \piencodf{ \lambda x . (M'[ {\widetilde{x}} \leftarrow  {x}]) }_v \{ \widetilde{w}_1 / \widetilde{y}_{{1_i}} \}  ) \cdots   \{ \widetilde{w}_p / \widetilde{y}_{{p_i}} \}  )
            \\
            \end{aligned}
            \]
            Hence we represent $\piencodf{M}_v$ as:

            \[
            \begin{aligned}
            \piencodf{M}_v &= \res{\unvar{z}_q,  \cdots , \unvar{z}_1} ( \res{\widetilde{w}_p}( \gsome{\widetilde{w}_p}{\llfv{C_p}};\piencodf{ C_p }_{\widetilde{w}_p }  \| \\
                    & \qquad \qquad \bignd_{\widetilde{y}_{p_i} \in \perm{\widetilde{y}_p}} \cdots   \res{\widetilde{w}_1}( \gsome{\widetilde{w}_1}{\llfv{C_1}};\piencodf{ C_1 }_{\widetilde{w}_1 }  \| \\
                    & \qquad \qquad \bignd_{\widetilde{y}_{1_i} \in \perm{\widetilde{y}_1}}  \piencodf{ \lambda x . (M'[ {\widetilde{x}} \leftarrow  {x}]) }_v \{ \widetilde{w}_1 / \widetilde{y}_{{1_i}} \}  ) \cdots   \{ \widetilde{w}_p / \widetilde{y}_{{p_i}} \}  ) \\
                    & \qquad \qquad \|   ~ \guname{\unvar{z}_1}{z_1}; \piencodf{ U_1 }_{z_1}  \cdots   \|   ~ \guname{\unvar{z}_q}{z_q}; \piencodf{ U_q }_{z_q} )
            \\
            \end{aligned}
            \]
            Finally we shall simplify the process to become:
            \[
                \piencodf{M}_v  = \res{\widetilde{z} , \widetilde{w} } ( \bignd_{i \in I}  \piencodf{ \lambda x . (M'[ {\widetilde{x}} \leftarrow  {x}]) }_v \{ \widetilde{w} / \widetilde{y}_{i} \}  \|  Q'' )
            \\
            \]
            With this shape for $M$, we then have the following:
            {
            \small
            \[
            \begin{aligned}
            \piencodf{N}_u & = \piencodf{M(C \bagsep U)}_u= \res{v} (\piencodf{M}_v \| \gsome{v}{u , \llfv{C}};\pname{v}{x}; ( \piencodf{C \bagsep U}_x  \| \pfwd{v}{u}  ) )\\
            &= \res{v} (\res{\widetilde{z} , \widetilde{w} } ( \bignd_{i \in I}  \piencodf{ \lambda x . (M'[ {\widetilde{x}} \leftarrow  {x}]) }_v \{ \widetilde{w} / \widetilde{y}_{i} \}  \|  Q'' )  \| \gsome{v}{u , \llfv{C}};\pname{v}{x}; ( \piencodf{C \bagsep U}_x  \| \pfwd{v}{u}  ) )\\
            &= \res{v} (\res{\widetilde{z} , \widetilde{w} } ( \bignd_{i \in I}  \psome{v};\gname{v}{x};  \psome{x};\gname{x}{\linvar{x}}; \gname{x}{\unvar{x}};  \gclose{x} ; \piencodf{M[ {\widetilde{x}} \leftarrow  {x}]}_v \{ \widetilde{w} / \widetilde{y}_{i} \}  \|  Q'' ) \\
            & \qquad \qquad \| \gsome{v}{u , \llfv{C}};\pname{v}{x}; ( \piencodf{C \bagsep U}_x  \| \pfwd{v}{u}  ) )\\
            & \redtwo \res{v} (\res{\widetilde{z} , \widetilde{w} } ( \bignd_{i \in I}  \gname{v}{x};  \psome{x};\gname{x}{\linvar{x}}; \gname{x}{\unvar{x}};  \gclose{x} ; \piencodf{M[ {\widetilde{x}} \leftarrow  {x}]}_v \{ \widetilde{w} / \widetilde{y}_{i} \}  \|  Q'' ) & = Q_1 \\
            & \qquad \qquad \| \pname{v}{x}; ( \piencodf{C \bagsep U}_x  \| \pfwd{v}{u}  ) )\\
            & \redtwo \res{v} (   \pfwd{v}{u} \| \res{x}( \res{\widetilde{z} , \widetilde{w} } ( \bignd_{i \in I}  \psome{x};\gname{x}{\linvar{x}}; \gname{x}{\unvar{x}};  \gclose{x} ; \piencodf{M[ {\widetilde{x}} \leftarrow  {x}]}_v \{ \widetilde{w} / \widetilde{y}_{i} \}  \|  Q'' ) & = Q_2\\
            & \qquad \qquad \|  \piencodf{C \bagsep U}_x     ) )\\
            & \redtwo  \res{x}( \res{\widetilde{z} , \widetilde{w} } ( \bignd_{i \in I}  \psome{x};\gname{x}{\linvar{x}}; \gname{x}{\unvar{x}};  \gclose{x} ; \piencodf{M[ {\widetilde{x}} \leftarrow  {x}]}_u \{ \widetilde{w} / \widetilde{y}_{i} \}  \|  Q'' )  \|  \piencodf{C \bagsep U}_x     ) & = Q_3\\
            \end{aligned}
            \]
            }
            We also have that
                                \[
                                \begin{aligned}
                                    N &=(\lambda x . (M'[ {\widetilde{x}} \leftarrow  {x}]))  \linexsub{C_1 / \widetilde{ y}_1} \cdots \linexsub{C_p / \widetilde{ y}_p} \unexsub{U_1 / \unvar{z}_1} \cdots \unexsub{U_q / \unvar{z}_q}  (C \bagsep U) \\
                                    & \equivlam  (\lambda x . (M'[ {\widetilde{x}} \leftarrow  {x}]) (C \bagsep U))  \linexsub{C_1 / \widetilde{ y}_1} \cdots \linexsub{C_p / \widetilde{ y}_p} \unexsub{U_1 / \unvar{z}_1} \cdots \unexsub{U_q / \unvar{z}_q}  \\
                                & \red   M'[ {\widetilde{x}} \leftarrow  {x}] \esubst{(C \bagsep U)}{x}  \linexsub{C_1 / \widetilde{ y}_1} \cdots \linexsub{C_p / \widetilde{ y}_p} \unexsub{U_1 / \unvar{z}_1} \cdots \unexsub{U_q / \unvar{z}_q}  = {M}
                                \end{aligned}
                                \]
            Furthermore, we have:
            \[
                                \begin{aligned}
                                    \piencodf{M}_u &= \piencodf{M'[ {\widetilde{x}} \leftarrow  {x}] \esubst{(C \bagsep U)}{x}  \linexsub{C_1 / \widetilde{ y}_1} \cdots \linexsub{C_p / \widetilde{ y}_p} \unexsub{U_1 / \unvar{z}_1} \cdots \unexsub{U_q / \unvar{z}_q} }_u \\
                                    & = \res{x}( \res{\widetilde{z} , \widetilde{w} } ( \bignd_{i \in I}  \psome{x};\gname{x}{\linvar{x}}; \gname{x}{\unvar{x}};  \gclose{x} ; \piencodf{M[ {\widetilde{x}} \leftarrow  {x}]}_u \{ \widetilde{w} / \widetilde{y}_{i} \}  \|  Q'' ) \\
                                    & \qquad \qquad \|  \piencodf{C \bagsep U}_x
                                \end{aligned}
                                \]

                    We consider different possibilities for $n \geq 1$; in all  the cases, the result follows.
                                    \smallskip

            \noindent  {\bf When $n = 1$:}
                We have $Q = Q_1$, $ \piencodf{N}_u \redtwo^1 Q_1$.
                    We also have that
                    \begin{itemize}
                    \item  $Q_1 \redtwo^2 Q_3 = Q'$ ,
                    \item ${N} \red^1 M'[ {\widetilde{x}} \leftarrow  {x}] \esubst{(C \bagsep U)}{x}  \linexsub{C_1 / \widetilde{ y}_1} \cdots \linexsub{C_p / \widetilde{ y}_p} \unexsub{U_1 / \unvar{z}_1} \cdots \unexsub{U_q / \unvar{z}_q} ={N}'$
                    \item and $\piencodf{M'[ {\widetilde{x}} \leftarrow  {x}] \esubst{(C \bagsep U)}{x}  \linexsub{C_1 / \widetilde{ y}_1} \cdots \linexsub{C_p / \widetilde{ y}_p} \unexsub{U_1 / \unvar{z}_1} \cdots \unexsub{U_q / \unvar{z}_q}}_u = Q_3$.
                    \end{itemize}

                                            \smallskip

            \noindent  {\bf When $n = 2$:} the analysis is similar.

            \noindent {\bf When $n \geq 3$:}
            We have $ \piencodf{{N}}_u \redtwo^3 Q_3 \redtwo^l Q$, for $l \geq 0$. We also know that ${N} \red {M}$, $Q_3 = \piencodf{{M}}_u$. By the IH, there exist $ Q' , {N}'$ such that $Q \redtwo^i Q'$, ${M} \red^j {N}'$ and $\piencodf{{N}'}_u = Q'$ . Finally, $\piencodf{{N}}_u \redtwo^3 Q_3 \redtwo^l Q \redtwo^i Q'$ and ${N} \rightarrow {M}  \red^j {N}'$.

            \item $M = \fail^{\widetilde{z}}$.

            Then,                     \(
                        \begin{aligned}
                            \piencodf{M}_v &= \piencodf{\fail^{\widetilde{z}}}_v = \pnone{ v}  \| \pnone{ \widetilde{z}} .
                        \end{aligned}
                    \)
                    With this shape for $M$, we have:

                    \[
                    \begin{aligned}
                        \piencodf{{N}}_u & = \piencodf{(M\ (C \bagsep U))}_u =\res{v} (\piencodf{M}_v \| \gsome{v}{u , \llfv{C}};\pname{v}{x}; ( \piencodf{C \bagsep U}_x  \| \pfwd{v}{u}  ) )
                        \\
                        & =\res{v} (\pnone{ v}  \| \pnone{ \widetilde{z}} \| \gsome{v}{u , \llfv{C}};\pname{v}{x}; ( \piencodf{C \bagsep U}_x  \| \pfwd{v}{u}  ) )
                        \\
                        & \redtwo  \pnone{ \widetilde{z}} \| \pnone{u} \| \pnone{ \llfv{C} }
                        \\
                        \end{aligned}
                    \]

                    \end{enumerate}

                    We also have that
                    \(  {N} = \fail^{\widetilde{z}}\ C \bagsep U \red   \fail^{\widetilde{z} \cup \llfv{C}}  = {M}.  \)
                    Furthermore,
                    \[
                     \begin{aligned}
                          \piencodf{{M}}_u &= \piencodf{ \fail^{\widetilde{z} \cup \llfv{C}  } }_u
                          = \pnone{ \widetilde{z}} \| \pnone{u} \| \pnone{ \llfv{C} }\\
                    \end{aligned}
                    \]

  \item When $m \geq 1$ and $ n \geq 0$, we distinguish two cases:

 \begin{enumerate}
\item When $n = 0$:

Then, $ \res{v} ( R \| \gsome{v}{u , \llfv{C}};\pname{v}{x}; ( \piencodf{C \bagsep U}_x  \| \pfwd{v}{u}  ) )) =  Q $ and $\piencodf{M}_u \redtwo^m R$ where $m \geq 1$. Then by the IH there exist $R'$  and ${M}' $ such that $R \redtwo^i R'$, $M \red^j {M}'$, and $\piencodf{{M}'}_u = R'$.  Hence we have that

            \[
                \begin{aligned}
                   \piencodf{{N}}_u
                   & = \res{v} (\piencodf{M}_v \| \gsome{v}{u , \llfv{C}};\pname{v}{x}; ( \piencodf{C \bagsep U}_x  \| \pfwd{v}{u}  ) )\\
                   & \redtwo^m  \res{v} ( R \| \gsome{v}{u , \llfv{C}};\pname{v}{x}; ( \piencodf{C \bagsep U}_x  \| \pfwd{v}{u}  ) )  = Q
                \end{aligned}
             \]
            We also know that
            \[
            \begin{aligned}
              Q & \redtwo^i  \res{v} ( R \| \gsome{v}{u , \llfv{C}};\pname{v}{x}; ( \piencodf{C \bagsep U}_x  \| \pfwd{v}{u}  ) ) = Q'\\
            \end{aligned}
            \]
            and so the \lamcoldetsh term can reduce as follows: ${N} = (M\ ( C \bagsep U )) \red^j M'\ ( C \bagsep U ) = {N}'$ and  $\piencodf{ {N}'}_u = Q'$.

                        \item When $n \geq 1$:

                            Then  $R$ has an occurrence of an unguarded $ \psome{v} $ or $\pnone{v}$, hence it is of the form

                            $ \piencodf{(\lambda x . (M'[ {\widetilde{x}} \leftarrow  {x}])) \linexsub{C_1 / \widetilde{ y}_1} \cdots \linexsub{C_p / \widetilde{ y}_p} \unexsub{U_1 / \unvar{z}_1} \cdots \unexsub{U_q / \unvar{z}_q}  }_v $ or $ \piencodf{\fail^{\widetilde{x}}}_v. $
              This case follows by IH.
                    \end{enumerate}

            \end{enumerate}

        \end{enumerate}

        This concludes the analysis for the case ${N} = (M \, ( C \bagsep U ))$.

        \item ${N} = M[ {\widetilde{x}} \leftarrow  {x}]$.

    The sharing variable $ {x}$ is not free and the result follows by vacuity.

        \item ${N} = M[ {\widetilde{x}} \leftarrow  {x}] \esubst{ C \bagsep U }{ x}$. Then we have

            \[
                \begin{aligned}
                    \piencodf{{N}}_u &=\piencodf{ M[ {\widetilde{x}} \leftarrow  {x}] \esubst{ C \bagsep U }{ x} }_u= \res{x}( \psome{x}; \gname{x}{\linvar{x}}; \gname{x}{\unvar{x}};  \gclose{x} ;\piencodf{ M[ {\widetilde{x}} \leftarrow  {x}]}_u \| \piencodf{ C \bagsep U}_x )
                \end{aligned}
            \]

            Let us consider three cases.

            \begin{enumerate}
                \item When $\size{ {\widetilde{x}}} = \size{C}$.
                    Then let us consider the shape of the bag $ C$.

  \begin{enumerate}
  \item When $C = \oneb$.

                              We have the following
                             \[
                             \begin{aligned}
                             \piencodf{{N}}_u
                             &= \res{x}( \psome{x}; \gname{x}{\linvar{x}}; \gname{x}{\unvar{x}};  \gclose{x} ;\piencodf{ M[  \leftarrow  {x}]}_u \| \piencodf{ \oneb \bagsep U}_x )\\
                             &= \res{x}( \psome{x}; \gname{x}{\linvar{x}}; \gname{x}{\unvar{x}};  \gclose{x} ;\piencodf{ M[  \leftarrow  {x}]}_u \| \\
                             & \qquad \qquad \gsome{x}{\llfv{C}};  \pname{x}{\linvar{x}}; \big( \piencodf{ \oneb }_{\linvar{x}} \|  \pname{x}{\unvar{x}}; ( \guname{\unvar{x}}{x_i}; \piencodf{ U }_{x_i} \| \pclose{x} ) \big) )\\
                             & \redtwo \res{x}(  \gname{x}{\linvar{x}}; \gname{x}{\unvar{x}};  \gclose{x} ;\piencodf{ M[  \leftarrow  {x}]}_u \| \pname{x}{\linvar{x}}; \big( \piencodf{ \oneb }_{\linvar{x}} \|  \pname{x}{\unvar{x}}; ( \guname{\unvar{x}}{x_i}; \piencodf{ U }_{x_i} \| \pclose{x} ) \big) )  & = Q_1  \\
                             & \redtwo \res{x}(  \pname{x}{\unvar{x}}; ( \guname{\unvar{x}}{x_i}; \piencodf{ U }_{x_i} \| \pclose{x} ) \|  \res{\linvar{x}} (  \gname{x}{\unvar{x}};  \gclose{x} ;\piencodf{ M[  \leftarrow  {x}]}_u \| \piencodf{ \oneb }_{\linvar{x}}   ))  & = Q_2  \\
                             & \redtwo \res{x}( \pclose{x} \|  \res{\unvar{x}} ( \guname{\unvar{x}}{x_i}; \piencodf{ U }_{x_i}  \|  \res{\linvar{x}} (   \gclose{x} ;\piencodf{ M[  \leftarrow  {x}]}_u \| \piencodf{ \oneb }_{\linvar{x}}   )))  & = Q_3  \\
                             & \redtwo  \res{\unvar{x}} ( \guname{\unvar{x}}{x_i}; \piencodf{ U }_{x_i}  \|  \res{\linvar{x}} (  \piencodf{ M[  \leftarrow  {x}]}_u \| \piencodf{ \oneb }_{\linvar{x}}   )))  & = Q_4  \\
                             & =  \res{\unvar{x}} ( \guname{\unvar{x}}{x_i}; \piencodf{ U }_{x_i}  \|  \res{\linvar{x}} (  \psome{\linvar{x}}; \pname{\linvar{x}}{y_i}; ( \gsome{y_i}{ u,\llfv{M} }; \gclose{ y_{i} } ;\piencodf{M}_u \| \pnone{ \linvar{x} } ) \| \\
                             & \qquad \qquad  \gsome{\linvar{x}}{\emptyset};\gname{x}{y_i};  ( \psome{ y_i}; \pclose{y_i}  \| \gsome{\linvar{x}}{\emptyset}; \pnone{\linvar{x}} )   )))  \\
                             & \redtwo  \res{\unvar{x}} ( \guname{\unvar{x}}{x_i}; \piencodf{ U }_{x_i}  \|
                             \res{\linvar{x}} ( \pname{\linvar{x}}{y_i}; ( \gsome{y_i}{ u,\llfv{M} }; \gclose{ y_{i} } ;\piencodf{M}_u \| \pnone{ \linvar{x} } ) \| \\
                             & \qquad \qquad  \gname{x}{y_i};  ( \psome{ y_i}; \pclose{y_i}  \| \gsome{\linvar{x}}{\emptyset}; \pnone{\linvar{x}} )   )))   & = Q_5 \\
                             & \redtwo  \res{\unvar{x}} ( \guname{\unvar{x}}{x_i}; \piencodf{ U }_{x_i}  \|
                             \res{\linvar{x}} ( \pnone{ \linvar{x} } \| \res{y_i} ( \gsome{y_i}{ u,\llfv{M} }; \gclose{ y_{i} } ;\piencodf{M}_u   \| \\
                             & \qquad \qquad  \psome{ y_i}; \pclose{y_i}  \| \gsome{\linvar{x}}{\emptyset}; \pnone{\linvar{x}} )   ))   & = Q_6 \\
                             & \redtwo  \res{\unvar{x}} ( \guname{\unvar{x}}{x_i}; \piencodf{ U }_{x_i}  \|
                              \res{y_i} ( \gsome{y_i}{ u,\llfv{M} }; \gclose{ y_{i} } ;\piencodf{M}_u   \|   \psome{ y_i}; \pclose{y_i}  ) )   & = Q_7 \\
                              & \redtwo  \res{\unvar{x}} ( \guname{\unvar{x}}{x_i}; \piencodf{ U }_{x_i}  \|
                              \res{y_i} ( \ \gclose{ y_{i} } ;\piencodf{M}_u   \|    \pclose{y_i}  ) )   & = Q_8 \\
                              & \redtwo  \res{\unvar{x}} ( \guname{\unvar{x}}{x_i}; \piencodf{ U }_{x_i}  \|
                              \piencodf{M}_u   ) =  \piencodf{M \unexsub{U / \unvar{x}}}_u   & = Q_9 \\
                            \end{aligned}
                            \]
                            Notice how $Q_6$ has a choice however the $\linvar{x}$ name can be closed at any time so for simplicity we perform communication across this name first followed by all other comunications that can take place.

                        Now we proceed by induction on the number of reductions $\piencodf{{N}}_u \redtwo^k Q$.

                            \begin{enumerate}

                                \item When $k = 0$, the result follows trivially. Just take ${N}={N}'$ and $\piencodf{{N}}_u=Q=Q'$.

                                \item When $k = 1$.

                                    We have $Q = Q_1$, $ \piencodf{{N}}_u \redtwo^1 Q_1$
                                    We also have that $Q_1 \redtwo^8 Q_9 = Q'$ , ${N} \red M \unexsub{U / \unvar{x}} = N'$ and $\piencodf{ N' }_u = Q_9$

                                \item When $2 \leq  k \leq 8$.

                                    Proceeds similarly to the previous case

                                \item When $k \geq 9$.

                                We have $ \piencodf{{N}}_u \redtwo^9 Q_9 \redtwo^l Q$, for $l \geq 0$. Since $Q_9 = \piencodf{ M \unexsub{U / \unvar{x}} }_u$ we apply the induction hypothesis we have that  there exist $ Q' , {N}' \ s.t. \ Q \redtwo^i Q' ,  M \unexsub{U / \unvar{x}}  \red^j {N}'$ and $\piencodf{{N}'}_u = Q'$.                                    Then,  $ \piencodf{{N}}_u \redtwo^5 Q_5 \redtwo^l Q \redtwo^i Q'$ and by the contextual reduction rule it follows that ${N} = (M[ \leftarrow x])\esubst{ 1 }{ x } \red  M \unexsub{U / \unvar{x}} \red^j  {N}' $ and the case holds.

\end{enumerate}

\item When $C = \bag{N_1} \cdot \cdots \cdot \bag{N_l}$, for $l \geq 1$.
                    Then,

 \[
   \begin{aligned}
   \piencodf{{N}}_u &=\piencodf{ M[ {\widetilde{x}} \leftarrow  {x}] \esubst{ C \bagsep U }{x} }_u\\
   & =  \res{x}( \psome{x}; \gname{x}{\linvar{x}}; \gname{x}{\unvar{x}};  \gclose{x} ;\piencodf{ M[ {\widetilde{x}} \leftarrow  {x}]}_u \| \piencodf{ C \bagsep U}_x )
    \\
   & \redtwo^4  \res{\unvar{x}} ( \guname{\unvar{x}}{x_i}; \piencodf{ U }_{x_i}  \|  \res{\linvar{x}} (  \piencodf{ M[ {\widetilde{x}} \leftarrow  {x}]}_u \| \piencodf{ C }_{\linvar{x}}   )))   \\
   & =  \res{\unvar{x}} ( \guname{\unvar{x}}{x_i}; \piencodf{ U }_{x_i}  \|  \res{\linvar{x}} (
    \psome{\linvar{x}}; \pname{\linvar{x}}{y_1}; \big( \gsome{y_1}{ \emptyset }; \gclose{ y_{1} } ; \0   \\
    & \qquad \qquad \|\psome{\linvar{x}}; \gsome{\linvar{x}}{u, \llfv{M} \setminus  \widetilde{x} }; \bignd_{x_{i_1} \in \widetilde{x}} \gname{x}{{x}_{i_1}};
    \cdots
    \psome{\linvar{x}}; \pname{\linvar{x}}{y_l}; \big( \gsome{y_l}{ \emptyset }; \gclose{ y_{l} } ; \0   \\
    & \qquad \qquad \|\psome{\linvar{x}}; \gsome{\linvar{x}}{u, \llfv{M} \setminus  (\widetilde{x} \setminus x_1, \cdots , x_{i_{l-1}} )}; \bignd_{x_{i_l} \in (\widetilde{x} \setminus x_1, \cdots , x_{i_{l-1}})} \gname{x}{{x}_{i_l}};\piencodf{M[  \leftarrow  {x}]}_u \big)
    \cdots
    \big)
    \\
    & \qquad \qquad \| \gsome{\linvar{x}}{\llfv{C} }; \gname{x}{y_1}; \gsome{\linvar{x}}{y_1, \llfv{C}}; \psome{\linvar{x}}; \pname{\linvar{x}}{z_1}; \\
    & \qquad \qquad  ( \gsome{z_1}{\llfv{M_1}};  \piencodf{M_1}_{z_1} \| \pnone{y_1} \|
    \cdots \\
    & \qquad \qquad \gsome{\linvar{x}}{\llfv{M_l} }; \gname{x}{y_l}; \gsome{\linvar{x}}{y_l, \llfv{M_l}}; \psome{\linvar{x}}; \pname{\linvar{x}}{z_l}; \\
    & \qquad \qquad  ( \gsome{z_l}{\llfv{M_l}};  \piencodf{M_l}_{z_l} \| \pnone{y_l} \| \piencodf{\oneb}_{\linvar{x}}  ) \cdots  )     ))   \\
\end{aligned}
\]
We shall now perform multiple non-committing reductions at once. Notice that non-determinism guards the same prefixes denying the use of the reduction $\rredtwo{\piprecong{x}}$ hence denying the commitment of non-determinism.

\[
   \begin{aligned}
    & \redtwo^{6l}  \res{\unvar{x}} ( \guname{\unvar{x}}{x_i}; \piencodf{ U }_{x_i}  \|  \res{\linvar{x}} ( \piencodf{\oneb}_{\linvar{x}} \|
    \\
    & \qquad \qquad \res{z_1}  (  \gsome{z_1}{\llfv{M_1}};  \piencodf{M_1}_{z_1} \| \cdots
    \res{z_l} ( \gsome{z_l}{\llfv{M_l}};  \piencodf{M_l}_{z_l} \| \\
    & \qquad \qquad  \bignd_{x_{i_1} \in \widetilde{x}} \bignd_{x_{i_l} \in (\widetilde{x} \setminus x_1, \cdots , x_{i_{l-1}})} \piencodf{M[  \leftarrow  {x}]}_u \{ z_1 / x_{i_1} \} \cdots \{ z_l / x_{i_l} \} )
    \cdots    )     ))   \\
    & =  \res{\unvar{x}} ( \guname{\unvar{x}}{x_i}; \piencodf{ U }_{x_i}  \|  \res{\linvar{x}} ( \gsome{\linvar{x}}{\emptyset};\gname{x}{y_{l+1}};  ( \psome{ y_{l+1}}; \pclose{y_{l+1}}  \| \gsome{\linvar{x}}{\emptyset}; \pnone{\linvar{x}} ) \|
    \\
    & \qquad \qquad \res{z_1}  (  \gsome{z_1}{\llfv{M_1}};  \piencodf{M_1}_{z_1} \| \cdots
    \res{z_l} ( \gsome{z_l}{\llfv{M_l}};  \piencodf{M_l}_{z_l} \| \\
    & \qquad \qquad  \bignd_{x_{i_1} \in \widetilde{x}} \bignd_{x_{i_l} \in (\widetilde{x} \setminus x_1, \cdots , x_{i_{l-1}})} \psome{\linvar{x}}; \\
    & \qquad \qquad \pname{\linvar{x}}{y_{l+1}}; ( \gsome{y_{l+1}}{ u,\llfv{M} }; \gclose{ y_{{l+1}} } ;\piencodf{M}_u \{ z_1 / x_{i_1} \} \cdots \{ z_l / x_{i_l} \} \| \pnone{ \linvar{x} } )  )
    \cdots    )     ))   \\
    & \redtwo^{5}  \res{\unvar{x}} ( \guname{\unvar{x}}{x_i}; \piencodf{ U }_{x_i}  \|
    \\
    & \qquad \qquad \res{z_1}  (  \gsome{z_1}{\llfv{M_1}};  \piencodf{M_1}_{z_1} \| \cdots
    \res{z_l} ( \gsome{z_l}{\llfv{M_l}};  \piencodf{M_l}_{z_l} \| \\
    & \qquad \qquad  \bignd_{x_{i_1} \in \widetilde{x}} \bignd_{x_{i_l} \in (\widetilde{x} \setminus x_1, \cdots , x_{i_{l-1}})} \piencodf{M}_u  \{ z_1 / x_{i_1} \} \cdots \{ z_l / x_{i_l} \} )  \cdots    )     )   \\
   & = \piencodf{ M\linexsub{\bag{M_1} \cdot \cdots \cdots \bag{M_l}/ {x_1}, \cdots , x_1}  \unexsub{U /\unvar{x} }}_{u}= Q_{6l + 9}\\
  \end{aligned}
 \]

    The proof follows by induction on the number of reductions $\piencodf{{N}}_u \redtwo^k Q$.

\begin{enumerate}
\item When $k = 0$, the result follows trivially. Just take ${N}={N}'$ and $\piencodf{{N}}_u=Q=Q'$.

 \item When $1 \leq k \leq 6l + 9$.

 Let $Q_k$ such that $ \piencodf{{N}}_u \redtwo^k Q_k$.
    We also have that $Q_k \redtwo^{6l + 9 - k} Q_{6l + 9} = Q'$ ,

    ${N} \red^1 M\linexsub{\bag{M_1} \cdot \cdots \cdots \bag{M_l}/ {x_1}, \cdots , x_1}  \unexsub{U /\unvar{x} } = {N}'$ and

    $\piencodf{M\linexsub{\bag{M_1} \cdot \cdots \cdots \bag{M_l}/ {x_1}, \cdots , x_1}  \unexsub{U /\unvar{x} }}_u = Q_{6l + 9}$.

\item When $k > 6l + 9$.

Then,  $ \piencodf{{N}}_u \redtwo^{6l + 9} Q_{6l + 9} \redtwo^n Q$ for $n \geq 1$. Also,

\(
\begin{aligned}
&  {N} \red^1 M\linexsub{\bag{M_1} \cdot \cdots \cdots \bag{M_l}/ {x_1}, \cdots , x_1}  \unexsub{U /\unvar{x} } \text { and } \\
&    Q_{6l + 9} = \piencodf{M\linexsub{\bag{M_1} \cdot \cdots \cdots \bag{M_l}/ {x_1}, \cdots , x_1}  \unexsub{U /\unvar{x} }}_u.
    \end{aligned}
\)

By the induction hypothesis, there exist $ Q'$ and ${N}'$ such that
$\ Q \redtwo^i Q'$,

$M\linexsub{\bag{M_1} \cdot \cdots \cdots \bag{M_l}/ {x_1}, \cdots , x_1}  \unexsub{U /\unvar{x} } \red^j {N}'$ and $\piencodf{{N}'}_u = Q'$.

Finally, $\piencodf{{N}}_u \redtwo^{6l + 9} Q_{6l + 9} \redtwo^n Q \redtwo^i Q'$ and $$ {N} \red M\linexsub{\bag{M_1} \cdot \cdots \cdots \bag{M_l}/ {x_1}, \cdots , x_1}  \unexsub{U /\unvar{x} }  \red^j {N}'. $$

                            \end{enumerate}

                    \end{enumerate}

                \item When $\size{\widetilde{x}} > \size{C}$.

                    Then we have
                    ${N} = M[ {x}_1, \cdots ,  {x}_k \leftarrow  {x}]\ \esubst{ C \bagsep U }{x}$ with $C = \bag{N_1}  \cdots  \bag{N_l} \quad k > l$. ${N} \red   \fail^{\widetilde{z}} = {M}'$ and $ \widetilde{z} =  (\llfv{M} \setminus \{   {x}_1, \cdots ,  {x}_k \} ) \cup \llfv{C} $. On the one hand, we have:
                    Hence $k = l + m$ for some $m \geq 1$

        \begin{equation}\label{eq:compl_sh_faillin2}
        \begin{aligned}
            \piencodf{N}_u
            &= \piencodf{M[ {x}_1, \!\cdots\! ,  {x}_k \leftarrow  {x}]\ \esubst{C \bagsep U}{ x } }_u\\
            &= \res{x}( \psome{x}; \gname{x}{\linvar{x}}; \gname{x}{\unvar{x}};  \gclose{x} ;\piencodf{ M[ {\widetilde{x}} \leftarrow  {x}]}_u \| \piencodf{ C \bagsep U}_x )\\
            &= \res{x}( \psome{x}; \gname{x}{\linvar{x}}; \gname{x}{\unvar{x}};  \gclose{x} ;\piencodf{ M[ {\widetilde{x}} \leftarrow  {x}]}_u \| \\
            & \qquad \qquad \gsome{x}{\llfv{C}};  \pname{x}{\linvar{x}}; \big( \piencodf{ C }_{\linvar{x}} \|  \pname{x}{\unvar{x}}; ( \guname{\unvar{x}}{x_i}; \piencodf{ U }_{x_i} \| \pclose{x} ) \big) )\\
            & \redtwo^4  \res{\unvar{x}} ( \res{\linvar{x} } \big( \piencodf{ M[ {\widetilde{x}} \leftarrow  {x}]}_u \|  \piencodf{ C }_{\linvar{x}}   \big) \|  \guname{\unvar{x}}{x_i}; \piencodf{ U }_{x_i}  )  \\
            & = \res{\unvar{x}} ( \res{\linvar{x} } \big( \psome{\linvar{x}}; \pname{\linvar{x}}{y_1}; \big( \gsome{y_1}{ \emptyset }; \gclose{ y_{1} } ; \0   \\
            & \qquad \qquad \|\psome{\linvar{x}}; \gsome{\linvar{x}}{u, \llfv{M} \setminus  \widetilde{x} }; \bignd_{x_{i_1} \in \widetilde{x}} \gname{x}{{x}_{i_1}}; \cdots \\
            & \qquad \qquad \psome{\linvar{x}}; \pname{\linvar{x}}{y_k}; \big( \gsome{y_k}{ \emptyset }; \gclose{ y_{k} } ; \0   \\
            & \qquad \qquad \|\psome{\linvar{x}}; \gsome{\linvar{x}}{u, \llfv{M} \setminus  (\widetilde{x} \setminus x_{i_1} , \cdots , x_{i_{k-1}}   )}; \bignd_{x_{i_k} \in (\widetilde{x} \setminus x_{i_1} , \cdots , x_{i_{k-1}}  )} \gname{x}{{x}_{i_k}};\piencodf{M[  \leftarrow  {x}]}_u \big)
            \big) \| \\
            & \qquad \qquad  \gsome{\linvar{x}}{\llfv{C} }; \gname{x}{y_1}; \gsome{\linvar{x}}{y_1, \llfv{C}}; \psome{\linvar{x}}; \pname{\linvar{x}}{z_1}; \\
            & \qquad \qquad  ( \gsome{z_1}{\llfv{M_1}};  \piencodf{M_1}_{z_1} \| \pnone{y_1}  \| \cdots  \gsome{\linvar{x}}{\llfv{C} }; \gname{x}{y_l}; \gsome{\linvar{x}}{y_l, \llfv{M_l}}; \psome{\linvar{x}}; \pname{\linvar{x}}{z_l}; \\
            & \qquad \qquad   ( \gsome{z_l}{\llfv{M_l}};  \piencodf{M_l}_{z_l} \| \piencodf{ \oneb }_{\linvar{x}} \| \pnone{y_l} ) \cdots   )   \big) \|  \guname{\unvar{x}}{x_i}; \piencodf{ U }_{x_i}  ) \qquad
            (:= Q_4) \\
            \end{aligned}
            \end{equation}

            we reduce $Q_4$ arbitrarily synchronising along channels $\linvar{x} , y_1, \cdots y_l$.

            \begin{equation*}
            \begin{aligned}
                Q_4
                & \redtwo^{6l} \res{\unvar{x}} ( \res{\linvar{x} } \big( \piencodf{ \oneb }_{\linvar{x}} \|
                \res{z_1} (\gsome{z_1}{\llfv{M_1}};  \piencodf{M_1}_{z_1}   \| \cdots
                \res{z_l} (\gsome{z_l}{\llfv{M_l}};  \piencodf{M_l}_{z_l} \| \\
                &  \qquad \qquad \bignd_{x_{i_1} \in \widetilde{x}} \cdots \bignd_{x_{i_l} \in (\widetilde{x} \setminus x_{i_1} , \cdots , x_{i_{l-1}} )}  \psome{\linvar{x}}; \pname{\linvar{x}}{y_{l+1}}; \big( \gsome{y_{l+1}}{ \emptyset }; \gclose{ y_{{l+1}} } ; \0   \\
                & \qquad \qquad \|\psome{\linvar{x}}; \gsome{\linvar{x}}{u, \llfv{M} \setminus  (\widetilde{x} \setminus x_{i_1} , \cdots , x_{i_{l}}  )}; \bignd_{x_{i_{l+1}} \in (\widetilde{x} \setminus x_{i_1} , \cdots , x_{i_{l}} )} \gname{x}{{x}_{i_{l+1}}}; \cdots   \\
                & \qquad \qquad \psome{\linvar{x}}; \pname{\linvar{x}}{y_k}; \big( \gsome{y_k}{ \emptyset }; \gclose{ y_{k} } ; \0   \\
                & \qquad \qquad \|\psome{\linvar{x}}; \gsome{\linvar{x}}{u, \llfv{M} \setminus  (\widetilde{x} \setminus x_{i_1} , \cdots , x_{i_{k-1}}   )}; \bignd_{x_{i_k} \in (\widetilde{x} \setminus x_{i_1} , \cdots , x_{i_{k-1}}  )} \gname{x}{{x}_{i_k}};\\
                & \qquad \qquad \piencodf{M[  \leftarrow  {x}]}_u \{ z_1 / x_{i_1} \} \cdots \{ z_l / {x}_{i_l} \}
                \big) \cdots \big)  ) \cdots ) \big) \| \guname{\unvar{x}}{x_i}; \piencodf{ U }_{x_i}  )\\
                & = \res{\unvar{x}} ( \res{\linvar{x} } \big(
                 \gsome{\linvar{x}}{\emptyset}; \gname{x}{y_{l+1}};  ( \psome{ y_{l+1}}; \pclose{y_{l+1}}  \| \gsome{\linvar{x}}{\emptyset}; \pnone{\linvar{x}} ) \\
                & \qquad \qquad \|
                \res{z_1} (\gsome{z_1}{\llfv{M_1}};  \piencodf{M_1}_{z_1}   \| \cdots
                \res{z_l} (\gsome{z_l}{\llfv{M_l}};  \piencodf{M_l}_{z_l} \| \\
                &  \qquad \qquad \bignd_{x_{i_1} \in \widetilde{x}} \cdots \bignd_{x_{i_l} \in (\widetilde{x} \setminus x_{i_1} , \cdots , x_{i_{l-1}} )}\psome{\linvar{x}}; \pname{\linvar{x}}{y_{l+1}}; \big( \gsome{y_{l+1}}{ \emptyset }; \gclose{ y_{{l+1}} } ; \0   \\
                & \qquad \qquad \|\psome{\linvar{x}}; \gsome{\linvar{x}}{u, \llfv{M} \setminus  (\widetilde{x} \setminus x_{i_1} , \cdots , x_{i_{l}}  )}; \bignd_{x_{i_{l+1}} \in (\widetilde{x} \setminus x_{i_1} , \cdots , x_{i_{l}} )} \gname{x}{{x}_{i_{l+1}}}; \cdots   \\
                & \qquad \qquad \psome{\linvar{x}}; \pname{\linvar{x}}{y_k}; \big( \gsome{y_k}{ \emptyset }; \gclose{ y_{k} } ; \0   \\
                & \qquad \qquad \|\psome{\linvar{x}}; \gsome{\linvar{x}}{u, \llfv{M} \setminus  (\widetilde{x} \setminus x_{i_1} , \cdots , x_{i_{k-1}}   )}; \bignd_{x_{i_k} \in (\widetilde{x} \setminus x_{i_1} , \cdots , x_{i_{k-1}}  )} \gname{x}{{x}_{i_k}};\\
                & \qquad \qquad \piencodf{M[  \leftarrow  {x}]}_u \{ z_1 / x_{i_1} \} \cdots \{ z_l / {x}_{i_l} \}
                \big) \cdots \big)  ) \cdots ) \big) \| \guname{\unvar{x}}{x_i}; \piencodf{ U }_{x_i}  )\\
                & \redtwo^5 \res{\unvar{x}} ( \res{\linvar{x} } \big( \pnone{\linvar{x}}  \|
                \res{z_1} (\gsome{z_1}{\llfv{M_1}};  \piencodf{M_1}_{z_1}   \| \cdots
                \res{z_l} (\gsome{z_l}{\llfv{M_l}};  \piencodf{M_l}_{z_l} \| \\
                &  \qquad \qquad \bignd_{x_{i_1} \in \widetilde{x}} \cdots \bignd_{x_{i_l} \in (\widetilde{x} \setminus x_{i_1} , \cdots , x_{i_{l-1}} )} \gsome{\linvar{x}}{u, \llfv{M} \setminus  (\widetilde{x} \setminus x_{i_1} , \cdots , x_{i_{l}}  )}; \bignd_{x_{i_{l+1}} \in (\widetilde{x} \setminus x_{i_1} , \cdots , x_{i_{l}} )} \gname{x}{{x}_{i_{l+1}}}; \cdots   \\
                & \qquad \qquad \psome{\linvar{x}}; \pname{\linvar{x}}{y_k}; \big( \gsome{y_k}{ \emptyset }; \gclose{ y_{k} } ; \0   \\
                & \qquad \qquad \|\psome{\linvar{x}}; \gsome{\linvar{x}}{u, \llfv{M} \setminus  (\widetilde{x} \setminus x_{i_1} , \cdots , x_{i_{k-1}}   )}; \bignd_{x_{i_k} \in (\widetilde{x} \setminus x_{i_1} , \cdots , x_{i_{k-1}}  )} \gname{x}{{x}_{i_k}};\\
                & \qquad \qquad \piencodf{M[  \leftarrow  {x}]}_u \{ z_1 / x_{i_1} \} \cdots \{ z_l / {x}_{i_l} \}
                \big) \cdots \big)  ) \cdots ) \big) \| \guname{\unvar{x}}{x_i}; \piencodf{ U }_{x_i}  )\\
                & \redtwo \res{\unvar{x}} (
                \res{z_1} (\gsome{z_1}{\llfv{M_1}};  \piencodf{M_1}_{z_1}   \| \cdots
                \res{z_l} (\gsome{z_l}{\llfv{M_l}};  \piencodf{M_l}_{z_l} \| \\
                &  \qquad \qquad \bignd_{x_{i_1} \in \widetilde{x}} \cdots \bignd_{x_{i_l} \in (\widetilde{x} \setminus x_{i_1} , \cdots , x_{i_{l-1}} )} \pnone{u } \| \pnone{(\llfv{M} \setminus  \widetilde{x}) } \|  \pnone{(z_1 , \cdots ,z_l ) }
                ) \cdots ) \| \\
                & \qquad \qquad  \guname{\unvar{x}}{x_i}; \piencodf{ U }_{x_i}  )\\
                & \redtwo^l \res{\unvar{x}} (
                 \bignd_{x_{i_1} \in \widetilde{x}} \cdots \bignd_{x_{i_l} \in (\widetilde{x} \setminus x_{i_1} , \cdots , x_{i_{l-1}} )} \pnone{u } \| \pnone{(\llfv{M} \setminus  \widetilde{x}) }  \| \pnone{ \llfv{C}  } \|\\
                & \qquad \qquad  \guname{\unvar{x}}{x_i}; \piencodf{ U }_{x_i}  )\\
                & \equiv \pnone{u } \| \pnone{(\llfv{M} \setminus  \widetilde{x}) } \| \pnone{ \llfv{C}  }
                = \piencodf{ \fail^{\widetilde{z}}}_u = Q_{7l + 10}  \\
        \end{aligned}
        \end{equation*}

The rest of the proof is by induction on the number of reductions $\piencodf{{N}}_u \redtwo^j Q$.

                            \begin{enumerate}
                                \item When $j = 0$, the result follows trivially. Just take ${N}={N}'$ and $\piencodf{{N}}_u=Q=Q'$.
  \item When $1 \leq j \leq 7l + 10$.

Let $Q_j$ be such that $ \piencodf{{N}}_u \redtwo^j Q_j$.
By the steps above one has

\(\begin{aligned}
  &Q_j \redtwo^{7l + 10 - j} Q_{7l + 6} = Q',\\ &{N} \red^1   \fail^{\widetilde{z}} = {N}';\text{ and} \piencodf{ \fail^{\widetilde{z}}}_u = Q_{7l + 10}.
\end{aligned}
\)
\item When $j > 7l + 10$.

In this case, we have
$ \piencodf{{N}}_u \redtwo^{7l + 10} Q_{7l + 10} \redtwo^n Q,$ for $n \geq 1$.
We also know that
${N} \redtwo^1  \fail^{\widetilde{z}}$. However no further reductions can be performed.

                            \end{enumerate}

                \item When $\size{\widetilde{x}} < \size{C}$, the proof  proceeds similarly to the previous case.

            \end{enumerate}

        \item  ${N} =  M \linexsub{C /  \widetilde{x}}$.

           In this case let us consider $ C = \bag{M_1} \cdot \cdots \cdot \bag{M_k} $,
            \[
            \begin{aligned}
                \piencodf{ M \linexsub{C /  \widetilde{x}}}_u &=  \res{z_1}( \gsome{z_1}{\llfv{M_{1}}};\piencodf{ M_{1} }_{ {z_1}}  \|  \cdots \res{z_k} ( \gsome{z_k}{\llfv{M_{k}}};\piencodf{ M_{k} }_{ {z_k}} \\
                & \qquad \qquad  \| \bignd_{x_{i_1} \in \{ x_1 ,\cdots , x_k  \}} \cdots \bignd_{x_{i_k} \in \{ x_1 ,\cdots , x_k \setminus x_{i_1} , \cdots , x_{i_{k-1}}  \}} \piencodf{ M }_u \{ z_1 / x_{i_1} \} \cdots \{ z_k / x_{i_k} \} ) \cdots )
            \end{aligned}
            \]
            Therefore,
            \[
            \begin{aligned}
               \piencodf{{N}}_u
                &=  \res{z_1}( \gsome{z_1}{\llfv{M_{1}}};\piencodf{ M_{1} }_{ {z_1}}  \|  \cdots \res{z_k} ( \gsome{z_k}{\llfv{M_{k}}};\piencodf{ M_{k} }_{ {z_k}} \\
                & \qquad \qquad  \| \bignd_{x_{i_1} \in \{ x_1 ,\cdots , x_k  \}} \cdots \bignd_{x_{i_k} \in \{ x_1 ,\cdots , x_k \setminus x_{i_1} , \cdots , x_{i_{k-1}}  \}} \piencodf{ M }_u \{ z_1 / x_{i_1} \} \cdots \{ z_k / x_{i_k} \} ) \cdots )\\
                & \redtwo^m  \res{z_1}( \gsome{z_1}{\llfv{M_{1}}};\piencodf{ M_{1} }_{ {z_1}}  \|  \cdots \res{z_k} ( \gsome{z_k}{\llfv{M_{k}}};\piencodf{ M_{k} }_{ {z_k}} \\
                & \qquad \qquad  \| \bignd_{x_{i_1} \in \{ x_1 ,\cdots , x_k  \}} \cdots \bignd_{x_{i_k} \in \{ x_1 ,\cdots , x_k \setminus x_{i_1} , \cdots , x_{i_{k-1}}  \}} R \{ z_1 / x_{i_1} \} \cdots \{ z_k / x_{i_k} \} ) \cdots )\\
                &  \redtwo^n  Q,\\
            \end{aligned}
            \]

            for some process $R$. Where $\redtwo^n$ is a reduction that  initially synchronizes with $ \gsome{z_i}{\llfv{M_{i}}}$ for some $i \leq k$ when $n \geq 1$, $n + m = k \geq 1$. Type preservation in \clpi ensures reducing $\piencodf{ M}_v \redtwo^m$ does not consume possible synchronizations with $\psome{z_i} $, if they occur. Let us consider the the possible sizes of both $m$ and $n$.

            \begin{enumerate}
                \item For $m = 0$ and $n \geq 1$.

                    We have that $R = \piencodf{M}_u$ as $\piencodf{M}_u \redtwo^0 \piencodf{M}_u$.

                    Notice that there are two possibilities of having an unguarded $\psome{z_i}$ or $\pnone{z_i}$ without internal reductions:

                    \begin{enumerate}
                        \item $M = \fail^{ \widetilde{x}, \widetilde{y}}$.
  \[
  \begin{aligned}
  \piencodf{ {N}}_u
    & = \res{z_1}( \gsome{z_1}{\llfv{M_{1}}};\piencodf{ M_{1} }_{ {z_1}}  \|  \cdots \res{z_k} ( \gsome{z_k}{\llfv{M_{k}}};\piencodf{ M_{k} }_{ {z_k}} \\
        &  \qquad  \| \bignd_{x_{i_1} \in \{ x_1 ,\cdots , x_k  \}} \cdots \bignd_{x_{i_k} \in \{ x_1 ,\cdots , x_k \setminus x_{i_1} , \cdots , x_{i_{k-1}}  \}} \piencodf{ M }_u \{ z_1 / x_{i_1} \} \cdots \{ z_k / x_{i_k} \} ) \cdots )\\
  & = \res{z_1}( \gsome{z_1}{\llfv{M_{1}}};\piencodf{ M_{1} }_{ {z_1}}  \|  \cdots \res{z_k} ( \gsome{z_k}{\llfv{M_{k}}};\piencodf{ M_{k} }_{ {z_k}} \\
        &  \qquad  \| \bignd_{x_{i_1} \in \{ x_1 ,\cdots , x_k  \}} \cdots \bignd_{x_{i_k} \in \{ x_1 ,\cdots , x_k \setminus x_{i_1} , \cdots , x_{i_{k-1}}  \}} \piencodf{\fail^{ \widetilde{x}, \widetilde{y}} }_u \{ z_1 / x_{i_1} \} \cdots \{ z_k / x_{i_k} \} ) \cdots )\\
  & = \res{z_1}( \gsome{z_1}{\llfv{M_{1}}};\piencodf{ M_{1} }_{ {z_1}}  \|  \cdots \res{z_k} ( \gsome{z_k}{\llfv{M_{k}}};\piencodf{ M_{k} }_{ {z_k}} \\
        &  \qquad  \| \bignd_{x_{i_1} \in \{ x_1 ,\cdots , x_k  \}} \cdots \bignd_{x_{i_k} \in \{ x_1 ,\cdots , x_k \setminus x_{i_1} , \cdots , x_{i_{k-1}}  \}}  \pnone{\widetilde{x}} \| \pnone{\widetilde{y}} \| \pnone{u}  \{ z_1 / x_{i_1} \} \cdots \{ z_k / x_{i_k} \}\\
        &  \qquad   ) \cdots )\\
  & \equiv  \res{z_1}( \gsome{z_1}{\llfv{M_{1}}};\piencodf{ M_{1} }_{ {z_1}}  \| \\
        &  \qquad \cdots \res{z_k} ( \gsome{z_k}{\llfv{M_{k}}};\piencodf{ M_{k} }_{ {z_k}}  \|   \pnone{z_1} \| \cdots \| \pnone{z_k} \| \pnone{\widetilde{y}} \| \pnone{u}  ) \cdots )\\
  & \redtwo^k \pnone{\llfv{C}} \| \pnone{\widetilde{y}} \| \pnone{u} =Q',
  \end{aligned}
    \]
and  no further reductions can be performed.
We also have that  ${N} \red \fail^{ \widetilde{y} \cup \llfv{C} } = {N}'$ and $\piencodf{ \fail^{\widetilde{y} \cup \llfv{C}} }_u = Q'$.

                        \item $\headf{M} =  {x}_i$.

                            Then we have the following
 \[
 \begin{aligned}
 \piencodf{{N}}_u
  &=  \res{z_1}( \gsome{z_1}{\llfv{M_{1}}};\piencodf{ M_{1} }_{ {z_1}}  \|  \cdots \res{z_k} ( \gsome{z_k}{\llfv{M_{k}}};\piencodf{ M_{k} }_{ {z_k}} \\
                & \qquad \qquad  \| \bignd_{x_{i_1} \in \{ x_1 ,\cdots , x_k  \}} \cdots \bignd_{x_{i_k} \in \{ x_1 ,\cdots , x_k \setminus x_{i_1} , \cdots , x_{i_{k-1}}  \}} \res{\widetilde{y}} (\piencodf{  {x}_i }_{j} \| P) \{ z_1 / x_{i_1} \} \cdots \{ z_k / x_{i_k} \} ) \cdots )\\
 &=  \res{z_1}( \gsome{z_1}{\llfv{M_{1}}};\piencodf{ M_{1} }_{ {z_1}}  \|  \cdots \res{z_k} ( \gsome{z_k}{\llfv{M_{k}}};\piencodf{ M_{k} }_{ {z_k}} \\
                & \qquad \qquad  \| \bignd_{x_{i_1} \in \{ x_1 ,\cdots , x_k  \}} \cdots \bignd_{x_{i_k} \in \{ x_1 ,\cdots , x_k \setminus x_{i_1} , \cdots , x_{i_{k-1}}  \}} \\
                & \qquad \qquad \qquad \res{\widetilde{y}} ( \psome{x_i}; \pfwd{x_i}{j} \| P) \{ z_1 / x_{i_1} \} \cdots \{ z_k / x_{i_k} \} ) \cdots )\\
 \end{aligned}
 \]

 Notice that multiple reductions can take place along any of the channels $z_l$. Let us consider for simplicity that $i = l = k$.

  \[
 \begin{aligned}
 \piencodf{{N}}_u
 &=  \res{z_1}( \gsome{z_1}{\llfv{M_{1}}};\piencodf{ M_{1} }_{ {z_1}}  \|  \cdots \res{z_k} ( \gsome{z_k}{\llfv{M_{k}}};\piencodf{ M_{k} }_{ {z_k}} \\
                & \qquad  \| \bignd_{x_{i_1} \in \{ x_1 ,\cdots , x_k  \}} \cdots \bignd_{x_{i_k} \in \{ x_1 ,\cdots , x_k \setminus x_{i_1} , \cdots , x_{i_{k-1}}  \}}  \res{\widetilde{y}} ( \psome{x_k}; \pfwd{x_k}{j} \| P) \{ z_1 / x_{i_1} \} \cdots \{ z_k / x_{i_k} \} ) \cdots )\\
\end{aligned}
\]
We show the full process of the reduction in this case for correctness. Applying multiple reductions of the form $\rredtwo{\nu}$ we obtain:

{
\small
\begin{prooftree}
    \hspace*{\dimexpr-\leftmargini-\leftmarginii-\leftmarginiii}
        \AxiomC{$\begin{aligned}
                &\res{z_k} ( \gsome{z_k}{\llfv{M_{k}}};\piencodf{ M_{k} }_{ {z_k}} \\
                & \qquad   \| \bignd_{x_{i_1} \in \{ x_1 ,\cdots , x_k  \}} \cdots \bignd_{x_{i_k} \in \{ x_1 ,\cdots , x_k \setminus x_{i_1} , \cdots , x_{i_{k-1}}  \}} \\
                & \qquad  \res{\widetilde{y}} ( \psome{x_k}; \pfwd{x_k}{j} \| P) \{ z_1 / x_{i_1} \} \cdots \{ z_k / x_{i_k} \} )
                \end{aligned} \redtwo_{z_k} R_{k}
                $}
        \LeftLabel{$ \rredtwo{\nu} $}
        \UnaryInfC{$ \vdots $}
        \noLine
        \UnaryInfC{$
        \begin{aligned}
            &\res{z_2}( \gsome{z_2}{\llfv{M_{2}}};\piencodf{ M_{2} }_{ {z_2}}  \| \\
                & \qquad  \cdots \res{z_k} ( \gsome{z_k}{\llfv{M_{k}}};\piencodf{ M_{k} }_{ {z_k}} \\
                & \qquad   \| \bignd_{x_{i_1} \in \{ x_1 ,\cdots , x_k  \}} \cdots \bignd_{x_{i_k} \in \{ x_1 ,\cdots , x_k \setminus x_{i_1} , \cdots , x_{i_{k-1}}  \}} \\
                & \qquad  \res{\widetilde{y}} ( \psome{x_k}; \pfwd{x_k}{j} \| P) \{ z_1 / x_{i_1} \} \cdots \{ z_k / x_{i_k} \} ) \cdots )\\
        \end{aligned} \redtwo_{z_k} R_1$}
        \LeftLabel{$ \rredtwo{\nu} $}
        \UnaryInfC{$
        \begin{aligned}
            &\res{z_1}( \gsome{z_1}{\llfv{M_{1}}};\piencodf{ M_{1} }_{ {z_1}}  \| \\
                & \quad \cdots \res{z_k} ( \gsome{z_k}{\llfv{M_{k}}};\piencodf{ M_{k} }_{ {z_k}} \\
                & \quad   \| \bignd_{x_{i_1} \in \{ x_1 ,\cdots , x_k  \}} \cdots \bignd_{x_{i_k} \in \{ x_1 ,\cdots , x_k \setminus x_{i_1} , \cdots , x_{i_{k-1}}  \}} \\
                & \quad  \res{\widetilde{y}} ( \psome{x_k}; \pfwd{x_k}{j} \| P) \{ z_1 / x_{i_1} \} \cdots \{ z_k / x_{i_k} \} ) \cdots )\\
        \end{aligned} \redtwo_{z_k} \res{z_1}( \gsome{z_1}{\llfv{M_{1}}};\piencodf{ M_{1} }_{ {z_1}}  \| R_1 )$}
\end{prooftree}
}

where $R_i = \res{z_i}( \gsome{z_i}{\llfv{M_{i}}};\piencodf{ M_{i} }_{ {z_i}}  \| R_{i+1} ) $ for $ i < k$

Hence we wish to show the following reduction:

\[
\begin{aligned}
                &\res{z_k} ( \gsome{z_k}{\llfv{M_{k}}};\piencodf{ M_{k} }_{ {z_k}} \\
                & \qquad   \| \bignd_{x_{i_1} \in \{ x_1 ,\cdots , x_k  \}} \cdots \bignd_{x_{i_k} \in \{ x_1 ,\cdots , x_k \setminus x_{i_1} , \cdots , x_{i_{k-1}}  \}} \\
                & \qquad  \res{\widetilde{y}} ( \psome{x_k}; \pfwd{x_k}{j} \| P) \{ z_1 / x_{i_1} \} \cdots \{ z_k / x_{i_k} \} )
\end{aligned} \redtwo_{z_k} R_{k}
\]

We apply the reduction rule

\begin{prooftree}
        \AxiomC{$ P \piprecong{z_k} P' $}
        \AxiomC{$ Q \piprecong{z_k} Q' $}
        \AxiomC{$ \res{z_k}(P' \| Q') \redtwo_{z_k} R_k $}
        \TrinaryInfC{$ \res{z_k}(P \| Q) \redtwo_{z_k} R_k $}
\end{prooftree}

Where we take $P = \gsome{z_k}{\llfv{M_{k}}};\piencodf{ M_{k} }_{ {z_k}} \piprecong{z_k } \gsome{z_k}{\llfv{M_{k}}};\piencodf{ M_{k} }_{ {z_k}} = P'$
and
Notice that when $ x_{i_k} = x_k$ we have that the substitution $\{ z_k / x_{i_k} \} $ takes place
\[
\begin{aligned}
    Q & = \bignd_{x_{i_1} \in \{ x_1 ,\cdots , x_k  \}} \cdots \bignd_{x_{i_k} \in \{ x_1 ,\cdots , x_k \setminus x_{i_1} , \cdots , x_{i_{k-1}}  \}}  \res{\widetilde{y}} ( \psome{x_k}; \pfwd{x_k}{j} \| P) \{ z_1 / x_{i_1} \} \cdots \{ z_k / x_{i_k} \} \\
    & = \bignd_{x_{i_1} \in \{ x_1 ,\cdots , x_{k-1}  \}} \cdots \bignd_{x_{i_{k-1}} \in \{ x_1 ,\cdots , x_{k-1} \setminus x_{i_1} , \cdots , x_{i_{k-2}}  \}}  \res{\widetilde{y}} ( \psome{z_k}; \pfwd{z_k}{j} \| P) \{ z_1 / x_{i_1} \} \cdots \{ z_{k-1} / x_{i_{k-1}} \} \\
    & \nd \bignd_{x_{i_k} \in \{ x_1 ,\cdots , x_{k-1} \} }\bignd_{x_{i_1} \in \{ x_1 ,\cdots , x_k \setminus x_{i_k} \}} \cdots \bignd_{x_{i_{k-1}} \in \{ x_1 ,\cdots , x_{k-1} \setminus x_{i_1} , \cdots , x_{i_{k-2}} , x_{i_k} \}}  \\
    & \qquad \res{\widetilde{y}} ( \psome{x_k}; \pfwd{x_k}{j} \| P) \{ z_1 / x_{i_1} \} \cdots \{ z_k / x_{i_k} \} \\
    & \piprecong{z_k}  \bignd_{x_{i_1} \in \{ x_1 ,\cdots , x_{k-1}  \}} \cdots \bignd_{x_{i_{k-1}} \in \{ x_1 ,\cdots , x_{k-1} \setminus x_{i_1} , \cdots , x_{i_{k-2}}  \}}  \res{\widetilde{y}} ( \psome{z_k}; \pfwd{z_k}{j} \| P) \{ z_1 / x_{i_1} \} \cdots \{ z_{k-1} / x_{i_{k-1}} \} \\
    & = Q'
\end{aligned}
\]

Hence we wish to show the following reduction
\(
\res{z_k}(\gsome{z_k}{\llfv{M_{k}}};\piencodf{ M_{k} }_{ {z_k}} \| Q' ) \redtwo_{z_k} R_k
\)
We do this via the rule $\rredtwo{\some}$ to obtain

$
R_k = \res{z_k}\Big(\piencodf{ M_{k} }_{ {z_k}} \|
\begin{aligned}
    &\bignd_{x_{i_1} \in \{ x_1 ,\cdots , x_{k-1}  \}} \cdots \bignd_{x_{i_{k-1}} \in \{ x_1 ,\cdots , x_{k-1} \setminus x_{i_1} , \cdots , x_{i_{k-2}}  \}} \\
    & \qquad \res{\widetilde{y}} (  \pfwd{z_k}{j} \| P) \{ z_1 / x_{i_1} \} \cdots \{ z_{k-1} / x_{i_{k-1}} \}
\end{aligned}
\Big)
$

Hence we continue with the following reductions:
\[
 \begin{aligned}
 & \redtwo  \res{z_1}( \gsome{z_1}{\llfv{M_{1}}};\piencodf{ M_{1} }_{ {z_1}}  \| \cdots \res{z_k} ( \piencodf{ M_{k} }_{ {z_k}} \\
                & \qquad \qquad  \| \bignd_{x_{i_1} \in \{ x_1 ,\cdots , x_{k-1}  \}} \cdots \bignd_{x_{i_{k-1}} \in \{ x_1 ,\cdots , x_{k-1} \setminus x_{i_1} , \cdots , x_{i_{k-2}}  \}} \\
                & \qquad \qquad \qquad \res{\widetilde{y}} (  \pfwd{x_k}{j} \| P) \{ z_1 / x_{i_1} \} \cdots \{ z_{k-1} / x_{i_{k-1}} \} ) \cdots ) & = Q_1\\
 & \redtwo \res{z_1}( \gsome{z_1}{\llfv{M_{1}}};\piencodf{ M_{1} }_{ {z_1}}  \| \cdots \res{z_{k-1}} ( \piencodf{ M_{k-1} }_{ {z_{k-1}}} \\
                & \qquad \qquad  \| \bignd_{x_{i_1} \in \{ x_1 ,\cdots , x_{k-1}  \}} \cdots \bignd_{x_{i_{k-1}} \in \{ x_1 ,\cdots , x_{k-1} \setminus x_{i_1} , \cdots , x_{i_{k-2}}  \}} \\
                & \qquad \qquad \qquad \res{\widetilde{y}} (  \piencodf{ M_{k} }_{ {j}} \| P) \{ z_1 / x_{i_1} \} \cdots \{ z_{k-1} / x_{i_{k-1}} \} ) \cdots )  = Q_2\\
 \end{aligned}
 \]

In addition,
\(
 {N} = M \linexsub{C /  \widetilde{x}} \red  M \headlin{ M_k / x_k }\linexsub{C \setminus M_k / \widetilde{x} \setminus x_k}  = {M}'\).
Finally,
\[
\begin{aligned}
\piencodf{{M}'}_u &= \piencodf{M \headlin{ M_k / x_k }\linexsub{C \setminus M_k / \widetilde{x} \setminus x_k}}_u\\
&=  \res{z_1}( \gsome{z_1}{\llfv{M_{1}}};\piencodf{ M_{1} }_{ {z_1}}  \|  \cdots \res{z_{k-1}} ( \piencodf{ M_{k-1} }_{ {z_{k-1}}} \\
                & \qquad \qquad  \| \bignd_{x_{i_1} \in \{ x_1 ,\cdots , x_{k-1}  \}} \cdots \bignd_{x_{i_{k-1}} \in \{ x_1 ,\cdots , x_{k-1} \setminus x_{i_1} , \cdots , x_{i_{k-2}}  \}} \\
                & \qquad \qquad \qquad \res{\widetilde{y}} (  \piencodf{ M_{k} }_{ {j}} \| P) \{ z_1 / x_{i_1} \} \cdots \{ z_{k-1} / x_{i_{k-1}} \} ) \cdots )
     = Q_2
\end{aligned}
\]

\begin{enumerate}
\item When $n = 1$:

Then, $Q = Q_1$ and  $ \piencodf{{N}}_u \redtwo^1 Q_1$. Also,

$Q_1 \redtwo^1 Q_2 = Q'$, ${N} \red^1 M \headlin{ M_k / x_k }\linexsub{C \setminus M_k / \widetilde{x} \setminus x_k} = {N}'$ and\\
 $\piencodf{M \headlin{ M_k / x_k }\linexsub{C \setminus M_k / \widetilde{x} \setminus x_k}}_u = Q_2$.
\item When $n \geq 2$:

Then  $ \piencodf{{N}}_u \redtwo^2 Q_2 \redtwo^l Q$, for $l \geq 0$.  Also,
${N} \red {M}'$, $Q_2 = \piencodf{{M}'}_u$. By the induction hypothesis, there exist $ Q'$ and ${N}'$ such that $ Q \redtwo^i Q'$, ${M}' \red^j {N}'$ and $\piencodf{{N}'}_u = Q'$. Finally, $\piencodf{{N}}_u \redtwo^2 Q_2 \redtwo^l Q \redtwo^i Q'$ and ${N} \rightarrow {M}'  \red^j {N}'$.

                            \end{enumerate}

                    \end{enumerate}
 \item  For $m \geq 1$ and $ n \geq 0$.

            \begin{enumerate}
            \item When $n = 0$.

               Then
               \[
                \begin{aligned}
                   Q& =  \res{z_1}( \gsome{z_1}{\llfv{M_{1}}};\piencodf{ M_{1} }_{ {z_1}}  \| \cdots \res{z_k} ( \gsome{z_k}{\llfv{M_{k}}};\piencodf{ M_{k} }_{ {z_k}} \\
                & \qquad \qquad  \| \bignd_{x_{i_1} \in \{ x_1 ,\cdots , x_k  \}} \cdots \bignd_{x_{i_k} \in \{ x_1 ,\cdots , x_k \setminus x_{i_1} , \cdots , x_{i_{k-1}}  \}} R \{ z_1 / x_{i_1} \} \cdots \{ z_k / x_{i_k} \} ) \cdots )\\
                \end{aligned}
               \]
                and $\piencodf{M}_u \redtwo^m R$ where $m \geq 1$. By the IH there exist $R'$  and ${M}' $ such that $R \redtwo^i R'$, $M \red^j {M}'$ and $\piencodf{{M}'}_u = R'$. Thus,
              \[
               \begin{aligned}
                   \piencodf{{N}}_u & \redtwo^m \res{z_1}( \gsome{z_1}{\llfv{M_{1}}};\piencodf{ M_{1} }_{ {z_1}}  \| \cdots \res{z_k} ( \gsome{z_k}{\llfv{M_{k}}};\piencodf{ M_{k} }_{ {z_k}} \\
                    & \qquad \qquad  \| \bignd_{x_{i_1} \in \{ x_1 ,\cdots , x_k  \}} \cdots \bignd_{x_{i_k} \in \{ x_1 ,\cdots , x_k \setminus x_{i_1} , \cdots , x_{i_{k-1}}  \}} R \{ z_1 / x_{i_1} \} \cdots \{ z_k / x_{i_k} \} ) \cdots ) & = Q\\
                    & \redtwo^i \res{z_1}( \gsome{z_1}{\llfv{M_{1}}};\piencodf{ M_{1} }_{ {z_1}}  \| \cdots \res{z_k} ( \gsome{z_k}{\llfv{M_{k}}};\piencodf{ M_{k} }_{ {z_k}} \\
                    & \qquad \qquad  \| \bignd_{x_{i_1} \in \{ x_1 ,\cdots , x_k  \}} \cdots \bignd_{x_{i_k} \in \{ x_1 ,\cdots , x_k \setminus x_{i_1} , \cdots , x_{i_{k-1}}  \}} R' \{ z_1 / x_{i_1} \} \cdots \{ z_k / x_{i_k} \} ) \cdots ) & = Q' \\
                \end{aligned}
                 \]
                Also, ${N} = M \linexsub{C /  \widetilde{x}} \red^j  M' \linexsub{C /  \widetilde{x}}  = {N}'$ and  $\piencodf{{N}'}_u = Q'$

            \item When $n \geq 1$.
                Then  $R$ has an occurrence of an unguarded $\psome{x_i}$ or $\pnone{x_i}$, this case follows by IH.

                    \end{enumerate}
            \end{enumerate}

            \item  ${N} =  M \unexsub{U / \unvar{x}}$.

            In this case,
            \(
            \begin{aligned}
                \piencodf{M \unexsub{U / \unvar{x}}}_u &=   \res{\unvar{x}} ( \piencodf{ M }_u \|   ~ \guname{\unvar{x}}{x_i}; \piencodf{ U }_{x_i} ).
            \end{aligned}
            \)
            Then,
            \[
            \begin{aligned}
               \piencodf{{N}}_u & =   \res{\unvar{x}} ( \piencodf{ M }_u \|   ~ \guname{\unvar{x}}{x_i}; \piencodf{ U }_{x_i} )  \redtwo^m  \res{\unvar{x}} ( R \|   ~ \guname{\unvar{x}}{x_i}; \piencodf{ U }_{x_i} ) \redtwo^n  Q.
            \end{aligned}
            \]
            for some process $R$. Where $\redtwo^n$ is a reduction initially synchronises with $\guname{\unvar{x}}{x_i}$ when $n \geq 1$, $n + m = k \geq 1$. Type preservation in \clpi ensures reducing $\piencodf{ M}_v \redtwo^m$ doesn't consume possible synchronisations with $ \guname{ \unvar{x} }{ x_i }  $ if they occur. Let us consider the the possible sizes of both $m$ and $n$.

            \begin{enumerate}
                \item For $m = 0$ and $n \geq 1$.

                   In this case,  $R = \piencodf{M}_u$ as $\piencodf{M}_u \red^0 \piencodf{M}_u$.

                    Notice that the only possibility of having an unguarded $ \puname{ \unvar{x} }{ x_i }$ without internal reductions is when   $\headf{M} =  {x}[j]$ for some index $j$.
                            Then we have the following:

                            \[
                            \begin{aligned}
                            \piencodf{{N}}_u
                            & =\res{\unvar{x}} ( \res{\widetilde{y}} ( \piencodf{ x[j] }_k \| P ) \|   ~ \guname{\unvar{x}}{x_i}; \piencodf{ U }_{x_i} )\\
                            & =\res{\unvar{x}} ( \res{\widetilde{y}} ( \puname{\unvar{x}}{{x_i}}; \psel{x_i}{j}; \pfwd{x_i}{k}  \| P ) \|   ~ \guname{\unvar{x}}{x_i}; \piencodf{ U }_{x_i} )\\
                            & \redtwo \res{\unvar{x}} (  \res{x_i}(  \res{\widetilde{y}} ( \psel{x_i}{j}; \pfwd{x_i}{k}  \| P ) \| \piencodf{ U }_{x_i}  ) \|   ~ \guname{\unvar{x}}{x_i}; \piencodf{ U }_{x_i} ) & = Q_1\\
                            & = \res{\unvar{x}} (  \res{x_i}(  \res{\widetilde{y}} ( \psel{x_i}{j}; \pfwd{x_i}{k}  \| P ) \| \gsel{x_i}\{i:\piencodf{ U_i }_{x_i} \}_{U_i \in U}  ) \|   ~ \guname{\unvar{x}}{x_i}; \piencodf{ U }_{x_i} )\\
                            & \redtwo \res{\unvar{x}} (  \res{x_i}(  \res{\widetilde{y}} ( \ \pfwd{x_i}{k}  \| P ) \| \piencodf{ U_j }_{x_i}  ) \|   ~ \guname{\unvar{x}}{x_i}; \piencodf{ U }_{x_i} ) & = Q_2\\
                            & \redtwo \res{\unvar{x}} (    \res{\widetilde{y}} ( \piencodf{ U_j }_{k}   \| P )  \|   ~ \guname{\unvar{x}}{x_i}; \piencodf{ U }_{x_i} ) & = Q_3\\
                            \end{aligned}
                            \]

                We consider the two cases of the form of $U_{j}$ and show that the choice of $U_{j}$ is inconsequential

                \begin{itemize}
                \item When $ U_j = \unvar{\bag{N}}$:

                In this case,
                \(
                \begin{aligned}
                {N} &=M \unexsub{U / \unvar{x}}\red M \headlin{ N /\unvar{x} }\unexsub{U / \unvar{x}} = {M}'.
                \end{aligned}
                \)
                 and
                \[
                 \begin{aligned}
                 \piencodf{{M}'}_u
                 &= \piencodf{M \headlin{ N /\unvar{x} }\unexsub{U / \unvar{x}}}_u=  \res{\unvar{x}} (    \res{\widetilde{y}} ( \piencodf{ N }_{k}   \| P )  \|   ~ \guname{\unvar{x}}{x_i}; \piencodf{ U }_{x_i} )  & = Q_3
                                            \end{aligned}
                 \]

                \item When $ U_i = \unvar{\oneb} $:

                  In this case,
                        \(
                        \begin{aligned}
                            {N} &=M \unexsub{U / \unvar{x}} \red M \headlin{ \fail^{\emptyset} /\unvar{x} } \unexsub{U /\unvar{x} } = {M}'.
                        \end{aligned}
                        \)

                        Notice that $\piencodf{\unvar{\oneb}}_{k} =  \pnone{k}$ and that $\piencodf{\fail^{\emptyset}}_k = \pnone{k} $. In addition,

                            \[
                            \begin{aligned}
                               \piencodf{{M}'}_u= \piencodf{M \headlin{ \fail^{\emptyset} /\unvar{x} } \unexsub{U /\unvar{x} }}_u&=  \res{\unvar{x}} (    \res{\widetilde{y}} ( \piencodf{\fail^{\emptyset}}_{k}   \| P )  \|   ~ \guname{\unvar{x}}{x_i}; \piencodf{ U }_{x_i} )\\
                               &=  \res{\unvar{x}} (    \res{\widetilde{y}} ( \piencodf{\unvar{\oneb}}_{k}   \| P )  \|   ~ \guname{\unvar{x}}{x_i}; \piencodf{ U }_{x_i} ) & = Q_3
                            \end{aligned}
                            \]
                \end{itemize}

                Both choices give an ${M}$ that are equivalent to $Q_3$.

    \begin{enumerate}
    \item When $n \leq 2$.

   In this case, $Q = Q_n$ and  $ \piencodf{{N}}_u \redtwo^n Q_n$.

Also, $Q_n \redtwo^{3-n} Q_3 = Q'$, ${N} \red^1 {M}' = {N}'$ and $\piencodf{{M}' }_u = Q_2$.

     \item When $n \geq 3$.

     We have $ \piencodf{{N}}_u \redtwo^3 Q_3 \redtwo^l Q$ for $l \geq 0$. We also know that ${N} \rightarrow {M}'$, $Q_3 = \piencodf{{M}'}_u$. By the IH, there exist $ Q$ and ${N}'$ such that $Q \redtwo^i Q'$, ${M}' \red^j {N}'$ and $\piencodf{{N}'}_u = Q'$. Finally, $\piencodf{{N}}_u \redtwo^3 Q_3 \redtwo^l Q \redtwo^i Q'$ and ${N} \rightarrow {M}'  \red^j {N}' $.

                    \end{enumerate}
 \item For $m \geq 1$ and $ n \geq 0$.

            \begin{enumerate}
            \item When $n = 0$.

               Then $ \res{\unvar{x}} ( R \|   ~ \guname{\unvar{x}}{x_i}; \piencodf{ U }_{x_i} )  = Q$ and $\piencodf{M}_u \redtwo^m R$ where $m \geq 1$. By the IH there exist $R'$  and ${M}' $ such that $R \redtwo^i R'$, $M \red^j {M}'$ and $\piencodf{{M}'}_u = R'$.
              Hence,
              \[
               \begin{aligned}
                   \piencodf{{N}}_u & =   \res{\unvar{x}} ( \piencodf{M}_u \|   ~ \guname{\unvar{x}}{x_i}; \piencodf{ U }_{x_i} )\redtwo^m   \res{\unvar{x}} ( R \|   ~ \guname{\unvar{x}}{x_i}; \piencodf{ U }_{x_i} )
                   & = Q.
                \end{aligned}
                 \]
                In addition,
                \(
                   Q  \redtwo^i  \res{\unvar{x}} ( R' \|   ~ \guname{\unvar{x}}{x_i}; \piencodf{ U }_{x_i} ) = Q \), and the term can reduce as follows: $ {N} = M \unexsub{U / \unvar{x}} \red^j  M' \unexsub{U / \unvar{x}} = {N}'$ and  $\piencodf{{N}'}_u = Q'$.

            \item When $n \geq 1$.

            Then $R$ has an occurrence of an unguarded $\guname{\unvar{x}}{x_i}$, and the case follows by IH.
            \qedhere
\end{enumerate}
            \end{enumerate}
               \end{enumerate}
\end{proof}

\subsection{Success Sensitivity}\label{a:tsuccess}

    \begin{proposition}[Preservation of Success]
        \label{Prop:checkprespiunres}
        For all closed $M\in \lamcoldetsh$, the following hold:
        \begin{enumerate}
            \item $ \headf{M} = \checkmark \implies \piencodf{M} =  \res{ \widetilde{x} }  (P   \| \checkmark) $
            \item $ \piencodf{M}_u =  P   \| \checkmark \nd Q \implies \headf{M} = \checkmark$
        \end{enumerate}

        \end{proposition}

        \begin{proof}

        Proof of both cases by induction on the structure of $M$.

        \begin{enumerate}
        \item We only need to consider terms of the following form:

            \begin{enumerate}

                \item  $ M = \checkmark $:

                This case is immediate.

                \item $M = N\ (C \bagsep U)$:

                Then, $\headf{N \ (C \bagsep U)} = \headf{N}$. If $\headf{N} = \checkmark$, then  $$ \piencodf{M (C \bagsep U)}_u =   \res{ v }  (\piencodf{M}_v   \|  \gsome{ v }{ u , \llfv{C} }; \pname{v}{x} . (\pfwd{v}{u}   \| \piencodf{C \bagsep U}_x ) ).$$
                By the IH,  $\checkmark$ is unguarded in $\piencodf{N}_u$.

                \item $M = M' \linexsub{C /  x_1 , \cdots , x_k}$

                Then we have that $\headf{M' \linexsub{C /  x_1 , \cdots , x_k}}  = \headf{M'} = \checkmark$. Then

                \[
                \begin{aligned}
                    \piencodf{ M' \linexsub{C /  x_1 , \cdots , x_k}  }_u    & =
                    \res{z_1}( \gsome{z_1}{\llfv{M_{1}}};\piencodf{ M_{1} }_{ {z_1}}  \| \cdots \res{z_k} ( \gsome{z_k}{\llfv{M_{k}}};\piencodf{ M_{k} }_{ {z_k}} \\
                    & \qquad  \| \bignd_{x_{i_1} \in \{ x_1 ,\cdots , x_k  \}} \cdots \bignd_{x_{i_k} \in \{ x_1 ,\cdots , x_k \setminus x_{i_1} , \cdots , x_{i_{k-1}}  \}} \piencodf{ M' }_u \{ z_1 / x_{i_1} \} \cdots \{ z_k / x_{i_k} \} ) \cdots )
                \end{aligned}
                \]

                and by the IH $\checkmark$ is unguarded in $\piencodf{M'}_u$ hence ungaurded in every summand in $\piencodf{ M \linexsub{C /  x_1 , \cdots , x_k}  }_u$.

                \item $M = M' \unexsub{U / \unvar{x}}$

                Then we have that $\headf{M' \unexsub{U / \unvar{x}}}  = \headf{M'} = \checkmark$. Then $\piencodf{ M' \unexsub{U / \unvar{x}}  }_u   =    \res{ \unvar{x} }  ( \piencodf{ M' }_u   \|    \guname{ \unvar{x} }{ x_i } ; \piencodf{ U }_{x_i} ) $ and by the IH  $\checkmark$ is unguarded in $\piencodf{M'}_u$.

            \end{enumerate}

           \item We only need to consider terms of the following form:

            \begin{enumerate}

                \item {\bf Case $M = \checkmark$:}

                Then,
                $\piencodf{\checkmark}_u = \checkmark$
                which is an unguarded occurrence of $\checkmark$ and that $\headf{\checkmark} = \checkmark$.

                \item {\bf Case $M = N (C \bagsep U)$:}

                Then, $\piencodf{N (C \bagsep U)}_u =  \res{ v }  (\piencodf{N}_v   \| \gsome{ v }{ u , \llfv{C} };   \pname{v}{x} . (\pfwd{v}{u}   \| \piencodf{C \bagsep U}_x ) )$. The only occurrence of an unguarded $\checkmark$ can occur is within $\piencodf{N}_v$. By the IH, $\headf{N} = \checkmark$ and finally $\headf{N \ (C \bagsep U)} = \headf{N}$.

                \item {\bf Case $M = M' \linexsub{C /  x_1 , \cdots , x_k}$:}

                Then:
                \[
                \begin{aligned}
                    \piencodf{ M' \linexsub{C /  x_1 , \cdots , x_k}  }_u    & =
                    \res{z_1}( \gsome{z_1}{\llfv{M_{1}}};\piencodf{ M_{1} }_{ {z_1}}  \|  \cdots \res{z_k} ( \gsome{z_k}{\llfv{M_{k}}};\piencodf{ M_{k} }_{ {z_k}} \\
                & \quad \qquad  \| \bignd_{x_{i_1} \in \{ x_1 ,\cdots , x_k  \}} \cdots \bignd_{x_{i_k} \in \{ x_1 ,\cdots , x_k \setminus x_{i_1} , \cdots , x_{i_{k-1}}  \}} \piencodf{ M' }_u \{ z_1 / x_{i_1} \} \cdots \{ z_k / x_{i_k} \} ) \cdots )
                \end{aligned}
                \]
                 An unguarded occurrence of $\checkmark$ can only occur within $\piencodf{ M' }_u $. By the IH,  $\headf{M'} = \checkmark$ and hence $\headf{ M' \linexsub{C /  x_1 , \cdots , x_k}}  = \headf{M'}$.

                \item {\bf Case $M = M' \unexsub{U / \unvar{x}}$:}
                This case is analogous to the previous.
                \qedhere
            \end{enumerate}
        \end{enumerate}
        \end{proof}

\begin{restatable}[Success Sensitivity (Under $\redtwo$)]{theorem}{thmEncLazySucc}\label{proof:successsenscetwo}
    $\succp{{M}}{\checkmark}$ if and only if ${\piencodf{{M}}_u}\succtwo_{\checkmark}$ for well-formed closed terms $M$.
\end{restatable}

\begin{proof}
We proceed with the proof in two parts.

\begin{enumerate}

    \item Suppose that  ${M} \Downarrow_{\checkmark} $. We will prove that $\piencodf{{M}} \succtwo_{\checkmark}$.

    By \defref{def:app_Suc3unres}, there exists  $ {M}'$ such that $ M \red^* {M}'$ and
    $\headf{M'} = \checkmark$.
    By completeness, if $ M \red M'$ then there exist $Q, Q'$ such that $\piencodf{M}_u \equiv Q \redtwo^* Q'$ and $\piencodf{M'}_u \equiv Q' $.

    We wish to show that there exists $Q''$such that $Q' \redtwo^* Q''$ and $Q''$ has an unguarded occurrence of $\checkmark$.

    By Proposition \ref{Prop:checkprespiunres} (1) we have that $\headf{M'} = \checkmark \implies \piencodf{M'}_u =   \res{ \widetilde{x} } (P   \| \checkmark)$. Finally $\piencodf{M'}_u  =   \res{ \widetilde{x} } (P   \| \checkmark) \equiv Q'$. Hence $Q$ reduces to a process that has an unguarded occurrence of $\checkmark$.

    \item Suppose that $\piencodf{{M}}_u \succtwo_{\checkmark}$. We will prove that $ {M} \Downarrow_{\checkmark}$.

    By operational soundness we have that if  $\piencodf{{M}}_u \redtwo^* Q$ then there exist ${M}'$ and $Q'$ such that:
    (i)~${M} \red^* {M}'$
    and
    (ii)~$Q \redtwo^* Q' $ with
    $ \piencodf{{M}'}_u \equiv Q'$.

   Since $\piencodf{{M}}_u \redtwo^*  Q$, and $Q= Q'' \| \checkmark$ and $Q \redtwo^* Q'$ we must have that $Q'' \redtwo^* Q''' $ with $ Q' = Q''' \| \checkmark $. As $ \piencodf{{M}'}_u \equiv Q'$ we have that $ \piencodf{{M}'}_u  = $. Finally applying Proposition \ref{Prop:checkprespiunres} (2) we have that $\piencodf{{M}'}_u  = Q''' \| \checkmark$ is itself a term with unguarded $\checkmark$, then ${M}$ is itself headed with $\checkmark$.
   \qedhere

\end{enumerate}
\end{proof}

%% file: appendix/eager-encod.tex
\section{Proof of Loose Correctness of the Translation under the Eager Semantics}\label{a:secloose}

\subsection{Completeness}\label{a:looscompleteness}

\thmEncLWCompl*

\begin{proof}
By induction on the reduction rule applied to infer $N\red M$.
We have five cases.

 \begin{enumerate}
        \item  Case $\redlab{RS:Beta}$:
        Then  $ N= (\lambda x . (M'[ {\widetilde{x}} \leftarrow  {x}])) B  \red (M'[ {\widetilde{x}} \leftarrow  {x}])\esubst{ B }{ x }  = M$ , where $B = C \bagsep U$. The result folows easily, since

\begin{equation}\label{eq:compl_lsbeta1failunres}
\begin{aligned}
\piencodf{N}_u &=    \res{ v }  (\piencodf{\lambda x . (M'[ {\widetilde{x}} \leftarrow  {x}])}_v   \|  \gsome{ v }{ u , \llfv{C} };  \pname{v}{x} . (\pfwd{v}{u}   \| \piencodf{C \bagsep U}_x ) )  \\
&=   \res{ v }  (\psome{v}; \gname{v}{x}; \psome{x}; \gname{x}{\linvar{x}}; \gname{x}{\unvar{x}};\gclose{ x } ; \piencodf{M'[ {\widetilde{x}} \leftarrow  {x}]}_v   \|  \\
& \qquad \gsome{ v }{ u , \llfv{C} };  \pname{v}{x} . (\pfwd{v}{u}   \| \piencodf{C \bagsep U}_x ) )  \\
& \redone   \res{ v }  ( \gname{v}{x}; \psome{x}; \gname{x}{\linvar{x}}; \gname{x}{\unvar{x}};\gclose{ x } ; \piencodf{M'[ {\widetilde{x}} \leftarrow  {x}]}_v   \|  \pname{v}{x} . (\pfwd{v}{u}   \| \piencodf{C \bagsep U}_x ) )  \\
& \redone   \res{ x }  ( \res{v}( \psome{x}; \gname{x}{\linvar{x}}; \gname{x}{\unvar{x}}; \gclose{ x } ; \piencodf{M'[ {\widetilde{x}} \leftarrow  {x}]}_v   \|  \pfwd{v}{u} )  \| \piencodf{C \bagsep U}_x  )  \\
& \redone   \res{ x }  (  \psome{x}; \gname{x}{\linvar{x}}; \gname{x}{\unvar{x}};  \gclose{ x } ; \piencodf{M'[ {\widetilde{x}} \leftarrow  {x}]}_u   \|  \piencodf{C \bagsep U}_x  )  =   \piencodf{M}_u\\
\end{aligned}
\end{equation}

\item  Case $ \redlab{RS:Ex \dash Sub}$: Then $ N =M'[ {x}_1, \!\cdots\! ,  {x}_k \leftarrow  {x}]\esubst{ C \bagsep U }{ x }$, with $C = \bag{M_1}
\cdots  \bag{M_k}$, $k\geq 0$ and $M'\not= \fail^{\widetilde{y}}$.

\begin{equation}\label{eq:compl_lsbeta1failunres2}
    \begin{aligned}
    \piencodf{N}_u &=    \res{ v }  (\piencodf{\lambda x . (M'[ {\widetilde{x}} \leftarrow  {x}])}_v   \|  \gsome{ v }{ u , \llfv{C} };  \pname{v}{x} . (\pfwd{v}{u}   \| \piencodf{C \bagsep U}_x ) )  \\
    &=   \res{ v }  (\psome{v}; \gname{v}{x}; \psome{x}; \gname{x}{\linvar{x}}; \gname{x}{\unvar{x}};\gclose{ x } ; \piencodf{M'[ {\widetilde{x}} \leftarrow  {x}]}_v   \|  \\
    & \qquad \gsome{ v }{ u , \llfv{C} };  \pname{v}{x} . (\pfwd{v}{u}   \| \piencodf{C \bagsep U}_x ) )  \\
    & \redone   \res{ v }  ( \gname{v}{x}; \psome{x}; \gname{x}{\linvar{x}}; \gname{x}{\unvar{x}};\gclose{ x } ; \piencodf{M'[ {\widetilde{x}} \leftarrow  {x}]}_v   \|  \pname{v}{x} . (\pfwd{v}{u}   \| \piencodf{C \bagsep U}_x ) )  \\
    & \redone   \res{ x }  ( \res{v}( \psome{x}; \gname{x}{\linvar{x}}; \gname{x}{\unvar{x}}; \gclose{ x } ; \piencodf{M'[ {\widetilde{x}} \leftarrow  {x}]}_v   \|  \pfwd{v}{u} )  \| \piencodf{C \bagsep U}_x  )  \\
    & \redone   \res{ x }  (  \psome{x}; \gname{x}{\linvar{x}}; \gname{x}{\unvar{x}};  \gclose{ x } ; \piencodf{M'[ {\widetilde{x}} \leftarrow  {x}]}_u   \|  \piencodf{C \bagsep U}_x  )  =   \piencodf{M}_u\\
    \end{aligned}
    \end{equation}

\item  Case $ \redlab{RS:Ex \dash Sub}$: Then $ N =M'[ {x}_1, \!\cdots\! ,  {x}_k \leftarrow  {x}]\esubst{ C \bagsep U }{ x }$, with $C = \bag{M_1}
    \cdots  \bag{M_k}$, $k\geq 0$ and $M'\not= \fail^{\widetilde{y}}$.

    The reduction is $N = M'[ {x}_1, \!\cdots\! ,  {x}_k \leftarrow  {x}]\esubst{ C \bagsep U }{ x } \red  M' \linexsub{C  /  x_1 , \cdots , x_k} \unexsub{U / \unvar{x} } = M.$

    We detail the translations of $\piencodf{N}_u$ and $\piencodf{M}_u$. To simplify the proof, we  consider $k=2$ (the case in which $k> 2$ is follows analogously. Similarly the case of $k =0,1$ it contained within $k = 2$). On the one hand, we have:

    \begin{equation}\label{eq:compl_lsbeta3failunres0}
    \begin{aligned}
    \piencodf{N}_u &= \piencodf{M'[ {x}_1 \leftarrow  {x}]\esubst{ C \bagsep U }{ x }}_u\\
    &= \res{ x }  ( \psome{x}; \gname{x}{\linvar{x}}; \gname{x}{\unvar{x}}; \gclose{ x } ; \piencodf{ M'[ x_1, x_2 \leftarrow  {x}]}_u   \| \piencodf{ C \bagsep U}_x )   \\
    &= \res{ x }  ( \psome{x}; \gname{x}{\linvar{x}}; \gname{x}{\unvar{x}};  \gclose{ x } ;\piencodf{ M'[ x_1, x_2 \leftarrow  {x}]}_u   \|  \gsome{ x }{ \llfv{C} };  \pname{x}{\linvar{x}}.\big( \piencodf{ C }_{\linvar{x}}\\
    &  \qquad \qquad   \| \pname{x}{\unvar{x}} .(  \guname{ \unvar{x} }{ x_i } ;  \piencodf{ U }_{x_i}   \|  \pclose{ x } ) \big) )   \qquad (:= P_{\mathbb{N}}) \\[4pt]
    \end{aligned}
    \end{equation}

The reduction is $N = M'[ {x}_1, \!\cdots\! ,  {x}_k \leftarrow  {x}]\esubst{ C \bagsep U }{ x } \red  M' \linexsub{C  /  x_1 , \cdots , x_k} \unexsub{U / \unvar{x} } = M.$

We detail the translations of $\piencodf{N}_u$ and $\piencodf{M}_u$. To simplify the proof, we  consider $k=2$ (the case in which $k> 2$ is follows analogously. Similarly the case of $k =0,1$ it contained within $k = 2$). On the one hand, we have:

\begin{equation}\label{eq:compl_lsbeta3failunres}
\begin{aligned}
\piencodf{N}_u &= \piencodf{M'[ {x}_1 \leftarrow  {x}]\esubst{ C \bagsep U }{ x }}_u\\
&= \res{ x }  ( \psome{x}; \gname{x}{\linvar{x}}; \gname{x}{\unvar{x}}; \gclose{ x } ; \piencodf{ M'[ x_1, x_2 \leftarrow  {x}]}_u   \| \piencodf{ C \bagsep U}_x )   \\
&= \res{ x }  ( \psome{x}; \gname{x}{\linvar{x}}; \gname{x}{\unvar{x}};  \gclose{ x } ;\piencodf{ M'[ x_1, x_2 \leftarrow  {x}]}_u   \|  \gsome{ x }{ \llfv{C} };  \pname{x}{\linvar{x}}.\big( \piencodf{ C }_{\linvar{x}}\\
&  \qquad \qquad   \| \pname{x}{\unvar{x}} .(  \guname{ \unvar{x} }{ x_i } ;  \piencodf{ U }_{x_i}   \|  \pclose{ x } ) \big) )   \qquad (:= P_{\mathbb{N}}) \\[4pt]
\end{aligned}
\end{equation}

        \begin{equation*}
        \begin{aligned}
            P_{\mathbb{N}}& \redone^*  \res{  \unvar{x} } ( \res{\linvar{x}} (  \piencodf{ M'[ x_1, x_2 \leftarrow  {x}]}_u   \|  \piencodf{\bag{M_1} \cdot \bag{M_2} }_{\linvar{x}}  ) \|   \guname{ \unvar{x} }{ x_i } ;  \piencodf{ U }_{x_i}   ) \\
            & =  \res{  \unvar{x} } (
            \res{\linvar{x}} (
            \psome{\linvar{x}}; \pname{\linvar{x}}{y_1}; \big( \gsome{y_1}{ \emptyset }; \gclose{ y_{1} } ; \0   \\
            & \qquad \qquad \| \psome{\linvar{x}}; \gsome{\linvar{x}}{u, \llfv{M'} \setminus  \widetilde{x} }; \bignd_{x_{i_1} \in x_1, x_2} \gname{x}{{x}_{i_1}};
            \psome{\linvar{x}}; \pname{\linvar{x}}{y_2}; \big( \gsome{y_2}{ \emptyset }; \gclose{ y_{2} } ; \0   \\
            & \qquad \qquad \| \psome{\linvar{x}}; \gsome{\linvar{x}}{u, \llfv{M'} \setminus  \widetilde{x} }; \bignd_{x_{i_2} \in (x_1, x_2 \setminus x_{i_1} )} \gname{x}{{x}_{i_2}};\\
            & \qquad \qquad \psome{\linvar{x}}; \pname{\linvar{x}}{y_3}; ( \gsome{y_3}{ u, \llfv{M'} }; \gclose{ y_{3} } ;\piencodf{M'}_u \| \pnone{ \linvar{x} } )
            \big)
            \big)
            \\
            & \qquad \qquad\| \\
            & \qquad \qquad
            \gsome{\linvar{x}}{\llfv{C} }; \gname{x}{y_1}; \gsome{\linvar{x}}{y_1, \llfv{C}}; \psome{\linvar{x}}; \pname{\linvar{x}}{z_1}; \\
            & \qquad \qquad   ( \gsome{z_1}{\llfv{M_j}};  \piencodf{M_j}_{z_1} \|  \pnone{y_1} \| \\
            & \qquad \qquad \gsome{\linvar{x}}{\llfv{C} }; \gname{x}{y_2}; \gsome{\linvar{x}}{y_2, \llfv{C}}; \psome{\linvar{x}}; \pname{\linvar{x}}{z_2}; \\
            & \qquad \qquad   ( \gsome{z_2}{\llfv{M_j}};  \piencodf{M_j}_{z_2} \| \pnone{y_2} \| \\
            & \qquad \qquad \gsome{\linvar{x}}{\emptyset};\gname{x}{y_3};  ( \psome{ y_3}; \pclose{y_3}  \| \gsome{\linvar{x}}{\emptyset}; \pnone{\linvar{x}} )
            )
            )
            )\|   \guname{ \unvar{x} }{ x_i } ;  \piencodf{ U }_{x_i}   ) \\
            & \redone^*  \res{  \unvar{x} }  (
            \res{z_1}( \gsome{z_1}{\llfv{M_{1}}};\piencodf{ M_{1} }_{ {z_1}}  \|  \res{z_2} ( \gsome{z_2}{\llfv{M_{2}}};\piencodf{ M_{2} }_{ {z_2}} \| \\
            & \qquad \qquad  \piencodf{ M'}_u \{ z_1 / x_{i_1} \} \{ z_2 / x_{i_2} \} )  )
            \|  \guname{ \unvar{x} }{ x_i } ;  \piencodf{ U }_{x_i} )  \\
        \end{aligned}
        \end{equation*}

        Where $x_{i_1}, x_{i_2}$ is an arbitrary permutation of $x_1,x_2$. On the other hand, we have:

        \begin{equation}\label{eq:compl_lsbeta4failunres}
        \begin{aligned}
            \piencodf{M}_u &= \piencodf{ M' \linexsub{M_1/ {x_1}} \unexsub{U /\unvar{x} } }_u \\
            &=   \res{  \unvar{x} }  (
            \res{z_1}( \gsome{z_1}{\llfv{M_{1}}};\piencodf{ M_{1} }_{ {z_1}}  \| \res{z_2} ( \gsome{z_2}{\llfv{M_{2}}};\piencodf{ M_{2} }_{ {z_2}} \\
        & \qquad \qquad  \| \bignd_{x_{i_1} \in \{ x_1 , x_2  \}}  \bignd_{x_{i_2} \in \{ x_1 , x_2 \} \setminus \{ x_{i_1}  \}} \piencodf{ M' }_u \{ z_1 / x_{i_1} \} \{ z_2 / x_{i_2} \} )  )
            \|  \guname{ \unvar{x} }{ x_i } ;  \piencodf{ U }_{x_i} )  \\
            & \premat   \res{  \unvar{x} }  (
            \res{z_1}( \gsome{z_1}{\llfv{M_{1}}};\piencodf{ M_{1} }_{ {z_1}}  \| \\
            &\qquad \qquad   \res{z_2} ( \gsome{z_2}{\llfv{M_{2}}};\piencodf{ M_{2} }_{ {z_2}}           \|  \piencodf{ M' }_u \{ z_1 / x_{i_1} \} \{ z_2 / x_{i_2} \} )  )
            \|  \guname{ \unvar{x} }{ x_i } ;  \piencodf{ U }_{x_i} )
        \end{aligned}
        \end{equation}
        Therefore, by \eqref{eq:compl_lsbeta3failunres}
        and  \eqref{eq:compl_lsbeta4failunres} the result follows.

        \item Case $\redlab{RS{:}Fetch^{\ell}}$:
        Then, we have
        $N = M' \linexsub{C /  x_1 , \cdots , x_k} $ with $\headf{M'} =  {x}_j$, $C = M_1 , \cdots , M_k$ and $N \red  M' \headlin{ M_i / x_j }  \linexsub{(C \setminus M_i ) /  x_1 , \cdots , x_k \setminus x_j }  = M$, for some $M_i \in C$.
        On the one hand, we have:
            {
            \small
            \begin{equation}\label{eq:compl_lsbeta5failunres0}
            \begin{aligned}
            \piencodf{N}_u &= \res{z_1}( \gsome{z_1}{\llfv{M_{1}}};\piencodf{ M_{1} }_{ {z_1}}  \|
              \cdots \res{z_k} ( \gsome{z_k}{\llfv{M_{k}}};\piencodf{ M_{k} }_{ {z_k}} \\
                & \qquad \qquad  \| \bignd_{x_{i_1} \in \{ x_1 ,\cdots , x_k  \}} \cdots \bignd_{x_{i_k} \in \{ x_1 ,\cdots , x_k \setminus x_{i_1} , \cdots , x_{i_{k-1}}  \}} \piencodf{ M' }_u \{ z_1 / x_{i_1} \} \cdots \{ z_k / x_{i_k} \} ) \cdots )
               \\
            &= \res{z_1}( \gsome{z_1}{\llfv{M_{1}}};\piencodf{ M_{1} }_{ {z_1}}  \|
              \cdots \res{z_k} ( \gsome{z_k}{\llfv{M_{k}}};\piencodf{ M_{k} }_{ {z_k}} \\
                & \qquad \qquad  \| \bignd_{x_{i_1} \in \{ x_1 ,\cdots , x_k  \}} \cdots \bignd_{x_{i_k} \in \{ x_1 ,\cdots , x_k \setminus x_{i_1} , \cdots , x_{i_{k-1}}  \}} \res{ \widetilde{y} }  (\piencodf{  {x}_{j} }_v   \| P) \{ z_1 / x_{i_1} \} \cdots \{ z_k / x_{i_k} \} ) \cdots ) \quad (*)
               \\
            &= \res{z_1}( \gsome{z_1}{\llfv{M_{1}}};\piencodf{ M_{1} }_{ {z_1}}  \|
             \cdots \res{z_k} ( \gsome{z_k}{\llfv{M_{k}}};\piencodf{ M_{k} }_{ {z_k}} \\
                & \qquad \qquad  \| \bignd_{x_{i_1} \in \{ x_1 ,\cdots , x_k  \}} \cdots \bignd_{x_{i_k} \in \{ x_1 ,\cdots , x_k \setminus x_{i_1} , \cdots , x_{i_{k-1}}  \}} \\
                & \qquad \qquad \res{ \widetilde{y} } (\psome{x_j};  \pfwd{x_j}{v}   \| P)  \{ z_1 / x_{i_1} \} \cdots \{ z_k / x_{i_k} \} ) \cdots )
               \\
            \end{aligned}
            \end{equation}
            }
        where $(*)$ is inferred via Proposition~\ref{prop:correctformfailunres}.
        Let us consider the case when $x_{i_k} = x_j$ the other cases proceed similarly. Then, we have  reduction:

        \begin{equation}\label{eq:compl_lsbeta5failunres}
            \begin{aligned}
            &\redone \res{z_1}( \gsome{z_1}{\llfv{M_{1}}};\piencodf{ M_{1} }_{ {z_1}}  \| \\
            & \qquad \cdots \res{z_k} ( \piencodf{ M_{k} }_{ {z_k}}   \|  \res{ \widetilde{y} } (  \pfwd{x_k}{v}   \| P)  \{ z_1 / x_{i_1} \} \cdots \{ z_{k-1} / x_{i_{k-1}} \} ) \cdots )
               \\
            &\redone \res{z_1}( \gsome{z_1}{\llfv{M_{1}}};\piencodf{ M_{1} }_{ {z_1}}  \| \cdots \res{z_{k-1}}( \gsome{z_{k-1}}{\llfv{M_{{k-1}}}};\piencodf{ M_{{k-1}} }_{ {z_{k-1}}}  \| \\
            & \qquad \res{ \widetilde{y} } (  \piencodf{ M_{k} }_{ {v}}   \| P)  \{ z_1 / x_{i_1} \} \cdots \{ z_{k-1} / x_{i_{k-1}} \} ) \cdots )
               \\
            \end{aligned}
            \end{equation}

        for some permutation $ x_{i_1}, \cdots , x_{i_{k-1}} $ of the bag $x_1, \cdots , x_{k-1}$.
         On the other hand, we have:
        \begin{equation}\label{eq:compl_lsbeta6failunres}
        \begin{aligned}
        \piencodf{M}_u &= \piencodf{M' \headlin{ M_k / x_k }  \linexsub{(C \setminus M_k ) /  x_1 , \cdots , x_{k-1} } }_u \\
        & = \res{z_1}( \gsome{z_1}{\llfv{M_{1}}};\piencodf{ M_{1} }_{ {z_1}}  \| \cdots \res{z_{k-1}} ( \gsome{z_{k-1}}{\llfv{M_{k-1}}};\piencodf{ M_{k-1} }_{ {z_{k-1}}} \\
                & \qquad \qquad  \| \bignd_{x_{i_1} \in \{ x_1 ,\cdots , x_{k-1}  \}} \cdots \bignd_{x_{i_{k-1}} \in \{ x_1 ,\cdots , x_{k-1} \setminus x_{i_1} , \cdots , x_{i_{k-2}}  \}} \\
                & \qquad \qquad \res{ \widetilde{y} } (  \piencodf{ M_{k} }_{ {v}}   \| P)  \{ z_1 / x_{i_1} \} \cdots \{ z_{k-1} / x_{i_{k-1}} \} ) \cdots )
               \\
        & \premat  \res{z_1}( \gsome{z_1}{\llfv{M_{1}}};\piencodf{ M_{1} }_{ {z_1}}  \|  \cdots \res{z_{k-1}} ( \gsome{z_{k-1}}{\llfv{M_{k-1}}};\piencodf{ M_{k-1} }_{ {z_{k-1}}} \\
                & \qquad \qquad  \| \res{ \widetilde{y} } (  \piencodf{ M_{k} }_{ {v}}   \| P)  \{ z_1 / x_{i_1} \} \cdots \{ z_{k-1} / x_{i_{k-1}} \} ) \cdots )
               \\
        \end{aligned}
        \end{equation}
                Therefore, by \eqref{eq:compl_lsbeta5failunres}
        and  \eqref{eq:compl_lsbeta6failunres} the result follows.

        \item Case $ \redlab{RS{:} Fetch^!}$:
        Then we have
        $N = M' \unexsub{U / \unvar{x}}$ with $\headf{M'} = \unvar{x}[k]$, $U_k = \unvar{\bag{N}}$ and $N \red  M' \headlin{ N /\unvar{x} }\unexsub{U / \unvar{x}} = M$.
    The result follows easily from
            \begin{equation}\label{eq:compl_lsbeta5failunresunres}
            \begin{aligned}
            \piencodf{N}_u &= \piencodf{M' \unexsub{U / \unvar{x}}}_u
            =  \res{ \unvar{x} } ( \piencodf{ M' }_u   \|    \guname{ \unvar{x} }{ x_k } ;\piencodf{ U }_{x_k}  ) \\
            & \redone^* \res{ \unvar{x} } (  \res{ \widetilde{y} }  (\piencodf{ \unvar{x}[k] }_{j}   \| P)   \|   \guname{ \unvar{x} }{ x_k } ;  \piencodf{ U }_{x_k}  ) \qquad (*)
            \\
            &  =  \res{ \unvar{x} }   (  \res{ \widetilde{y} }  (\puname{ \unvar{x} }{ x_k }; \psel{ {x}_k }{ k }; \pfwd{x_k}{j}   \| P)   \|   \guname{ \unvar{x} }{ x_k } ;  \piencodf{ U }_{x_k}  )
            \\
            & \redone  \res{ \unvar{x} }  (  \res{ \widetilde{y} }  (   \res{ x_k } ( \psel{ {x}_k }{ k }; \pfwd{x_k}{j}   \| \piencodf{ U }_{x_k})   \| P)   \|   \guname{ \unvar{x} }{ x_k } ;  \piencodf{ U }_{x_k}  )
            \\
            & =  \res{ \unvar{x} }  (  \res{ \widetilde{y} } (   \res{ x_k } ( \psel{ {x}_k }{ k }; \pfwd{x_k}{j}   \| \gsel{x_k}\{i:\piencodf{ U_i }_{x} \}_{U_i \in U} )   \| P)   \|   \guname{ \unvar{x} }{ x_k } ;  \piencodf{ U }_{x_k}  )
            \\
            & \redone  \res{ \unvar{x} }   (  \res{ \widetilde{y} }  (  \piencodf{\unvar{\bag{N}}}_{j} )   \| P)   \|   \guname{ \unvar{x} }{ x_k } ;  \piencodf{ U }_{x_k}  )\\
            &=  \res{ \unvar{x} }   (  \res{ \widetilde{y} } (  \piencodf{N}_{j} )   \| P)   \|   \guname{ \unvar{x} }{ x_k } ;  \piencodf{ U }_{x_k}  ) = \piencodf{M}_u
            \end{aligned}
            \end{equation}
       where the reductions denoted by $(*)$ are inferred via Proposition~\ref{prop:correctformfailunres}.

        \item Cases $\redlab{RS:TCont}$ :
        This case follows by IH.

        \item Case $\redlab{RS{:}Fail^{\ell}}$:
        Then we have
        $N = M'[ {x}_1, \!\cdots\! ,  {x}_k \leftarrow  {x}]\ \esubst{C \bagsep U}{ x } $ with $k \neq \size{C}$ and
        $N \red  \fail^{\widetilde{y}} = M$, where $\widetilde{y} = (\llfv{M'} \setminus \{   {x}_1, \cdots ,  {x}_k \} ) \cup \llfv{C}$. Let $ \size{C} = l$ and we assume that $k > l$ and we proceed similarly for $k > l$. Hence $k = l + m$ for some $m \geq 1$

        {
        \small
        \begin{equation}\label{eq:compl_fail1-failunres}
        \begin{aligned}
            \piencodf{N}_u &= \piencodf{M'[ {x}_1, \!\cdots\! ,  {x}_k \leftarrow  {x}]\ \esubst{C \bagsep U}{ x } }_u
            =    \res{ x }  ( \psome{x}; \gname{x}{\linvar{x}}; \gname{x}{\unvar{x}};  \gclose{ x } ;\piencodf{ M'[\widetilde{x} \leftarrow  {x}]}_u   \| \piencodf{ C \bagsep U}_x )
            \\
            & =   \res{  x }  ( \psome{x}; \gname{x}{\linvar{x}}; \gname{x}{\unvar{x}};  \gclose{ x } ;\piencodf{ M'[\widetilde{x} \leftarrow  {x}]}_u   \| \\
            & \hspace{2.5cm}  \gsome{ x }{ \llfv{C} };  \pname{x}{\linvar{x}} .( \piencodf{ C }_{\linvar{x}}   \| \pname{x}{\unvar{x}} .(  \guname{ \unvar{x} }{ x_i } ;  \piencodf{ U }_{x_i}   \| \pclose{ x } ) ) )
            \\
            & \redone^*    \res{ \unvar{x} } ( \res{\linvar{x}} ( \piencodf{ M'[\widetilde{x} \leftarrow  {x}]}_u   \| \piencodf{ C }_{\linvar{x}} )  \|  \guname{ \unvar{x} }{ x_i } ;  \piencodf{ U }_{x_i}  )\\
            & =    \res{ \unvar{x} } (
            \res{\linvar{x}} (
            \psome{\linvar{x}}; \pname{\linvar{x}}{y_1}; \big( \gsome{y_1}{ \emptyset }; \gclose{ y_{1} } ; \0   \\
            & \qquad \qquad \| \psome{\linvar{x}}; \gsome{\linvar{x}}{u, \llfv{M'} \setminus  \widetilde{x} }; \bignd_{x_{i_1} \in \widetilde{x}} \gname{x}{{x}_{i_1}}; \cdots   \psome{\linvar{x}}; \pname{\linvar{x}}{y_k}; \big( \gsome{y_k}{ \emptyset }; \gclose{ y_{k} } ; \0   \\
            & \qquad \qquad \| \psome{\linvar{x}}; \gsome{\linvar{x}}{u, \llfv{M'} \setminus  \widetilde{x} }; \bignd_{x_{i_k} \in (\widetilde{x} \setminus x_{i_1} , \cdots , x_{i_{k-1}}    )} \gname{x}{{x}_{i_k}}; \\
            & \qquad \qquad \psome{\linvar{x}}; \pname{\linvar{x}}{y_{k+1}}; ( \gsome{y_{k+1}}{ u, \llfv{M'} }; \gclose{ y_{{k+1}} } ;\piencodf{M'}_u \| \pnone{ \linvar{x} } )
            \big) \cdots
            \big)
            \\
            & \qquad \qquad \| \gsome{\linvar{x}}{\llfv{C} }; \gname{x}{y_1}; \gsome{\linvar{x}}{y_1, \llfv{C}}; \psome{\linvar{x}}; \pname{\linvar{x}}{z_1}; \\
            & \qquad \qquad   ( \gsome{z_1}{\llfv{M_1}};  \piencodf{M_1}_{z_1} \| \pnone{y_1}  \| \\
            & \qquad \qquad \cdots \gsome{\linvar{x}}{\llfv{M_l} }; \gname{x}{y_l}; \gsome{\linvar{x}}{y_l, \llfv{C}}; \psome{\linvar{x}}; \pname{\linvar{x}}{z_l}; \cdots\\
            & \qquad \qquad  ( \gsome{z_l}{\llfv{M_l}};  \piencodf{M_l}_{z_l} \| \pnone{y_l}  \| \\
            & \qquad \qquad  \gsome{\linvar{x}}{\emptyset};\gname{x}{y_{l+1}};  ( \psome{ y_{l+1}}; \pclose{y_{l+1}}  \| \gsome{\linvar{x}}{\emptyset}; \pnone{\linvar{x}} )
            ) \cdots
            )
            )
            \|
            \guname{ \unvar{x} }{ x_i } ;  \piencodf{ U }_{x_i}
            ) \qquad
            (:= P_\mathbb{N})
               \\
            \end{aligned}
        \end{equation}
        }

            we reduce $P_\mathbb{N}$ arbitrarily discarding non-deterministic sums.

            \begin{equation*}
            \begin{aligned}
P_\mathbb{N}
& \redone^* \res{ \unvar{x} } (
            \res{\linvar{x}} ( \res{z_1} ( \gsome{z_1}{\llfv{M_1}};  \piencodf{M_1}_{z_1} \| \cdots  \res{z_k} ( \gsome{z_k}{\llfv{M_k}};  \piencodf{M_k}_{z_k} \| \\
            & \qquad \qquad \psome{\linvar{x}}; \pname{\linvar{x}}{y_{k+1}}; ( \gsome{y_{k+1}}{ u, \llfv{M'} }; \gclose{ y_{{k+1}} } ;\piencodf{M'}_u\{ z_1 / x_{i_1} \} \cdots \{ z_k / x_{i_k} \} \| \pnone{ \linvar{x} } )) \cdots   )
            \\
            & \qquad \qquad \| \gsome{\linvar{x}}{\llfv{C} \setminus (M_1 , \cdots , M_k) }; \gname{x}{y_{k+1}}; \gsome{\linvar{x}}{y_{k+1}, \llfv{C}\setminus (M_1 , \cdots , M_k)}; \psome{\linvar{x}}; \pname{\linvar{x}}{z_{k+1}}; \\
            & \qquad \qquad   ( \gsome{z_{k+1}}{\llfv{M_{k+1}}};  \piencodf{M_{k+1}}_{z_{k+1}} \| \pnone{y_1}  \| \\
            & \qquad \qquad \cdots \gsome{\linvar{x}}{\llfv{M_l} }; \gname{x}{y_l}; \gsome{\linvar{x}}{y_l, \llfv{M_l}}; \psome{\linvar{x}}; \pname{\linvar{x}}{z_l}; \cdots\\
            & \qquad \qquad  ( \gsome{z_l}{\llfv{M_l}};  \piencodf{M_l}_{z_l} \| \pnone{y_l}  \| \\
            & \qquad \qquad  \gsome{\linvar{x}}{\emptyset};\gname{x}{y_{l+1}};  ( \psome{ y_{l+1}}; \pclose{y_{l+1}}  \| \gsome{\linvar{x}}{\emptyset}; \pnone{\linvar{x}} )
            ) \cdots
            )
            )
            \|
            \guname{ \unvar{x} }{ x_i } ;  \piencodf{ U }_{x_i}
            )
\\
& \redone^* \res{ \unvar{x} } (
              \res{z_1} ( \gsome{z_1}{\llfv{M_1}};  \piencodf{M_1}_{z_1} \| \cdots  \res{z_k} ( \gsome{z_k}{\llfv{M_k}};  \piencodf{M_k}_{z_k} \| \\
            & \qquad \qquad   \pnone{u} \| \pnone{(\llfv{M'} \setminus \widetilde{x})} \|  \pnone{z_1} \| \cdots \| \pnone{z_k} \| \pnone{( \llfv{C} \setminus (M_1 , \cdots , M_k) )}  ) \cdots   )
            \\
            & \qquad \qquad \|
            \guname{ \unvar{x} }{ x_i } ;  \piencodf{ U }_{x_i}
            )
\\
& \redone^* \res{ \unvar{x} } ( ( \pnone{u} \| \pnone{(\llfv{M'} \setminus \widetilde{x})} \|   \| \pnone{ \llfv{C} }  )     \|
            \guname{ \unvar{x} }{ x_i } ;  \piencodf{ U }_{x_i}
            )
\\
& \equiv  \pnone{u} \| \pnone{(\llfv{M'} \setminus \widetilde{x})} \|   \| \pnone{ \llfv{C} }=   \piencodf{M}_u   \\
        \end{aligned}
        \end{equation*}

       \item Case $\redlab{RS{:}Fail^!}$:         Then we have
        $N = M' \unexsub{U /\unvar{x}}  $ with $\headf{M'} =  {x}[i]$, $U_i = \unvar{\oneb} $ and
        $N \red   M' \headlin{ \fail^{\emptyset} /\unvar{x} } \unexsub{U /\unvar{x} } $.
The result follows easily from

        \begin{equation}\label{eq:compl_unfail-failunres}
        \begin{aligned}
            \piencodf{N}_u &= \piencodf{ M' \unexsub{U /\unvar{x}} }_u  =   \res{ \unvar{x} }  ( \piencodf{ M' }_u   \|   \guname{ \unvar{x} }{ x_i } ;  \piencodf{ U }_{x_i}  )  \\
            & \redone^* \res{ \unvar{x} }  (  \res{ \widetilde{y} } (\piencodf{  {x}[i] }_{j}   \| P)   \|   \guname{ \unvar{x} }{ x_k } ;  \piencodf{ U }_{x_k}  ) \qquad (*)
            \\
            &  =  \res{ \unvar{x} }   (  \res{ \widetilde{y} } (\puname{ \unvar{x} }{ x_k };\psel{ {x}_k }{ i }; \pfwd{x_k}{j}   \| P)   \|   \guname{ \unvar{x} }{ x_k } ;  \piencodf{ U }_{x_k}  ) \qquad (*)
            \\
            & \redone  \res{ \unvar{x} }  (  \res{ \widetilde{y} }(   \res{ x_k } ( x_k.l_{i}; \pfwd{x_k}{j}   \| \piencodf{ U }_{x_k})   \| P)   \|   \guname{ \unvar{x} }{ x_k } ;  \piencodf{ U }_{x_k}  )
            \\
            & =  \res{ \unvar{x} } (  \res{ \widetilde{y} } (   \res{  x_k }  ( x_k.l_{i}; \pfwd{x_k}{j}   \| x_k. case( i.\piencodf{U_i}_{x} ))   \| P)   \|   \guname{ \unvar{x} }{ x_k } ;  \piencodf{ U }_{x_k}  )
            \\
            & \redone  \res{ \unvar{x} }  (  \res{ \widetilde{y} }  (  \piencodf{\unvar{\oneb}}_{j}    \| P)   \|   \guname{ \unvar{x} }{ x_k } ; \piencodf{ U }_{x_k}  )
           =  \res{ \unvar{x} }  (  \res{ \widetilde{y} }  (   \pnone{ j }    \| P)   \|   \guname{ \unvar{x} }{ x_k } ;  \piencodf{ U }_{x_k}  ) = \piencodf{M}_u
        \end{aligned}
        \end{equation}

        \item Case $\redlab{RS:Cons_1}$:
        Then we have
        $N = \fail^{\widetilde{x}}\ C \bagsep U$ and $N \red \fail^{\widetilde{x} \cup \widetilde{y}}  = M$ where $ \widetilde{y} = \llfv{C}$. The result follows easily from

        \begin{equation}\label{eq:compl_cons1-failunres}
        \begin{aligned}
            \piencodf{N}_u &= \piencodf{ \fail^{\widetilde{x}}\ C \bagsep U }_u=   \res{ v }  (\piencodf{\fail^{\widetilde{x}}}_v   \|  \gsome{ v }{ u , \llfv{C} };  \pname{v}{x} . (\pfwd{v}{u}   \| \piencodf{C \bagsep U}_x ) ) \\
            &=  \res{ v } (   \pnone{ v }    \|   \pnone{ \widetilde{x} }    \|  \gsome{ v }{ u , \llfv{C} };  \pname{v}{x} . (\pfwd{v}{u}   \| \piencodf{C \bagsep U}_x ) ) \\
            & \redone    \pnone{ u }   \|   \pnone{ \widetilde{x} }    \|  \pnone{ \widetilde{y} }= \piencodf{M}_u   \\
        \end{aligned}
        \end{equation}

        \item Cases $\redlab{RS:Cons_2}$ and $\redlab{RS:Cons_3}$: These cases follow by IH similarly to Case 7.

        \item Case $\redlab{RS{:}Cons_4}$:
        Then we have
        $N =  \fail^{\widetilde{y}} \unexsub{U / \unvar{x}} $ and $N \red \fail^{\widetilde{y}}  = M$. The result follows easily from

        \begin{align}\label{eq:compl_cons4-failunres}
            \piencodf{N}_u &= \piencodf{ \fail^{\widetilde{y}} \unexsub{U / \unvar{x}}}_u
            =  \res{ \unvar{x} }  ( \piencodf{ \fail^{\widetilde{y}} }_u   \|   \guname{ \unvar{x} }{ x_i } ;  \piencodf{ U }_{x_i} )  \\
            &=  \res{ \unvar{x} } (   \pnone{ u }    \|   \pnone{ \widetilde{x} }    \|    \guname{ \unvar{x} }{ x_i } ; \piencodf{ U }_{x_i} )
             \equiv    \pnone{ u }   \|   \pnone{ \widetilde{x} }=   \piencodf{M}_u
             \tag*{\qedhere}
         \end{align}

    \end{enumerate}

\end{proof}

\subsection{Soundness} \label{a:loossoundness}

We define $P \red \{P_i\}_{i \in I}$, for a fixed finite set $I$ where $ I = \{ i \ s.t. \  P \red P_i \}$. Similarly we define $P \red^* \{P_i\}_{i \in I}$ to be defined inductively by $P \red^* \{P_i\}_{i \in I} $ and $P_i \red \{P_j\}_{j \in J_i}$ for each $i \in I$ then $P \red^* \{P_j\}_{j \in J} $ with $J = \cup_{ i \in I} J_i  $

\propSoundextra*

\begin{proof}

Proof by induction on the precongruence rules.

\begin{itemize}
    \item When $P =P \premat P = Q$ then all reductions in $P$ are matched in $Q$.

    \item When $ P = P_1 \nd P_2 \premat Q$ with $P_i \premat Q \quad i \in \{ 1 , 2 \} $. Let us take $i = 1$ Then by the rule:

    \begin{prooftree}
            \infAss{
                $P_1\redone P_1' $
            }
            \infUn{
                $P_1 \nd P_2 \redone P_1' \nd P_2 $
            }{$\rredone{\nd}$}
    \end{prooftree}

    we have $P_1 \nd P_2 \redone \{P'_{1_i}\}_{i \in I_1}  \nd P_2 $  and $P_1 \nd P_2\redone P_1 \nd \{ P'_{2_i} \}_{i \in I_2} $
    for some $I_1, I_2$ Hence we have that $ P_1 \nd P_2\redone \{ P'_i \}_{i \in I_1 \cup I_2}$ where $P_i' = P'_{1_i} \nd P_2$ if $i \in I_1$ and $P_i' = P_{1} \nd P_{2_i}'$ if $i \in I_2$ . By the induction hypothesis $P_1 \premat Q$ and $P_1\redone \{ P_{1_i} \}_{i \in I_1}' $ imply $\exists J', \{ Q_j \}_{j \in J'} \ s.t. \  Q \redone^* \{Q_j\}_{j \in J'} $ , $ J' \subseteq I_1 $ and $P_j \premat Q_j \  , \ \forall j \in J'$. We take $J = J'$ and hence we can deduce $ P \redone \{P'_{i}\}_{i \in I_1 \cup I_2}$ , $Q \redone^* \{Q_j\}_{j \in J}$. Finally we have that $\forall j \in J$.

    \begin{prooftree}
        \AxiomC{$ P_{1_j}' \premat Q_j  $}
        \UnaryInfC{$ P_{1_j}' \nd P_2 \premat Q$}
    \end{prooftree}

    \item When $P = P_1   \| P_2  $ with $ P_1   \| P_2 \premat Q $. Then by the rule:

        \begin{prooftree}
            \infAss{
                $P_1\redone P_1'$
            }
            \infUn{
                $P_1   \| P_2\redone P_1'   \| P_2 $
            }{$\rredone{  \| }$}
        \end{prooftree}

    we have $P_1   \| P_2\redone \{P'_{1_{i}}\}_{i  \in I_1}   \| P_2 $  and $P_1   \| P_2\redone P_1   \| \{P'_{2_{i}}\}_{i \in I_1} $ for some $I_1, I_2$ Hence we have that $ P_1   \| P_2\redone \{P'_i \}_{i \in I_1 \cup I_2}$ where $P_i' = P_{1_i}'   \| P_2$ if $i \in I_1$ and $P_i' = P_{1}   \| P'_{2_i}$. By the induction hypothesis $P_1   \| P_2 \premat Q$
    and $P_1\redone \{P'_{1_i} \}_{i \in I_1} $ imply $\exists J_1', \{Q_j\}_{j \in J'_1} \ s.t. \  Q \redone^* \{Q_j\}_{j \in J_1'} $ ,
    $ J_1' \subseteq I_1 $ and $P'_{1_j} \premat Q_j \  , \ \forall j \in J_1'$. Similarly we have that $P_2\redone \{P'_2\}_{i \in I_2} $ imply $\exists J_2', \{Q_j\}_{j \in J'_2} \ s.t. \  Q \redone^* \{Q_j\}_{j \in J_2'} $ , $ J_2' \subseteq I_2 $ and $P_{2_j} \premat Q_j \  , \ \forall j \in J_2'$.
    We take $Q_{j \in J} = P_{1}   \| \{P'_{2_{i}}\}_{i \in I_2} \cup \{P'_{1_{i}}\}{i \in I_1}   \| P_{2}$ and hence we can deduce $ P\redone \{P'_{i}\}_{i \in I_1 \cup I_2}$ , $Q \redone^* \{Q_j\}_{j \in J}$. Finally we have that $\forall j \in J$ $ P_j \premat Q_j $

    \item When $ P =  \res{  x }  P_1 $ with $   \res{  x }   P_1 \premat Q $. Then by the rule:

        \begin{prooftree}
            \infAss{
                $P_1\redone P'$
            }
            \infUn{
                $ \res{ x }  P_1\redone  \res{ x }  P'$
            }{$\rredone{ \nu }$}
        \end{prooftree}

    we have $ \res{ x }  P_1 \redone  \res{ x }   \{P'_i\}_{{i \in I}} $ for some $I$. By the induction hypothesis $ \res{ x }   P_1 \ \premat Q$
    and $ \res{ x }  P_1\redone  \res{ x }  \{P'_i\}_{{i \in I}} $ imply $\exists J', \{Q_j\}_{j \in J'} \ s.t. \  Q \redone^* \{Q_j\}_{j \in J'} $ ,
    $ J' \subseteq I $ and $P'_{j} \premat Q_j \  , \ \forall j \in J'$.
     We take $J = J'$ and hence we can deduce $ P\redone \{P'_i\}_{i \in I}$ , $Q \redone^* Q_{j \in J}$. Finally we have that $\forall j \in J$

    \begin{prooftree}
            \AxiomC{$ P_j' \premat Q_j $}
            \UnaryInfC{$  \res{ x }   P_j' \premat  \res{ x }   Q_j  $}
    \end{prooftree}
    \qedhere

\end{itemize}

\end{proof}

\thmOpsoundone*

\begin{proof}
By induction on the structure of $N $ and then induction on the number of reductions of $\piencodf{N} \redone^* Q$.

\begin{enumerate}
    \item {\bf Base case:} $N =  {x}$, $N =  {x}[j]$, $N = \fail^{\emptyset}$ and $N = \lambda x . (M'[ {\widetilde{x}} \leftarrow  {x}])$.
.

    No reductions can take place, and the result follows trivially. Take $I = \{ a \}$,

    $Q =  \piencodf{N}_u \redone^0 \piencodf{N}_u = Q_a$ and $ {x} \red^0  {x} = N'$.

    \item $N =  M'(C \bagsep U) $.

        Then,
        $ \piencodf{M'(C \bagsep U)}_u =   \res{ v }  (\piencodf{M'}_v   \|  \gsome{ v }{ u , \llfv{C} };  \pname{v}{x} . (\pfwd{v}{u}   \| \piencodf{C \bagsep U}_x ) )$, and we are able to perform the  reductions from $\piencodf{M'(C \bagsep U)}_u$.

        We now proceed by induction on $k$, with  $\piencodf{N}_u \redone^k Q$. There are two main cases:
        \begin{enumerate}
        \item When $k = 0$ the thesis follows easily:

            We have $i = \{ a \} $,
    $Q =  \piencodf{M'(C \bagsep U)}_u \redone^0 \piencodf{M'(C \bagsep U)}_u = Q_a$ and $M'(C \bagsep U) \red^0 M'(C \bagsep U) = N'$.

            \item The interesting case is when $k \geq 1$.

            Then, for some process $R$ and $n, m$ such that $k = n+m$, we have the following:
            \[
            \begin{aligned}
               \piencodf{N}_u & =    \res{ v } (\piencodf{M'}_v   \|  \gsome{ v }{ u , \llfv{C} }; \pname{v}{x} . (\pfwd{v}{u}   \| \piencodf{C \bagsep U}_x ) )\\
               & \redone^m   \res{ v }  (R   \|  \gsome{ v }{ u , \llfv{C} };  \pname{v}{x} . (\pfwd{v}{u}   \| \piencodf{C \bagsep U}_x ) ) \redone^n  Q\\
            \end{aligned}
            \]
            Thus, the first $m \geq 0$ reduction steps are  internal to $\piencodf{ M'}_v$; type preservation in \clpi ensures that, if they occur,  these reductions  do not discard the possibility of synchronizing with $\psome{v}$. Then, the first of the $n \geq 0$ reduction steps towards $Q$ is a synchronization between $R$ and $ \gsome{ v }{ u, \llfv{C} }$.

            We consider two sub-cases, depending on the values of  $m$ and $n$:

\noindent{\bf (b.1) Case $m = 0$ and $n \geq 1$:}

Then $R = \piencodf{M}_v$ as $\piencodf{M}_v \redone^0 \piencodf{M}_v$.
 Notice that there are two possibilities of having an unguarded:

\begin{enumerate}
\item $M'=  (\lambda x . (M''[ {\widetilde{x}} \leftarrow  {x}])) \linexsub{C_1 / \widetilde{y_1}} \cdots \linexsub{C_p / \widetilde{y_p}} \unexsub{U_1 / \unvar{z}_1} \cdots \unexsub{U_q / \unvar{z}_q}   \quad (p, q \geq 0)$

   \[
   \begin{aligned}
   \piencodf{M'}_v &= \piencodf{ (\lambda x . (M''[ {\widetilde{x}} \leftarrow  {x}])) \linexsub{C_1 / \widetilde{y_1}} \cdots \linexsub{C_p / \widetilde{y_p}} \unexsub{U_1 / \unvar{z}_1} \cdots \unexsub{U_q / \unvar{z}_q} }_v \\
          \end{aligned}
    \]

    For simplicity we shall denote
    $$\piencodf{ (\lambda x . (M''[ {\widetilde{x}} \leftarrow  {x}])) \linexsub{C_1 / \widetilde{y_1}} \cdots \linexsub{C_p / \widetilde{y_p}} \unexsub{U_1 / \unvar{z}_1} \cdots \unexsub{U_q / \unvar{z}_q} }_v = \res{ \widetilde{y} ,\widetilde{z} } ( \piencodf{\lambda x . (M''[ {\widetilde{x}} \leftarrow  {x}])}_v   \| Q'' )$$
     where $\widetilde{y} = \widetilde{y_1} , \cdots , \widetilde{y_p}$. $\widetilde{z} = \unvar{z}_1, \cdots ,\unvar{z}_q$ and we continue the evaluation as:

      \[
      \begin{aligned}
      &=  \res{ \widetilde{y} ,\widetilde{z} } ( \piencodf{\lambda x . (M''[ {\widetilde{x}} \leftarrow  {x}])}_v   \| Q'' )\\
      &=  \res{ \widetilde{y},\widetilde{z} }  ( \psome{v}; \gname{v}{x}; \psome{x}; \gname{x}{\linvar{x}}; \gname{x}{\unvar{x}}; \gclose{ x } ;  \piencodf{M''[ {\widetilde{x}} \leftarrow  {x}]}_v   \| Q'' )
        \end{aligned}
        \]
  \noindent

With this shape for $M$, we then have the following:
 \[
 \begin{aligned}
 \piencodf{N}_u & = \piencodf{(M'\ B)}_u=   \res{ v }  (\piencodf{M'}_v   \|  \gsome{ v }{ u , \llfv{C} }; \pname{v}{x} . (\pfwd{v}{u}   \| \piencodf{C \bagsep U}_x ) )\\
 & \redone  \res{ v  }( \res{ \widetilde{y},\widetilde{z}} (  \gname{v}{x}; \psome{x}; \gname{x}{\linvar{x}}; \gname{x}{\unvar{x}}; \gclose{ x } ;  \piencodf{M''[ {\widetilde{x}} \leftarrow  {x}]}_v \\
 & \hspace{.5cm}   \| Q'' )   \| \pname{v}{x} . (\pfwd{v}{u}   \| \piencodf{C \bagsep U}_x ) ) & = Q_1 \\
& \redone  \res{ x}  ( \res{ v }(  \res{ \widetilde{y},\widetilde{z}}( \psome{x}; \gname{x}{\linvar{x}}; \gname{x}{\unvar{x}}; \gclose{ x } ;  \piencodf{M''[ {\widetilde{x}} \leftarrow  {x}]}_v  \\
 & \hspace{.5cm}  \| Q'' )   \|  \pfwd{v}{u}  ) \| \piencodf{C \bagsep U}_x ) & = Q_2 \\
 & \redone   \res{ x } ( \res{ \widetilde{y},\widetilde{z}}(\psome{x}; \gname{x}{\linvar{x}}; \gname{x}{\unvar{x}};  \gclose{ x } ; \piencodf{M''[ {\widetilde{x}} \leftarrow  {x}]}_u   \| Q'')\\
 & \hspace{.5cm}  \|  \piencodf{C \bagsep U}_x ) & = Q_3 \\
\end{aligned}
     \]
We also have that
\[
\begin{aligned}
    N &=(\lambda x . (M''[ {\widetilde{x}} \leftarrow  {x}])) \linexsub{C_1 /  \widetilde{y_1}} \cdots \linexsub{C_p / \widetilde{y_p}} \unexsub{U_1 / \unvar{z}_1} \cdots \unexsub{U_q / \unvar{z}_q} (C \bagsep U) \\
    & \equivlam (\lambda x . (M''[ {\widetilde{x}} \leftarrow  {x}]) (C \bagsep U)) \linexsub{C_1 / \widetilde{y_1}} \cdots \linexsub{C_p / \widetilde{y_p}} \unexsub{U_1 / \unvar{z}_1} \cdots \unexsub{U_q / \unvar{z}_q} \\
    & \red   M''[ {\widetilde{x}} \leftarrow  {x}] \esubst{(C \bagsep U)}{x} \linexsub{C_1 / \widetilde{y_1}} \cdots \linexsub{C_p / \widetilde{y_p}} \unexsub{U_1 / \unvar{z}_1} \cdots \unexsub{U_q / \unvar{z}_q} = M
\end{aligned}
\]
Furthermore, we have:
\[
\begin{aligned}
&\piencodf{M}_u = \piencodf{M''[ {\widetilde{x}} \leftarrow  {x}] \esubst{(C \bagsep U)}{x} \linexsub{C_1 / \widetilde{y_1}} \cdots \linexsub{C_p / \widetilde{y_p}} \unexsub{U_1 / \unvar{z}_1} \cdots \unexsub{U_q / \unvar{z}_q}}_u \\
& =  \res{  x } ( \res{ \widetilde{y},\widetilde{z} } ( \psome{x}; \gname{x}{\linvar{x}}; \gname{x}{\unvar{x}}; \gclose{ x } ;  \piencodf{M''[ {\widetilde{x}} \leftarrow  {x}]}_u  \| Q'' )  \|  \piencodf{C \bagsep U}_x    )
\end{aligned}
\]

We consider different possibilities for $n \geq 1$; in all  the cases, the result follows.
        \smallskip

 \noindent  {\bf When $n = 1$:}

We have $I = \{a \}$, $Q = Q_1$, $ \piencodf{N}_u \redone^1 Q_1$.
We also have that
\begin{itemize}
\item  $Q_1 \redone^2 Q_3 = Q_a$ ,
\item $N \red^1 M''[ {\widetilde{x}} \leftarrow  {x}] \esubst{(C \bagsep U)}{x} \linexsub{C_1 / \widetilde{y_1}} \cdots \linexsub{C_p / \widetilde{y_p}} \unexsub{U_1 / \unvar{z}_1} \cdots \unexsub{U_q / \unvar{z}_q} = N'$
\item and $\piencodf{M''[ {\widetilde{x}} \leftarrow  {x}] \esubst{(C \bagsep U)}{x} \linexsub{C_1 / \widetilde{y_1}} \cdots \linexsub{C_p / \widetilde{y_p}} \unexsub{U_1 / \unvar{z}_1} \cdots \unexsub{U_q / \unvar{z}_q}}_u = Q_3$.
\end{itemize}

                    \smallskip

 \noindent  {\bf When $n = 2,3$:} the analysis is similar.

\noindent {\bf When $n \geq 4$:}

 We have $ \piencodf{N}_u \redone^3 Q_3 \redone^*_{} Q$. We also know that $N \red M$, $Q_3 = \piencodf{M}_u$. By the IH, there exist $\{Q_i\}_{i \in I}$ , $ N'$ such that $Q \redone^* \{ Q_i \}_{i \in I}$, $M \red^* N'$ and $\piencodf{N'}_u \premat Q_i \quad \forall i \in I $ . Finally, $\piencodf{N}_u \redone^3 Q_3 \redone^l Q \redone_I Q_i$ and $N \red M  \red^* N'$.

\item $M'= \fail^{\widetilde{z}}$.

Then,                     \(
\begin{aligned}
\piencodf{M'}_v &= \piencodf{\fail^{\widetilde{z}}}_v =   \pnone{ v }    \|   \pnone{ \widetilde{z} } .
\end{aligned}
\)
With this shape for $M$, we have:

\[
\begin{aligned}
\piencodf{N}_u & = \piencodf{(M'\ (C \bagsep U))}_u=   \res{ v } (\piencodf{M'}_v   \|  \gsome{ v }{ u , \llfv{C} };  \pname{v}{x} . (\pfwd{v}{u}   \| \piencodf{C \bagsep U}_x ) )\\
& =  \res{  v }  (  \pnone{ v }    \|   \pnone{ \widetilde{z} }    \|  \gsome{ v }{ u , \llfv{C} };  \pname{v}{x} . (\pfwd{v}{u}   \| \piencodf{C \bagsep U}_x ) )\\
& \redone   \pnone{ u }     \|  \pnone{ \widetilde{z} }      \|   \pnone{ \llfv{C} }  \\
\end{aligned}
\]

\end{enumerate}

We also have that
\(  N = \fail^{\widetilde{x}}\ C \bagsep U \red  \fail^{\widetilde{x} \cup \llfv{C}}  = N'.  \)
Furthermore,
\[
\begin{aligned}
\piencodf{N'}_u &= \piencodf{ \fail^{\widetilde{z} \cup \llfv{C}  } }_u
= \piencodf{ \fail^{\widetilde{z} \cup \llfv{C} }}_u
=    \pnone{ u }   \|   \pnone{ \widetilde{z} }    \|    \pnone{ \llfv{C} } .
\end{aligned}
\]

\noindent{ \bf (b.2) Case  $m \geq 1$ and $ n \geq 0$:}

We distinguish two cases:

 \begin{enumerate}
\item When $n = 0$:

Then, $  \res{ v }  (R   \|  \gsome{ v }{ u , \llfv{C} };  \pname{v}{x} . (\pfwd{v}{u}   \| \piencodf{C \bagsep U}_x ) ) =  Q $ and $\piencodf{M'}_u \redone^m R$ where $m \geq 1$. Then by the IH there exist  $\{R_i'\}_{i \in I}$  and $M'' $ such that $R \redone^* \{ R_i' \}_{i \in I}$, $M'\red^* M''$, and $\piencodf{M''}_u \premat R_i \quad \forall i \in I $.  Hence we have that

            \[
            \begin{aligned}
                \piencodf{N}_u & =   \res{ v } (\piencodf{M'}_v   \|  \gsome{ v }{ u , \llfv{C} };  \pname{v}{x} . (\pfwd{v}{u}   \| \piencodf{C \bagsep U}_x ) )\\
                    & \redone^m   \res{ v }  (R   \|  \gsome{ v }{ u , \llfv{C} };  \pname{v}{x} . (\pfwd{v}{u}   \| \piencodf{C \bagsep U}_x ) )  = Q
            \end{aligned}
             \]
            We also know that
            \[
            \begin{aligned}
              Q & \redone^* \{  \res{ v } (R'_i   \|  \gsome{ v }{ u , \llfv{C} };  \pname{v}{x} . (\pfwd{v}{u}   \| \piencodf{C \bagsep U}_x ) ) \}_{i \in I} = \{ Q_i \}_{i \in I}\\
            \end{aligned}
            \]

            and so the \lamcoldetsh term can reduce as follows: $N = (M'\ ( C \bagsep U )) \red^* M''\ ( C \bagsep U ) = N'$ and  $\piencodf{N'}_u \premat Q_i \quad \forall i \in I$ via the $\premat$ rules.

                \item When $n \geq 1$:

                    Then $R$ has an occurrence of an unguarded $\psome{v}$ or $  \pnone{ v } $, hence it is of the form

                    $ \piencodf{(\lambda x . (M''[ {\widetilde{x}} \leftarrow  {x}]))  \linexsub{N_1 / y_1} \cdots \linexsub{N_p / y_p} \unexsub{U_1 / \unvar{z}_1} \cdots \unexsub{U_q / \unvar{z}_q} }_v $ or $ \piencodf{\fail^{\widetilde{x}}}_v. $

                    \end{enumerate}

        \end{enumerate}

        This concludes the analysis for the case $N = (M'\, ( C \bagsep U ))$.

        \item $N = M'[ {\widetilde{x}} \leftarrow  {x}]$.
    The sharing variable $ {x}$ is not free and the result follows by vacuity.

        \item $N = M'[ {\widetilde{x}} \leftarrow  {x}] \esubst{ C \bagsep U }{ x}$. Then we have

            \[
                \begin{aligned}
                    \piencodf{N}_u &=\piencodf{ M'[ {\widetilde{x}} \leftarrow  {x}] \esubst{ C \bagsep U }{ x} }_u=  \res{ x } ( \psome{x}; \gname{x}{\linvar{x}}; \gname{x}{\unvar{x}};  \gclose{ x } ;\piencodf{ M'[ {\widetilde{x}} \leftarrow  {x}]}_u   \| \piencodf{ C \bagsep U}_x )
                \end{aligned}
            \]

            Let us consider three cases.

            \begin{enumerate}
                \item When $\size{ {\widetilde{x}}} = \size{C}$.
                    Then let us consider the shape of the bag $ C$.

  \begin{enumerate}
  \item When $C = \oneb$.

  We have the following
 \[
 \begin{aligned}
 \piencodf{N}_u  &=   \res{ x } ( \psome{x}; \gname{x}{\linvar{x}}; \gname{x}{\unvar{x}};  \gclose{ x } ;\piencodf{ M'[ \leftarrow  {x}]}_u   \| \piencodf{ \oneb \bagsep U}_x )\\
 &=   \res{ x }  ( \psome{x}; \gname{x}{\linvar{x}}; \gname{x}{\unvar{x}}; \gclose{ x } ; \piencodf{ M'[ \leftarrow  {x}]}_u   \|  \gsome{ x }{ \llfv{C} };  \pname{x}{\linvar{x}} . \\
 & \hspace{1cm}( \piencodf{ \oneb }_{\linvar{x}}   \|\pname{x}{\unvar{x}} .(  \guname{ \unvar{x} }{ x_i } ;  \piencodf{ U }_{x_i}   \|  \pclose{ x } ) ) )\\
&\redone  \res{ x }  (  \gname{x}{\linvar{x}}; \gname{x}{\unvar{x}}; \gclose{ x } ; \piencodf{ M'[ \leftarrow  {x}]}_u   \|
 \pname{x}{\linvar{x}} .( \piencodf{ \oneb }_{\linvar{x}}   \| \pname{x}{\unvar{x}} .  (  \guname{ \unvar{x} }{ x_i } ;  \piencodf{ U }_{x_i}  \|  \pclose{ x } ) ) ) & = Q_1
                              \\
  &\redone  \res{ x,\linvar{x} } (  \gname{x}{\unvar{x}}; \gclose{ x } ; \piencodf{ M'[ \leftarrow  {x}]}_u   \| \piencodf{ \oneb }_{\linvar{x}}   \|\pname{x}{\unvar{x}} .(  \guname{ \unvar{x} }{ x_i } ;  \piencodf{ U }_{x_i}   \|  \pclose{ x } ) ) & = Q_2
                              \\
  &\redone  \res{ x,\linvar{x}, \unvar{x} } (  \gclose{ x } ; \piencodf{ M'[ \leftarrow  {x}]}_u   \| \piencodf{ \oneb }_{\linvar{x}}   \|  \guname{ \unvar{x} }{ x_i } ;  \piencodf{ U }_{x_i}   \|  \pclose{ x } ) & = Q_3   \\
 &\redone  \res{ \linvar{x}, \unvar{x} } (  \piencodf{ M'[ \leftarrow  {x}]}_u   \| \piencodf{ \oneb }_{\linvar{x}}   \|  \guname{ \unvar{x} }{ x_i } ;  \piencodf{ U }_{x_i} ) & = Q_4\\
    & =  \res{ \linvar{x}, \unvar{x} } ( \psome{\linvar{x}}; \pname{\linvar{x}}{y_i} . (  \gsome{ y_i }{ u,\llfv{M'} };  \gclose{ y_{i} } ; \piencodf{M'}_u   \|   \pnone{ \linvar{x} } )   \| \\
    & \qquad  \gsome{ \linvar{x} }{ \emptyset };  \gname{\linvar{x}}{y_n}; ( \psome{y_n}; \pclose{ y_n }    \|  \gsome{ \linvar{x} }{ \emptyset };   \pnone{ \linvar{x} } )    \|  \guname{ \unvar{x} }{ x_i } ;  \piencodf{ U }_{x_i} )
                            \\
      & \redone  \res{ \linvar{x}, \unvar{x} }  (  \pname{\linvar{x}}{y_i} . ( \gsome{  y_i }{ u,\llfv{M'} };  \gclose{ y_{i} } ; \piencodf{M'}_u   \|   \pnone{ \linvar{x} } )   \| \\
                              & \qquad \gname{\linvar{x}}{y_n}; ( \psome{y_n}; \pclose{ y_n }    \|  \gsome{ \linvar{x} }{ \emptyset };   \pnone{  \linvar{x} } )    \|  \guname{ \unvar{x} }{ x_i } ;  \piencodf{ U }_{x_i} )  & = Q_5
                            \\
                            & \redone  \res{ \linvar{x}, \unvar{x} , y_i } (   \gsome{ y_i }{ u,\llfv{M'} }; \gclose{ y_{i} } ; \piencodf{M'}_u   \|   \pnone{ \linvar{x} }    \|  \psome{y_i}; \pclose{ y_i }   \\
 & \hspace{1cm}  \|  \gsome{ \linvar{x} }{ \emptyset };   \pnone{ \linvar{x} }     \|  \guname{ \unvar{x} }{ x_i } ;  \piencodf{ U }_{x_i} )  & = Q_6
                            \\
                            & \redone  \res{ \linvar{x}, \unvar{x} , y_i }  (  \gclose{ y_{i} } ;  \piencodf{M'}_u   \|   \pnone{  \linvar{x} }   \|   \pclose{ y_i }    \|  \gsome{ \linvar{x} }{ \emptyset };   \pnone{ \linvar{x} }    \|  \guname{ \unvar{x} }{ x_i } ;  \piencodf{ U }_{x_i} )  & = Q_7
                            \\
                            & \redone  \res{ \linvar{x}, \unvar{x} } (  \piencodf{M'}_u   \|   \pnone{ \linvar{x} }    \|   \gsome{ \linvar{x} }{ \emptyset };    \pnone{ \linvar{x} }     \|  \guname{ \unvar{x} }{ x_i } ;  \piencodf{ U }_{x_i} )  & = Q_8
                            \\
                            & \redone  \res{ \unvar{x} }  (  \piencodf{M'}_u   \|  \guname{ \unvar{x} }{ x_i } ;  \piencodf{ U }_{x_i} )
                            =  \piencodf{M'\unexsub{U / \unvar{x}}}_u
                            & = Q_9
                            \end{aligned}
                            \]
                            Notice how $Q_8$ has a choice however the $\linvar{x}$ name can be closed at any time so for simplicity we only perform communication across this name once all other names have completed their reductions.

                        Now we proceed by induction on the number of reductions $\piencodf{N}_u \red^k Q$.

                            \begin{enumerate}

                                \item When $k = 0$, the result follows trivially. Just take $I = \{a\}$, $\mathbb{N}=\mathbb{N}'$ and $\piencodf{N}_u=Q=Q_a$.

                                \item When $k = 1$.

                                    We have $Q = Q_1$, $ \piencodf{N}_u \redone^1 Q_1$. Let us take $I = a$, we also have that $Q_1 \redone^8 Q_9 = Q_a$ , $N \red M'\unexsub{U / \unvar{x}} = M$ and $\piencodf{ M'}_u \premat Q_9$

                                \item When $2 \leq  k \leq 8$.

                                    Proceeds similarly to the previous case

                                \item When $k \geq 9$.

      We have $ \piencodf{N}_u \redone^9 Q_9 \redone^l Q$, for $l \geq 1$. Since $Q_9 = \piencodf{ M'}_u$ we apply the induction hypothesis we have that  there $\exists \{Q_i\}_{ i \in I } $, $N' \ s.t. \ Q \redone^* \{ Q_i\}_{ i \in I} ,  M'\red^* N'$ and $\piencodf{N'}_u \premat Q_i \quad \forall i \in I$.                                    Then,  $ \piencodf{N}_u \redone^9 Q_9 \redone^l Q \redone^* \{ Q_i \}_{i \in I}$ and by the contextual reduction rule it follows that $N = (M'[ \leftarrow x])\esubst{ 1 }{ x } \red^*  N' $ and the case holds.

\end{enumerate}

\item When $C = \bag{N_1} \cdot \cdots \cdot \bag{N_l}$, for $l \geq 1$.
                    Then,

 \[
   \begin{aligned}
   \piencodf{N}_u &=\piencodf{ M'[ {\widetilde{x}} \leftarrow  {x}] \esubst{ C \bagsep U }{x} }_u\\
   &=   \res{ x }  ( \psome{x}; \gname{x}{\linvar{x}}; \gname{x}{\unvar{x}}; \gclose{ x } ; \piencodf{ M'[ {\widetilde{x}} \leftarrow  {x}]}_u   \| \piencodf{ C \bagsep U}_x ) \\
  &\redone ^{4}  \res{ \unvar{x} }  ( \res{\linvar{x}} ( \piencodf{ M'[ {\widetilde{x}} \leftarrow  {x}]}_u   \| \piencodf{ C }_{\linvar{x}} )   \|  \guname{ \unvar{x} }{ x_i } ;  \piencodf{ U }_{x_i} )\\
  &=
  \res{ \unvar{x} }  (
  \res{\linvar{x}} (
  \psome{\linvar{x}}; \pname{\linvar{x}}{y_1}; \big( \gsome{y_1}{ \emptyset }; \gclose{ y_{1} } ; \0   \\
    & \qquad \qquad \| \psome{\linvar{x}}; \gsome{\linvar{x}}{u, \llfv{M'} \setminus  \widetilde{x} }; \bignd_{x_{i_1} \in \widetilde{x}} \gname{x}{{x}_{i_1}};
    \cdots \\
    & \qquad \qquad \psome{\linvar{x}}; \pname{\linvar{x}}{y_l}; \big( \gsome{y_l}{ \emptyset }; \gclose{ y_{l} } ; \0   \\
    & \qquad \qquad \| \psome{\linvar{x}}; \gsome{\linvar{x}}{u, \llfv{M'} \setminus  (\widetilde{x} \setminus (x_{i_1} , \cdots , x_{i_{l-1}}) )}; \bignd_{x_{i_l} \in \widetilde{x}} \gname{x}{{x}_{i_l}};\\
    & \qquad \qquad \psome{\linvar{x}}; \pname{\linvar{x}}{y_{l+1}}; ( \gsome{y_{l+1}}{ u , \llfv{M'} }; \gclose{ y_{l+1} } ;\piencodf{M'}_u \| \pnone{ \linvar{x} } )\big) \cdots
    \big)
  \\
  & \qquad \qquad \| \\
  & \qquad \qquad \gsome{\linvar{x}}{\llfv{C} }; \gname{x}{y_1}; \gsome{\linvar{x}}{y_1, \llfv{C}}; \psome{\linvar{x}}; \pname{\linvar{x}}{z_1}; \\
       & \qquad \qquad  ( \gsome{z_1}{\llfv{M_1}};  \piencodf{M_1}_{z_1} \| \pnone{y_1}  \| \\
  & \qquad \qquad \cdots \gsome{\linvar{x}}{\llfv{M_l} }; \gname{x}{y_l}; \gsome{\linvar{x}}{y_l, \llfv{M_l}}; \psome{\linvar{x}}; \pname{\linvar{x}}{z_l}; \\
       & \qquad \qquad  ( \gsome{z_l}{\llfv{M_l}};  \piencodf{M_l}_{z_l} \| \pnone{y_l}  \| \\
  & \qquad \qquad \gsome{\linvar{x}}{\emptyset};\gname{x}{y_{l+1}};  ( \psome{ y_{l+1}}; \pclose{y_{l+1}}  \| \gsome{\linvar{x}}{\emptyset}; \pnone{\linvar{x}} )) \cdots )
  )\|
  \guname{ \unvar{x} }{ x_i } ;  \piencodf{ U }_{x_i} )\\
 & \redone^{6l}
    \res{ \unvar{x} }  ( \res{\linvar{x}} ( \res{ z_1 }( \cdots \res{ z_l } ( \psome{\linvar{x}}; \pname{\linvar{x}}{y_{l+1}}. (  \gsome{ y_{l+1} }{ u,\llfv{M'} };   \\
    &\qquad  \gclose{ y_{l+1} } ;  \piencodf{M'}_u\{ z_1 / x_{i_1} \} \cdots \{ z_l / x_{i_l} \}   \|   \pnone{ \linvar{x} }  )\\
    & \qquad   \|  \gsome{ z_l }{ \llfv{M_{j_l}} };  \piencodf{M_{j_l}}_{z_l} ) \cdots \|  \gsome{ z_1 }{ \llfv{M_{j_1}} }; \piencodf{M_{j_1}}_{z_1} )       \|
    \\
    &\qquad  \gsome{ \linvar{x} }{ \emptyset };  \linvar{x}(y_{l+1}). ( \psome{y_{l+1}}; \pclose{ y_{l+1} }    \|  \gsome{ \linvar{x} }{ \emptyset };    \pnone{ \linvar{x} } ) )
   \|  \guname{ \unvar{x} }{ x_i } ; \piencodf{ U }_{x_i} )\\
& \redone^{5}
    \res{ \unvar{x} }  (  \res{ z_1 }(\gsome{ x_1 }{ \llfv{M_{j_1}} };  \piencodf{M_{j_1}}_{x_1} \| \cdots \\
    & \qquad \qquad \res{ z_l }  ( \gsome{ x_l }{ \llfv{M_{j_l}} };  \piencodf{M_{j_l}}_{x_l} \| \piencodf{M'}_u \{ z_1 / x_{i_1} \} \cdots \{ z_l / x_{i_l} \}  )  \cdots      )
    \|  \guname{ \unvar{x} }{ x_i } ;  \piencodf{ U }_{x_i} )\\
  \end{aligned}
 \]

                            The proof follows by induction on the number of reductions $\piencodf{N}_u \redone^k Q$.

\begin{enumerate}
\item When $k = 0$, the result follows trivially. Just take $I = \{ a\} $, $\mathbb{N}=\mathbb{N}'$ and $\piencodf{N}_u=Q=Q_a$.

 \item When $1 \leq k \leq 6l + 9$.

 Let $I = \{ a \}$, $Q_k$ be such that $ \piencodf{N}_u \redone^k Q_k$.
            We also have that $Q_k \redone^{6l + 9 - k} Q_{6l + 9} = Q_a$ ,

            $N \red  M'\linexsub{C  /  x_1 , \cdots , x_l} \unexsub{U / \unvar{x} }= N'$ and

            $
            \begin{aligned}
                \piencodf{  M'\linexsub{C  /  x_1 , \cdots , x_l} \unexsub{U / \unvar{x} } }_u & =  \res{  \unvar{x} }  \Big( \res{z_1}( \gsome{z_1}{\llfv{M_{1}}};\piencodf{ M_{1} }_{ {z_1}}  \| \\
                & \qquad  \cdots \res{z_l} ( \gsome{z_l}{\llfv{M_{l}}};\piencodf{ M_{l} }_{ {z_l}} \\
                & \qquad   \| \bignd_{x_{i_1} \in \{ x_1 ,\cdots , x_l  \}} \cdots \bignd_{x_{i_l} \in \{ x_1 ,\cdots , x_l \setminus x_{i_1} , \cdots , x_{i_{l-1}}  \}}\\
                & \qquad  \piencodf{ M'}_u \{ z_1 / x_{i_1} \} \cdots \{ z_l / x_{i_l} \} ) \cdots )
                     \|  \guname{ \unvar{x} }{ x_i } ;  \piencodf{ U }_{x_i} \Big )\\
            & \premat  \res{  \unvar{x} }  \Big( \res{z_1}( \gsome{z_1}{\llfv{M_{1}}};\piencodf{ M_{1} }_{ {z_1}}  \| \\
                & \qquad  \cdots \res{z_l} ( \gsome{z_l}{\llfv{M_{l}}};\piencodf{ M_{l} }_{ {z_l}} \\
                & \qquad   \| \piencodf{ M'}_u \{ z_1 / x_{i_1} \} \cdots \{ z_l / x_{i_l} \} ) \cdots )
                    \\
                   &  \qquad   \|  \guname{ \unvar{x} }{ x_i } ;  \piencodf{ U }_{x_i} \Big )
                    = Q_{6l + 9}
            \end{aligned}
            $.

\item When $k > 6l + 9$.

Then,  $ \piencodf{N}_u \redone^{6l + 9} Q_{6l + 9} \redone^n Q$ for $n \geq 1$. Also,

\(
\begin{aligned}
&N \red^1   M'\linexsub{C  /  x_1 , \cdots , x_l} \unexsub{U / \unvar{x} } \text { and } \\
&   \piencodf{ M'\linexsub{C  /  x_1 , \cdots , x_l} \unexsub{U / \unvar{x} } }_u \premat Q_{6l + 9} .
    \end{aligned}
\)

$\exists  \{ P_j \}_{ j \in J}, N', \ s.t. \ M'\linexsub{C  /  x_1 , \cdots , x_l} \unexsub{U / \unvar{x} } \red^* N',$ $ \piencodf{M'\linexsub{C  /  x_1 , \cdots , x_l} \unexsub{U / \unvar{x} }}$ $ \redone^n P \redone^* \{ P_j \}_{j \in J}  $ and
$ \piencodf{N'}_u \premat P_j \quad \forall j \in J $. We also have by Prop.\ref{prop:soundextra} that as $\piencodf{M'\linexsub{C  /  x_1 , \cdots , x_l} \unexsub{U / \unvar{x} }} \premat Q_{6l + 9} $ and $\piencodf{M'\linexsub{C  /  x_1 , \cdots , x_l} \unexsub{U / \unvar{x} }} \redone^* \{ Q_j \}_{j \in J} $ implies $\exists  \{Q_i\}_{i \in I} \ s.t. \  Q_{6l + 9} \redone^* Q_{i \in I} $ , $ I \subset J $ and $P_i \premat Q_i \  , \ \forall i \in I$

                        \end{enumerate}

                    \end{enumerate}

                \item When $\size{\widetilde{x}} > \size{C}$.

                    Then we have
                    $N = M'[ {x}_1, \cdots ,  {x}_k \leftarrow  {x}]\ \esubst{ C \bagsep U }{x}$ with $C = \bag{M_1}  \cdots  \bag{M_l} \quad k > l$, $N \red  \fail^{\widetilde{z}} = M$ and $ \widetilde{z} =  (\llfv{M'} \setminus \{   {x}_1, \cdots ,  {x}_k \} ) \cup \llfv{C} $. On the one hand, we have:
                    Hence $k = l + m$ for some $m \geq 1$

                    \[
                    \begin{aligned}
                        \piencodf{N}_u &= \piencodf{M'[ {x}_1, \cdots ,  {x}_k \leftarrow  {x}]\ \esubst{ C \bagsep U }{x}}_u \\
                        & =   \res{ x } ( \psome{x}; \gname{x}{\linvar{x}}; \gname{x}{\unvar{x}}; \gclose{ x } ; \piencodf{ M'[ {x}_1, \cdots ,  {x}_k \leftarrow  {x}]}_u   \| \piencodf{ C \bagsep U}_x ) \\
                              &\redone ^{4}  \res{ \unvar{x} } ( \res{\linvar{x}} ( \piencodf{M'[ {x}_1, \cdots ,  {x}_k \leftarrow  {x}]}_u   \| \piencodf{ C }_{\linvar{x}}  ) \|  \guname{ \unvar{x} }{ x_i } ;  \piencodf{ U }_{x_i} )\\
            & = \res{ \unvar{x} } (
            \res{\linvar{x}} (
            \psome{\linvar{x}}; \pname{\linvar{x}}{y_1}; \big( \gsome{y_1}{ \emptyset }; \gclose{ y_{1} } ; \0   \\
            & \qquad \qquad \|\psome{\linvar{x}}; \gsome{\linvar{x}}{u, \llfv{M'} \setminus  \widetilde{x} }; \bignd_{x_{i_1} \in \widetilde{x}} \gname{x}{{x}_{i_1}}; \cdots  \\
            & \qquad \qquad \psome{\linvar{x}}; \pname{\linvar{x}}{y_k}; \big( \gsome{y_k}{ \emptyset }; \gclose{ y_{k} } ; \0   \\
            & \qquad \qquad \| \psome{\linvar{x}}; \gsome{\linvar{x}}{u, \llfv{M'} \setminus  \widetilde{x} }; \bignd_{x_{i_k} \in (\widetilde{x} \setminus x_{i_1} , \cdots , x_{i_{k-1}}    )} \gname{x}{{x}_{i_k}}; \\
            & \qquad \qquad \psome{\linvar{x}}; \pname{\linvar{x}}{y_{k+1}}; ( \gsome{y_{k+1}}{ u, \llfv{M'} }; \gclose{ y_{{k+1}} } ;\piencodf{M'}_u \| \pnone{ \linvar{x} } )
            \big) \cdots
            \big)
            \\
            & \qquad \qquad \| \gsome{\linvar{x}}{\llfv{C} }; \gname{x}{y_1}; \gsome{\linvar{x}}{y_1, \llfv{C}}; \psome{\linvar{x}}; \pname{\linvar{x}}{z_1}; \\
            & \qquad \qquad   ( \gsome{z_1}{\llfv{M_1}};  \piencodf{M_1}_{z_1} \| \pnone{y_1}  \| \\
            & \qquad \qquad \cdots \gsome{\linvar{x}}{\llfv{M_l} }; \gname{x}{y_l}; \gsome{\linvar{x}}{y_l, \llfv{C}}; \psome{\linvar{x}}; \pname{\linvar{x}}{z_l}; \cdots\\
            & \qquad \qquad  ( \gsome{z_l}{\llfv{M_l}};  \piencodf{M_l}_{z_l} \| \pnone{y_l}  \| \\
            & \qquad \qquad  \gsome{\linvar{x}}{\emptyset};\gname{x}{y_{l+1}};  ( \psome{ y_{l+1}}; \pclose{y_{l+1}}  \| \gsome{\linvar{x}}{\emptyset}; \pnone{\linvar{x}} )
            ) \cdots
            )
            )
            \|
            \guname{ \unvar{x} }{ x_i } ;  \piencodf{ U }_{x_i}
            ) \\
            & \redone^{6k}
             \res{ \unvar{x} } (
            \res{\linvar{x}} ( \res{z_1} ( \gsome{z_1}{\llfv{M_1}};  \piencodf{M_1}_{z_1} \| \cdots  \res{z_k} ( \gsome{z_k}{\llfv{M_k}};  \piencodf{M_k}_{z_k} \| \\
            & \qquad \qquad \psome{\linvar{x}}; \pname{\linvar{x}}{y_{k+1}}; ( \gsome{y_{k+1}}{ u, \llfv{M'} }; \gclose{ y_{{k+1}} } ;\piencodf{M'}_u\{ z_1 / x_{i_1} \} \cdots \{ z_k / x_{i_k} \} \| \pnone{ \linvar{x} } )) \cdots   )
            \\
            & \qquad \qquad \| \gsome{\linvar{x}}{\llfv{C} \setminus (M_1 , \cdots , M_k) }; \gname{x}{y_{k+1}}; \gsome{\linvar{x}}{y_{k+1}, \llfv{C}\setminus (M_1 , \cdots , M_k)}; \psome{\linvar{x}}; \pname{\linvar{x}}{z_{k+1}}; \\
            & \qquad \qquad   ( \gsome{z_{k+1}}{\llfv{M_{k+1}}};  \piencodf{M_{k+1}}_{z_{k+1}} \| \pnone{y_1}  \| \\
            & \qquad \qquad \cdots \gsome{\linvar{x}}{\llfv{M_l} }; \gname{x}{y_l}; \gsome{\linvar{x}}{y_l, \llfv{M_l}}; \psome{\linvar{x}}; \pname{\linvar{x}}{z_l}; \cdots\\
            & \qquad \qquad  ( \gsome{z_l}{\llfv{M_l}};  \piencodf{M_l}_{z_l} \| \pnone{y_l}  \| \\
            & \qquad \qquad  \gsome{\linvar{x}}{\emptyset};\gname{x}{y_{l+1}};  ( \psome{ y_{l+1}}; \pclose{y_{l+1}}  \| \gsome{\linvar{x}}{\emptyset}; \pnone{\linvar{x}} )
            ) \cdots
            )
            )
            \|
            \guname{ \unvar{x} }{ x_i } ;  \piencodf{ U }_{x_i}
            )
\\
 & \redone^{6}  \res{ \unvar{x} } (
              \res{z_1} ( \gsome{z_1}{\llfv{M_1}};  \piencodf{M_1}_{z_1} \| \cdots  \res{z_k} ( \gsome{z_k}{\llfv{M_k}};  \piencodf{M_k}_{z_k} \| \\
            & \qquad \qquad   \pnone{u} \| \pnone{(\llfv{M'} \setminus \widetilde{x})} \|  \pnone{z_1} \| \cdots \| \pnone{z_k} \| \pnone{( \llfv{C} \setminus (M_1 , \cdots , M_k) )}  ) \cdots   )
            \\
            & \qquad \qquad \|
            \guname{ \unvar{x} }{ x_i } ;  \piencodf{ U }_{x_i}
            ) \\
  & \redone^{k}   \pnone{ u }     \|   \pnone{ (\llfv{M'} \setminus \{  x_1, \cdots , x_k \} ) }    \|   \pnone{ \llfv{C} }  \\
 &= \piencodf{  \fail^{\widetilde{z}}}_u = Q_{7l + 10}  \\
 \end{aligned}
 \]
The rest of the proof is by induction on the number of reductions $\piencodf{N}_u \redone^j Q$.

                            \begin{enumerate}
                                \item When $j = 0$, the result follows trivially. Just take $I = \{a\}$ $\mathbb{N}=\mathbb{N}'$ and $\piencodf{N}_u=Q=Q_a$.
  \item When $1 \leq j \leq 7k + 10$.

Let $I = \{ a\}$ and  $Q_j$ be such that $ \piencodf{N}_u \redone^j Q_j$.
By the steps above one has

\(\begin{aligned}
  &Q_j \redone^{7k + 10 - j} Q_{7k + 10} = Q_a,\\ &N \red^1 \fail^{\widetilde{z}} = N';\text{ and} \piencodf{ \fail^{\widetilde{z}}}_u = Q_{7k + 10}.
\end{aligned}
\)
\item When $j > 7k + 10$.

In this case, we have
$ \piencodf{N}_u \redone^{7k + 10} Q_{7k + 10} \redone^n Q,$ for $n \geq 1$.
We also know that

$N \red^1 \fail^{\widetilde{z}}$. However no further reductions can be performed.

                            \end{enumerate}

                \item When $\size{\widetilde{x}} < \size{C}$, the proof  proceeds similarly to the previous case.

            \end{enumerate}

        \item  $N =   M'\linexsub{C /  x_1 , \cdots , x_k} $.

           In this case we let $C = \bag{M_1} \cdot \cdots \cdot \bag{M_k}$,
            \[
            \begin{aligned}
               \piencodf{ M'\linexsub{C /  x_1 , \cdots , x_k} }_u =&  \res{z_1}( \gsome{z_1}{\llfv{M_{1}}};\piencodf{ M_{1} }_{ {z_1}}  \|  \cdots \res{z_k} ( \gsome{z_k}{\llfv{M_{k}}};\piencodf{ M_{k} }_{ {z_k}} \\
                & \qquad \qquad  \| \bignd_{x_{i_1} \in \{ x_1 ,\cdots , x_k  \}} \cdots \bignd_{x_{i_k} \in \{ x_1 ,\cdots , x_k \setminus x_{i_1} , \cdots , x_{i_{k-1}}  \}}\\
                & \qquad \qquad \piencodf{ M'}_u \{ z_1 / x_{i_1} \} \cdots \{ z_k / x_{i_k} \} ) \cdots ) \\
            \end{aligned}
            \]
            Therefore,
            \[
            \begin{aligned}
               \piencodf{N}_u & =   \res{z_1}( \gsome{z_1}{\llfv{M_{1}}};\piencodf{ M_{1} }_{ {z_1}}  \|
              \cdots \res{z_k} ( \gsome{z_k}{\llfv{M_{k}}};\piencodf{ M_{k} }_{ {z_k}} \\
                & \qquad \qquad  \| \bignd_{x_{i_1} \in \{ x_1 ,\cdots , x_k  \}} \cdots \bignd_{x_{i_k} \in \{ x_1 ,\cdots , x_k \setminus x_{i_1} , \cdots , x_{i_{k-1}}  \}} \piencodf{ M'}_u \{ z_1 / x_{i_1} \} \cdots \{ z_k / x_{i_k} \} ) \cdots ) \\
               & \redone^m   \res{z_1}( \gsome{z_1}{\llfv{M_{1}}};\piencodf{ M_{1} }_{ {z_1}}  \| \cdots \res{z_k} ( \gsome{z_k}{\llfv{M_{k}}};\piencodf{ M_{k} }_{ {z_k}} \\
                & \qquad \qquad  \| \bignd_{x_{i_1} \in \{ x_1 ,\cdots , x_k  \}} \cdots \bignd_{x_{i_k} \in \{ x_1 ,\cdots , x_k \setminus x_{i_1} , \cdots , x_{i_{k-1}}  \}} R \{ z_1 / x_{i_1} \} \cdots \{ z_k / x_{i_k} \} ) \cdots ) \\
               & \redone^n  Q\\
            \end{aligned}
            \]

            for some process $R$. Where $\redone^n$ is a reduction that  initially synchronizes with $  \gsome{ {x}_i }{ \llfv{M_{J_i}} } $ for some $ i \in \{ 1 , \cdots , k \}$, when $n \geq 1$, $n + m = k \geq 1$. Type preservation in \clpi ensures reducing $\piencodf{ M'}_v \redone^m$ does not consume possible synchronizations with $ \psome{{x}_i} $, if they occur. Let us consider the the possible sizes of both $m$ and $n$.

            \begin{enumerate}
                \item For $m = 0$ and $n \geq 1$.

                    We have that $R = \piencodf{M'}_u$ as $\piencodf{M'}_u \redone^0 \piencodf{M'}_u$.

                    Notice that there are two possibilities of having an unguarded $\psome{x_i}$ or $\pnone{ x_i }$ without internal reductions:

                    \begin{enumerate}
                        \item $M'= \fail^{ {x}_i, \widetilde{y}}$.
    \[
    \begin{aligned}
  \piencodf{N}_u & =
                \res{z_1}( \gsome{z_1}{\llfv{M_{1}}};\piencodf{ M_{1} }_{ {z_1}}  \|  \cdots \res{z_k} ( \gsome{z_k}{\llfv{M_{k}}};\piencodf{ M_{k} }_{ {z_k}} \\
                & \qquad \qquad  \| \bignd_{x_{i_1} \in \{ x_1 ,\cdots , x_k  \}} \cdots \bignd_{x_{i_k} \in \{ x_1 ,\cdots , x_k \setminus x_{i_1} , \cdots , x_{i_{k-1}}  \}} \piencodf{ M'}_u \{ z_1 / x_{i_1} \} \cdots \{ z_k / x_{i_k} \} ) \cdots ) \\
                & = \res{z_1}( \gsome{z_1}{\llfv{M_{1}}};\piencodf{ M_{1} }_{ {z_1}}  \|  \cdots \res{z_k} ( \gsome{z_k}{\llfv{M_{k}}};\piencodf{ M_{k} }_{ {z_k}} \\
                & \qquad \qquad  \| \bignd_{x_{i_1} \in \{ x_1 ,\cdots , x_k  \}} \cdots \bignd_{x_{i_k} \in \{ x_1 ,\cdots , x_k \setminus x_{i_1} , \cdots , x_{i_{k-1}}  \}} \piencodf{ \fail^{ {x}_i, \widetilde{y}}}_u \{ z_1 / x_{i_1} \} \cdots \{ z_k / x_{i_k} \} ) \cdots ) \\
                & = \res{z_1}( \gsome{z_1}{\llfv{M_{1}}};\piencodf{ M_{1} }_{ {z_1}}  \| \cdots \res{z_k} ( \gsome{z_k}{\llfv{M_{k}}};\piencodf{ M_{k} }_{ {z_k}} \\
                & \qquad \qquad  \| \bignd_{x_{i_1} \in \{ x_1 ,\cdots , x_k  \}} \cdots \bignd_{x_{i_k} \in \{ x_1 ,\cdots , x_k \setminus x_{i_1} , \cdots , x_{i_{k-1}}  \}} \\
                & \qquad \qquad \qquad
                \pnone{ u }    \|   \pnone{ {x}_i }     \|    \pnone{ \widetilde{y} }
                \{ z_1 / x_{i_1} \} \cdots \{ z_k / x_{i_k} \} ) \cdots ) \\
    \end{aligned}
    \]

    by type preservation we have that $\widetilde{y} = \{x_1 , \cdots , x_k \} \setminus x_i , \widetilde{y}' $ for some $\widetilde{y}'$

    \[
    \begin{aligned}
    & =\res{z_1}( \gsome{z_1}{\llfv{M_{1}}};\piencodf{ M_{1} }_{ {z_1}}  \|  \cdots \res{z_k} ( \gsome{z_k}{\llfv{M_{k}}};\piencodf{ M_{k} }_{ {z_k}} \\
                & \qquad \qquad  \| \bignd_{x_{i_1} \in \{ x_1 ,\cdots , x_k  \}} \cdots \bignd_{x_{i_k} \in \{ x_1 ,\cdots , x_k \setminus x_{i_1} , \cdots , x_{i_{k-1}}  \}}
                \pnone{ u }   \|    \pnone{ \widetilde{y}' } \\
                & \qquad \qquad \|   \pnone{ \{ x_1 , \cdots , x_k  \} }
                \{ z_1 / x_{i_1} \} \cdots \{ z_k / x_{i_k} \} ) \cdots ) \\
    & =\res{z_1}( \gsome{z_1}{\llfv{M_{1}}};\piencodf{ M_{1} }_{ {z_1}}  \| \cdots \res{z_k} ( \gsome{z_k}{\llfv{M_{k}}};\piencodf{ M_{k} }_{ {z_k}} \\
                & \qquad \qquad  \| \bignd_{x_{i_1} \in \{ x_1 ,\cdots , x_k  \}} \cdots \bignd_{x_{i_k} \in \{ x_1 ,\cdots , x_k \setminus x_{i_1} , \cdots , x_{i_{k-1}}  \}}
                \pnone{ u }   \|    \pnone{ \widetilde{y}' } \\
                & \qquad \qquad \|   \pnone{ \{ z_1 , \cdots , z_k  \} } ) \cdots ) \\
    &\redone^k   \pnone{ u }    \|   \pnone{ \widetilde{y}' }    \|   \pnone{ \llfv{C} }  \\
  \end{aligned}
    \]
  Notice that no further reductions can be performed.
  Thus we take $I = \{ a \}$ and,
 $$ \piencodf{N}_u \redone   \pnone{ u }   \|   \pnone{ \widetilde{y} }    \|   \pnone{ \lfv{C} }  = Q_a.$$

We also have that  $N \red \fail^{ \widetilde{y} \cup \llfv{C} } = N'$ and $\piencodf{ \fail^{\widetilde{y} \cup \llfv{C}} }_u = Q_a$.

                        \item $\headf{M'} =  {x}_i$ with $i \in \{ 1 , \cdots , k \}$.

 Then we have the following

 \[
    \begin{aligned}
  \piencodf{N}_u
  & =
                \res{z_1}( \gsome{z_1}{\llfv{M_{1}}};\piencodf{ M_{1} }_{ {z_1}}  \| \cdots \res{z_k} ( \gsome{z_k}{\llfv{M_{k}}};\piencodf{ M_{k} }_{ {z_k}} \\
                & \qquad \qquad  \| \bignd_{x_{i_1} \in \{ x_1 ,\cdots , x_k  \}} \cdots \bignd_{x_{i_k} \in \{ x_1 ,\cdots , x_k \setminus x_{i_1} , \cdots , x_{i_{k-1}}  \}} \piencodf{ M'}_u \{ z_1 / x_{i_1} \} \cdots \{ z_k / x_{i_k} \} ) \cdots ) \\
  & =
                \res{z_1}( \gsome{z_1}{\llfv{M_{1}}};\piencodf{ M_{1} }_{ {z_1}}  \|  \cdots \res{z_k} ( \gsome{z_k}{\llfv{M_{k}}};\piencodf{ M_{k} }_{ {z_k}} \\
                & \qquad \qquad  \| \bignd_{x_{i_1} \in \{ x_1 ,\cdots , x_k  \}} \cdots \bignd_{x_{i_k} \in \{ x_1 ,\cdots , x_k \setminus x_{i_1} , \cdots , x_{i_{k-1}}  \}}
                \res{ \widetilde{y} } (\piencodf{  {x_i} }_{j}   \| P)
                \{ z_1 / x_{i_1} \} \cdots \{ z_k / x_{i_k} \} ) \cdots ) \\
    & =
                \res{z_1}( \gsome{z_1}{\llfv{M_{1}}};\piencodf{ M_{1} }_{ {z_1}}  \| \cdots \res{z_k} ( \gsome{z_k}{\llfv{M_{k}}};\piencodf{ M_{k} }_{ {z_k}} \\
                & \qquad \qquad  \| \bignd_{x_{i_1} \in \{ x_1 ,\cdots , x_k  \}} \cdots \bignd_{x_{i_k} \in \{ x_1 ,\cdots , x_k \setminus x_{i_1} , \cdots , x_{i_{k-1}}  \}}\\
                & \qquad \qquad
                \res{ \widetilde{y} } (  \psome{{x_i}}; \pfwd{x_i}{j}    \| P)
                \{ z_1 / x_{i_1} \} \cdots \{ z_k / x_{i_k} \} ) \cdots ) \\
    \end{aligned}
    \]
 Let us consider a arbitrary sum where $x_{i_k} = x_i$ , other cases follow similarly.

 \[
 \begin{aligned}
    & =
                \res{z_1}( \gsome{z_1}{\llfv{M_{1}}};\piencodf{ M_{1} }_{ {z_1}}  \| \cdots \res{z_k} ( \gsome{z_k}{\llfv{M_{k}}};\piencodf{ M_{k} }_{ {z_k}} \\
                & \qquad \qquad  \| \bignd_{x_{i_1} \in \{ x_1 ,\cdots , x_k  \}} \cdots \bignd_{x_{i_k} \in \{ x_1 ,\cdots , x_k \setminus x_{i_1} , \cdots , x_{i_{k-1}}  \}} \\
                & \qquad \qquad
                \res{ \widetilde{y} } (  \psome{{x_i}}; \pfwd{x_i}{j}    \| P)
                \{ z_1 / x_{i_1} \} \cdots \{ z_k / x_{i_k} \} ) \cdots ) \\
     & \redone
     \res{z_1}( \gsome{z_1}{\llfv{M_{1}}};\piencodf{ M_{1} }_{ {z_1}}  \| \cdots \res{z_k} ( \piencodf{ M_{k} }_{ {z_k}} \\
                & \qquad \qquad  \|
                \res{ \widetilde{y} } (  \ \pfwd{x_k}{j}    \| P)
                \{ z_1 / x_{i_1} \} \cdots \{ z_{k-1} / x_{i_{k-1}} \} ) \cdots ) & = Q_1\\
    & \redone
     \res{z_1}( \gsome{z_1}{\llfv{M_{1}}};\piencodf{ M_{1} }_{ {z_1}}  \| \cdots \res{z_{k-1}}( \gsome{z_{k-1}}{\llfv{M_{{k-1}}}};\piencodf{ M_{{k-1}} }_{ {z_{k-1}}}    \\
                & \qquad \qquad \| \res{ \widetilde{y} } (  \piencodf{ M_{k} }_{j}\    \| P)
                \{ z_1 / x_{i_1} \} \cdots \{ z_{k-1} / x_{i_{k-1}} \} ) \cdots ) & = Q_2\\
 \end{aligned}
 \]

In addition,
\(
N = M'\linexsub{C /  x_1 , \cdots , x_k} \redone  M'\headlin{ C_i / x_k }  \linexsub{(C \setminus C_i ) /  x_1 , \cdots , x_{k-1}  } = M\).
Finally,
\[
\begin{aligned}
    \piencodf{M}_u &= \piencodf{ M'\headlin{ C_i / x_k }  \linexsub{(C \setminus C_i ) /  x_1 , \cdots , x_{k-1}  } }_u \\
    &=  \res{z_1}( \gsome{z_1}{\llfv{M_{1}}};\piencodf{ M_{1} }_{ {z_1}}  \|  \cdots \res{z_{k-1}}( \gsome{z_{k-1}}{\llfv{M_{{k-1}}}};\piencodf{ M_{{k-1}} }_{ {z_{k-1}}}  \|  \\
                & \qquad \qquad \bignd_{x_{i_1} \in \{ x_1 ,\cdots , x_{k-1}  \}} \cdots \bignd_{x_{i_{k-1}} \in \{ x_1 ,\cdots , x_{k-1} \setminus x_{i_1} , \cdots , x_{i_{k-2}}  \}} \\
                & \qquad \qquad \res{ \widetilde{y} } (  \piencodf{ M_{k} }_{j}\    \| P)
                \{ z_1 / x_{i_1} \} \cdots \{ z_{k-1} / x_{i_{k-1}} \} ) \cdots )\\
    & \premat Q_2.
\end{aligned}
\]

\begin{enumerate}
\item When $n = 1$:

Then, $I = \{ a \} $, $Q = Q_1$ and  $ \piencodf{N}_u \redone^1 Q_1$. Also,

$Q_1 \redone^1 Q_2 = Q_a$, $N \red^1  M'\headlin{ C_i / x_k }  \linexsub{(C \setminus C_i ) /  x_1 , \cdots , x_{k-1}  } = N'$ and

$\piencodf{M'\headlin{ C_i / x_k }  \linexsub{(C \setminus C_i ) /  x_1 , \cdots , x_{k-1}  }}_u \premat Q_a$.

\item When $n = 2$:

Then, $I = \{ a \} $, $Q = Q_2$ and  $ \piencodf{N}_u \redone^2 Q_2$. Also,

$Q_2 \redone^0 Q_2 = Q_a$, $N \red^1  M'\headlin{ C_i / x_k }  \linexsub{(C \setminus C_i ) /  x_1 , \cdots , x_{k-1}  } = N'$ and

$\piencodf{M'\headlin{ C_i / x_k }  \linexsub{(C \setminus C_i ) /  x_1 , \cdots , x_{k-1}  }}_u \premat Q_a$.

\item When $n > 2$:

Then  $ \piencodf{N}_u \redone^2 Q_2 \redone^l Q$, for $l \geq 1$.  Also,
$N \rightarrow^1 M$, $ \piencodf{M}_u \premat Q_2 $.
$\exists  \{P_j\}_{ j \in J}, N', \ s.t. M \red^* N', $ $ \ \piencodf{M} \redone^l P \redone^* \{P_j\}_{j \in J}  $ and
$ \piencodf{N'}_u \premat P_j \quad \forall j \in J $. We also have by Prop.\ref{prop:soundextra} that as $\piencodf{M} \premat Q_{2} $ and $\piencodf{M} \redone^* \{ Q_j \}_{j \in J}  $ implies $\exists \{ Q_i \}_{i \in I} \ s.t. \  Q_{2} \redone^* \{Q_i\}_{i \in I} $ , $ I \\subseteq J $ and $P_i \premat Q_i \  , \ \forall i \in I$

                            \end{enumerate}

                    \end{enumerate}
 \item  For $m \geq 1$ and $ n \geq 0$.

            \begin{enumerate}
            \item When $n = 0$.

               Then $   \res{ {x} }   ( R   \|   \gsome{ {x} }{ \llfv{N'} }; \piencodf{ N' }_{ {x}} )  = Q$ and $\piencodf{M'}_u \redone^m R$ where $m \geq 1$. By the IH $\exists  \{R_i\}_{ i \in I}, $ $ M'', \ s.t.\ M'\red^* M'', \  R \redone^* \{R_i\}_{i \in I}  $ and

               $ \piencodf{M''}_u \premat R_i \quad \forall i \in I $. Thus,
              \[
               \begin{aligned}
                   \piencodf{N}_u
                   & =   \res{z_1}( \gsome{z_1}{\llfv{M_{1}}};\piencodf{ M_{1} }_{ {z_1}}  \|  \cdots \res{z_k} ( \gsome{z_k}{\llfv{M_{k}}};\piencodf{ M_{k} }_{ {z_k}} \\
                & \qquad \qquad  \| \bignd_{x_{i_1} \in \{ x_1 ,\cdots , x_k  \}} \cdots \bignd_{x_{i_k} \in \{ x_1 ,\cdots , x_k \setminus x_{i_1} , \cdots , x_{i_{k-1}}  \}} \piencodf{ M'}_u \{ z_1 / x_{i_1} \} \cdots \{ z_k / x_{i_k} \} ) \cdots ) \\
               & \redone^m   \res{z_1}( \gsome{z_1}{\llfv{M_{1}}};\piencodf{ M_{1} }_{ {z_1}}  \|  \cdots \res{z_k} ( \gsome{z_k}{\llfv{M_{k}}};\piencodf{ M_{k} }_{ {z_k}} \\
                & \qquad \qquad  \| \bignd_{x_{i_1} \in \{ x_1 ,\cdots , x_k  \}} \cdots \bignd_{x_{i_k} \in \{ x_1 ,\cdots , x_k \setminus x_{i_1} , \cdots , x_{i_{k-1}}  \}} R \{ z_1 / x_{i_1} \} \cdots \{ z_k / x_{i_k} \} ) \cdots ) \\
               & \redone^n  Q\\
                \end{aligned}
                 \]
                Also,
                \(
               Q  \redone^*  \{ \res{  {x} }   ( R_i   \|  \gsome{ {x} }{ \llfv{N'} };  \piencodf{ N' }_{ {x}} ) \}_{i \in I} = Q_i,
                \)
                and the term can reduce as follows:

                $N = M'\linexsub{C /  x_1 , \cdots , x_k} \red^* M'' \linexsub{C /  x_1 , \cdots , x_k} = N'$ and  $\piencodf{N'}_u \premat Q_i$ $\forall i \in I$

            \item When $n \geq 1$.
                Then  $R$ has an occurrence of an unguarded $\psome{x}$ or $  \pnone{ x } $, this case follows by IH and applying Proposition \ref{prop:soundextra}.

                    \end{enumerate}
            \end{enumerate}

            \item  $N =  M'\unexsub{U / \unvar{x}}$.

            In this case,
            \(
            \begin{aligned}
                \piencodf{M'\unexsub{U / \unvar{x}}}_u &=   \res{ \unvar{x} }  ( \piencodf{ M'}_u   \|    \guname{ \unvar{x} }{ x_i } ; \piencodf{ U }_{x_i} ).
            \end{aligned}
            \)
            Then,
            \[
            \begin{aligned}
               \piencodf{N}_u & =   \res{ \unvar{x} }  ( \piencodf{ M'}_u   \|    \guname{ \unvar{x} }{ x_i } ; \piencodf{ U }_{x_i} )  \redone^m   \res{ \unvar{x} }  ( R   \|   \guname{ \unvar{x} }{ x_i } ;  \piencodf{ U }_{x_i} ) \redone^n  Q.
            \end{aligned}
            \]
            for some process $R$. Where $\redone^n$ is a reduction initially synchronises with $ \guname{ \unvar{x} }{ x_i }$ when $n \geq 1$, $n + m = k \geq 1$. Type preservation in \clpi ensures reducing $\piencodf{ M'}_v \redone^m$ doesn't consume possible synchronisations with $ \guname{ \unvar{x} }{ x_i } $ if they occur. Let us consider the the possible sizes of both $m$ and $n$.

            \begin{enumerate}
                \item For $m = 0$ and $n \geq 1$.

                   In this case,  $R = \piencodf{M'}_u$ as $\piencodf{M'}_u \redone^0 \piencodf{M'}_u$.

                    Notice that the only possibility of having an unguarded $ \puname{ \unvar{x} }{ x_i }$ without internal reductions is when   $\headf{M'} =  {x}[ind].$
                           By the diamond property  we will be reducing each non-deterministic choice of a process simultaneously.
                            Then we have the following:

                            \[
                            \begin{aligned}
                            \piencodf{N}_u  & =   \res{ \unvar{x} }  (  \res{ \widetilde{y} }  (\piencodf{  {x}[ind] }_{j}   \| P)   \|    \guname{ \unvar{x} }{ x_i } ; \piencodf{ U }_{x_i} ) \\
                            & =   \res{ \unvar{x} }   (  \res{ \widetilde{y} }  ( \puname{ \unvar{x} }{ x_i }; \psel{ {x}_i }{ ind }; \pfwd{x_i}{j}    \| P)   \|    \guname{ \unvar{x} }{ x_i } ; \piencodf{ U }_{x_i}  ) \\
                            & \redone   \res{ \unvar{x} }  (  \res{ \widetilde{y} }  (  \res{ x_i }  (\psel{ {x}_i }{ ind }; \pfwd{x_i}{j}   \| \piencodf{ U }_{x_i} )   \| P)   \|   \guname{ \unvar{x} }{ x_i } ;  \piencodf{ U }_{x_i}  ) &= Q_1 \\
                            & =   \res{ \unvar{x} }   (  \res{ \widetilde{y} }  (  \res{ x_i }   (\psel{ {x}_i }{ ind };\pfwd{x_i}{j}    \| x_i. case( ind.\piencodf{U_{ind}}_{x_i} ) )   \| P) \\
                            &  \|   \guname{ \unvar{x} }{ x_i } ; \piencodf{ U }_{x_i}  ) \\
                            & \redone   \res{ \unvar{x} }   (  \res{ \widetilde{y} } (  \res{ x_i } (\pfwd{x_i}{j}   \| \piencodf{U_{ind}}_{x_i} )   \| P)   \|    \guname{ \unvar{x} }{ x_i } ; \piencodf{ U }_{x_i}  ) &= Q_2\\
                            & \redone   \res{ \unvar{x} }   (  \res{ \widetilde{y} }  ( \piencodf{U_{ind}}_{j}    \| P)   \|  \guname{ \unvar{x} }{ x_i } ;   \piencodf{ U }_{x_i}  ) &= Q_3 \\
                            \end{aligned}
                            \]

                We consider the two cases of the form of $U_{ind}$ and show that the choice of $U_{ind}$ is inconsequential

                \begin{itemize}
                \item When $ U_{ind} = \unvar{\bag{N}}$:

                In this case,
                \(
                \begin{aligned}
                N &=M'\unexsub{U / \unvar{x}} \red M'\headlin{ N /\unvar{x} }\unexsub{U / \unvar{x}} = M.
                \end{aligned}
                \)
                 and
                \[
                 \begin{aligned}
                 \piencodf{M}_u= \piencodf{M'\headlin{ N /\unvar{x} }\unexsub{U / \unvar{x}}}_u &=   \res{ \unvar{x} }  ( \res{ \widetilde{y} }  ( \piencodf{\bag{N}}_{j}    \| P)   \|   \guname{ \unvar{x} }{ x_i } ; \piencodf{ U }_{x_i}  ) & = Q_3
                                            \end{aligned}
                 \]

                \item When $ U_i = \unvar{\oneb} $:

                  In this case,
                        \(
                        \begin{aligned}
                            N &=M'\unexsub{U / \unvar{x}} \red M'\headlin{ \fail^{\emptyset} /\unvar{x} } \unexsub{U /\unvar{x} } = M.
                        \end{aligned}
                        \)

                        Notice that $\piencodf{\unvar{\oneb}}_{j} =    \pnone{ j } $ and that $\piencodf{\fail^{\emptyset}}_j =   \pnone{ j } $. In addition,

                            \[
                            \begin{aligned}
                               \piencodf{M}_u&= \piencodf{M'\headlin{ \fail^{\emptyset} /\unvar{x} } \unexsub{U /\unvar{x} }}_u\\
                               &=   \res{ \unvar{x} }   (  \res{ \widetilde{y} } ( \piencodf{\fail^{\emptyset}}_{j}    \| P)   \|   \guname{ \unvar{x} }{ x_i } ;  \piencodf{ U }_{x_i}  ) \\
                               & =  \res{ \unvar{x} }   (  \res{ \widetilde{y} } ( \piencodf{\unvar{\oneb}}_{j}    \| P)   \|   \guname{ \unvar{x} }{ x_i } ;  \piencodf{ U }_{x_i}  ) & = Q_3
                            \end{aligned}
                            \]
                \end{itemize}

                Both choices give an $M$ that are equivalent to $Q_3$.

    \begin{enumerate}
    \item When $n \leq 2$.

   In this case, $Q = Q_n$ and  $ \piencodf{N}_u \redone^n Q_n$.

Also, $Q_n \redone^{3-n} Q_3 = Q'$, $N \red^1 M = N'$ and $\piencodf{M }_u = Q_2$.

\item When $n \geq 3$.

We have $ \piencodf{N}_u \redone^3 Q_3 \redone^l Q$ for $l \geq 0$. We also know that $N \rightarrow M$, $Q_3 = \piencodf{M}_u$.  By the IH, there exist $\{ Q_i \}_{i \in I}$ , $ N'$ such that $Q \redone^* \{Q_i\}_{i \in I}$, $M \red^* N'$ and $\piencodf{N'}_u \premat Q_i \quad \forall i \in I $ . Finally, $\piencodf{N}_u \redone^3 Q_3 \redone^l Q \redone \{Q_i\}_{i \in I}$ and $N \red M  \red^* N'$.

\end{enumerate}
 \item For $m \geq 1$ and $ n \geq 0$.

{\bf (b.1) Case $n = 0$.}

Then $    \res{ \unvar{x} }   ( R   \|   \guname{ \unvar{x} }{ x_i } ; \piencodf{ U }_{x_i} )  = Q$ and $\piencodf{M'}_u \redone^m R$ where $m \geq 1$. Then by the IH there exist $\{R_i'\}_{i \in I}$  and $M'' $ such that $R \redone^* \{R_i'\}_{i \in I}$, $M'\red^* M''$, and $\piencodf{M''}_u \premat R_i \quad \forall i \in I $.  Hence we have that

\[
\begin{aligned}
\piencodf{N}_u & =   \res{ \unvar{x} }  ( \piencodf{M'}_u   \|   \guname{ \unvar{x} }{ x_i } ; \piencodf{ U }_{x_i} ) \redone^m   \res{ \unvar{x} }  ( R   \|   \guname{ \unvar{x} }{ x_i } ; \piencodf{ U }_{x_i} )  = Q
\end{aligned}
\]
We also know that
\[
\begin{aligned}
Q & \redone^*  \{ \res{ \unvar{x} } ( R'_i   \|    \guname{ \unvar{x} }{ x_i } ; \piencodf{ U }_{x_i} )\}_{i \in I} = \{ Q_i\}_{i \in I} \\
\end{aligned}
\]
and so the \lamcoldetsh term can reduce as follows: $N =  M'\unexsub{U / \unvar{x}} \red^*  M'' \unexsub{U / \unvar{x}} = N'$ and  $\piencodf{N'}_u \premat Q_i \quad \forall i \in I$ via the $\premat$ rules

{\bf (b.2) Case  $n \geq 1$.}

Then $R$ has an occurrence of an unguarded $\puname{ \unvar{x} }{ x_i }$, and the case follows by IH.
\qedhere
\end{enumerate}
\end{enumerate}
\end{proof}

\thmEncLWSound*

\begin{proof}
Immediate from \Cref{thm:opsoundone}.
\end{proof}

\subsection{Success Sensitivity}\label{a:loosuccess}

\begin{restatable}[Success Sensitivity (Under $\redone$)]{theorem}{thmEncEagerSucc}\label{proof:successsenscetwounres}
    $\succp{{M}}{\sucs{\lambda}}$ iff ${\piencodf{{M}}_u}\succone{\sucs{\pi}}$ for well-formed closed terms $M$.
\end{restatable}

\begin{proof}
We proceed with the proof in two parts.

\begin{enumerate}

    \item Suppose that  ${M} \Downarrow_{\checkmark} $. We will prove that $\piencodf{{M}} \succone_{\checkmark}$.

    By \defref{def:app_Suc3unres}, there exists  $ {M}' $ such that $ M \red^* {M}'$ and
    $\headf{M'} = \checkmark$.
    By completeness, if $ M\red M'$ then there exist $Q, Q'$ such that $\piencodf{M}_u \equiv Q \redone^* Q'$ and $\piencodf{M'}_u \premat Q' $.

    We wish to show that there exists $Q''$such that $Q' \redone^* Q''$ and $Q''$ has an unguarded occurrence of $\checkmark$.

    By Proposition \ref{Prop:checkprespiunres} (1) we have that $\headf{M'} = \checkmark \implies \piencodf{M'}_u =   \res{ \widetilde{x} } (P   \| \checkmark)$. Finally $\piencodf{M'}_u  =   \res{ \widetilde{x} }  (P   \| \checkmark) \premat Q'$ hence $Q$  must be of the form $ \res{ \widetilde{x} } (P'   \| \checkmark) $ where $P \premat P' $ . Hence $Q$ reduces to a process that has an unguarded occurence of $\checkmark$.

    \item Suppose that $\piencodf{{M}}_u \succone_{\checkmark}$. We will prove that $ {M} \Downarrow_{\checkmark}$.

    By operational soundness we have that if  $\piencodf{{M}}_u \redone^* Q$ then there exist ${M}'$ and $\{Q_i\}_{i \in I}$ such that:
    (i)~${M} \red^* {M}'$
    and
    (ii)~$Q \redone^* \{Q_i\}_{i \in I}$ with
    $ \piencodf{{M}'}_u \premat Q_j$,  for all $j \in I$.

   Since $\piencodf{{M}}_u \redone^* P_1 \nd \ldots \nd P_k (:= Q)$, and $P_j= P_j'   \| \checkmark$ and $Q \redone^* \{Q_i\}_{i \in I}$ we must have that each $Q_i$ is of the form $P'_{i_1} \nd \ldots \nd P'_{i_k}$ with $P'_{i_j}= P''_{i_j}   \| \checkmark$. As $ \piencodf{{M}'}_u \premat Q_j$,  for all $j \in I$ we have that $ \piencodf{{M}'}_u  = (\bignd_{i \in I} Q_i  ) \nd R$ for some $R$. Finally applying Proposition \ref{Prop:checkprespiunres} (2) we have that $\piencodf{{M}'}_u$ is itself a term with unguarded $\checkmark$, then ${M}$ is itself headed with $\checkmark$.
    \qedhere
\end{enumerate}
\end{proof}

%% file: appendix/pi-comparison.tex
\section{Proof of Separation of Lazy and Eager Semantics}
\label{a:piBisim}

\begin{definition}[Dual Prefix]\label{d:readyPrefixDual}
    Given prefixes $\alpha$ and $\beta$ (\Cref{d:prefix}), we say $\alpha$ and $\beta$ are duals, denoted $\alpha \dualPrefix \beta$, if and only if
    $\sub{\alpha} \cap \sub{\beta} \neq \emptyset$.
\end{definition}

\begin{lemma}\label{l:readyPrefixDual}
    Given $P \vdash \emptyset$, if $P \readyPrefix{\alpha}$, then there exist $P',\beta$ such that $P \redone^\ast P' \readyPrefix{\beta}$ and $\alpha \dualPrefix \beta$.
\end{lemma}

\begin{proof}
    By well-typedness, there appears $\beta$ in $P$ with $\alpha \dualPrefix \beta$.
    However, we may have $P \nreadyPrefix{\beta}$, because $\beta$ is blocked by other prefixes.
    Hence, we need to find reductions from $P$ such that we unblock $\beta$.
    However, the prefixes blocking $\beta$ are connected to dual prefixes, that may be blocked themselves.
    The crux of this proof is thus to show that we can reduce $P$, such that we eventually unblock $\beta$.

    The proof is by induction on the number of names that may block $\beta$ (\ih{1}).
    Initially, this number corresponds to the total number of names appearing in $P$.
    Suppose $\beta$ is blocked by $n$ prefixes $\gamma_i$, where $\gamma_n$ blocks $\beta$, and $\gamma_1$ is not blocked.
    We apply another layer of induction on $n$ (\ih{2}).

    In the inductive case, $n \geq 1$.
    The goal is to perform a reduction that synchronizes $\gamma_1$ with its dual, say $\ol{\gamma_1}$.
    The prefix $\ol{\gamma_1}$ may be blocked by a number of prefixes itself.
    However, the type system of \clpi is based on \scc{Cut}, so $\ol{\gamma_1}$ appears in parallel with the duals of $\gamma_2,\ldots,\gamma_n$ and $\alpha$.
    We then may apply \ih{1} to find $P \redone^\ast P_0 \readyPrefix{\ol{\gamma_1}}$.
    We can then reduce $P_0$ by synchronizing between $\gamma_1$ and $\ol{\gamma_1}$: $P \redone^\ast P_0 \redone P_1$.
    In $P_1$, $\beta$ is blocked by one less prefix.
    Hence, by \ih{2}, $P \redone^\ast P_0 \redone P_1 \redone^\ast P' \readyPrefix{\beta}$, proving the thesis.

    In the base case, $\beta$ is not blocked: $P \readyPrefix{\beta}$.
    Let $P' := P$; trivially, $P \redone^\ast P' \readyPrefix{\beta}$, proving the thesis.
\end{proof}

\thmPiBisim*

\begin{proof}
    For~(i), we construct a relation $\mathbb{B}$ as follows:
    \begin{align*}
        \sff{Id}^\equiv
        &:=
        \{ (T,U) \mid T \equiv U \}
        \\
        \mathbb{B}'
        &:=
        \{
            (T,U) \mid \begin{array}[t]{@{}l@{}}
                T \equiv \pctx{M}[\beta_1; (V \nd W)] \vdash \emptyset
                \text{ and}
                \\
                U \equiv \pctx{M}[\beta_2; V \nd \beta_3; W] \vdash \emptyset
                \text{ and}
                \\
                V \nreadyPrefixBisim{L} W
                \text{ and}
                \\
                \beta_1 \relalpha \beta_2 \relalpha \beta_3
                \text{ and}
                \\
                \text{$\beta_1,\beta_2,\beta_3$ require a continuation}
            \}
        \end{array}
        \\
        \mathbb{B}
        &:=
        \sff{Id}^\equiv \cup \mathbb{B}'
    \end{align*}

    We prove that $\mathbb{B}$ is a strong ready-prefix bisimulation w.r.t.\ the lazy semantics by proving the three conditions of \Cref{d:readyPrefixBisim} for each $(T,U) \in \mathbb{B}$.
    We distinguish cases depending on whether $(T,U) \in \sff{Id}^\equiv$ or $(T,U) \in \mathbb{B}'$.
    \begin{itemize}
        \item
            $(T,U) \in \sff{Id}^\equiv$.
            The three conditions hold trivially.

        \item
            $(T,U) \in \mathbb{B}'$.
            Then $T \equiv \pctx{M}[\beta_1; (V \nd W)]$, $U \equiv \pctx{M}[\beta_2; V \nd \beta_3; W]$, $V \nreadyPrefixBisim{L} W$, and $\beta_1 \relalpha \beta_2 \relalpha \beta_3$.
            We prove each condition separately.
            \begin{enumerate}
                \item
                    Suppose $T \redtwo T'$.
                    Note that the hole in $\pctx{M}$ may appear inside a non-deterministic choice.
                    We distinguish three cases: (a)~the reduction is inside $\pctx{M}$ and maintains the branch with the hole, (b)~the reduction is inside $\pctx{M}$ and discards the branch with the hole, or (c)~the reduction synchronizes on $\beta_1$.
                    \begin{enumerate}
                        \item
                            The reduction is inside $\pctx{M}$ and maintains the branch with the hole.
                            Then $T' \equiv \pctx{M'}[\beta_1; (V \nd W)]$ and $U \redtwo U' \equiv \pctx{M'}[\beta_2; V \nd \beta_3; W]$.
                            Clearly, $(T',U') \in \mathbb{B}'$, so $(T',U') \in \mathbb{B}$.

                        \item
                            The reduction is inside $\pctx{M}$ and discards the branch with the hole.
                            Then there exists $U'$ such that $U \redtwo U' \equiv T'$, so $(T',U') \in \sff{Id}^\equiv$, and thus $(T',U') \in \mathbb{B}$.

                        \item
                            The reduction synchronizes on $\beta_1$.
                            Then $T' \equiv \pctx{M'}[V \nd W]$ and, since $\beta_1 \relalpha \beta_2 \relalpha \beta_3$, $U \redtwo U' \equiv \pctx{M'}[V \nd W]$.
                            Then $T' \equiv U'$, so $(T',U') \in \sff{Id}^\equiv$ and thus $(T',U') \in \mathbb{B}$.
                    \end{enumerate}

                \item
                    Suppose $U \redtwo U'$.
                    By reasoning analogous to above, $T \redtwo T'$ and $(T',U') \in \mathbb{B}$.

                \item
                    Suppose $T \readyPrefix{\gamma}$.
                    If the prefix $\gamma$ appears in $\pctx{M}$, then clearly also $U \readyPrefix{\gamma}$.
                    Otherwise, $\gamma = \beta_1$.
                    We have, e.g., $\gamma \relalpha \beta_2$ and clearly $U \readyPrefix{\beta_2}$.
                    The other direction is analogous.
            \end{enumerate}
    \end{itemize}

    It remains to show that $(R,S) \in \mathbb{B}$ which trivially holds.

    \medskip
    For~(ii), toward a contradiction, assume there exists a strong ready-prefix bisimulation w.r.t.\ \redone $\mathbb{B}$ where $(R,S) \in \mathbb{B}$.

    By \Cref{l:readyPrefixDual}, there exist $R',\beta_1$ such that $R \redone^\ast R' \readyPrefix{\beta_1}$, and $\alpha_1 \dualPrefix \beta_1$.
    By the well-typedness of $R$ and $S$, $\beta_1$ must appear in $\pctx{N}$, and the reduction $R \redone^\ast R'$ takes place in $\pctx{N}$.
    Take $x \in \sub{\alpha_1} \cap \sub{\beta_1}$ (which is non-empty by \Cref{l:readyPrefixDual}).
    Then $R' \equiv \pctx[\big]{N'_1}[\res{x}(\pctx{N'_2}[\beta_1; R'_2] \| \pctx{N'_3}[\alpha_1; (P \nd Q)])]$.
    Moreover, clearly $S \redone^\ast S'$ following the same reductions, resulting in $S' \equiv \pctx[\big]{N'_1}[\res{x}(\pctx{N'_2}[\beta_1; R'_2] \| \pctx{N'_3}[\alpha_2; P \nd \alpha_3; Q])]$; note that, since $\alpha_1 \relalpha \alpha_2 \relalpha \alpha_3$, also $\alpha_2 \dualPrefix \beta_1$ and $\alpha_3 \dualPrefix \beta_1$.
    At this point, we must have $(R',S') \in \mathbb{B}$.

    The synchronization between $\beta_1$ and $\alpha_1$ gives $R' \redone R'' \equiv \pctx{N''}[P \nd Q]$.
    Then by the bisimulation, there exists $S''$ such that $S' \redone S''$ with $(R'',S'') \in \mathbb{B}$.
    By clause~3 of the bisimulation, $R''$ and $S''$ must have the same ready-prefixes, so clearly the reduction $S' \redone S''$ results from a synchronization between $\beta_1$ and either of $\alpha_2$ and $\alpha_3$.
    W.l.o.g., let us assume this was $\alpha_3$.
    Then $S'' \equiv \pctx{N''}[Q]$.
    By assumption, $P \nreadyPrefixBisim{E} Q$ and thus $P \nd Q \nreadyPrefixBisim{E} Q$, so clearly $R'' \nreadyPrefixBisim{E} S''$.
    Hence, $\mathbb{B}$ cannot be a strong ready-prefix bisimulation w.r.t.\ \redone.
    In other words, $R \nreadyPrefixBisim{E} S$.
\end{proof}